\documentclass[12pt,a4paper]{article}

\usepackage{lineno,hyperref}

\usepackage{graphicx}
\usepackage[T1]{fontenc}
\usepackage[latin9]{inputenc}
\usepackage{textcomp}
\usepackage{amsmath}
\usepackage{amssymb}
\usepackage{eqnarray}
\usepackage[papersize={8.5in,11in}]{geometry}
\usepackage{bm}
\usepackage{authblk}

\def\be{\begin{equation}}
\def\ee{\end{equation}}

\newcommand{\bra}[1]{{\langle{#1}}}
\newcommand{\ket}[1]{{\vert{#1}\rangle}}
\newcolumntype{C}[1]{>{\centering\arraybackslash }b{#1}}
\usepackage[usenames]{color}
\usepackage{ulem}

\begin{document}


\title{Interplay between tensor force and deformation in even-even nuclei }

\author{R\'emi N. Bernard and Marta Anguiano}
\affil{Departamento de F\'{\i}sica At\'omica, Molecular y Nuclear, Universidad de Granada, E-18071, Granada, Spain}

\maketitle

\begin{abstract}
In this work we study the effect of the nuclear tensor force on properties related with deformation. 
We focus on isotopes in the Mg, Si, S, Ar, Sr and Zr chains within the Hartree-Fock-Bogoliubov theory
using the D1ST2a Gogny interaction. Contributions to the tensor energy in terms of saturated and unsaturated subshells are analyzed.
Like-particle and proton-neutron parts of the tensor term are independently examinated.
We found that the tensor term may considerably modify the potential energy landscapes and change the ground
state shape.
We analyze too how the pairing characteristics of the ground state change when the tensor force is included. 
\end{abstract}




\section{Introduction}

The tensor force get a major role in the nucleon-nucleon interaction. Indeed
in a boson exchange picture the nuclear interaction is generated at long range by a pion exchange
between two nucleons. The associated potential (One Pion
Exchange Potential) is composed by a central and a tensor term.
Besides the requirement to include a tensor term to the bare 
nucleon-nucleon interaction is supported by some well known experimental data such as
the none zero quadrupole moment of the deuteron \cite{RarPR37, RarPR51a, RarPR51b} or
the differential cross section of the p-p scattering.
Consequently all the most popular potentials used in the ab-initio approaches
as the Paris \cite{LacPRC80}, Bonn \cite{MacPR87, MacANP89, MacPRC01} or Argonne \cite{WirPRC95} 
potentials get a built-in tensor component.
Its impact on the shell structure properties has been studied in a large extent:
its contribution to the single particle energies depends on the filling of the shells;
it induces correlations which strongly influence the n-p pairs structures in light
nuclei \cite{SchPRL07}; the tensor force enables to get a convenient spectrum in the p-shell \cite{WirPRL02}.

In contrast the tensor term was initially neglected in self-consistent
mean field theories except for a few exceptions \cite{StaPLB77}. 
It should be emphasized here that in effective field theories 
some part of the bare tensor interaction is already taken into account in the central 
part of the effective interaction. As a consequence only the residual tensor
interaction was neglected in the usual Skyrme \cite{BeiNPA06, ChaNPA97} or Gogny 
\cite{DecPRC80, BerCPC91} effective interactions.

However the tensor force get a renewed interest over the past few years
in mean field theories. A lot of works recently aimed to determine the most
reliable tensor term built-in effective interaction.
It is now clear that the tensor term modifies the single particle energies \cite{OtsPRL05} and
the binding energies \cite{PudPRC97}, 
the multipoles giant resonances \cite{CaoPRC09, CoPRC12} 
and even may affect the magic numbers and the spin-orbit splitting in some cases \cite{OtsPRL06}.

The inclusion of the tensor term in the effective interaction has been done in 
perturbation from pre-existing parametrizations \cite{ColPLB07, BriPRC07, AngPRC12}
where all the other parameters remain unchanged. Alongside 
a full variational procedure is performed to get the new parametrizations for the 
Skyrme \cite{LesPRC07, BroPRC05} and Gogny \cite{OtsPRL06} interactions.
In all cases it rises the problem of the way to fit the parameters.
Otsuka {\it et al.} \cite{OtsPRL06} make the overall fit of the interaction using the properties of the AV8
potential \cite{PudPRC97}  to adjust the tensor strength.
Lesinski {\it et al.} \cite{LesPRC07} built 36 parametrizations of the zero range Skyrme interaction
including a tensor term. They are obtained by studying the structure properties
such as the spin-orbit splitting or the single particle energies on the Ca, Ni and
Sn chains. In the works of Zalewski {\it et al.} \cite{ZalPRC08, ZalPRC09} and Grasso  {\it et al.} \cite{GraPRC13} the spin-orbit 
strength is modified at the same time that the tensor parameters to reproduce some single particle properties,
(spin-orbit splittings) in the doubly magic $^{40}$Ca, $^{56}$Ni and $^{48}$Ca nuclei.
The work of Grasso {\it et al.} enables to reduce the number of parametrizations suggested by Lesinski {\it et al.}
excluding the ones whose parameter signs do not lead to satisfactory results.
In addition, while the tensor term is adjusted in some local structure properties some
efforts have been done to constraint the tensor strength from collectives excitations properties 
\cite{BaiPLB09, BaiPRC09, BaiPRL10, CaoPRC09} in the Skyrme Hartree--Fock+Random Phase Approximation framework.

In the present study we aim to analyze the impact of the
tensor term on the deformation of the even-even nuclei in the Hartree--Fock--Bogoliubov 
approximation. We use the D1ST2a Gogny interaction
proposed by Anguiano {\it et al.} and built on the finite range D1S Gogny interaction \cite{BerCPC91} with
a finite range tensor term \cite{AngPRC12}. This tensor term incorporates both a pure tensor
and a tensor--isospin contribution. Keeping the D1S parameters unchanged
the two tensor parameters were adjusted to reproduce the neutron single particle energies 
$1f_{5/2}$ and $1f_{7/2}$ in $^{48}$Ca.
Refitting a nuclear effective interaction in a global procedure as it is done in \cite{LesPRC07} for the Skyrme plus tensor interaction or in \cite{OtsPRL06} 
for the Gogny plus a tensor--isospin term interaction represents a considerable amount of work and is far beyond the scope of this study. 
In the present paper we are interested in the effect of tensor terms in the deformation of the ground state of various nuclei using the D1ST2a interaction. 
Keeping the D1S parameter set unchanged enables to isolate in a better way the specific tensor effects than in a consistent refit of the interaction.
Thus this study aims to isolate the situations in which tensor contribution is important, in order to choose the most pertinent observables 
to consider for a consistent refit of the interaction.

We encountered a few works in the literature on that matter.
The shape coexistence is analyzed for some $N = 28$ neutron-rich isotones in \cite{WanCPL14}. It is shown that
whereas the tensor force does not change the ground state the shape coexistence between prolate and oblate 
states vanishes for $^{42}$Si and $^{44}$S.
In the paper by Zalewski {\it et al.} \cite{ZalPRC09} the contribution of the isovector and isoscalar tensor terms to the binding
energy is depicted over the mass table. The impact of two Skyrme functionals on the superdeformed band heads is tested. 
SkO$_{\rm TX}$ is built on SkO \cite{ReiPRC99} plus a tensor term and SkO$_{{\rm T}'}$ get a modified spin-orbit
strength and the same tensor part. It is pointed out that deformation properties are very sensitive to the relative 
adjustment between the SO and tensor term for both spin-saturated and spin-unsaturated cases. 
 Bender {\it et al.} \cite{BenPRC09} studied the Skyrme parametrizations depicted in \cite{LesPRC07} with respect
to the deformation properties of some magic and semi--magic nuclei, doing 
an analysis in terms of the isoscalar and isovector parts of the tensor term.
They showed that since the effect of the tensor depends on the filling of the single particle orbitals, 
the tensor affects the nuclear chart in a different way. 
In particular Bender {\it et al.} observed that in doubly 
spin-saturated nuclei the tensor contribution to the total energy increases with the deformation whereas it is at sphericity that 
the tensor energy is the highest for doubly spin-unsaturated. 

Here we will attempt to deal with the tensor-deformation interplay 
by looking at the change of shape in specific regions, analyzing in terms of like--particle and proton--neutron contributions the tensor effects. 
We study too the impact on the pairing energy and spacial densities. We discuss too some differences found related with considering or not the Slater 
approximation in our calculations. Sec. \ref{gene} is dedicated to the introduction of the useful 
quantities and the numerical considerations of the theoretical framework.
Sec. \ref{ana} is devoted to the influence of the tensor term of the D1ST2a Gogny interaction on the potential energy curves and structures properties in 
the Zr, Mg, Si, S, Ar and Sr chains.
We also aim to look at the way the tensor energy is distributed in terms of isospin dependent contributions.
Conclusions are presented in Sec. \ref{conc}.

\section{Generalities}\label{gene}

\subsection{Computational Aspects}

All the calculations reported here are carried out at the Hartree--Fock or Hartree--
Fock--Bogoliubov
(HFB) approximation.
It is assumed that the simplex, axial and time reversal symmetries are preserved for all the calculations.
The HFBAXIAL code \cite{RobHFB02} 
uses a harmonic oscillator basis of the Fock space whose
dimension is chosen with respect to the nucleus under consideration.
In that context the HFB states are labeled by a set of
quantum numbers namely the radial $n_\perp$ the axial quantum number $n_z$  
(symmetry over the z-axis) the projection of the spacial and 
intrinsic angular momentum $m$, $s_z$ on the symmetry axis.
For all nuclei the number of major shells is chosen as $N_{\rm sh} = 10$. The oscillator length parameters are equal in each direction 
and get the fixed value $b=2.1$ fm. If not explicitly specified the Coulomb interaction is computed numerically with a Gauss Legendre integration, as it is 
explained in~\cite{ang01} and some calculations will be performed using the Slater approximation.
The two body kinetic energy coming from the center of mass correction is computed for both the Hartree--Fock 
and pairing fields. The local minima of the energy functional is determined within the second order gradient 
method. Details of the method are presented in \cite{RobPRC11} and references therein.

\subsection{Tensor interaction} \label{tenint}
The tensor part that is considered in the interaction D1ST2a writes:
\begin{eqnarray}
 V_{\rm TS}\, (\vec{r}_1,\vec{r}_2 )  = ( V_{\rm T1}+V_{\rm T2}\,  P_{12}^\tau ) 
 \left [ 3\, \frac{  (\vec{\sigma}_1 \cdot \vec{r}_{12})(\vec{\sigma}_2 \cdot 
  \vec{r}_{12})} {|\vec{r}_{12}|^2}   -\vec{\sigma}_1 \cdot \vec{\sigma}_2 \right ]  e^{-(\vec{r}_{12})^2/\mu_{\rm T}^2 } \, \, ,
  \label{D1ST2a}
\end{eqnarray}
where $\vec{r}_{12}= \vec{r}_{1}-\vec{r}_{2}$ , $P_{12}^\tau$ is the isospin exchange operator 
and $\vec{\sigma}$ is the usual 3-Dimension spin operator.
For even-even nuclei the only non zero contribution to the tensor energy comes from the exchange field.
The isospin dependence to the tensor matrix element  
can be decomposed as a summation of like-particle (lp) part and proton-neutron (pn) part:
\begin{equation}
\bra{\tau_a \tau_b}|V_{{\rm T}1}+P_{12}^{\tau}V_{{\rm T}2}\ket{\tau_b \tau_a}= \left ( V_{{\rm T}1}+V_{{\rm T}2} \right ) \delta_{\tau_a\tau_b} 
+ V_{{\rm T}2}\delta_{\tau_a-\tau_b} \, \, .
\label{parts}
\end{equation}
The two parameters $V_{{\rm T}1}$ and $V_{{\rm T}2}$ are $-135$ MeV and $+115$ MeV respectively.
The range $\mu_{\rm T}$ in Eq.~(\ref{D1ST2a})  is taken equal to $1.2$ fm, corresponding to the longest range in the D1S interaction.

Besides we can isolate the lp and pn contributions by modifying the values
of $V_{{\rm T}1}$ and $V_{{\rm T}2}$. Defining two new sets of parameters $\{V_{{\rm T}1}^{{\rm lp}}, V_{{\rm T}2}^{{\rm l p}} \}$ 
and $\{V_{{\rm T}1}^{{\rm p n}}, V_{{\rm T}2}^{{\rm p n}} \}$ such as:
\begin{equation}
 \begin{cases}
  V_{{\rm T}1}^{{\rm l p}} + V_{{\rm T}2}^{{\rm l p}} = -20 \, \,\mbox{MeV} \, \, , \\ 
  V_{{\rm T}2}^{{\rm l p}} = 0                 \,\mbox{MeV} \, \, ,
 \end{cases}
  \label{vlp}
\end{equation}
and 
\begin{equation}
 \begin{cases}
  V_{{\rm T}1}^{{\rm p n}} + V_{{\rm T}2}^{{\rm p n}} = 0 \,\, \mbox{MeV} \, \, , \\ 
  V_{{\rm T}2}^{{\rm p n}} = V_{{\rm T}2}      = +115 \, \,\mbox{MeV} \, \, ,
 \end{cases}
  \label{vpn}
\end{equation}
then replacing the $\{V_{{\rm T}1},V_{{\rm T}2}\}$ parameters in Eq.~(\ref{D1ST2a}) by the new coefficients
$\{V_{{\rm T}1}^{{\rm l p}}, V_{{\rm T}2}^{{\rm lp }} \}$
the proton-neutron part vanishes and the like-particle term is unchanged.   
The opposite conclusion applies for $\{V_{{\rm T}1}^{{\rm p n}}, V_{{\rm T}2}^{{\rm p n}} \}$. Let's note that adding these 
different new contributions give the initial parameters: $V_{{\rm T}1}^{{\rm l p}}+V_{{\rm T}1}^{{\rm p n}}=V_{{\rm T}1}$ and
$V_{{\rm T}2}^{{\rm l p}}+V_{{\rm T}2}^{{\rm p n}}=V_{{\rm T}2}$.

\subsection{Deformation parameters}

In this section we summarize all the experimental and theoretical quantities involved in the HFB calculations 
presented below dealing with the quadrupole deformation.

The spectroscopic quadrupole moment is defined as (see \cite{R&S80}, Appendix B):
\begin{equation}
 Q_{\rm S}  = \sqrt{ \frac{16\pi}{5} } \bra{I,M=I,K}|\hat{Q}^{\rm lab}_{20}\ket{I,M=I,K} \, \, ,
\end{equation}
where $\hat{Q}^{\rm lab}_{20}$ is the electric quadrupole operator. $M$ and $K$ are  the projection of the total angular momentum $I$ on the symmetry axis 
in the laboratory frame and in the body-fixed frame respectively. 
$Q_{\rm S}$ is defined in the laboratory frame and is measurable.
The electric quadrupole operator is defined as: 
\begin{equation}
 \hat{Q}^{\rm lab}_{20} = e  \int_{V} \rho_p(\vec{r})  r^{2} Y_{20}(\theta, \varphi) d\vec{r} \, \, ,
\end{equation}
where $\rho_p(\vec{r})$ is the proton density and $e$ the electric charge.  
The intrinsic quadrupole moment of the band $Q_0$ is defined as:
\begin{equation}
  Q_0 = \bra{I,M,K}|\hat{Q}^{\rm int}_{20}\ket{I,M,K} \, \, ,
\end{equation}
where  $\hat{Q}^{\rm int}_{20}$ is the quadrupole moment operator in the body-fixed frame.
This leads for an axially symmetric nucleus to (\cite{R&S80},  p25):
\begin{equation}
  Q_0  = \frac{3}{\sqrt{5\pi}} Ze R_0^ 2 \beta_2 \, \, .
 \label{Q0}
\end{equation}
$R_0$ is the radius of the sphere with the same volume than the nucleus, taken as
$R_0=1.2 A^{1/3}$~fm and  $\beta_2$ is one of the Hill-Wheeler coordinates \cite{HilPR53} 
($\gamma_2 = 0 $ in the axially symmetric case).

It is possible to relate the spectroscopic quadrupole moment $Q_{\rm S}$ with the intrinsic one when $M=I$ by the
following expression:
\begin{eqnarray}
 Q_{\rm S}  
     &=& \frac{3K^2 - I (I+1)}{(I+1)(2I+3)}Q_0 \, \, .
     \label{defQS}
\end{eqnarray} 
For the band head we usually have $K=I$ (\cite{R&S80}, p. 26).

The potential energy curves (PEC) presented below are plotted with respect to the total
deformation of the nucleus by the variable $q_{20}^{(\rm tot)}$ such as:
\begin{equation}
 q_{20}^{(\rm tot)} = \frac{3}{\sqrt{5\pi}} A R_0^2 \beta_2 \,\,  .
 \label{betaq}
\end{equation}
$q_{20}^{(\rm tot)}$ is the total quadrupole moment and is related to the proton and neutron quadrupole moments, 
$q_{20}^{(\rm p)}$ and $q_{20}^{(\rm n)}$ respectively, such as $q_{20}^{(\rm tot)}=q_{20}^{(\rm p)}+q_{20}^{(\rm n)}$.
These latter quantities are obtained as the average value 
of the operator $\hat{Q}_{20}= \sqrt{{\displaystyle  \frac{16\pi}{5} }} r^2 Y_{20}$ in the HFB state.
They can be expressed with respect to $\beta_2$ replacing $A$ by $Z$ and $N$ respectively 
in Eq.~(\ref{betaq}).
Consequently we get the relation (see Eq.~(\ref{Q0})):
\begin{equation}
 \frac{Q_0}{\rm e} = q_{20}^{(\rm p)} \, \, .
\end{equation} 
It is convenient to note that for all the nuclei presented below the PEC 
has been computed with a $\beta_2 = 0.05 $ step. 

Since we are only dealing with even-even nuclei all the states constituting the PEC  are $0^+$ states.
For all the minima in the above PEC the theoretical $Q_0$ is extracted.
Taking into account the definition of $Q_S$, it is clear that its value is zero for $0^+$ states.
The experimental $Q_S$ values in the following tables are such as $I=2$. The experimental $Q_0$ are extracted from 
the measured $Q_S$ using Eq.~(\ref{defQS}).
The theoretical $Q_0$  from $0^+$ states are compared to experimental $Q_0$  from $2^+$ states
assuming that the intrinsic quadrupole moment does not change in a pure K-band. 
For $I=K=2$ we get $Q_{S}=2/7 Q_{0}$ from Eq.~(\ref{defQS}), for $I=2,\,K=1$  $Q_{S}=-1/7 Q_{0}$ and for $I=2,\,K=0$  we have $Q_{S}=-2/7 Q_{0}$.

\section{Results}\label{ana}

The effect of the tensor interaction on some properties regarding deformation are analyzed for the Zr, Mg, Si, S, Ar, Kr and Sr isotopes chains
in the HFB framework.
The general feature that we find is that the tensor in most of the cases has a  repulsive effect 
in its contribution to the total HFB energy. It raises the HFB energy by a few MeV. The tensor contribution 
is zero for some cases and can be attractive for specific points of a few potential energy curves.

In the case of several local minima the tensor term tends towards stretching the local minima 
with respect to the deformation variable (i.e. oblate minima are more oblate,
prolate minima are more prolate for D1ST2a) and lower the energy difference between the minima.
If the minima are close in energy for D1S, D1ST2a can invert the ground state location. 
First of all, we analyze the effect of tensor by doing an energy decomposition.

\subsection{Tensor energy decomposition}\label{TED}

Spin-saturated shells occur when both $j=l+1/2$ and $j=l-1/2$ subshells are filled.
For instance $^{80}$Zr is both proton and neutron spin-saturated at sphericity: 
the proton and neutron subshells are filled up to the 
$2p_{1/2}$ one and its spin-orbit partner, $2p_{3/2}$, is also full.

\begin{table}[htb] \centering
 \begin{tabular}{|C{1cm}|c c|C{2cm}|C{2cm}|c|}
\hline	
              & $E_{\rm TS}(p-p)$ & $E_{\rm TS}(n-n)$ & $E_{\rm TS}(p-n)$ & $E_{\rm TS}$(tot) & p/n Shells  \\
\hline
  $^{36}$Ar   &      $-0.140$     &      $-0.135$     &      $+1.984$     &       $+1.637$    &    SU/SU    \\
\hline  
  $^{34}$Si   &      $-0.792$     &      $-0.005$     &      $+0.122$     &       $-0.675$    &    SU/SS    \\
\hline
 $^{40}$Ca    &      $-0.003$     &      $-0.003$     &      $+0.032$     &       $+0.027$    &    SS/SS    \\
\hline
 $^{48}$Ca    &      $-0.003$     &      $-1.248$     &      $+0.152$     &       $-1.099$    &    SS/SU    \\
\hline
 $^{60}$Ca    &      $-0.001$     &      $-0.057$     &      $+0.027$     &       $-0.031$    &    SS/SS    \\
\hline
 $^{60}$Ni    &      $-1.068$     &      $-0.537$     &      $+8.439$     &       $+6.834$    &    SU/SU    \\
\hline
 $^{78}$Sr    &      $-0.072$     &      $-0.010$     &      $-0.069$     &       $-0.151$    &    SU/SS    \\
\hline
 $^{88}$Sr    &      $-0.098$     &      $-1.498$     &      $+0.636$     &       $-0.960$    &    SU/SU    \\
\hline
 $^{80}$Zr    &      $-0.005$     &      $-0.006$     &      $+0.063$     &       $+0.052$    &    SS/SS    \\
\hline
 $^{90}$Zr    &      $-0.008$     &      $-1.449$     &      $+0.252$     &       $-1.205$    &    SS/SU    \\
\hline
 $^{108}$Zr   &      $-0.005$     &      $-0.262$     &      $+0.039$     &       $-0.228$    &    SS/SS    \\
\hline
 \end{tabular}
\caption{Tensor energy contributions in MeV at sphericity. The shell configuration is specified in the last column
(SS for spin-saturated, SU for spin-unsaturated). }
\label{TSenergy} 
\end{table}

In order to study how the tensor force behaves we look at the tensor energy contributions
for specific nuclei at sphericity. In Table~\ref{TSenergy} the total tensor contribution $E_{\rm TS}$(tot) to the D1ST2a HFB energy is 
presented. Using Eq.~(\ref{parts}) we separate $E_{\rm TS}$(tot) into its lp parts $E_{\rm TS}(p-p)$, $E_{\rm TS}(n-n)$ and its pn part 
$E_{\rm TS}(p-n)$.
For all the nuclei both lp parts of the energy are  attractive whereas 
the pn part is repulsive, with the only exception of  $^{78}$Sr for which the pn part is also attractive.
The components get very different orders of magnitude, from a few keV to a few MeV.
As a result the total tensor energy can be both repulsive and attractive and is most of the time dominated by a sole term.

In order to understand the way the energy arises we assume that only the valence subshells and their spin-orbit partners
contributes to the tensor energy, all the contributions from fully filled up spin-orbit partner doublets being neglected.
Thus there are three types of contributions to the tensor energies: the ones where particles are in the same subshell, 
the ones where particles comes from both spin-orbit partners and contributions where the particles are in uncorrelated subshells;
one of the two partners being the valence subshell in all cases.
However when two spin-orbit doublets are mixed up (typically the $2p$ and $1f$ subshells) we will also take into account the non valence subshell level 
whose spin-orbit partner is above the Fermi level.
Thus we separate the energy contributions $E_{\rm TS}(i-j)$ of isospin $i-j$ according to the subshells $\alpha, \beta ...$ involved:
\begin{eqnarray}
\begin{cases}
 E_{\rm TS}(i-j) = V_{ij} \Big[X^{ij}_{\alpha,\beta}+X^{ij}_{\gamma,\delta}+...  \Big], \hspace{2cm} i,j\in \{p,n\}\\
X^{ij}_{\alpha, \beta} \equiv {\displaystyle \sum_{a,c\in \alpha} \sum_{b,d \in \beta} \bra{ab}|V_{{\rm TS}}\ket{\widetilde{cd}}\Lambda^{(i)}_{db}\Lambda^{(j)}_{ca} } \, ,\\
 \end{cases}
\end{eqnarray}
where $V_{ij}$ is $V_{{\rm T}1}+V_{{\rm T}2}$ when $i=j$ and $2V_{{\rm T}2}$ for $i\neq j$.  $\bra{ab}| V_{\rm TS}\ket{\widetilde{cd}}$ is the antisymmetrized two body 
matrix element of the tensor interaction in the quasiparticle basis and $\Lambda^{(i)}$ is the density matrix of isospin $i$. 
$X^{ij}_{\alpha, \beta}$ is symmetric in its subshell/isospin indices: $X^{ij}_{\alpha, \beta}=X^{ji}_{\beta, \alpha}$.
In the Appendix, Hartree-Fock calculations are performed on  some selected nuclei. In this case
the quantity $X^{ij}_{\alpha, \beta}$ is no longer isospin dependent and reduces to a sum of two body matrix elements for which the particles run
over the subshells under consideration: 
\begin{eqnarray}
 X^{ij}_{\alpha, \beta}= X_{\alpha, \beta} = \sum_{a\in \alpha} \sum_{b \in \beta} \bra{ab}| V_{{\rm TS}}\ket{\widetilde{ab}} \, .
\end{eqnarray}
Calculations are done to determine the $X_{\alpha, \beta}$ for subshells from the $1d_{5/2}$ to the $2p_{1/2}$ one.
It is shown in the Appendix that $X_{\alpha, \beta}$ is positive when $\alpha$ and $\beta$ get
the same spin quantum number and is negative in the opposite configuration.
Within a doublet $\{ \alpha, \beta\}$ the order of magnitude is the same for all the terms 
$X_{\alpha, \alpha}\sim X_{\beta, \beta}\sim -X_{\alpha, \beta} $
though the cross term absolute value is slightly smaller.
Although most of the nuclei of Table~\ref{TSenergy} get pairing 
these Hartree-Fock properties enable to explain the tensor energy contributions qualitatively.

In Table~\ref{TSenergy} the simplest cases are $^{40}$Ca and $^{80}$Zr, two $N=Z$ nuclei for which the pairing is zero. 
The tensor energy comes from two spin-orbit saturated subshells: $1d_{3/2}$ with $1d_{5/2}$ and $2d_{1/2}$ with $2d_{3/2}$ respectively. 
In this case it is shown in the Appendix that  
the lp and pn components are built from the same sum of matrix elements: 
\begin{eqnarray}
 E_{\rm TS}(i-j) = V_{ij}\Big [ X_{\alpha,\alpha} + 2\, X_{\alpha,\beta} + X_{\beta,\beta}\Big ] \, .
 \label{SS1}
\end{eqnarray}
Consequently their energies in Table~\ref{TSenergy} only differ from each other 
by their respective parameters given in Eq.~(\ref{parts}): $V_{{\rm T}1}+V_{{\rm T}2}=-20$ MeV for the lp cases and 
$V_{{\rm T}2}=+115$ MeV for the pn one.
Besides $^{60}$Ca is also a SS/SS nucleus but for which the proton and neutron subshells involved are different.
 Like-particle contributions are also driven by Eq.~(\ref{SS1}).
As expected the value of $E_{\rm TS} (p-p)$, $-0.001$ MeV, is similar to the \ $^{40}$Ca one.
However $^{60}$Ca gets a non zero neutron pairing which modifies the $E_{\rm TS} (n-n)$ value 
from the corresponding $p-p$ or $n-n$ ones  in \ $^{80}$Zr, but stays small ($-0.057$ MeV). 
The value of $E_{\rm TS} (p-n)$ for  \ $^{60}$Ca, composed by two attractive and two repulsive $X$ is expected to be small. 
We get: 
\begin{eqnarray}
 E_{{\rm TS}}(p-n) = V_{pn} \left [ X^{pn}_{1d_{5/2},1f_{7/2}} + X^{pn}_{1d_{5/2},1f_{5/2}} + X^{pn}_{1d_{3/2},1f_{7/2}} +X^{pn}_{1d_{3/2},1f_{5/2}}\right ] 
 = +0.027 \,{\rm  MeV} .
 \label{SS2}
\end{eqnarray}
As a result, the $p-n$ contribution partially cancels the $n-n$ one and the total tensor energy is slightly attractive, contrary to \ $^{40}$Ca and \ $^{80}$Zr. 

For nuclei of SS/SU type in Table~\ref{TSenergy}  ($^{34}$Si, $^{48}$Ca, $^{90}$Zr and $^{78}$Sr) the total tensor energy is always attractive. The 
three first nuclei are in the same configuration where the $p-n$ contribution is repulsive 
and the total energy is dominated by the attractive SU subshell. Their like-particle SU energy $E_{\rm TS}(i-i)$ is directly proportional to a $X^{ii}_{\alpha,\alpha}$ term and $E_{{\rm TS}}(p-n)$ writes:
\begin{eqnarray}
 E_{{\rm TS}}(p-n) = 2\, V_{\rm T2} \left [ X^{pn}_{\alpha,\beta} +X^{pn}_{\alpha,\gamma} \right ] \, ,
 \label{EpnSSSU}
\end{eqnarray}
where $X^{pn}_{\alpha,\beta}$ and $X^{pn}_{\alpha,\gamma}$ get opposite signs. It explains qualitatively the fact that the absolute value of $E_{{\rm TS}}(p-n)$ 
is an order of magnitude smaller than the corresponding one of the SS subshell energy. The case of \ $^{78}$Sr is different because we need to take into account the 
$2p_{3/2}$ subshell for the $p-p$ and $p-n$ contributions. We have
\begin{eqnarray}
 E_{{\rm TS}}(p-n)=2\, V_{{\rm T}2} & \left[ X^{pn}_{2p_{1/2},1f_{5/2}} + X^{pn}_{2p_{3/2},1f_{7/2}} +  X^{pn}_{2p_{3/2},2p_{3/2}} \right .  \nonumber\\
+  & \left . X^{pn}_{2p_{3/2},1f_{5/2}} + X^{pn}_{2p_{1/2},2p_{3/2}} + X^{pn}_{2p_{1/2},1f_{7/2}} \right ] \, .
\label{EpnSr}
\end{eqnarray}
In the latter equation we expect all the $X$ on the first line give a positive contribution (same spin for both subshells) and all 
the $X$ on the second line give a negative contribution. Here $E_{{\rm TS}}(p-n)$ is attractive, with a value of $-0.069$ MeV. 
At the Hartree-Fock level most of the $X$ involved in Eq.~(\ref{EpnSr}) are much smaller than the ones for $^{34}$Si, $^{48}$Ca, $^{90}$Zr in Eq.~(\ref{EpnSSSU}) 
as depicted in Table~\ref{TableX},  and this value for $^{78}$Sr falls to $-0.011$ MeV.

The last type of configuration occurs when both neutron and proton subshells are spin unsaturated (SU/SU). 
Here the tensor energy can be either attractive or repulsive.
Like $^{40}$Ca and  $^{80}$Zr, $^{36}$Ar is a $N=Z$ nucleus and the ratio between the different contributions also reflects
the parameters but two orders of magnitude bigger than the doubly SS nuclei. This is due to the fact that the $1d_{3/2}$ subshell is half full.

As for $^{78}$Sr we need here to take into account the fact that the $2p$ and $1f$ shells are mixed for $^{60}$Ni and $^{88}$Sr.
These nuclei look more like SU/SS nuclei because their valence $1f_{5/2}$ subshell is saturated,
even if the $2p_{3/2}$ subshell must be considered without its partner in the tensor energy calculations. 
As expected  $E_{{\rm TS}}(p-p)$ from $^{78}$Sr and $^{88}$Sr are very close. The $^{88}$Sr total tensor energy is dominated by the attractive 
$E_{{\rm TS}}(n-n)$ from neutrons in the $1g_{9/2}$ subshell.
Contributions $X^{pn}_{1f_{7/2},1g_{9/2}}$ and $X^{pn}_{1f_{5/2},1g_{9/2}}$ should approximately cancel each other and let the $X^{pn}_{2p_{3/2},1g_{9/2}}$
term drive the pn tensor energy.
For $^{60}$Ni the repulsive pn contributions dominates the total tensor energy like a SU/SS nucleus. 
However we would expect from Hartree-Fock calculations in the Appendix to have a $n-n$ energy more attractive than the $p-p$ one.
This has to be related to the important pairing energy ($-9.517$ MeV with D1ST2a) for $^{60}$Ni in its spherical state.

Finally the neutron rich doubly spin-saturated $^{108}$Zr differs from the usual SS/SS case. 
Here, whereas the $E_{\rm TS}(p-p)$ value $p-p$ energy is equal to the $^{80}$Zr case, as expected, the 
$E_{\rm TS}(n-n)$ value is much bigger than what we
expect from a SS configuration. This can be  partially due to the strong neutron pairing energy, about $-14.627$ MeV. 
When the pairing is off the $E_{\rm TS}(n-n)$ value falls from $E_{\rm TS}(n-n) = -0.262$ MeV to $E_{\rm TS}(n-n) = -0.071$ MeV.

\subsection{Zirconium chain}

\begin{figure}[htb] \centering
\begin{tabular}{ccc}
   \includegraphics[width=4.6cm]{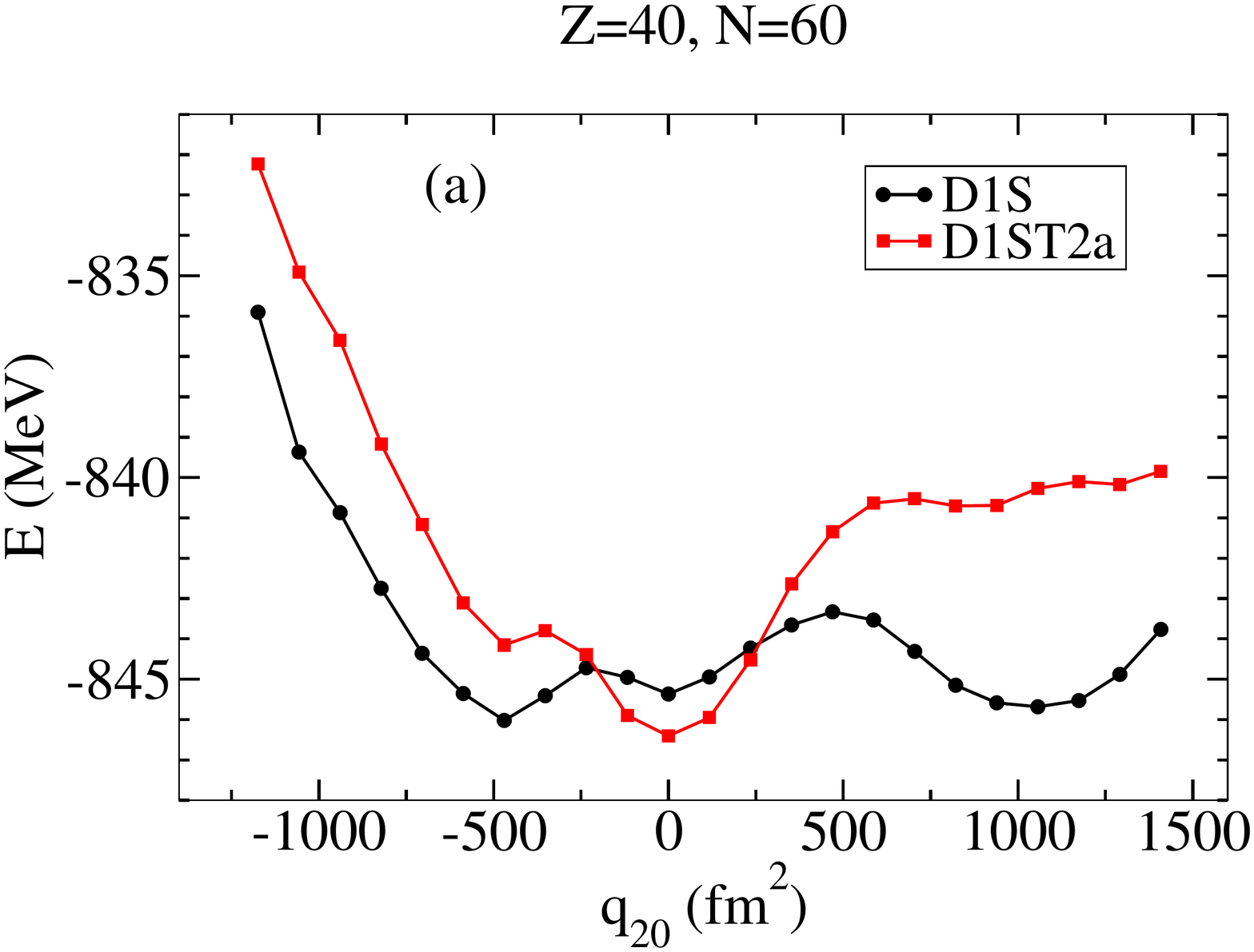} &
   \includegraphics[width=4.6cm]{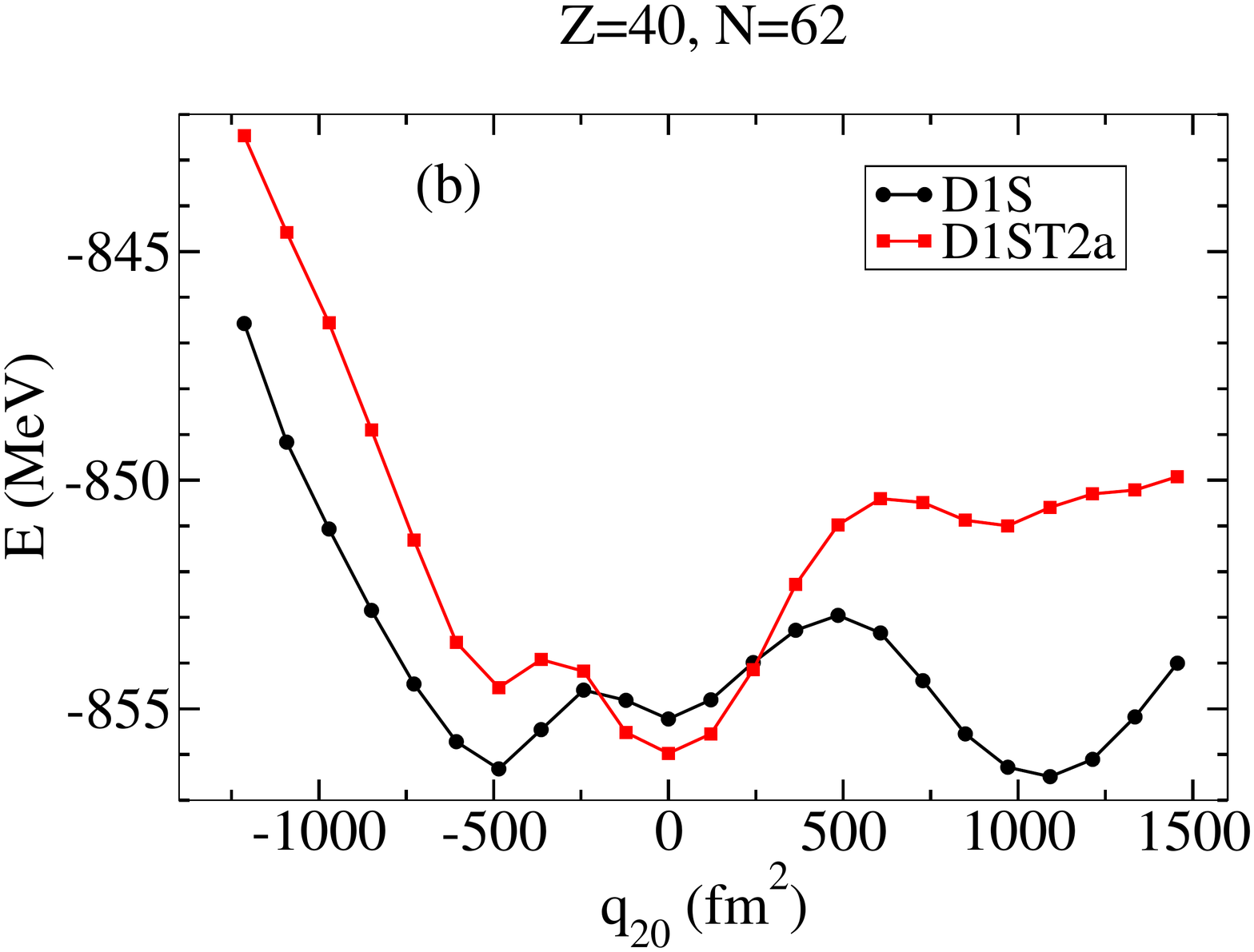} &
   \includegraphics[width=4.6cm]{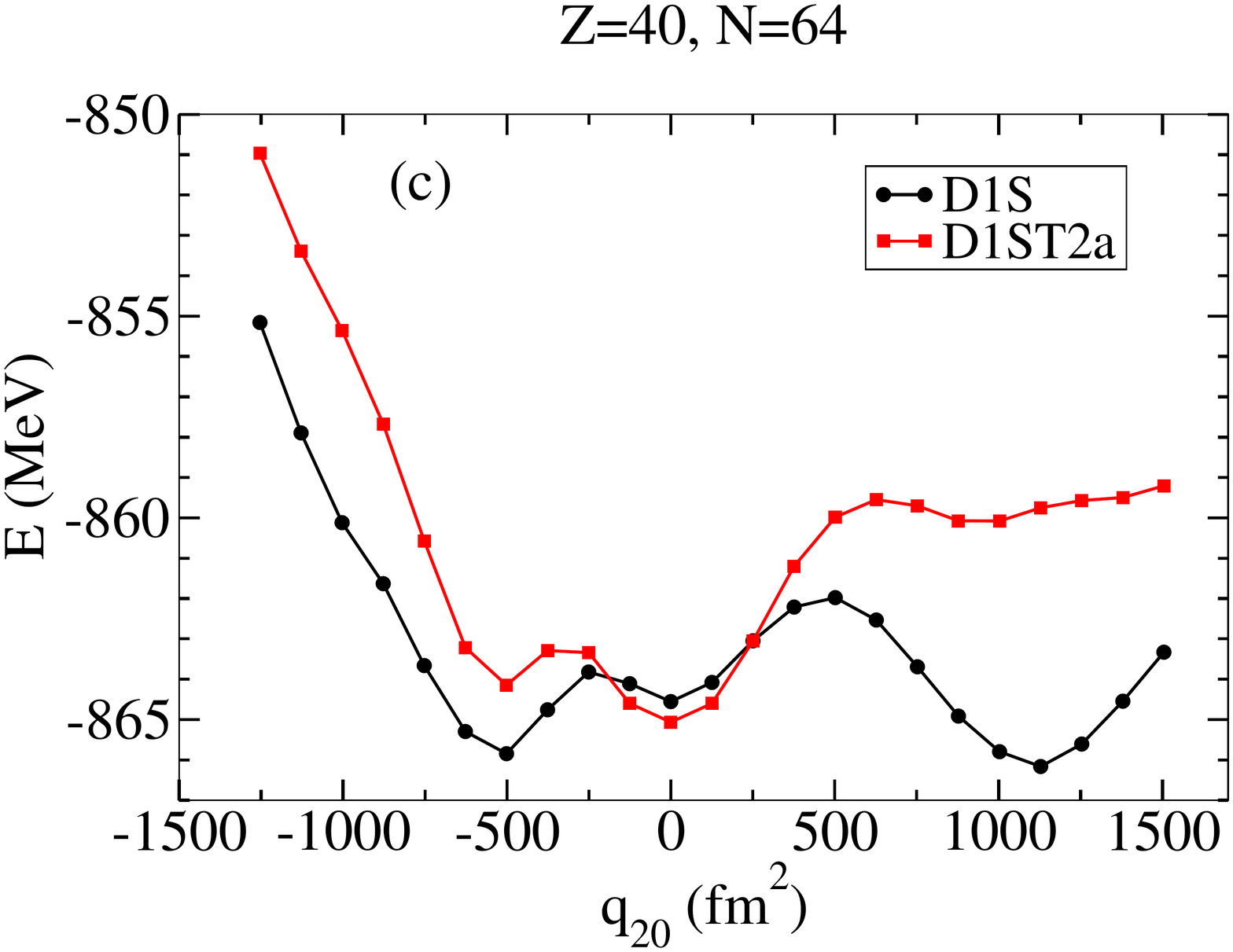} \\
   \includegraphics[width=4.6cm]{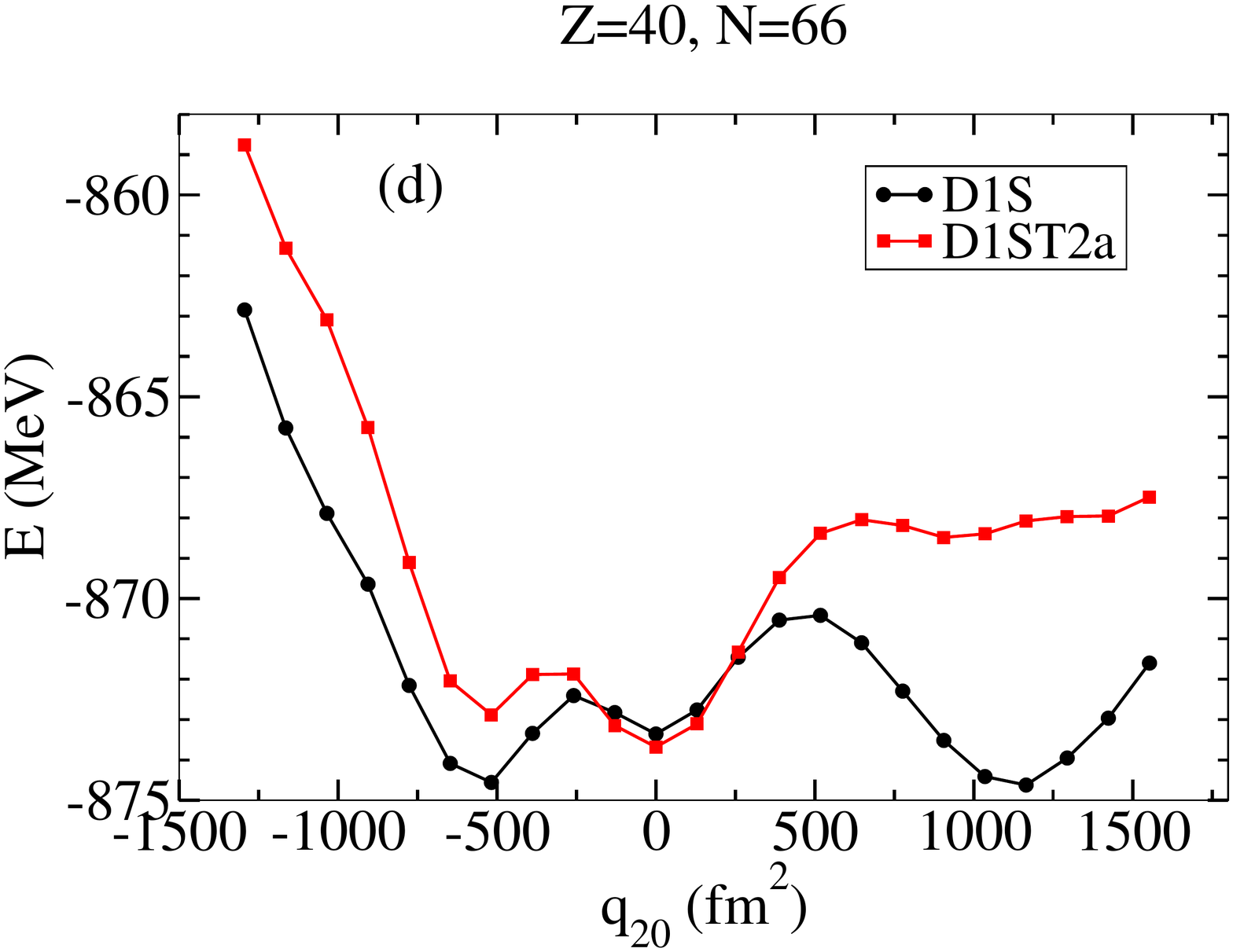} &
   \includegraphics[width=4.6cm]{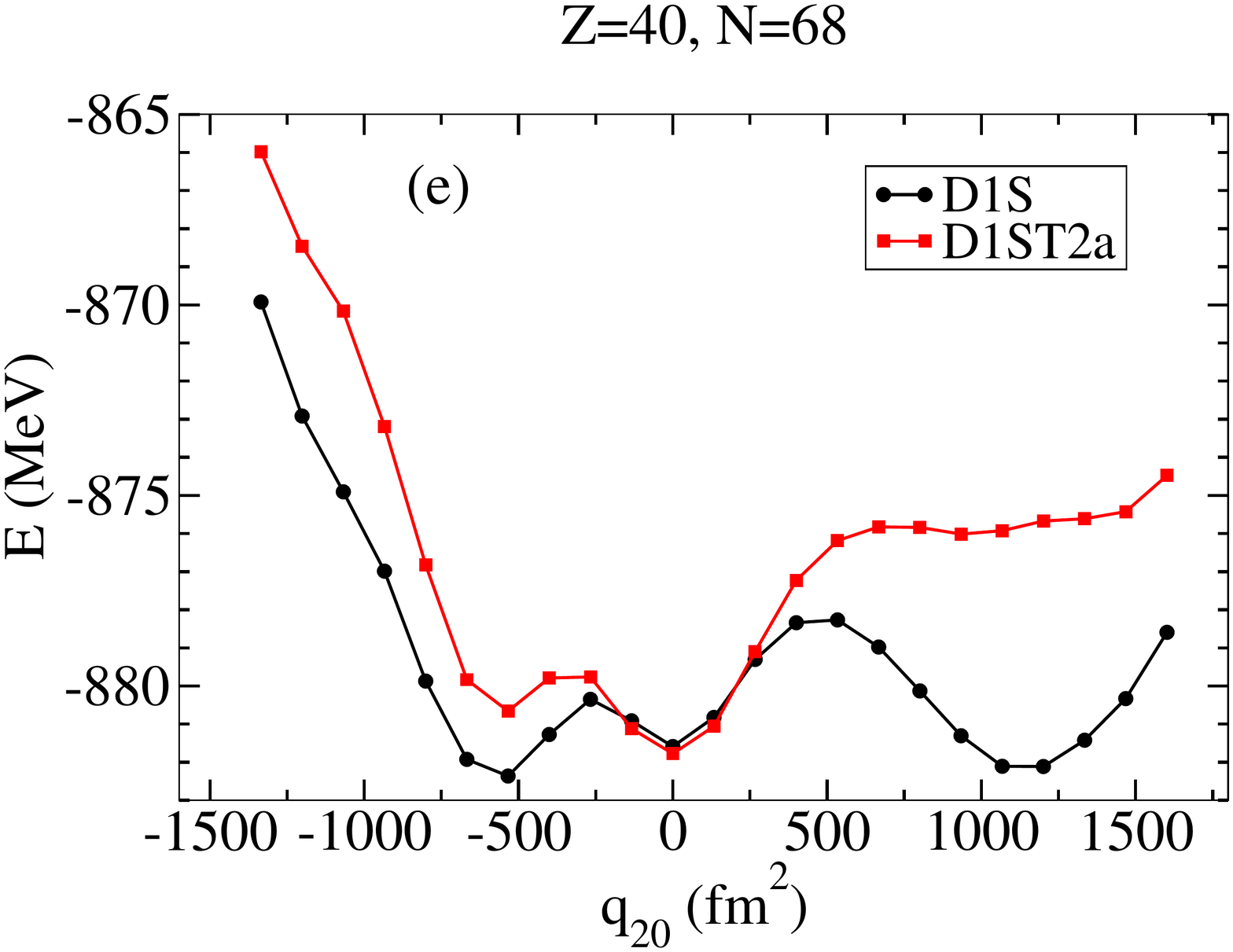} &
   \includegraphics[width=4.6cm]{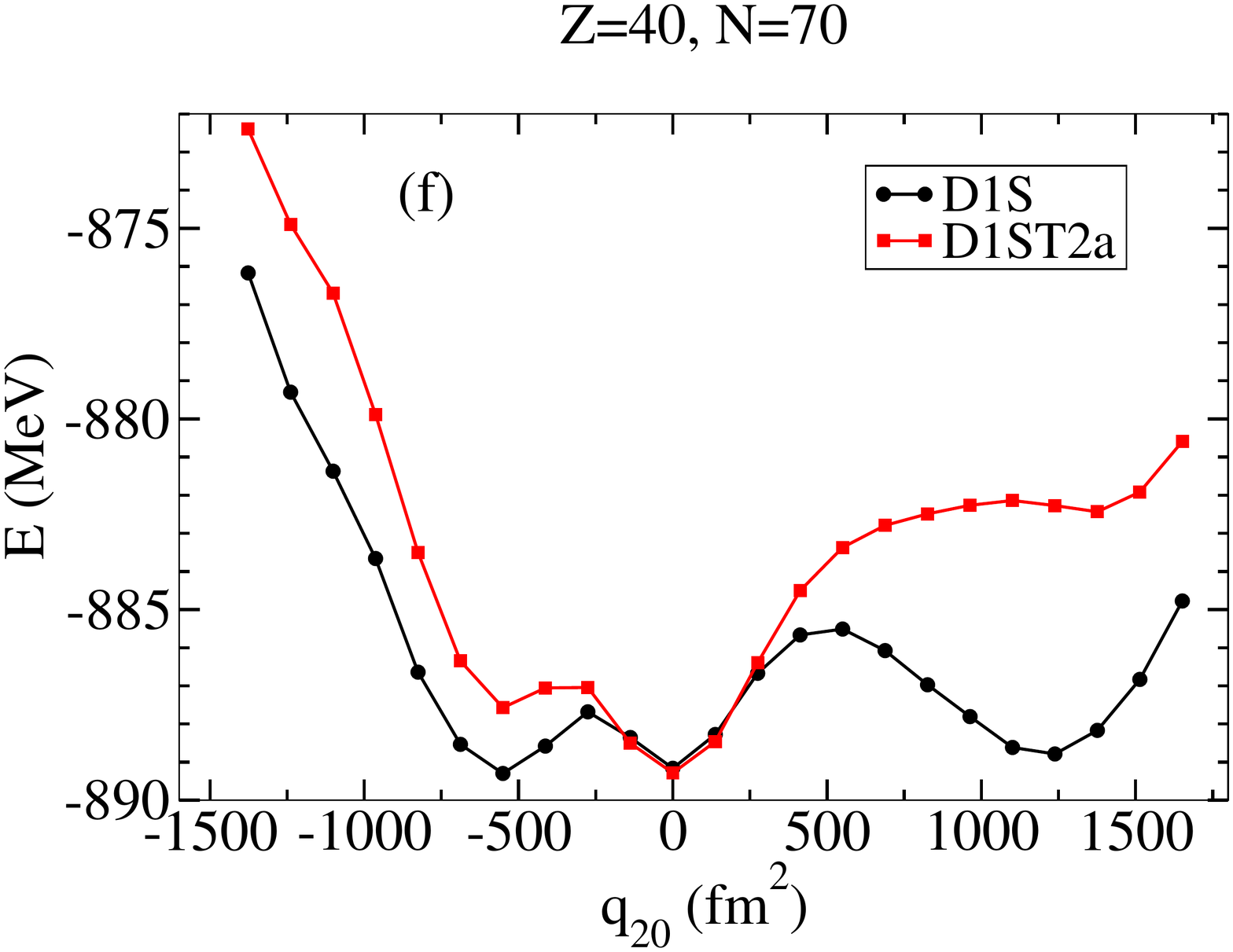} \\
   \includegraphics[width=4.6cm]{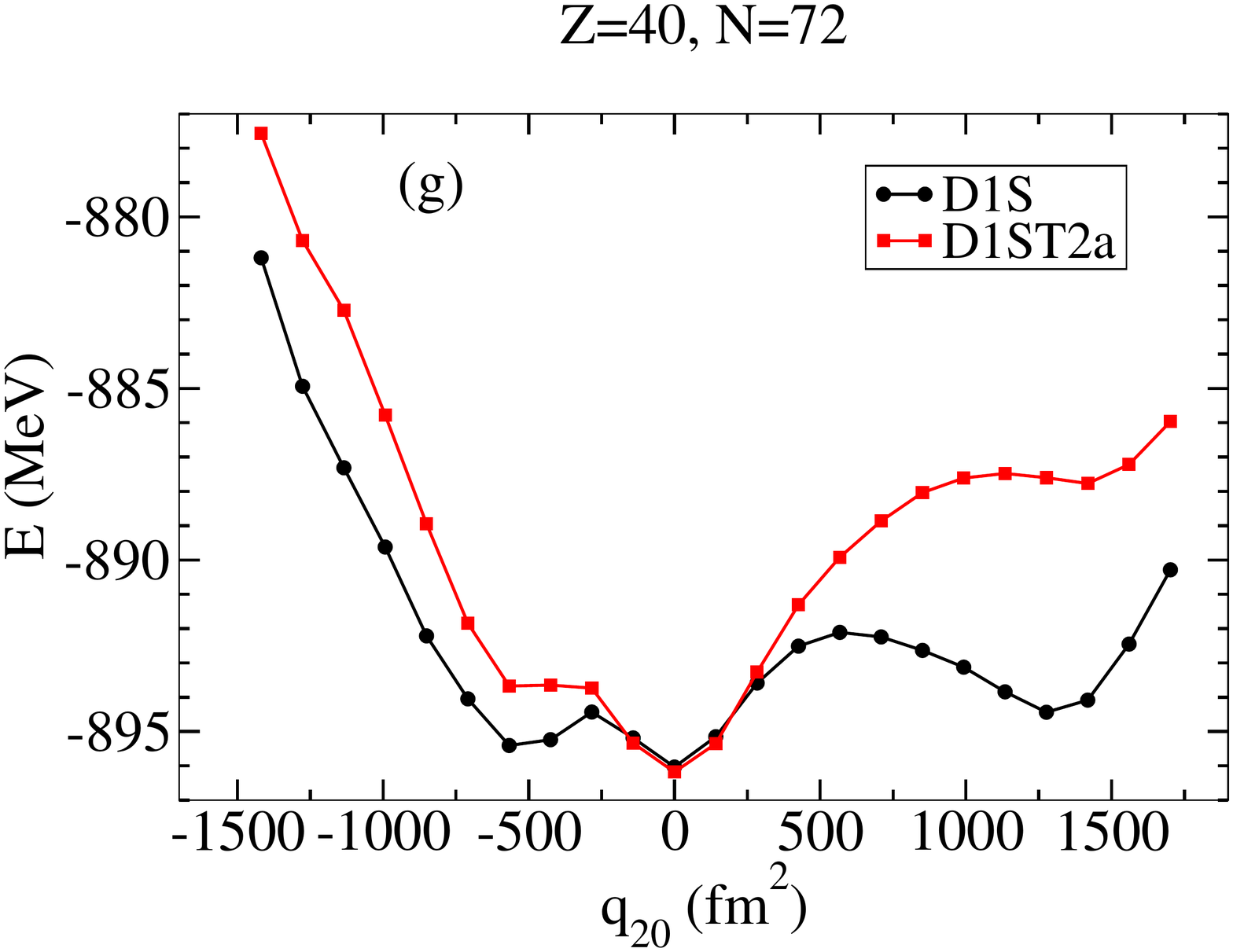} &
   \includegraphics[width=4.6cm]{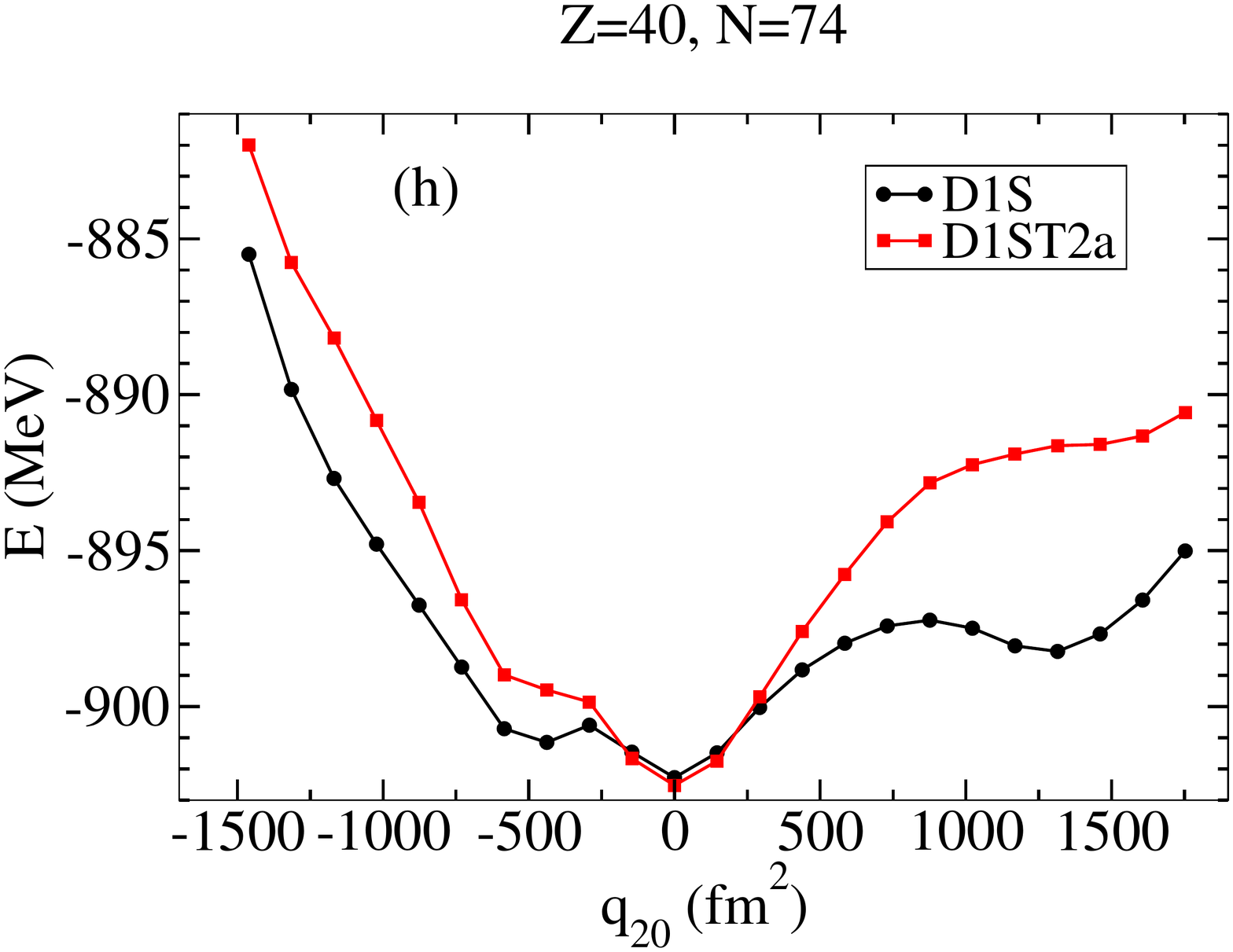} &
   \includegraphics[width=4.6cm]{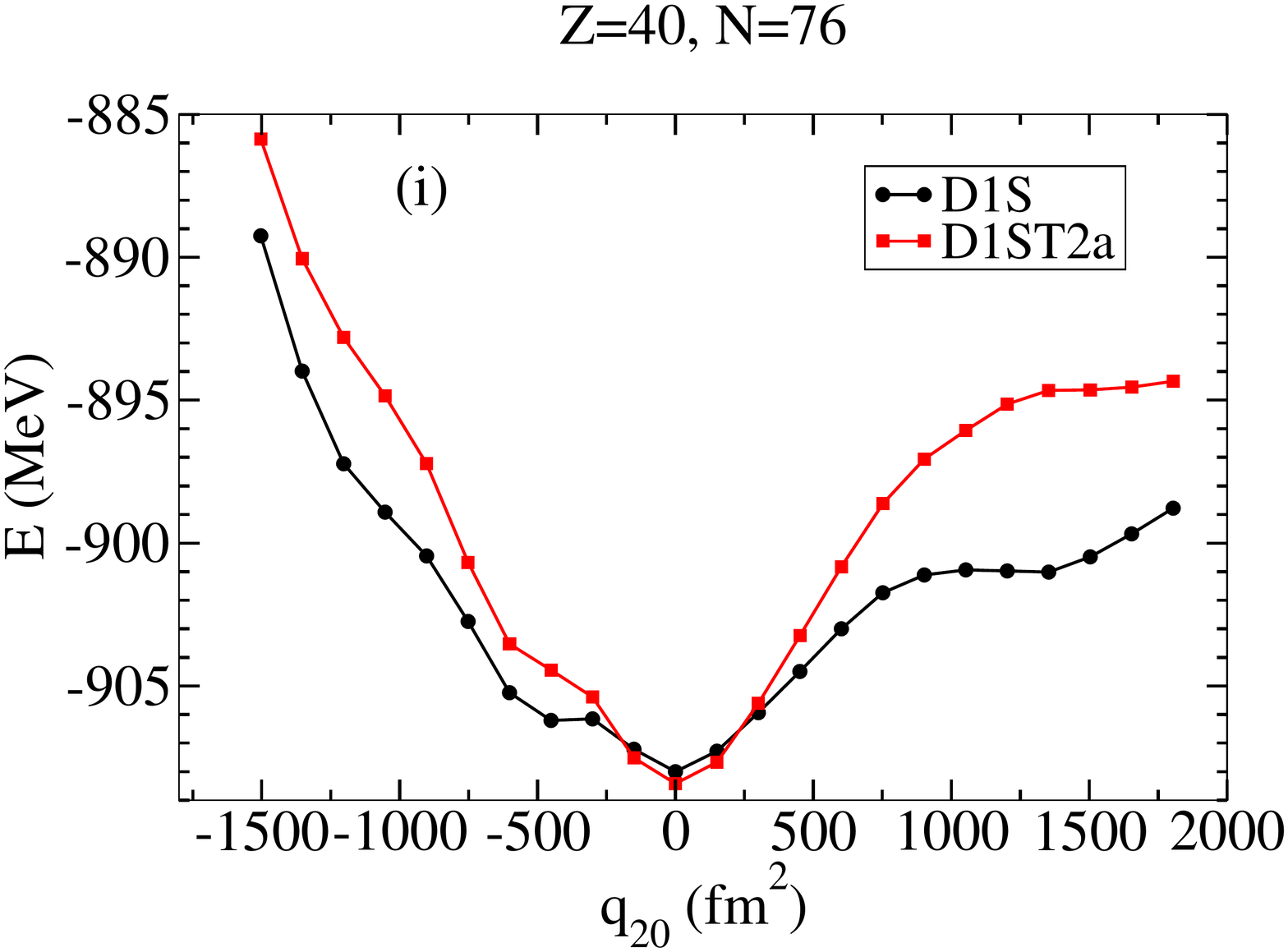} \\
\end{tabular}
\caption{Potential Energy Curves of the $Z=40$ chain 
for the D1S (black circles) and D1ST2a (red squares) interactions.} 
\label{PESZr2}
\end{figure}
In Fig. \ref{PESZr2} are depicted the PEC for Zirconium isotopes. $E$ is the HFB energy and $q_{20}$ the total quadrupole moment variable.
This chain has a proton spin-saturated configuration. 
We choose to restrict the Zr chain for isotopes from $A=100$ to $A=116$ to focus on the region where several minima (oblate/spherical/prolate) are
in competition. PEC obtained using D1S and D1ST2a interactions are quite different. Whereas the D1ST2a interaction is globally less attractive than the D1S one, 
the results are different around sphericity:  D1ST2a calculations give a more bound nucleus in the spherical minima
from $A=100$ up to $A=106$ and the tensor effect becomes negligible at sphericity from $A>106$.
Besides the prolate minima is always high in energy for D1ST2a whereas it can be the ground state for D1S 
($Z=102,\,104,\,106$). For $A>110$ the spherical minima is more and more dominating for both
D1S and D1ST2a. In Table~\ref{TableZr2} we show the values of the intrinsic quadrupole moment $Q_0$ for 
these isotopes, for the first minimum (ground state) and the second one (isomere state). We see that all the nuclei 
considered present a zero value for the intrinsic quadrupole moment in the case of the D1ST2a interaction
for the ground state. Both interactions provide very different results up to $A=110$: the first two minima 
are spherical and oblate for the D1ST2a interaction and are oblate and prolate for the D1S one. For $A=110$ the minima obtained with D1ST2a 
interaction are reversed with  respect to the D1S minima.  
For heavier isotopes the D1S prolate minimum becomes higher in energy and ground state locations obtained with both interactions 
become similar.

\begin{table}[htb] \centering
 \begin{tabular}{|c|c|c|c|c|c|c|c|c|c|}
\hline
           A    &   100 &   102 &   104 &   106&   108&   110 &   112 &   114 &  116   \\
\hline
 D1S gs         &-1.82  &+4.34  &+4.40  &+4.46 &-1.91 &-1.95  & 0.00  & 0.00  & 0.00   \\
 D1S isomere    &+4.26  &-1.84  &-1.86  &-1.88 &+4.51 & 0.00  &-2.00  &-1.59  &-1.62       \\
\hline
 D1ST2a gs      & 0.00  & 0.00  & 0.00  & 0.00 & 0.00 & 0.00  & 0.00  & 0.00  & 0.00   \\
 D1ST2a isomere &-1.83  &-1.85  &-1.87  &-1.89 &-1.92 &-1.96  &-2.01  &       &         \\
\hline
 \end{tabular}
\caption{Theoretical $Q_0$ (eb) for the $Z=40$ chain. The results for the ground state (gs) and for the first isomere state are
presented.}    
\label{TableZr2} 
\end{table}

\subsubsection{Like-particle and proton-neutron contributions}

\begin{figure}[htb] \centering
\begin{tabular}{ccc}
   \includegraphics[width=4.6cm]{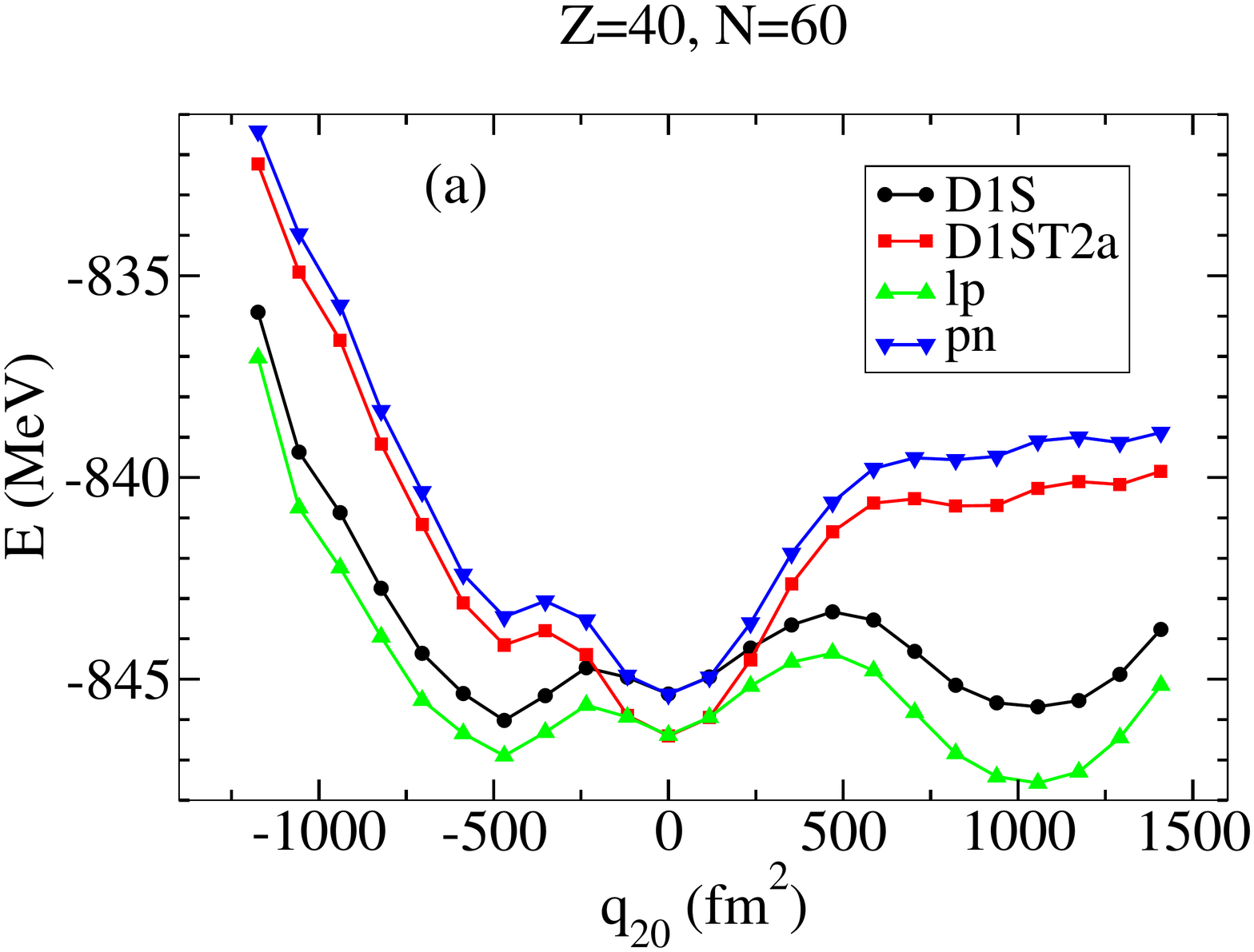} &
   \includegraphics[width=4.6cm]{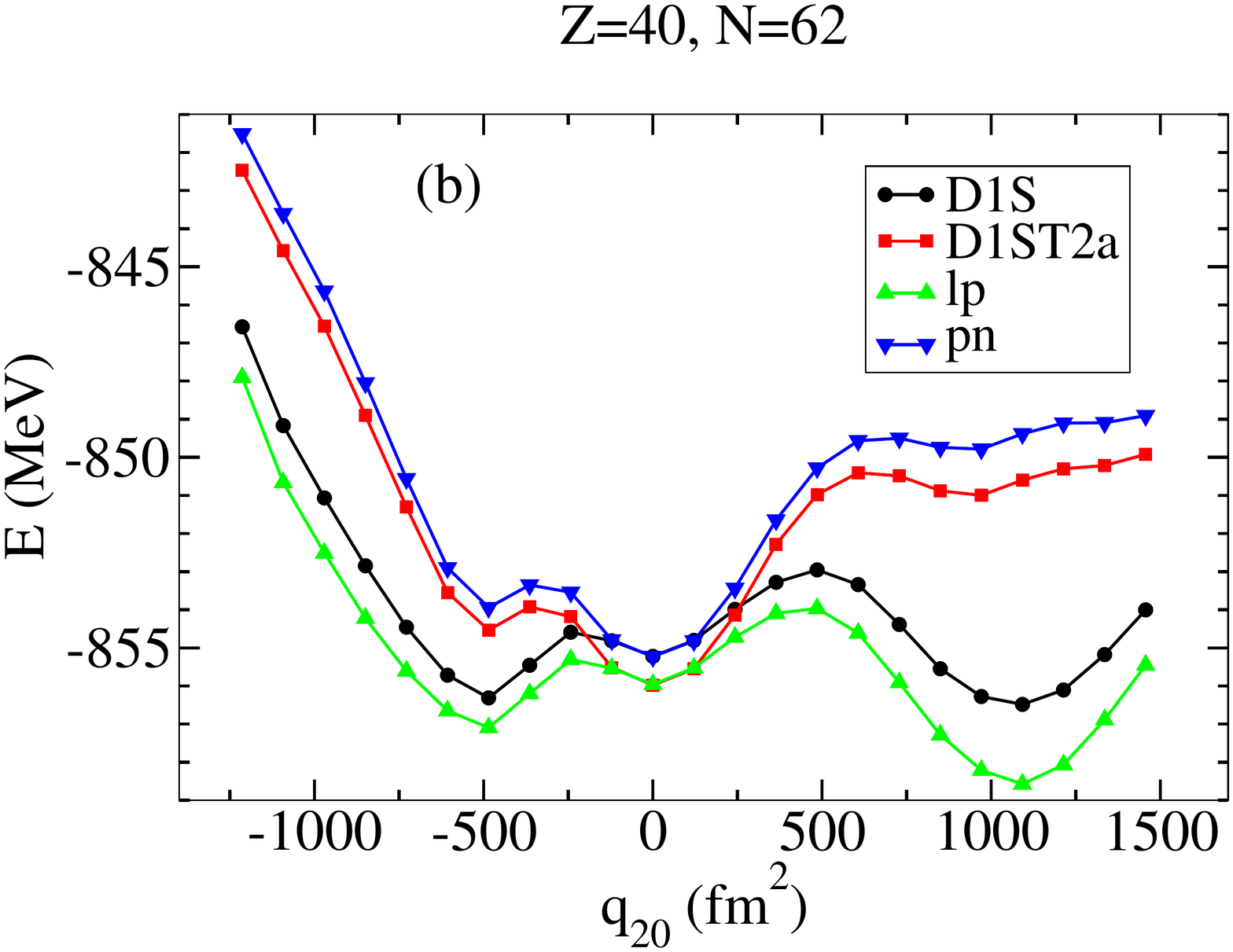} &
   \includegraphics[width=4.6cm]{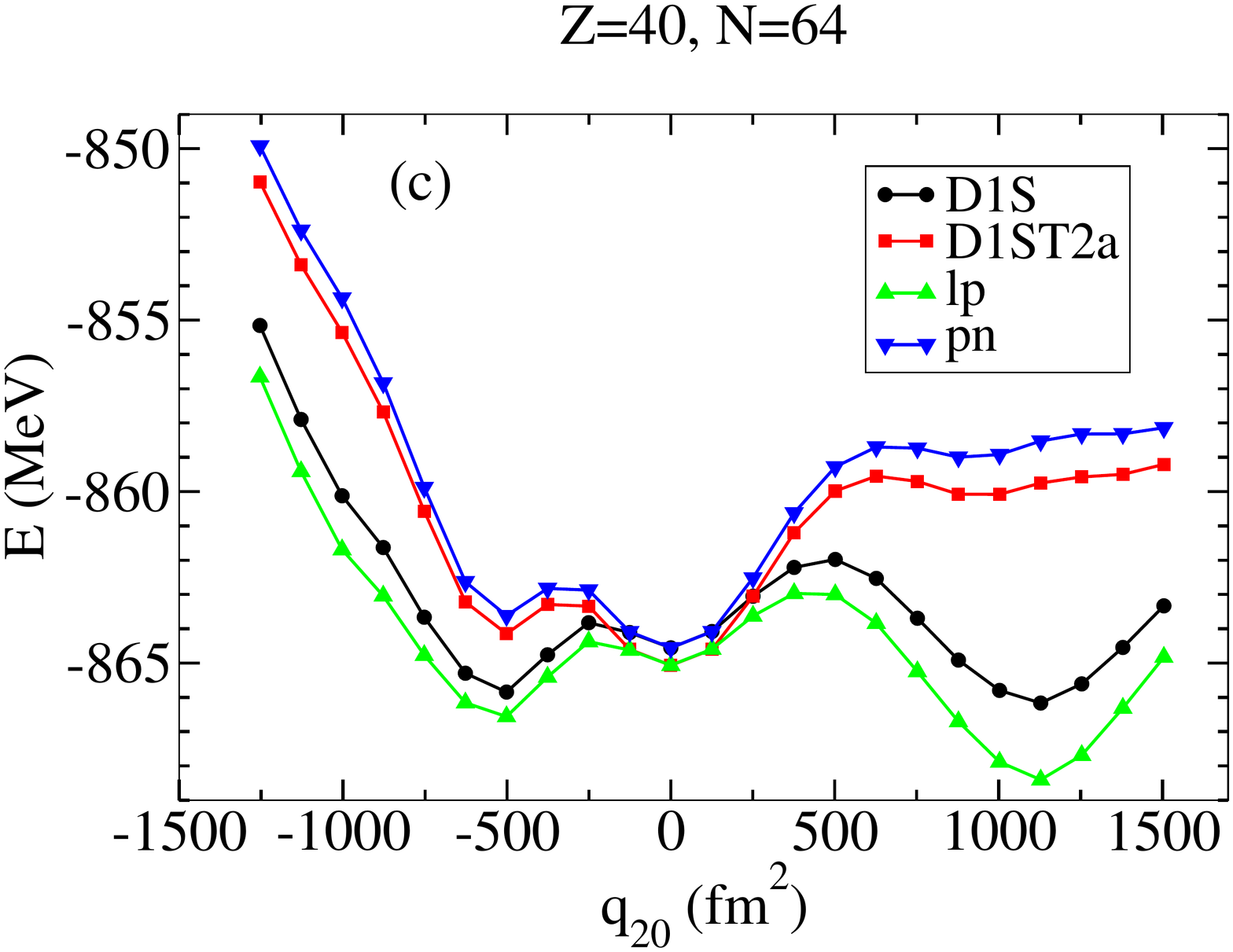} \\
   \includegraphics[width=4.6cm]{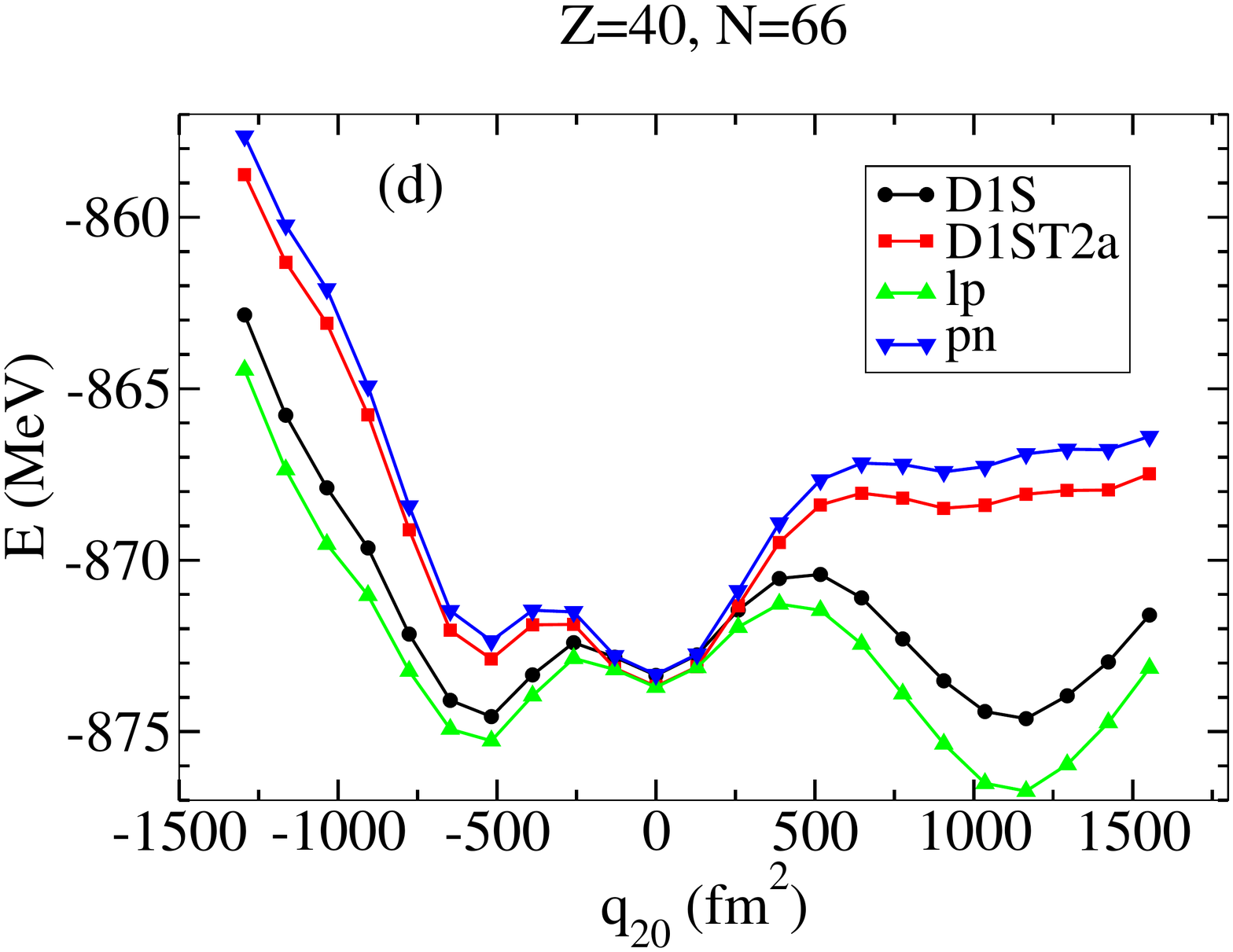} &
   \includegraphics[width=4.6cm]{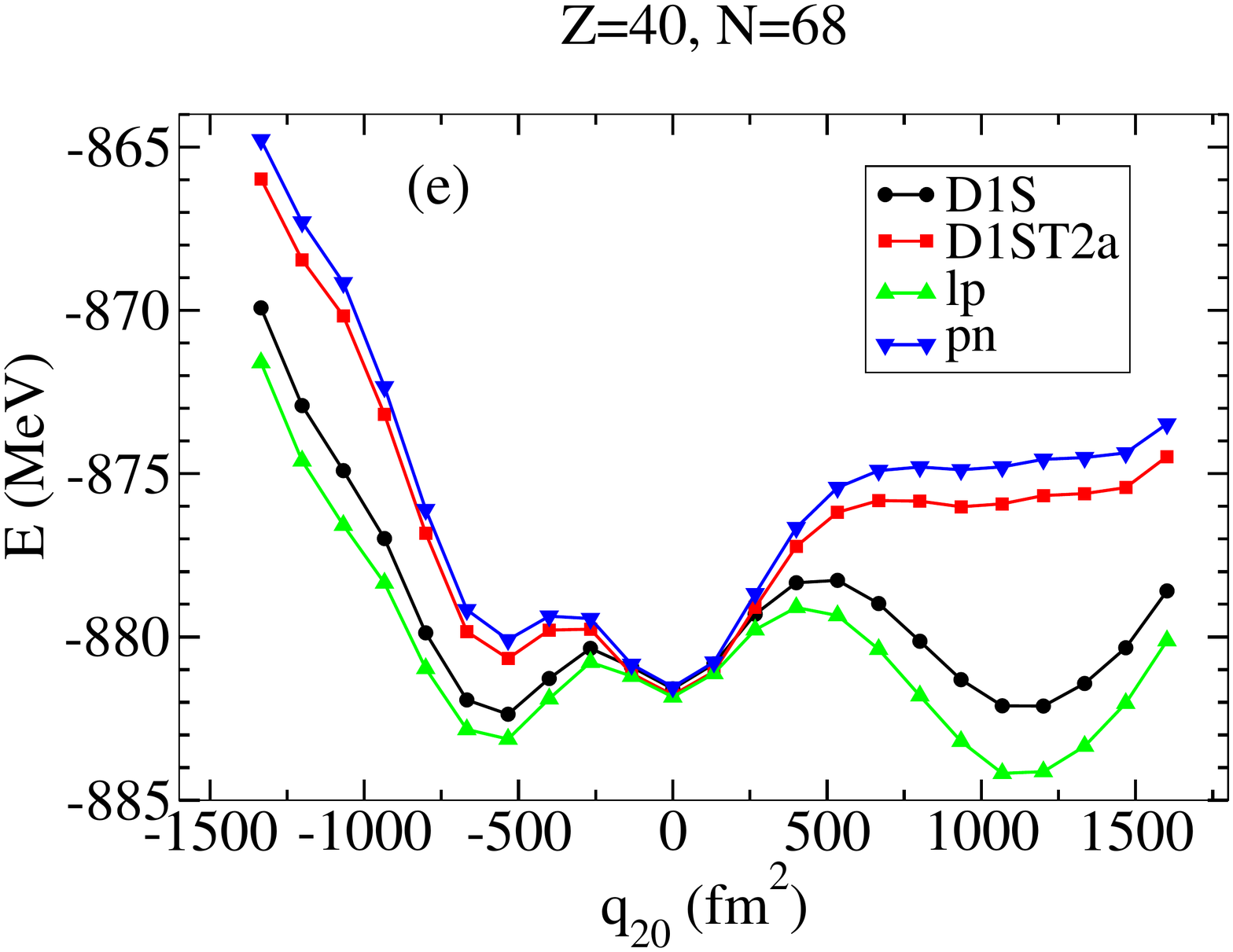} &
   \includegraphics[width=4.6cm]{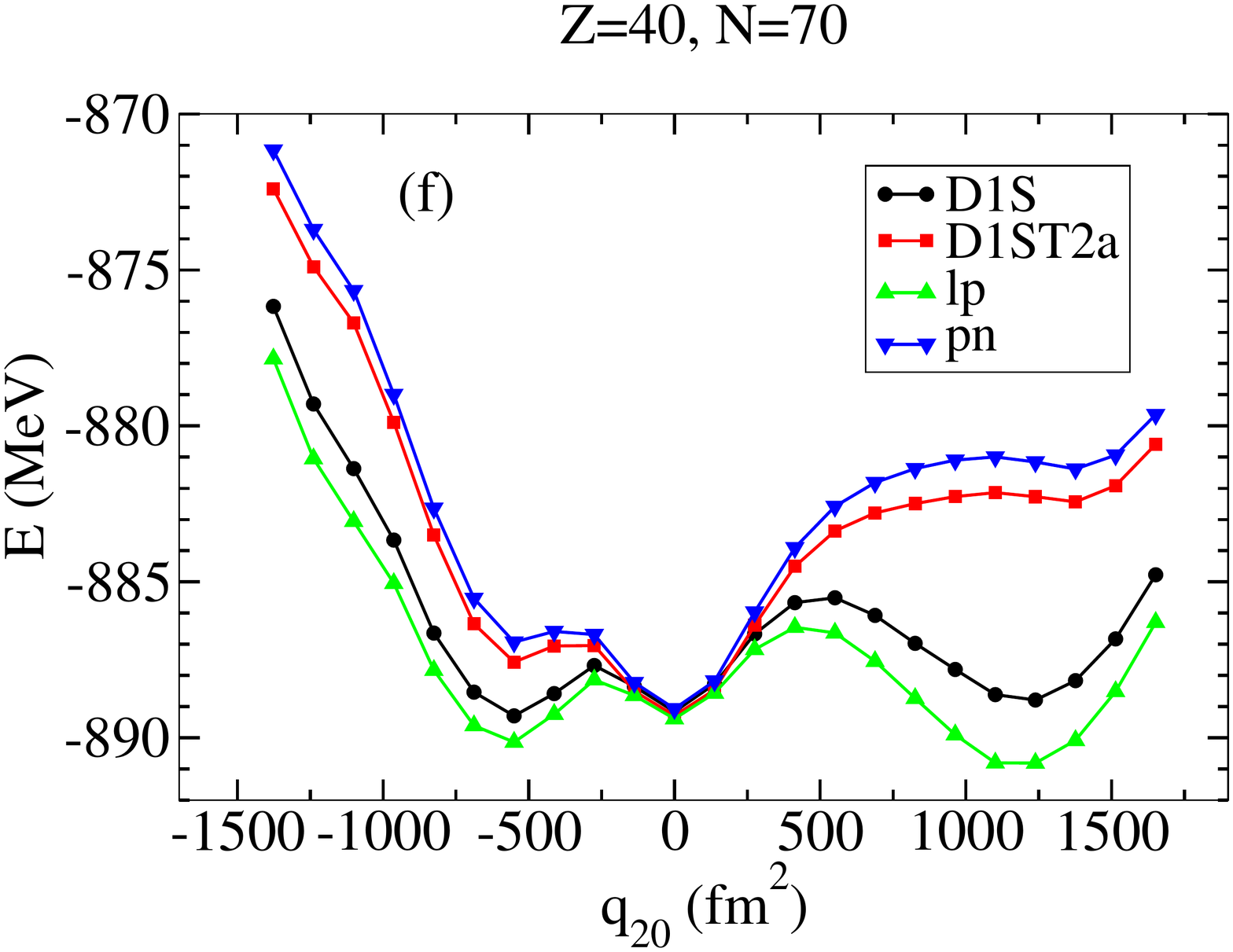} \\
   \includegraphics[width=4.6cm]{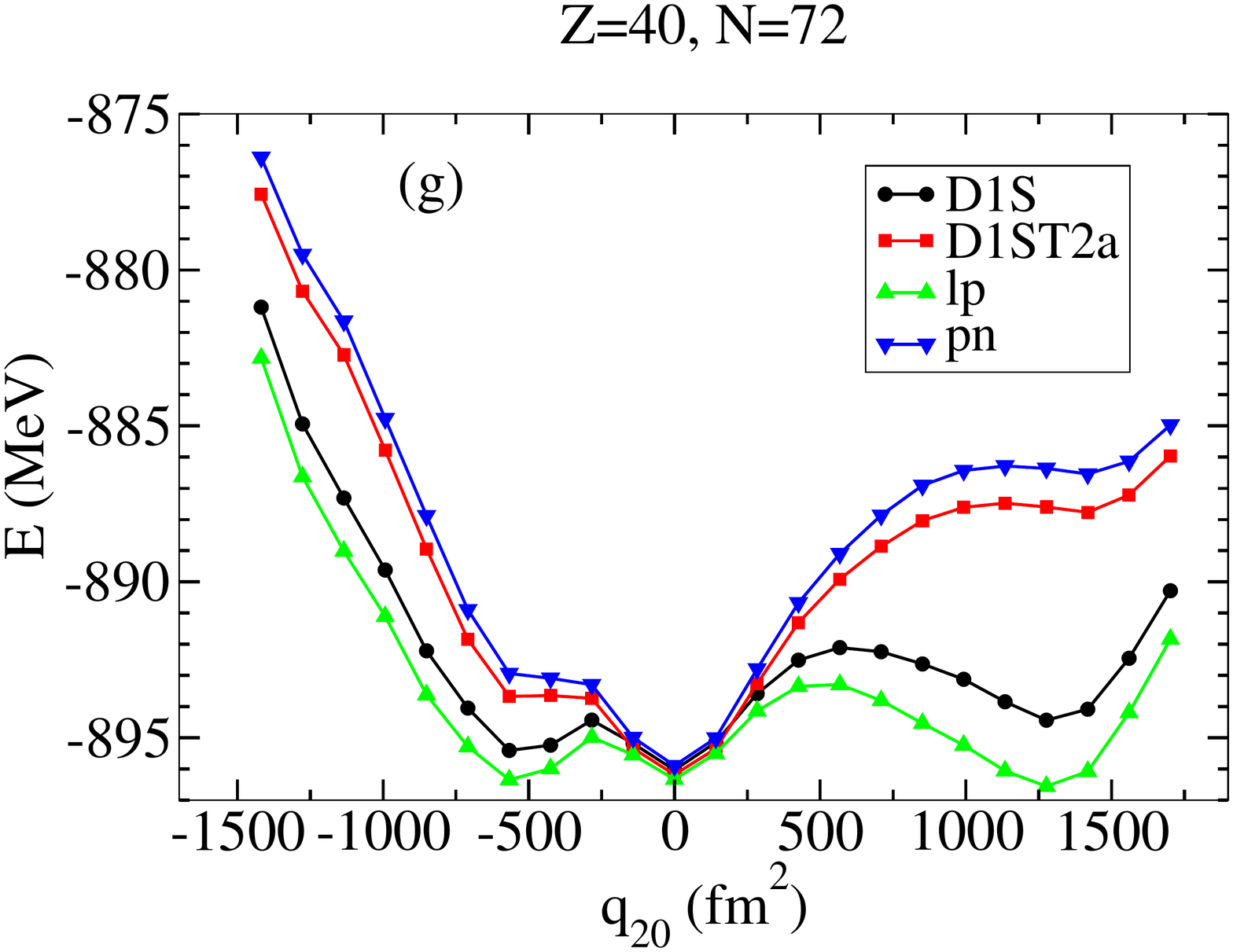} &
   \includegraphics[width=4.6cm]{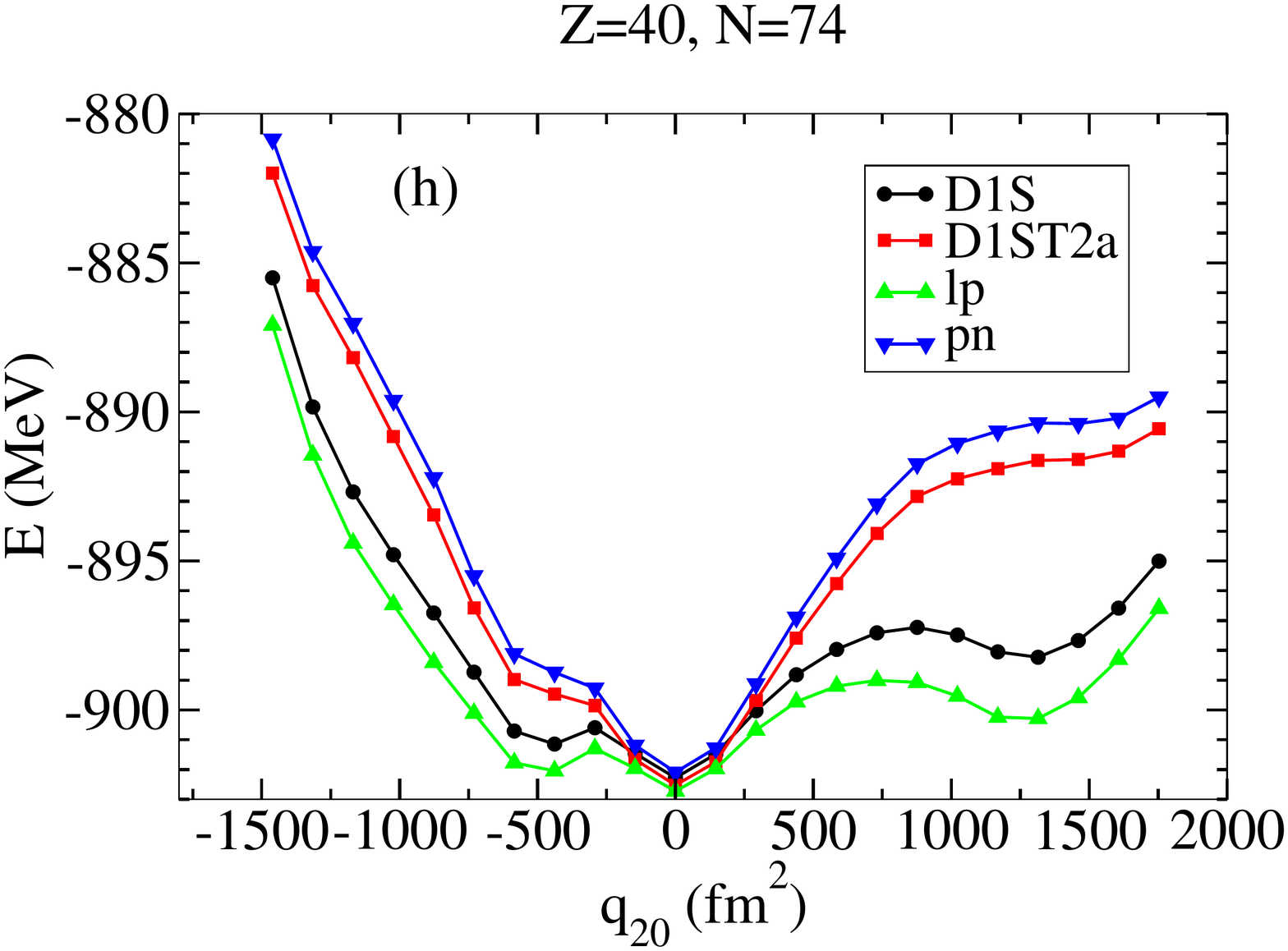} &
   \includegraphics[width=4.6cm]{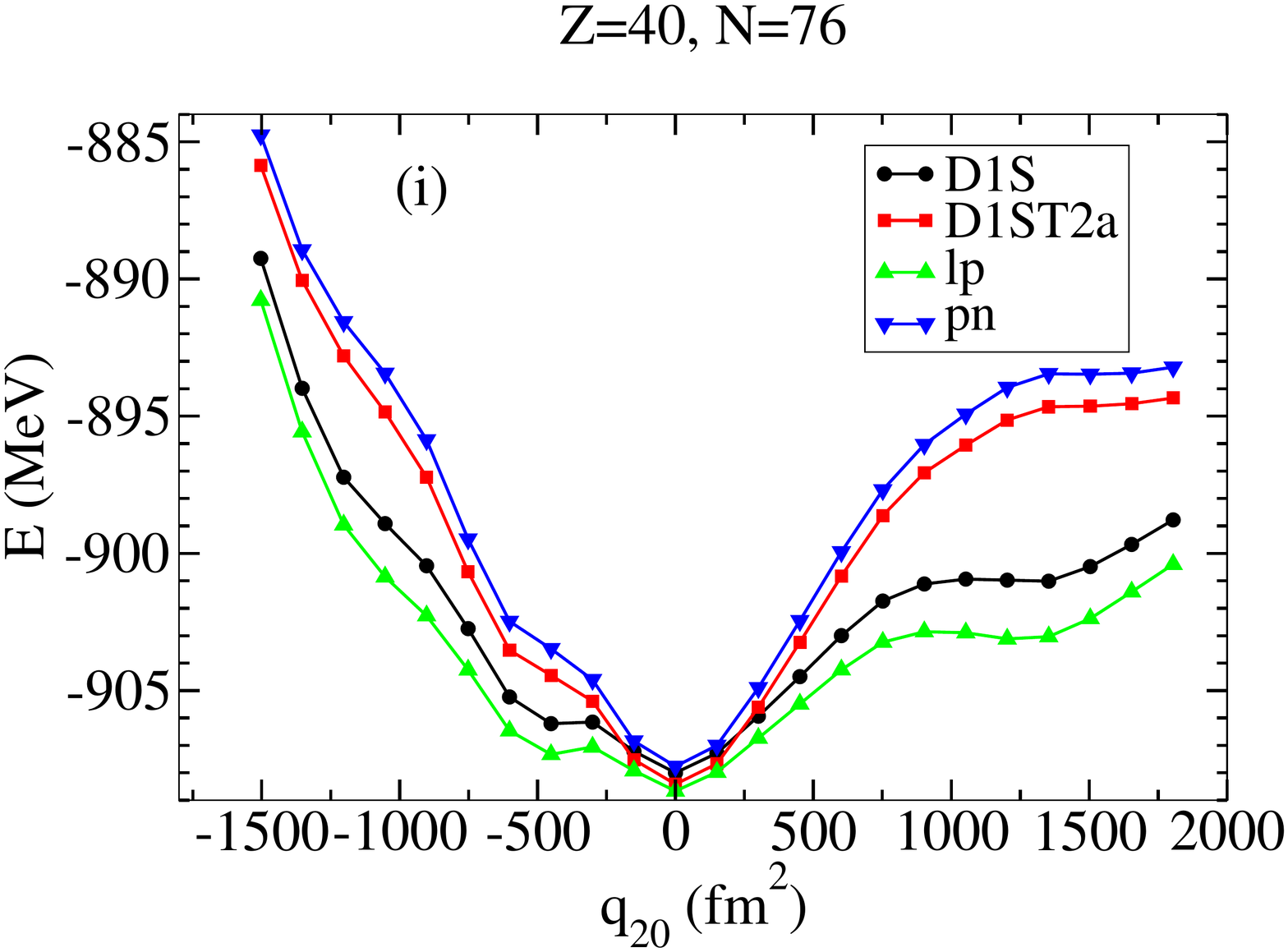} \\
\end{tabular}
\caption{Potential Energy Curves of the $Z=40$ chain 
for the D1S interaction (circles) and the D1ST2a one (squares). Triangles up correspond to the  D1S +
like-particle contribution and the triangles down to the D1S + 
pn one.} 
\label{Zrdec2}
\end{figure}

The contribution of the tensor interaction can be decomposed into two parts: the like-particle one and the proton-neutron one as explained in Section \ref{tenint}. The calculations achieved with the appropriate set of parameters are presented 
in Fig.~(\ref{Zrdec2}): the lp and pn contributions use the parameters defined in Eq.~(\ref{vlp}) and Eq.~(\ref{vpn}), 
respectively. All of them  are done in a self-consistent way.

For all the Zirconium isotopes the pn part dominates the effect of the tensor interaction. 
It is always repulsive and reproduces the trend of the D1ST2a curve 
for each Zr isotope. On the other hand the lp PEC is very close to the D1S one. Here the only non zero parameter is 
$V^{{\rm lp}}_{{\rm T}1}=-20$ MeV and is an order of magnitude smaller than $V^{\rm pn}_{{\rm T}1}$ and $V^{\rm pn}_{\rm T2}$. Its effect is slightly 
attractive ($V^{\rm lp}_{{\rm T}1}$ is negative) and is practically constant all over the nine isotopes.
It is worth to emphasize that the situation is quite different around sphericity. The pn part is zero (the D1S and pn curves are superimposed) 
for all the isotopes whereas the lp part remains slightly repulsive for the lighest isotopes ($A< 108$) before vanishing ($A\ge 108$) and increasing for $A\ge112$.
Here we use the numbers of the Zr isotopes in Table~\ref{TSenergy}. The Zr chain is proton spin-saturated.
For the first one, $^{100}$Zr, the $2d_{5/2}$ subshell becomes to be filled with two neutrons. The pn contribution to the tensor energy is negligible 
and the $n-n$ attractive part dominates the tensor energy as for the $^{90}$Zr case. 
Then the subshells are filled up to $N=68$ where the $2d_{3/2}$ is full, which makes $^ {108}$Zr a doubly spin-saturated nucleus.
The main contribution to the tensor energy then comes from the $n-n$ part as suggested by the $^{108}$Zr energy decomposition in Table~\ref{TSenergy}.
The next isotope, $^{110}$Zr, also doubly spin-saturated, is similar to $^ {108}$Zr. Then filling the $1h_{11/2}$ susbshell the pn part increases slightly
but the total tensor energy remains attractive.

\subsubsection{Particle number fluctuation and densities}

\begin{figure}[htb] \centering
\begin{tabular}{cc}
   \includegraphics[width=5cm]{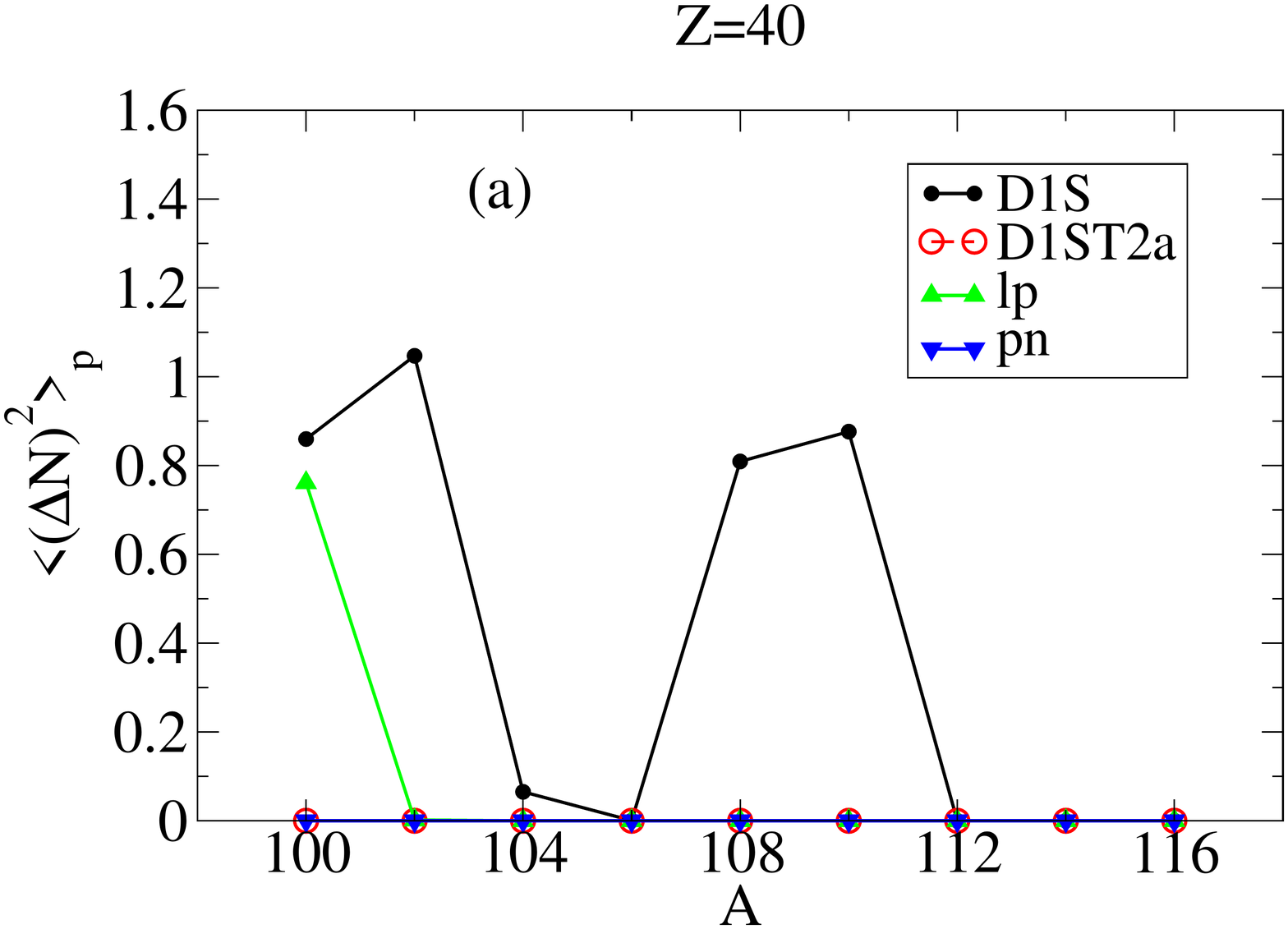} &
   \includegraphics[width=5cm]{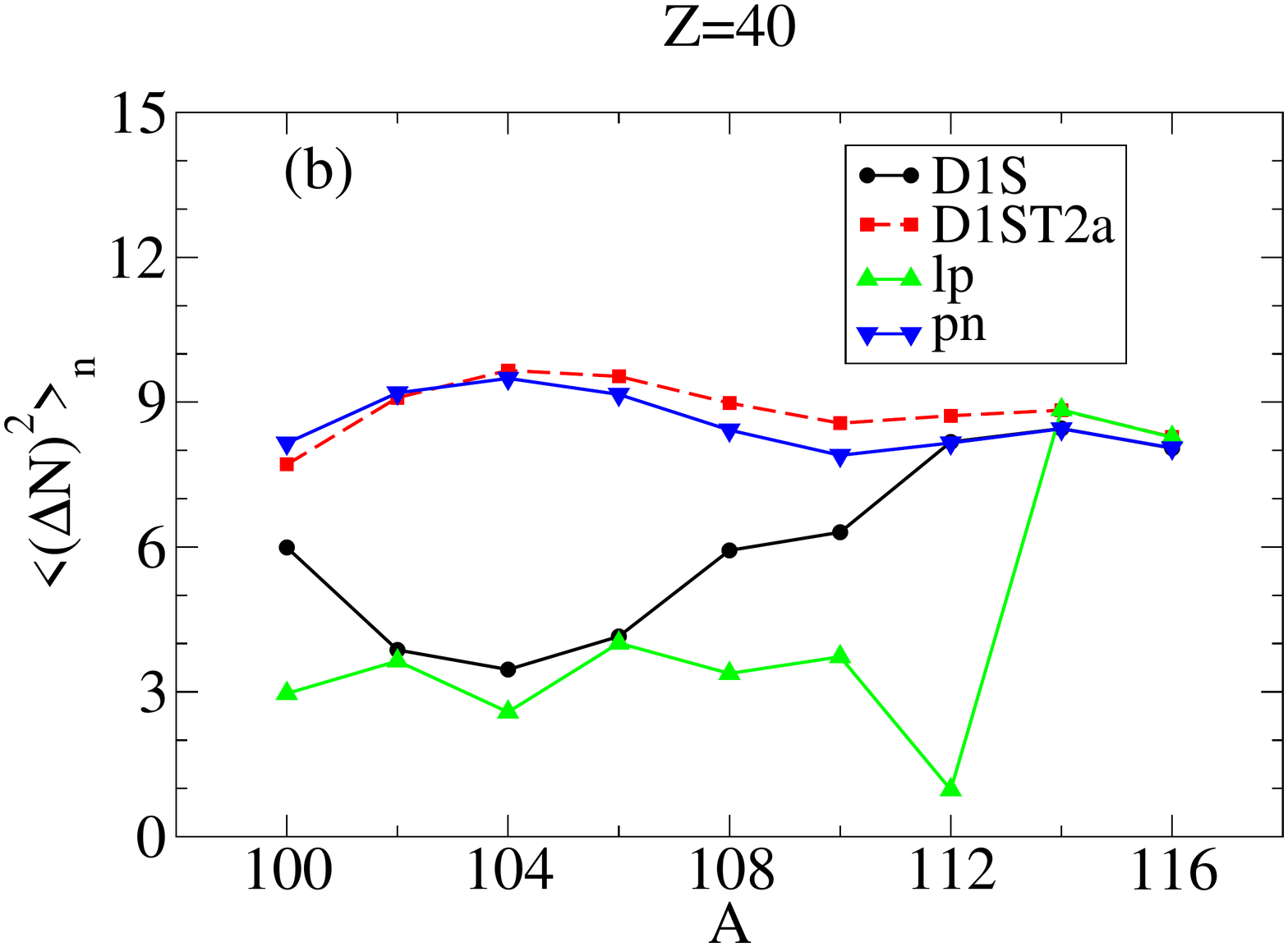} 
\end{tabular}
\caption{Proton and neutron particle number fluctuation for the $Z=40$ chain.}
\label{ZrDN2}
\end{figure}

In Fig. \ref{ZrDN2} we show the proton and neutron fluctuations numbers obtained for the ground state for all Zirconium
isotopes from $A=100$ to $A=116$. Red squares are the D1ST2a results and black circles are the D1S ones. On the other hand, green triangles up are the results 
considering only the lp part of the tensor interaction, and blue triangles down are those 
considering only the pn part. It is interesting to note that in what respect to proton fluctuations, 
we find practically zero for all the isotopes, using the D1ST2a interaction. 
With the D1S one, we find a non zero value for some isotopes, but in any case, the proton fluctuations are small, around $1.0$. 
On the other hand, the neutron fluctuations are always higher for the D1ST2a case, and from $A=112$ the results for D1S and D1ST2a 
are practically the same. This can be related with the fact that the ground state obtained with the interaction D1ST2a is always spherical, and in the case 
of the D1S one, only from $A=112$ is  spherical. 

Concerning the lp and  pn results  we see that the lp part gives 
very similar results to D1ST2a ones for both protons and neutrons. On the other hand the pn points are much closer to 
the D1S ones. This can be related to the fact that the ground state locations are most of the time the same for the lp (pn) 
curve and the D1ST2a (D1S) one in  Fig. \ref{Zrdec2}. For $A>110$ almost all the ground states are spherical  
and the tensor energy contributions are  small (see Fig. \ref{Zrdec2}). Consequently  the results 
are similar for $A>110$. Only the lp part for $A=112$ gets a prolate ground state associated to a small neutron number fluctuation.

 
\begin{figure}[htb] \centering
\begin{tabular}{ccc}
   \includegraphics[width=4.6cm]{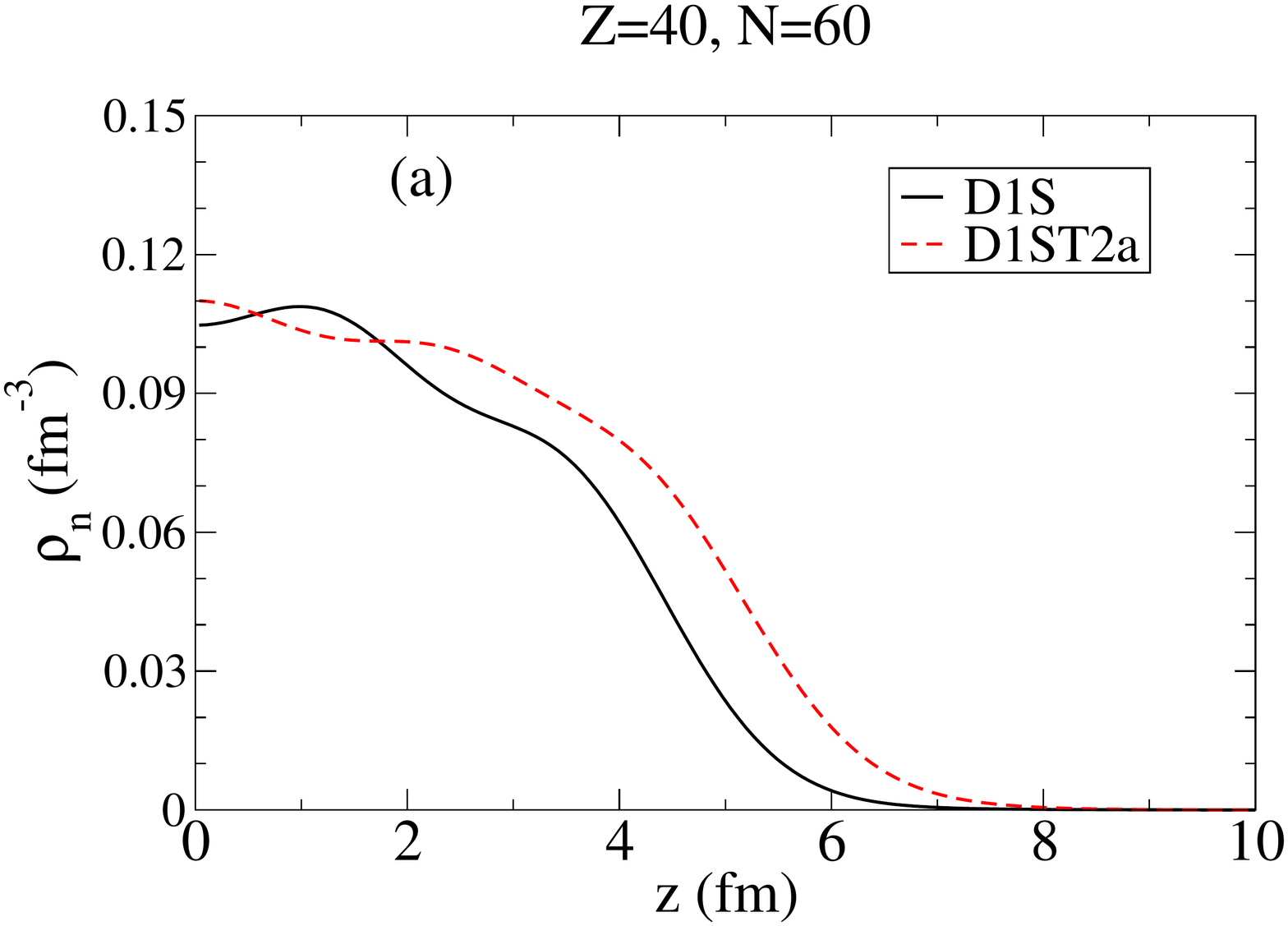} &
   \includegraphics[width=4.6cm]{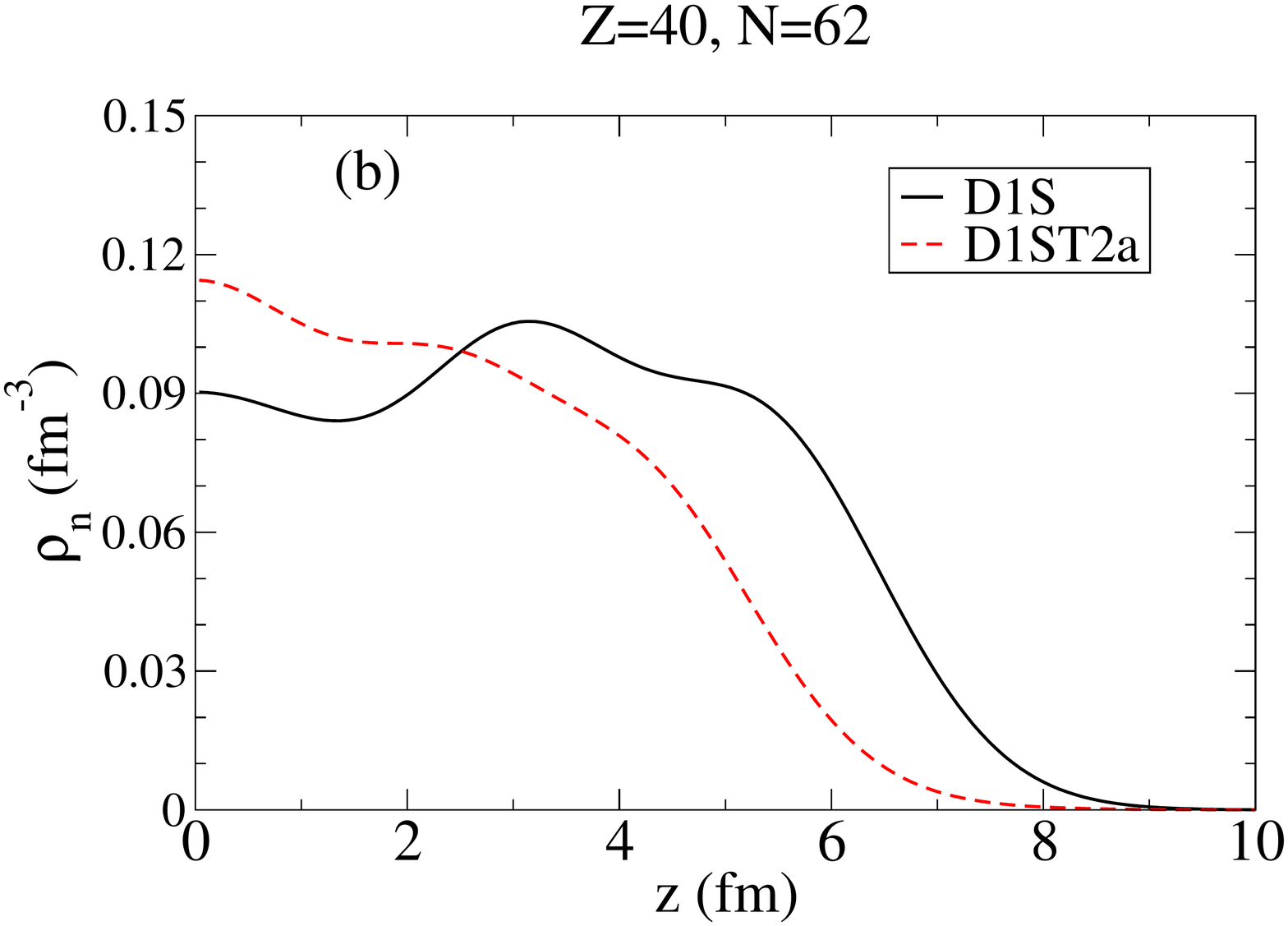} &
   \includegraphics[width=4.6cm]{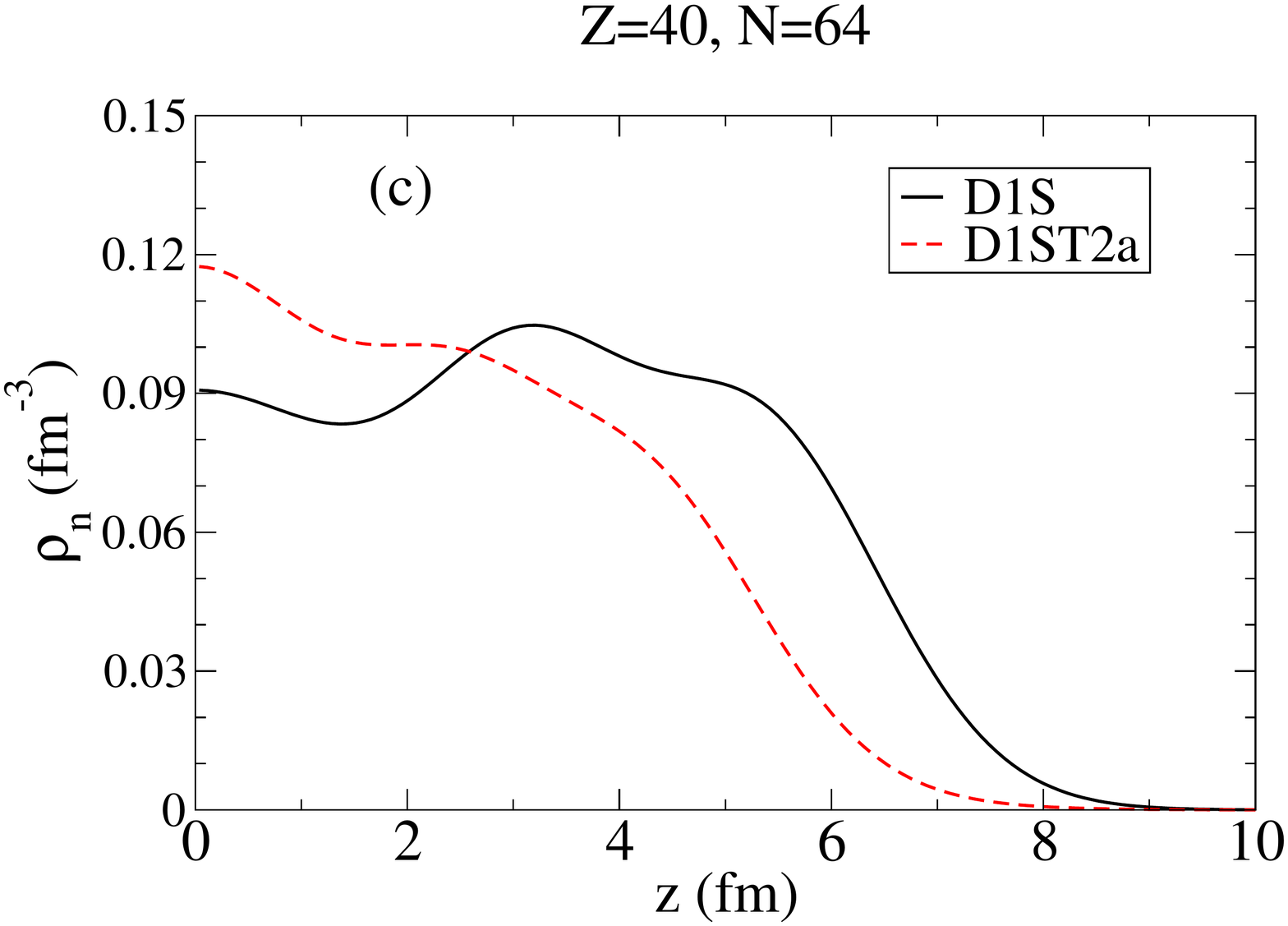} \\
   \includegraphics[width=4.6cm]{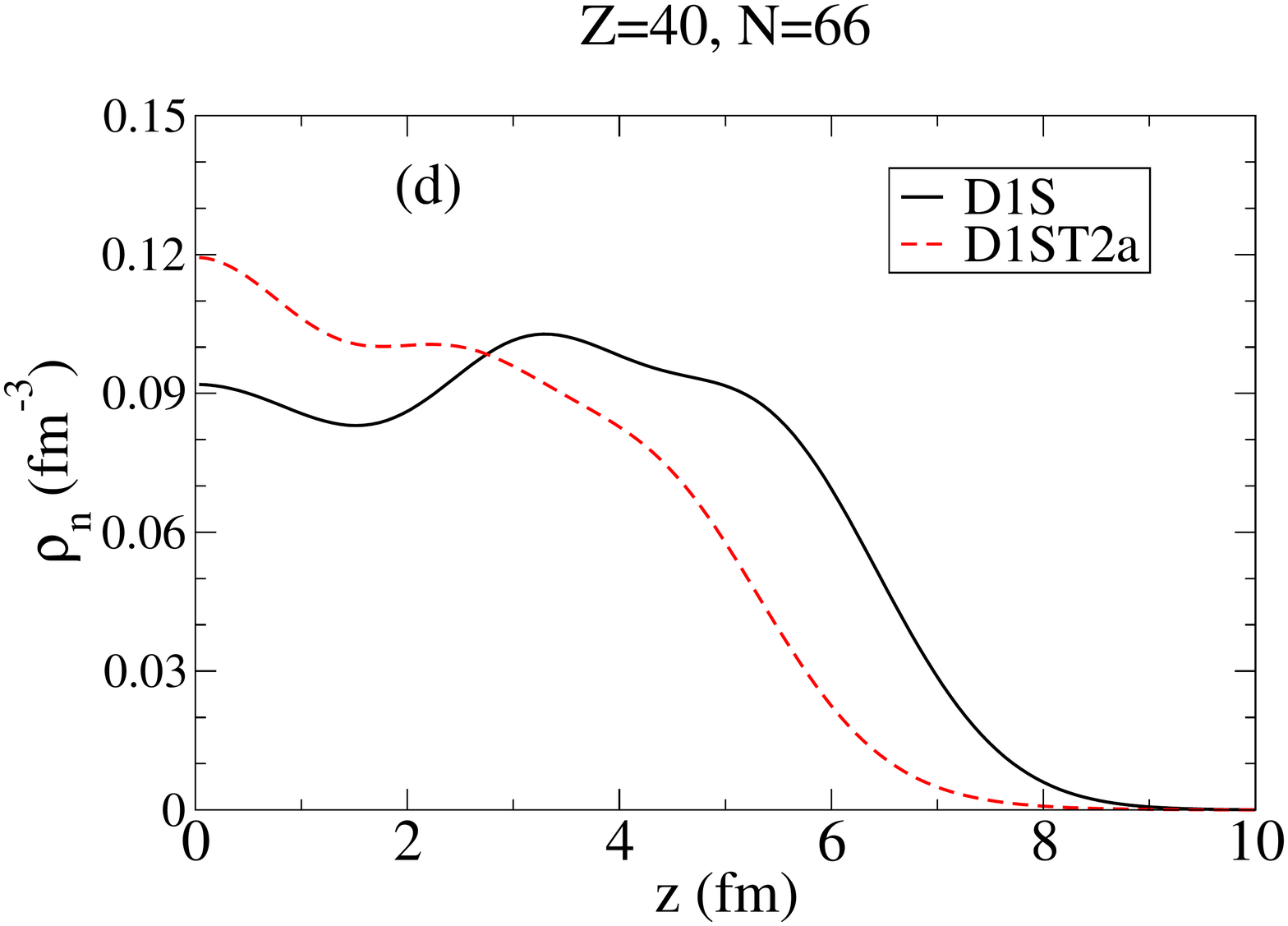} &
   \includegraphics[width=4.6cm]{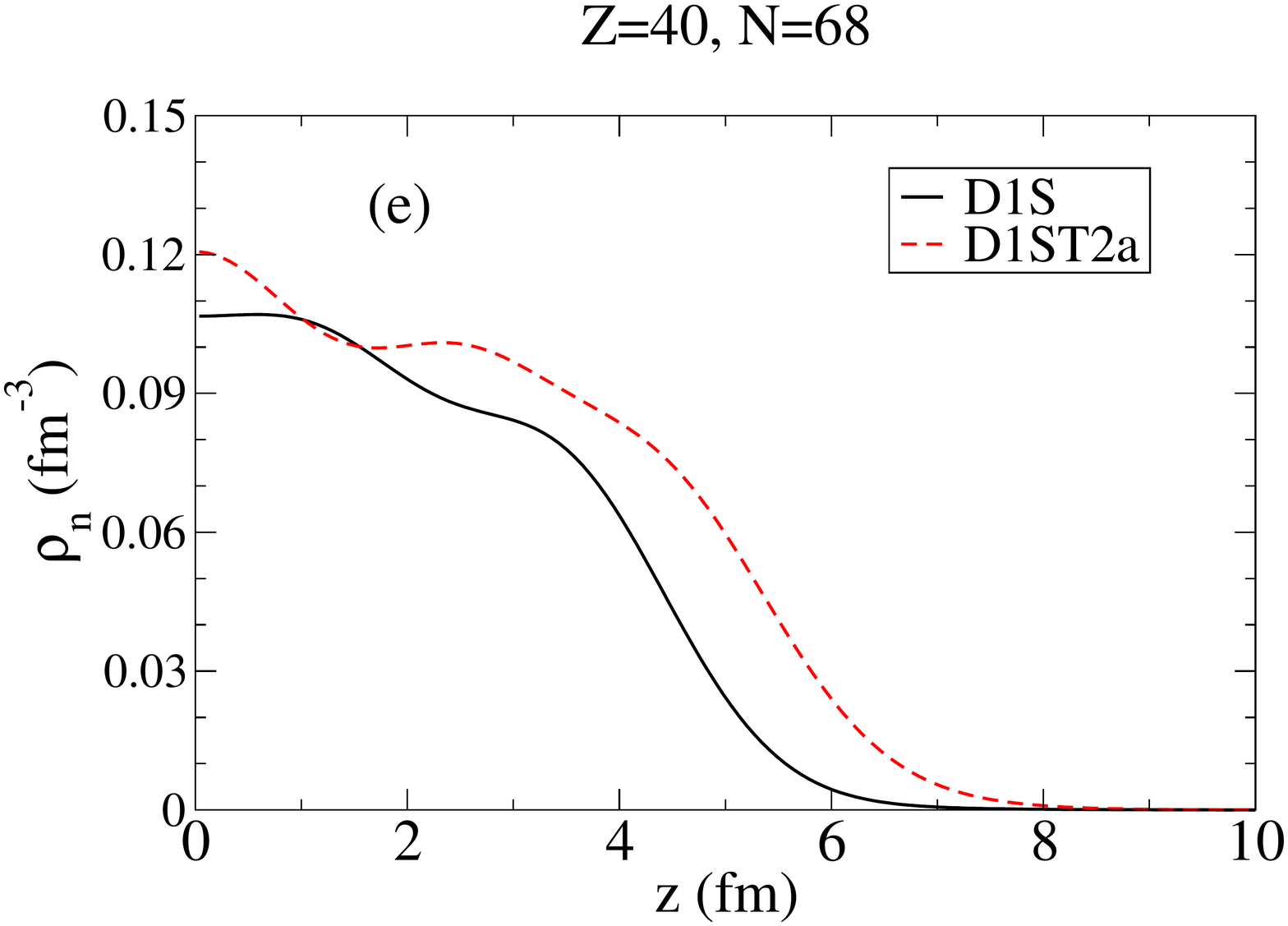} &
   \includegraphics[width=4.6cm]{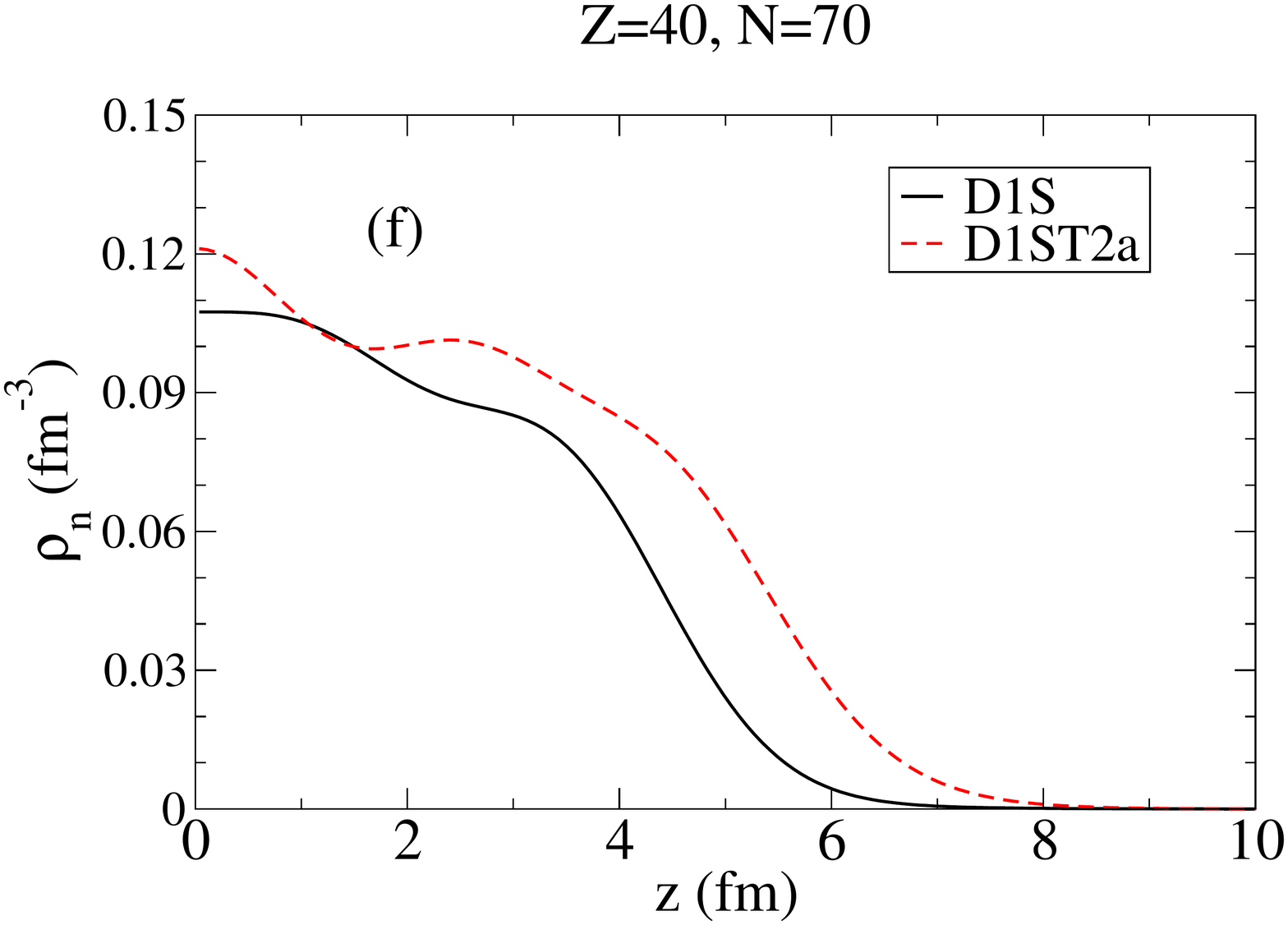} \\
\end{tabular}
\caption{Ground state neutron densities for the $Z=40$ chain for the D1S interaction (solid black lines) and  the D1ST2a one (red dashed lines).} 
\label{ZrDensn}
\end{figure}

\begin{figure}[htb] \centering
\begin{tabular}{ccc}
   \includegraphics[width=4.6cm]{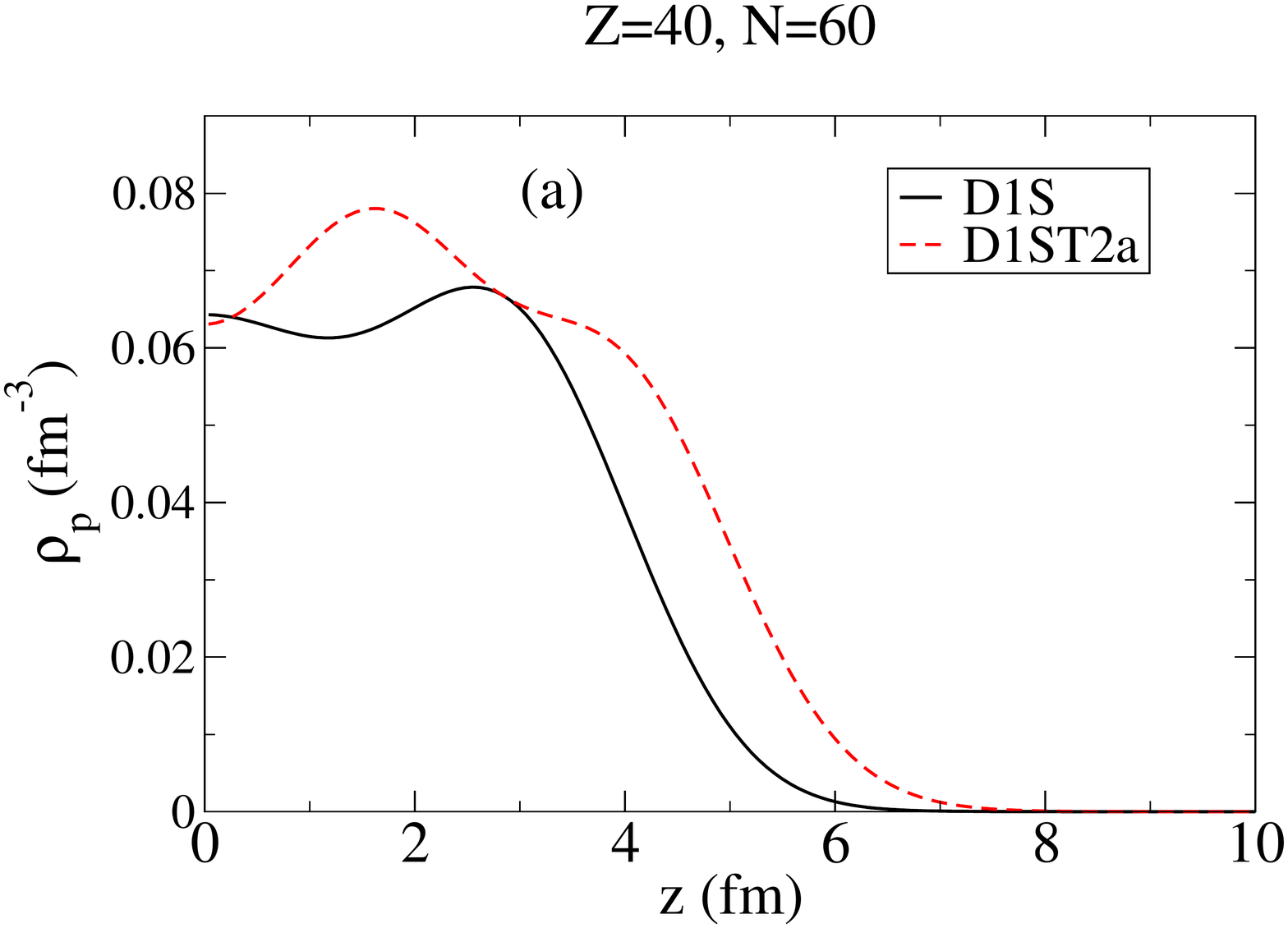} &
   \includegraphics[width=4.6cm]{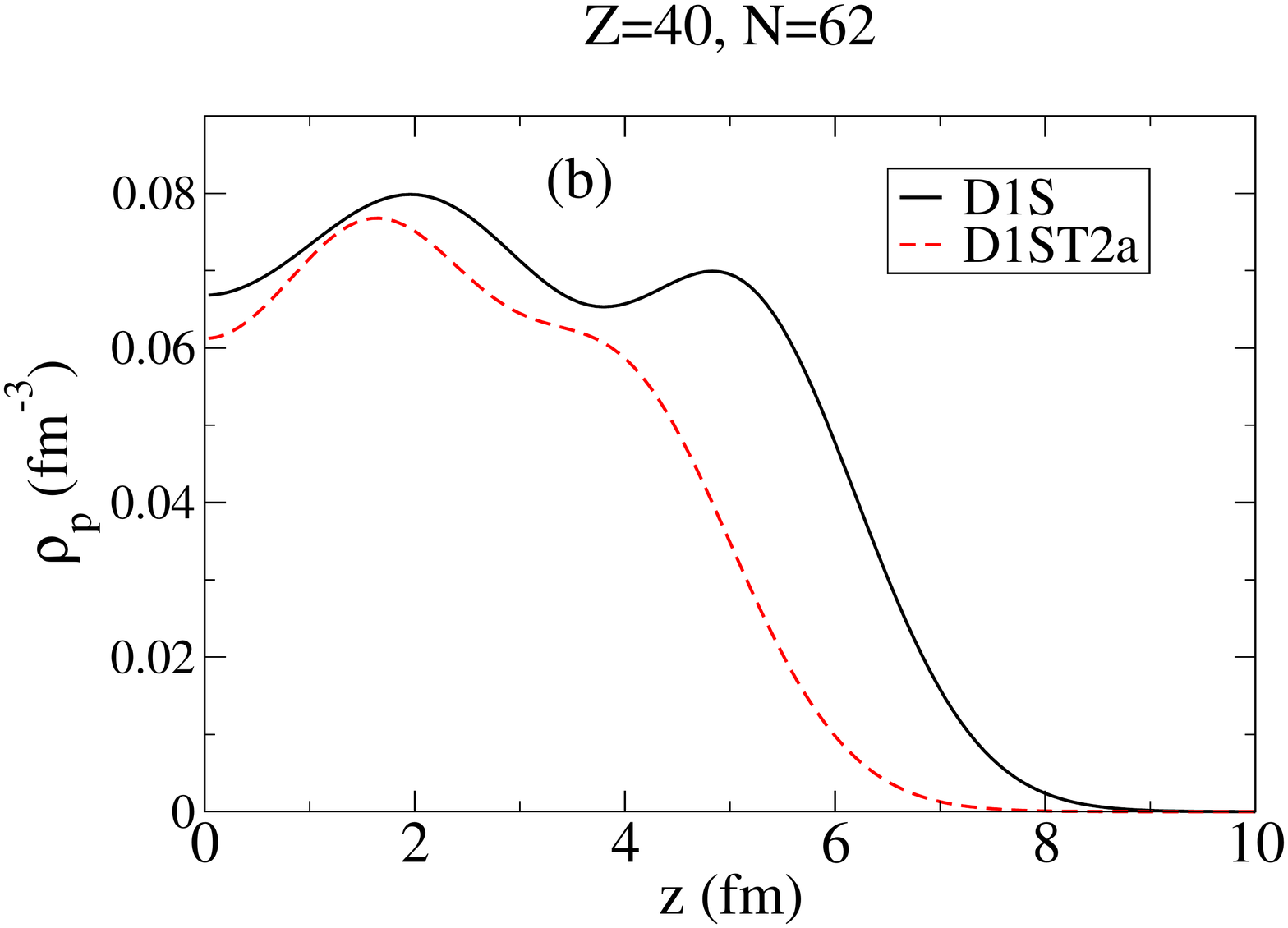} &
   \includegraphics[width=4.6cm]{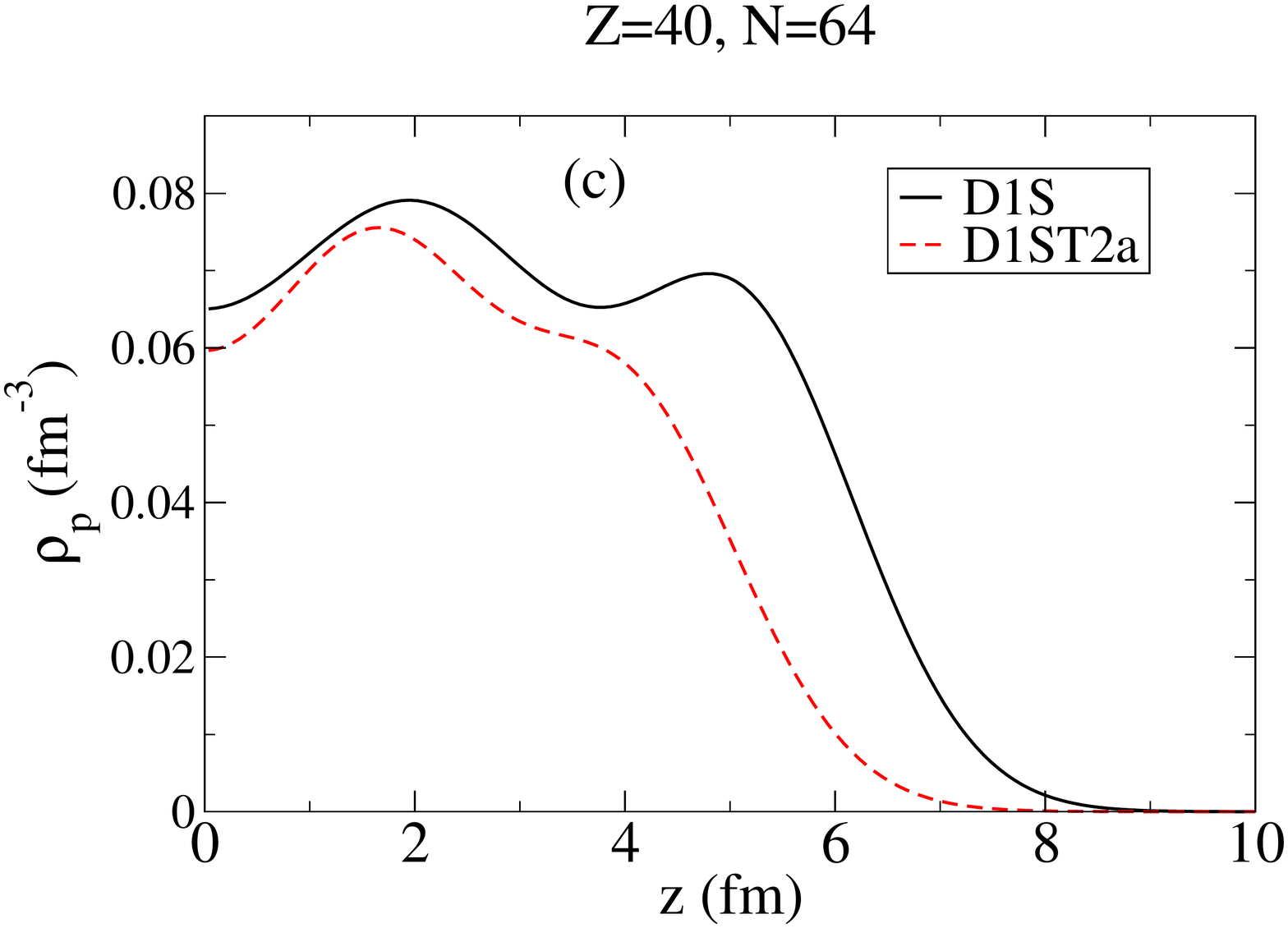} \\
   \includegraphics[width=4.6cm]{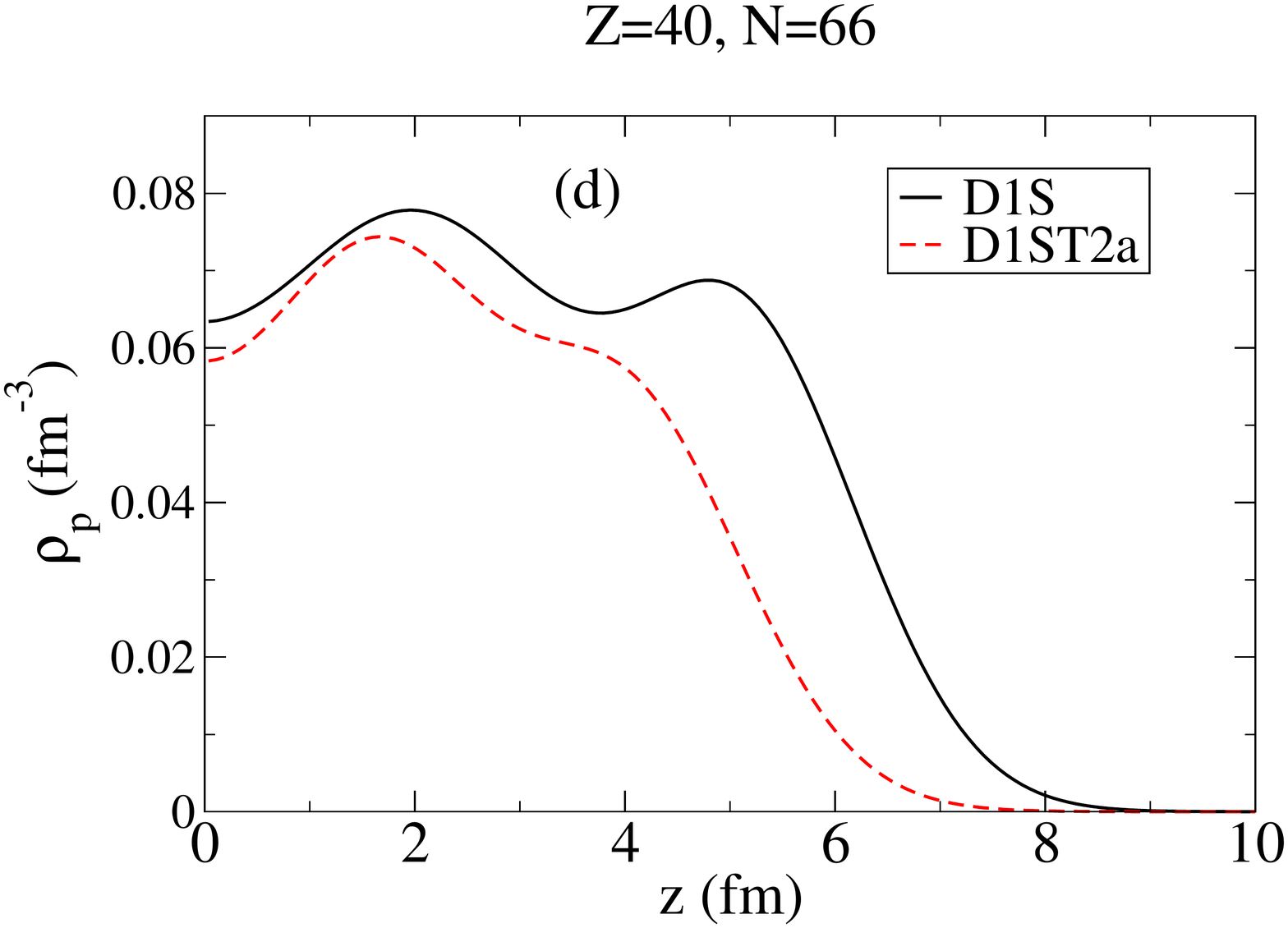} &
   \includegraphics[width=4.6cm]{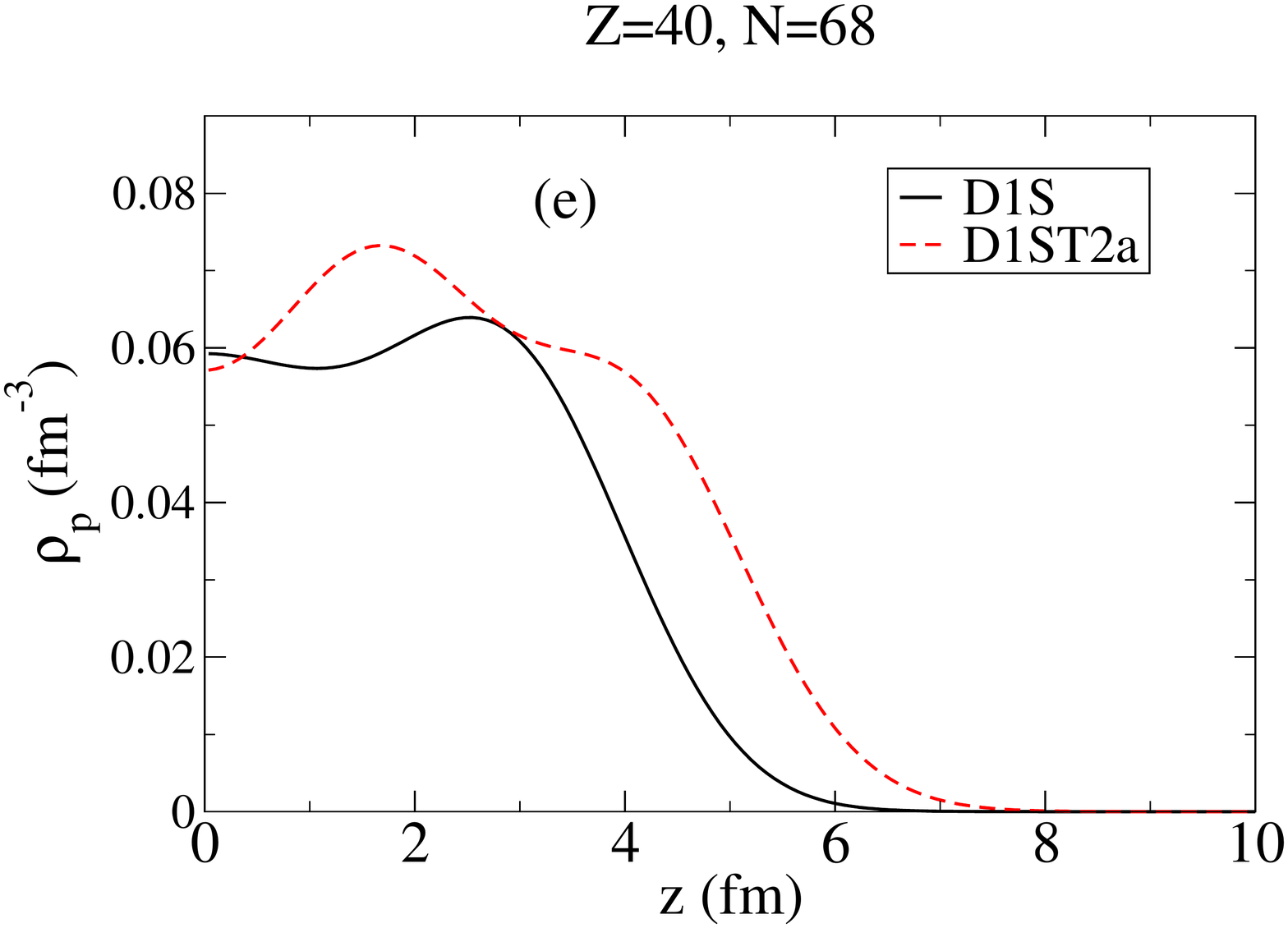} &
   \includegraphics[width=4.6cm]{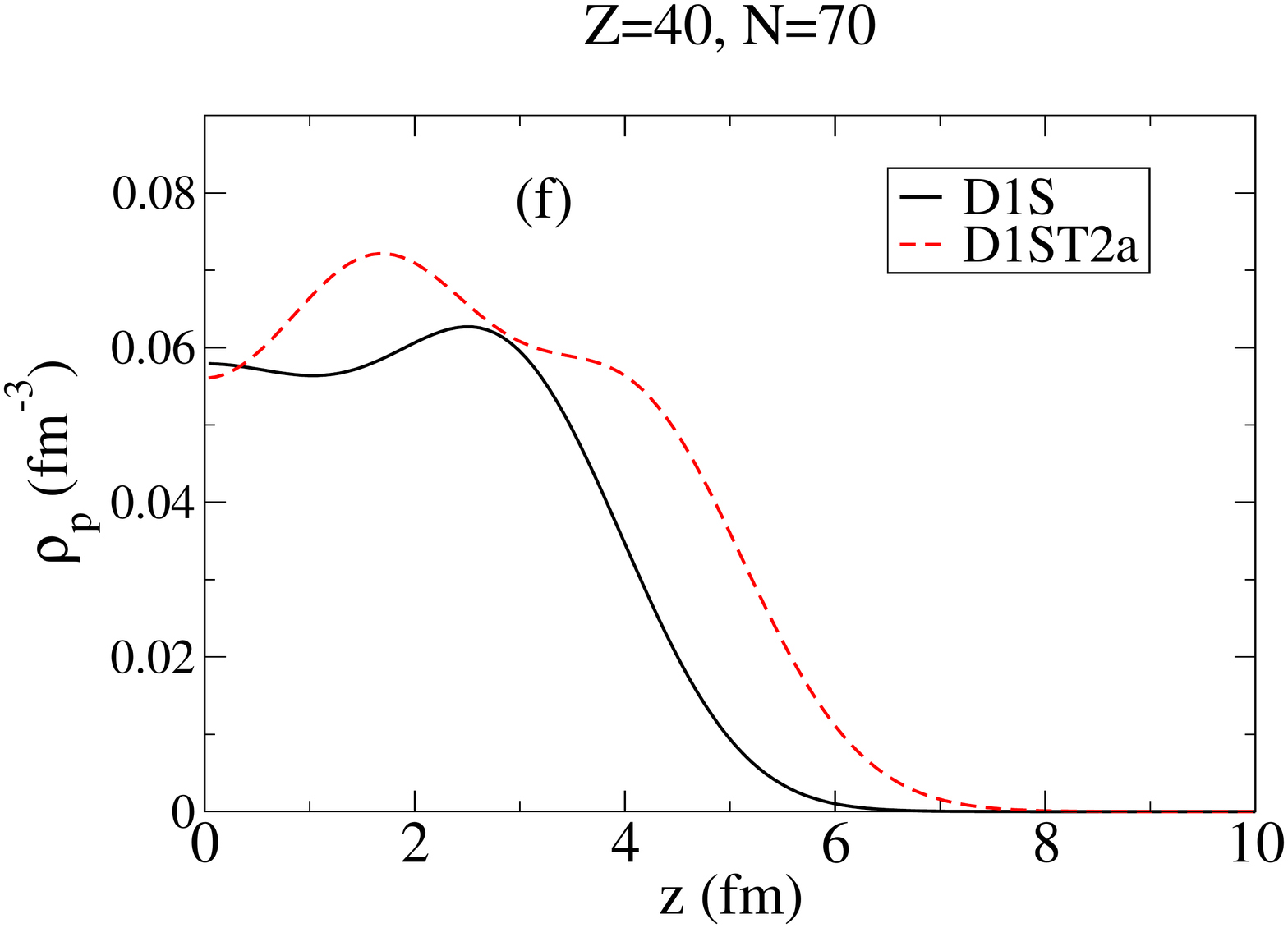} \\
\end{tabular}
\caption{Ground state proton densities for the $Z=40$ chain for the D1S interaction (solid black lines) and the D1ST2a one (red dashed lines).} 
\label{ZrDensp}
\end{figure}

In Fig.~\ref{ZrDensn} and \ref{ZrDensp} we plot the neutron and proton densities for some Zr isotopes along the  symetry axis $z$.
Densities obtained with D1S and D1ST2 are very different because minima for both interactions have not the same deformation character 
(see Table~\ref{TableZr2}).

\clearpage

\subsection{Magnesium chain}

\begin{figure}[htb] \centering
\begin{tabular}{ccc}
   \includegraphics[width=4.6cm]{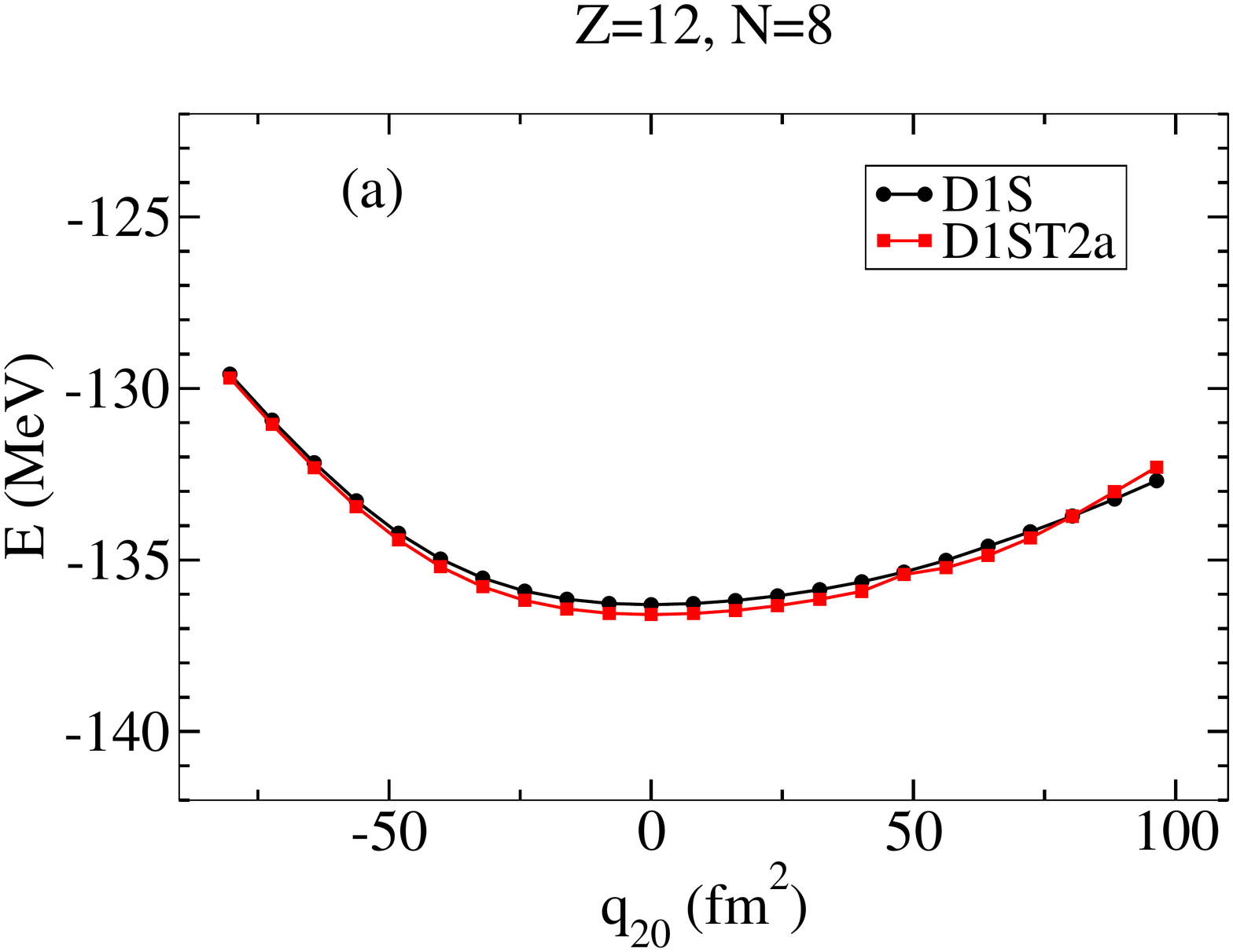} &
   \includegraphics[width=4.6cm]{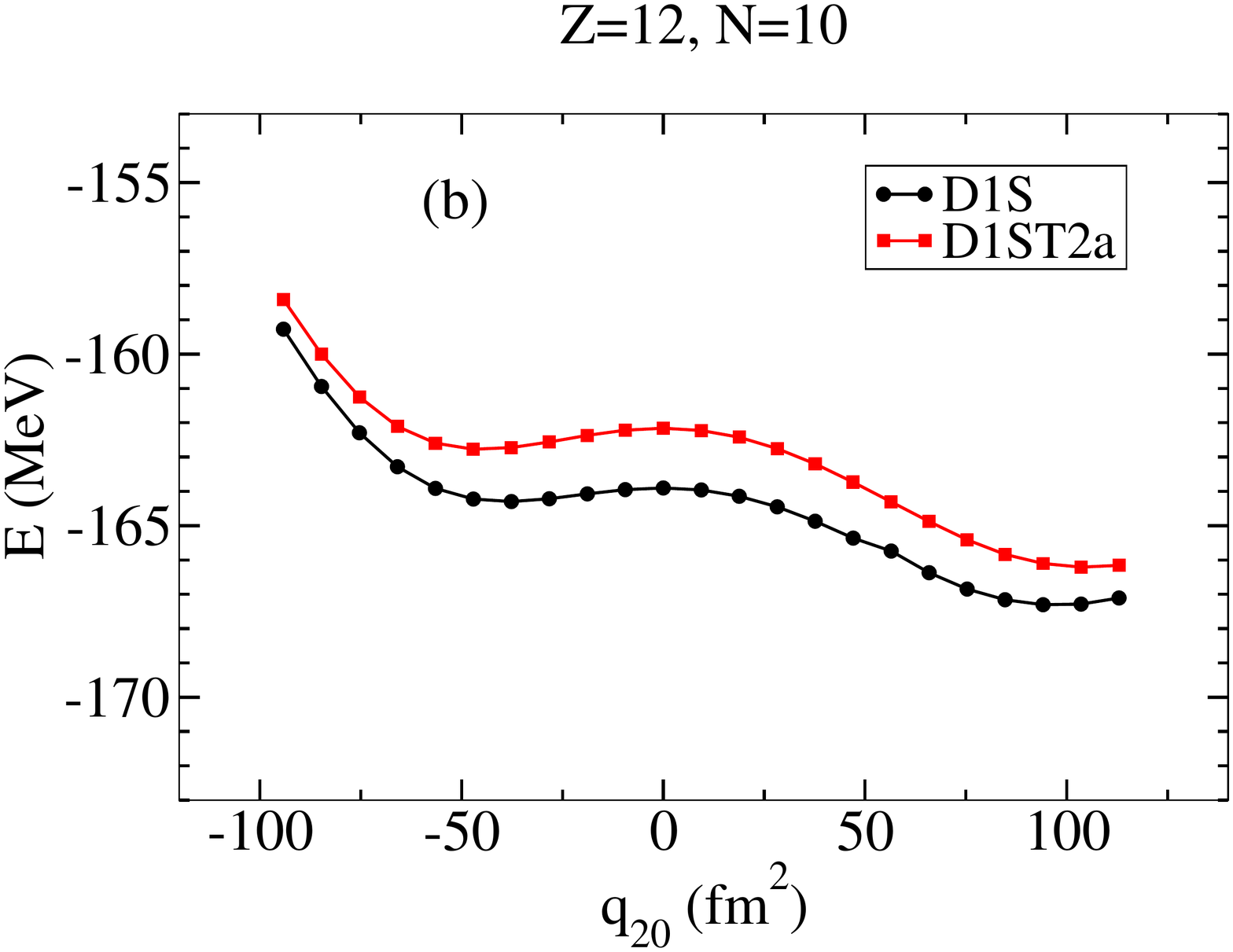} &
   \includegraphics[width=4.6cm]{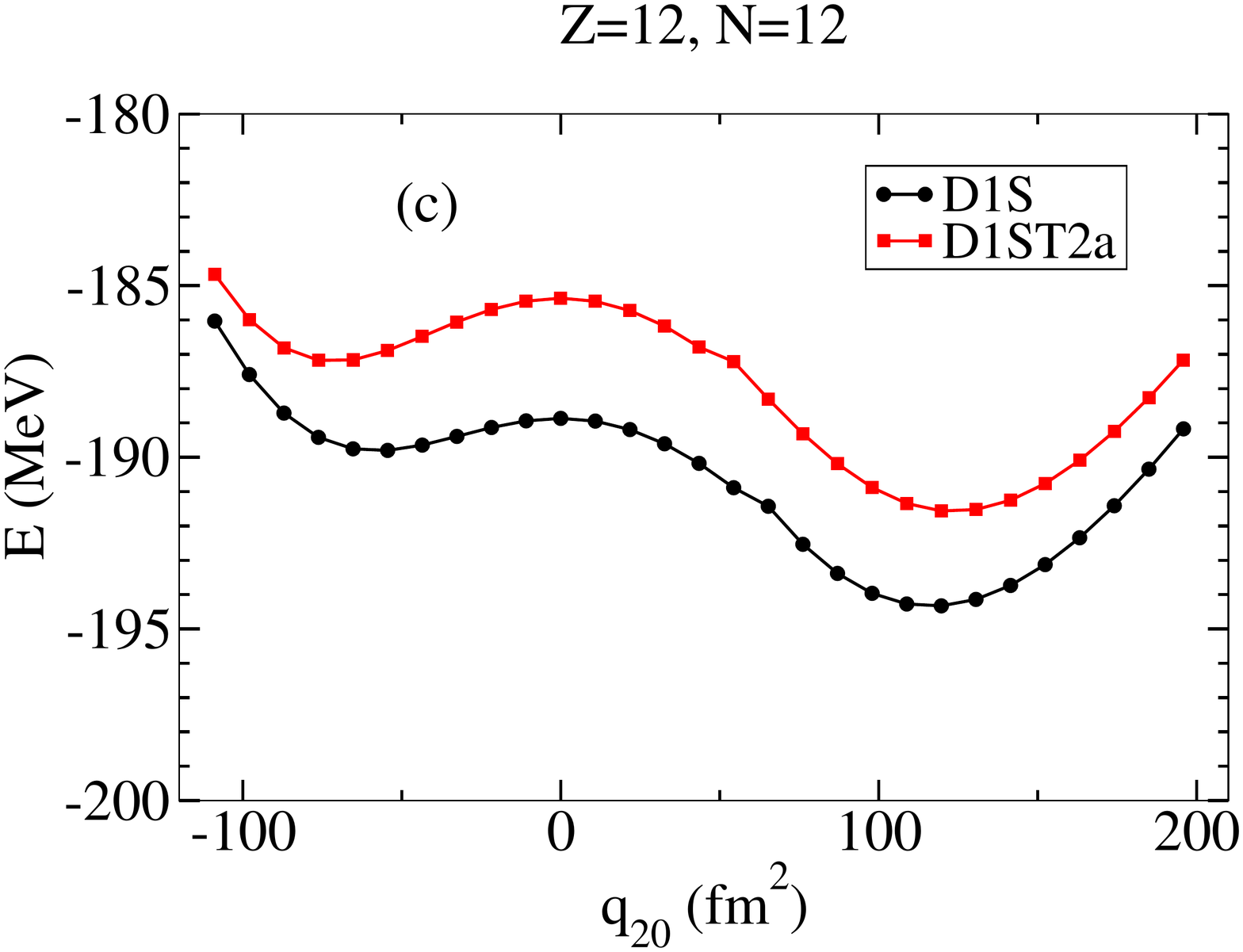} \\
   \includegraphics[width=4.6cm]{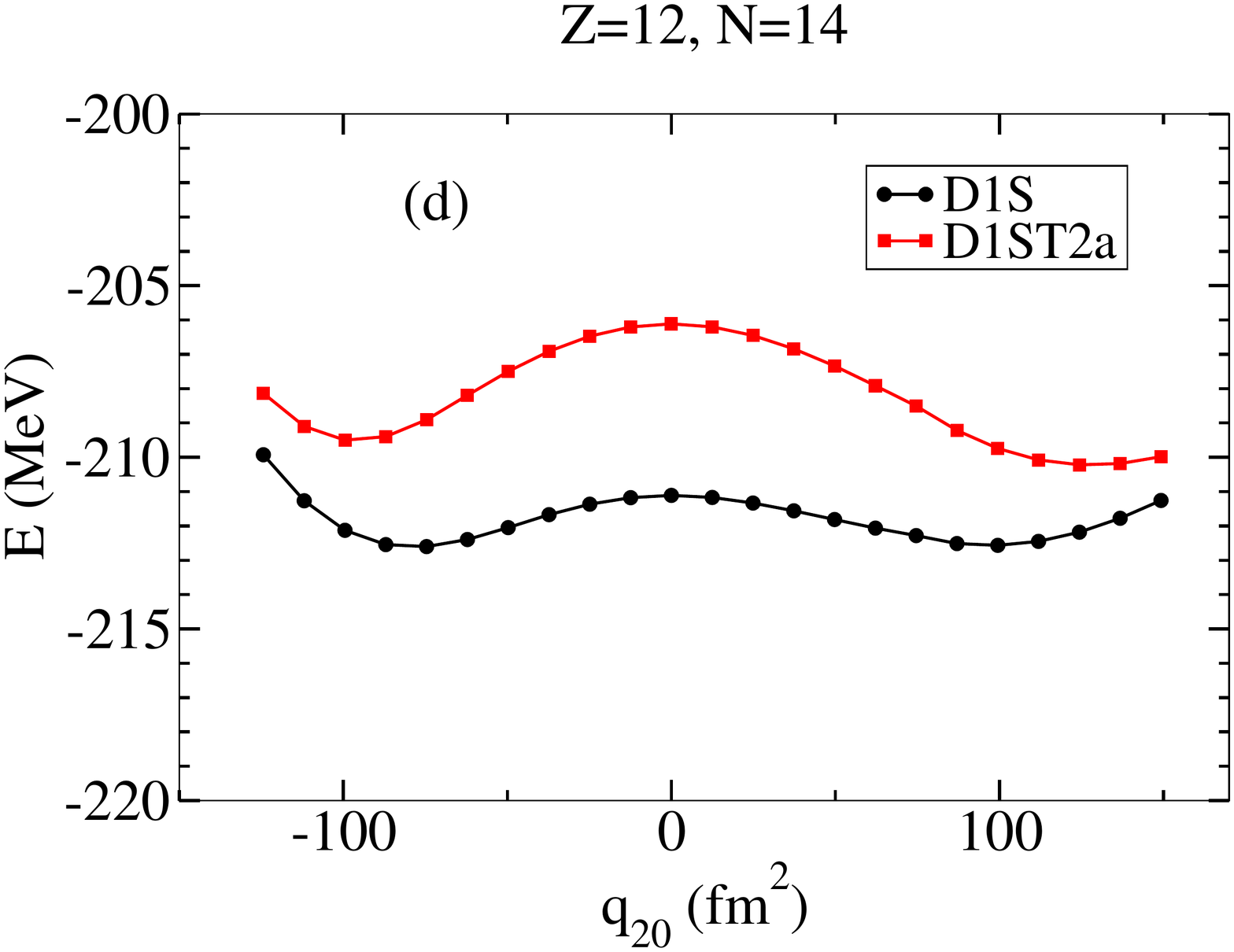} &
   \includegraphics[width=4.6cm]{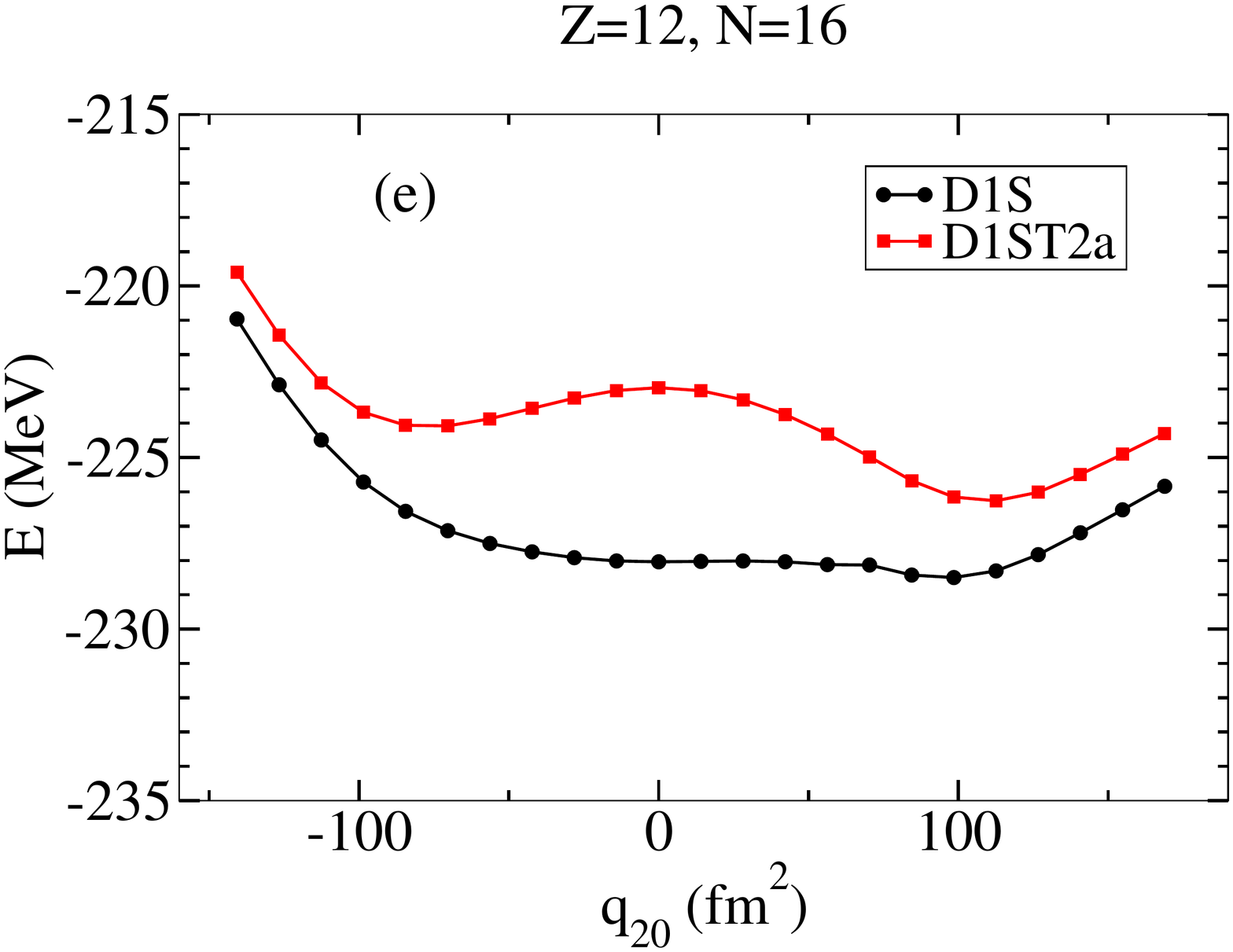} &
   \includegraphics[width=4.6cm]{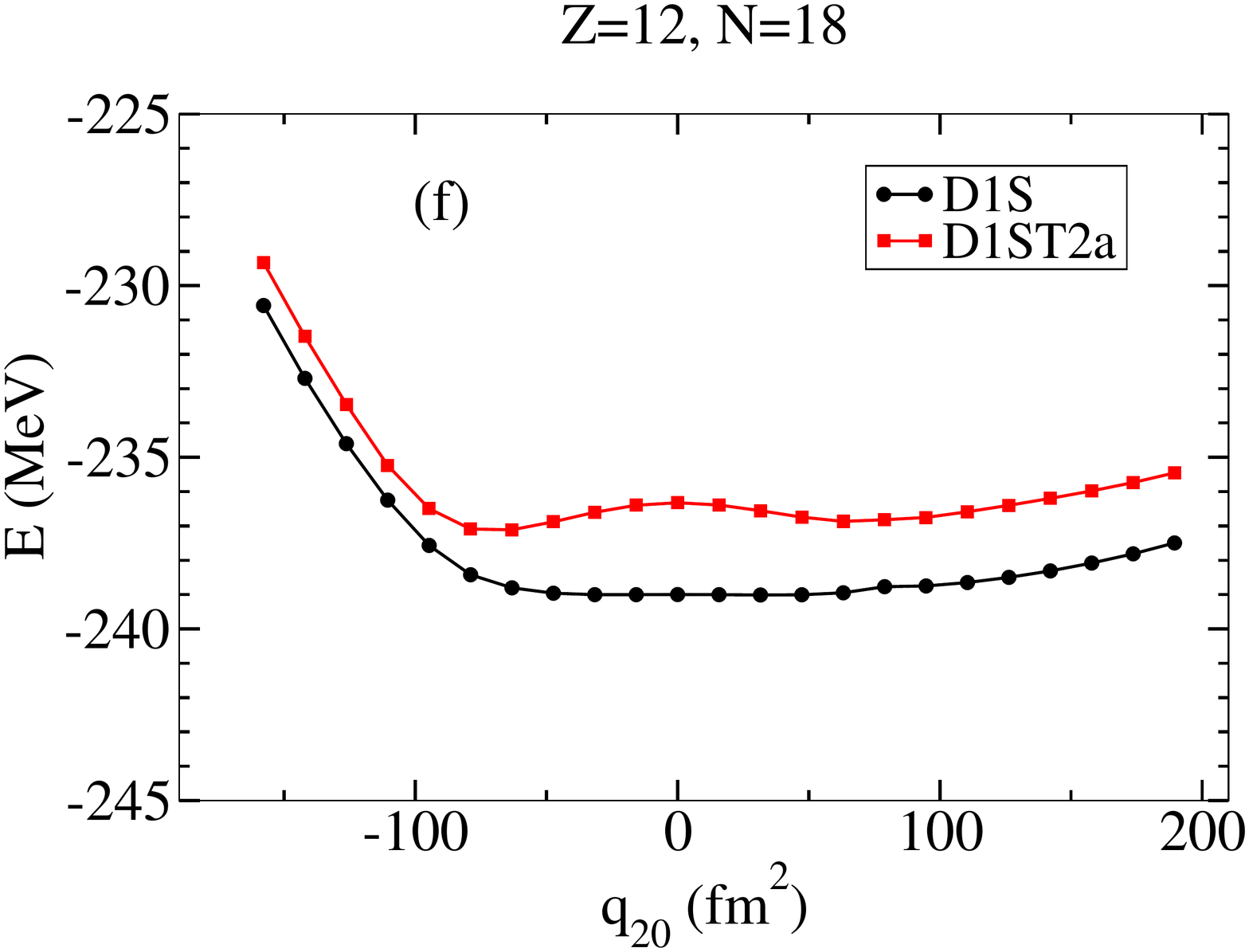} \\
   \includegraphics[width=4.6cm]{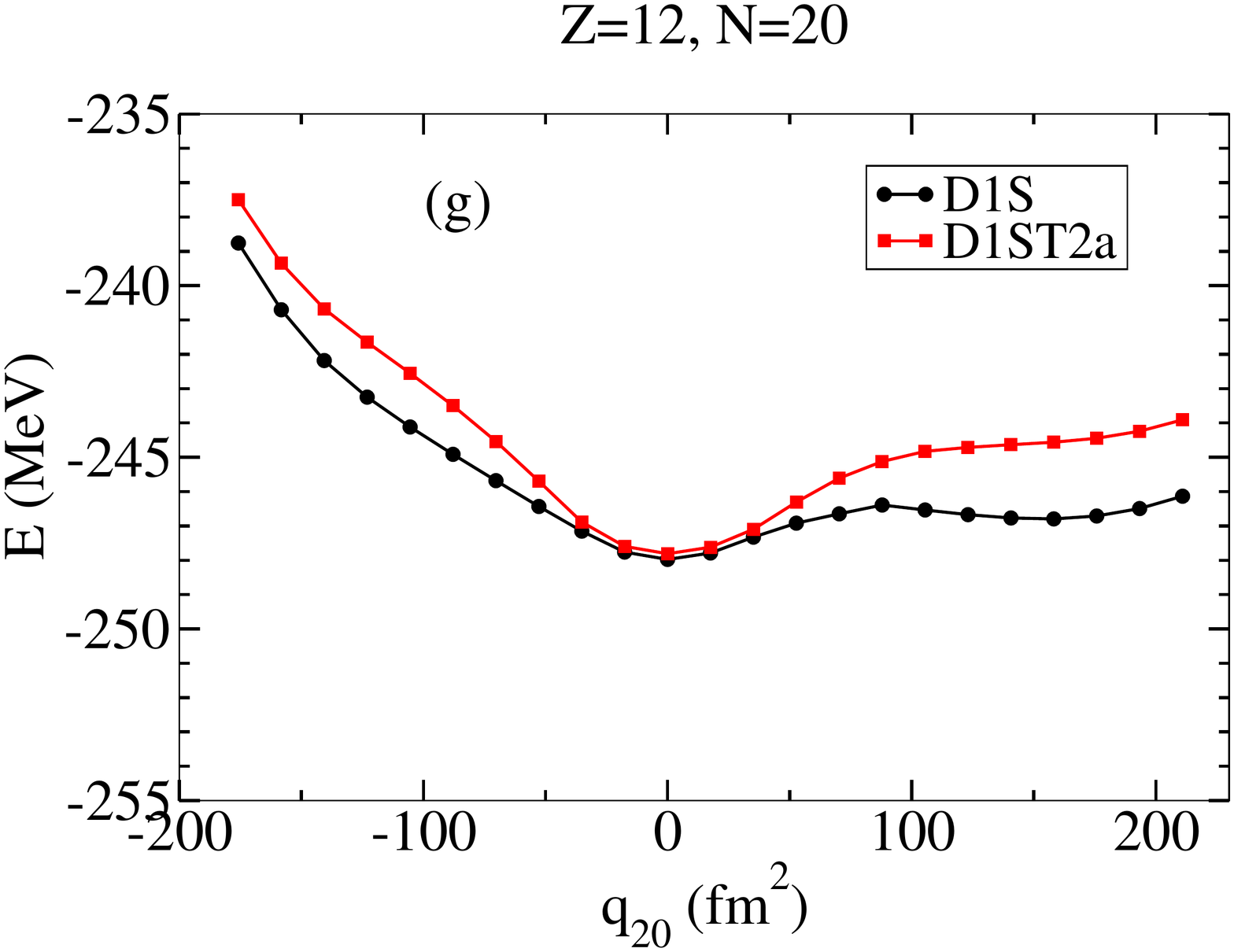} &
   \includegraphics[width=4.6cm]{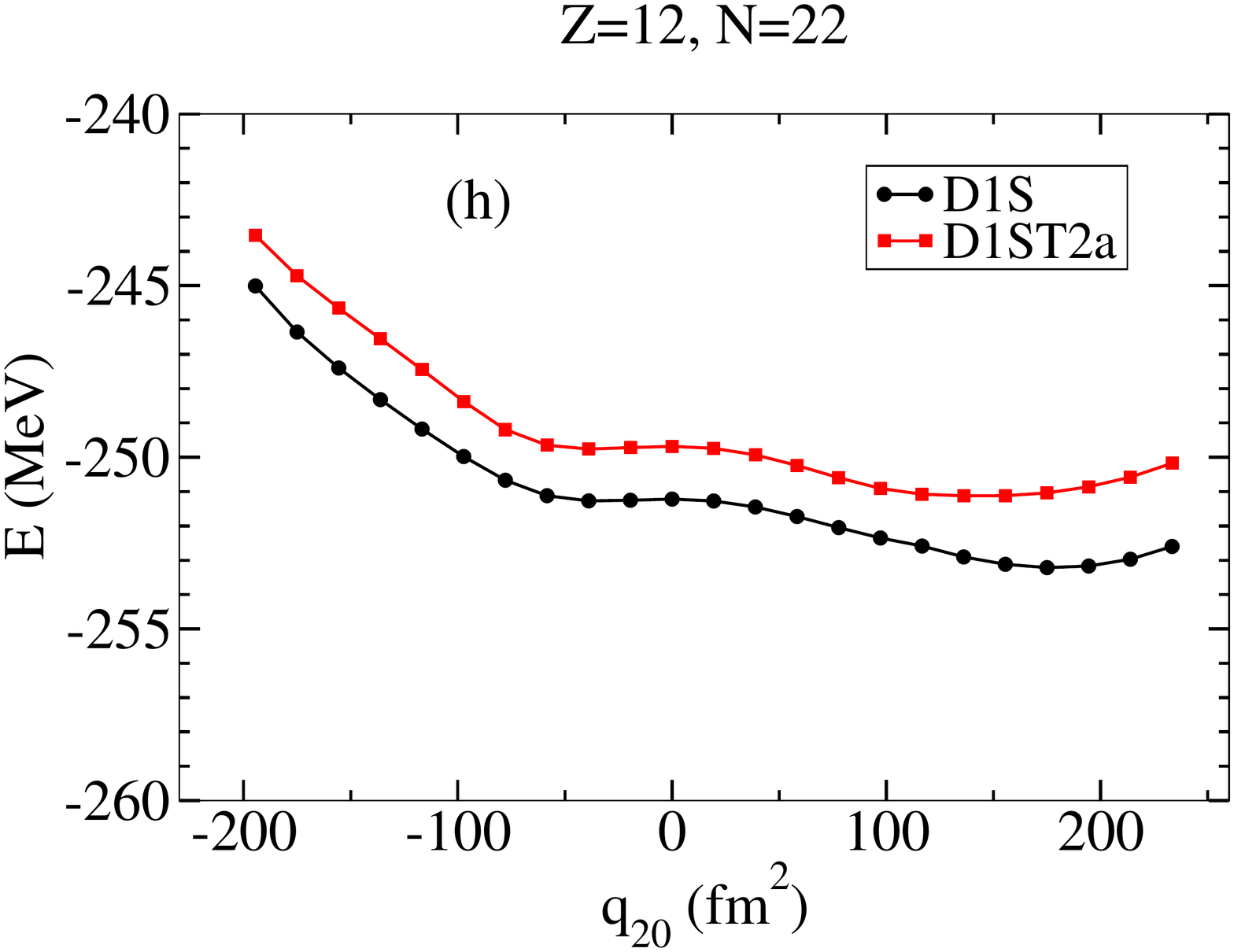} &
   \includegraphics[width=4.6cm]{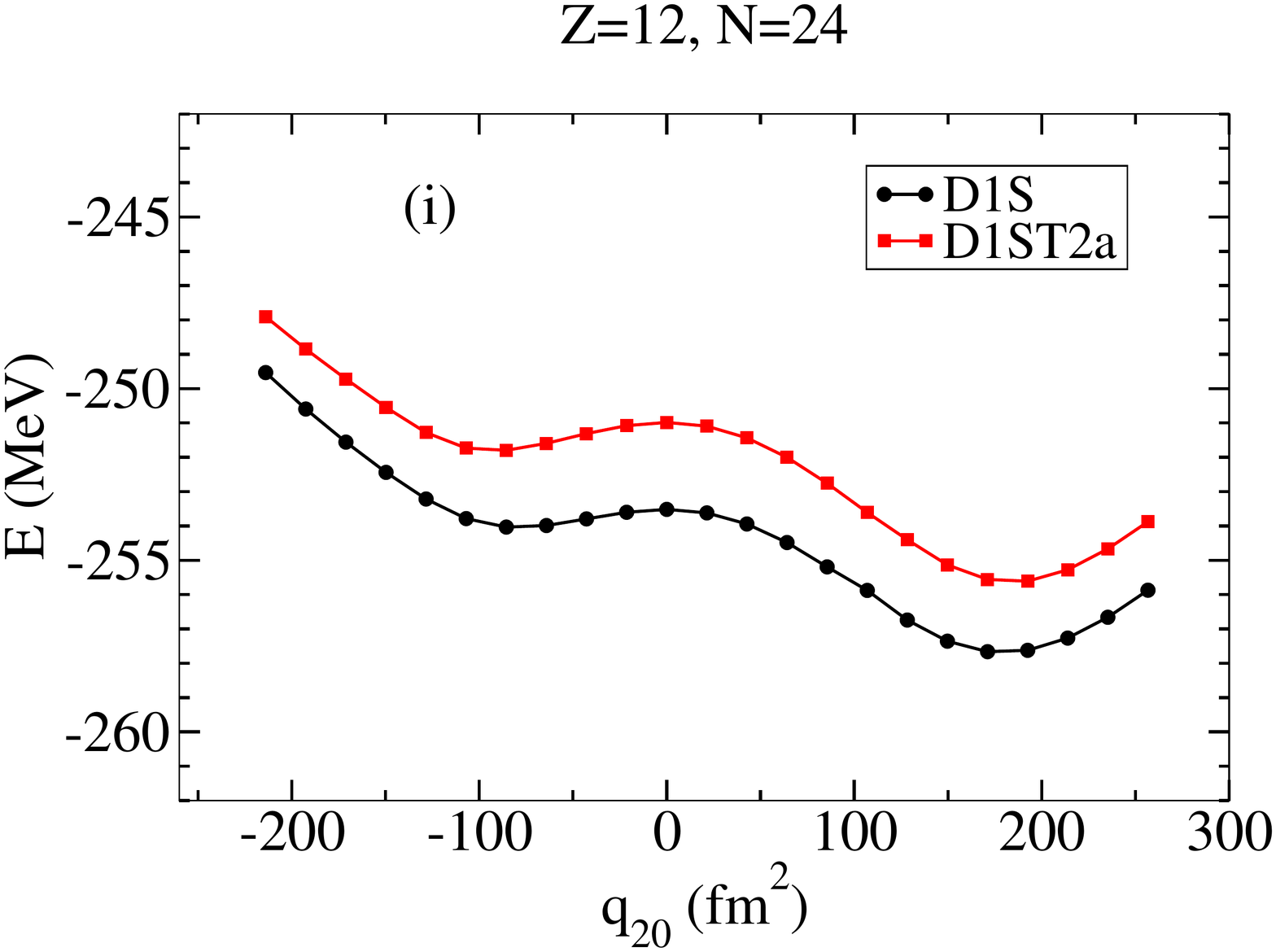} \\
   \includegraphics[width=4.6cm]{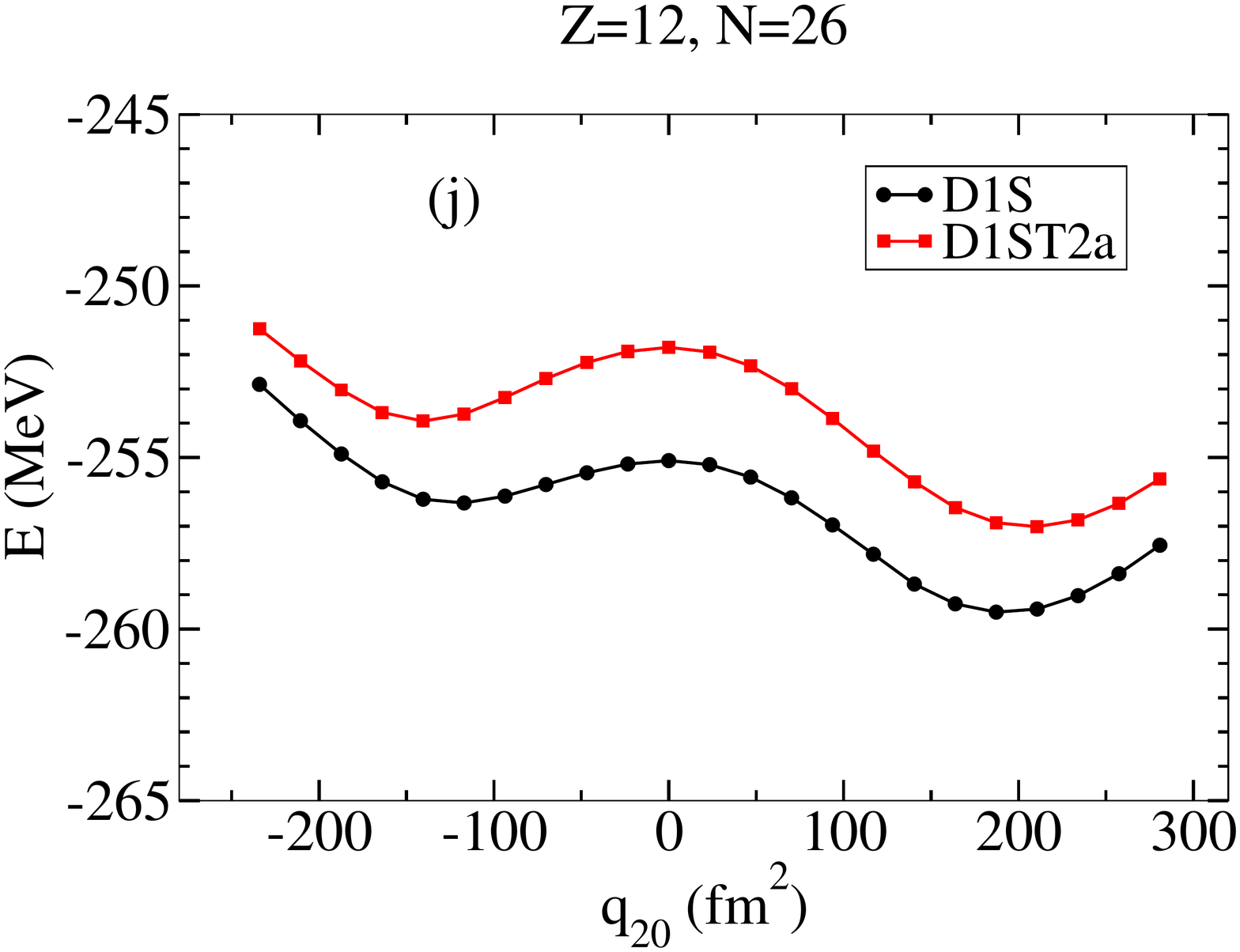} &
   \includegraphics[width=4.6cm]{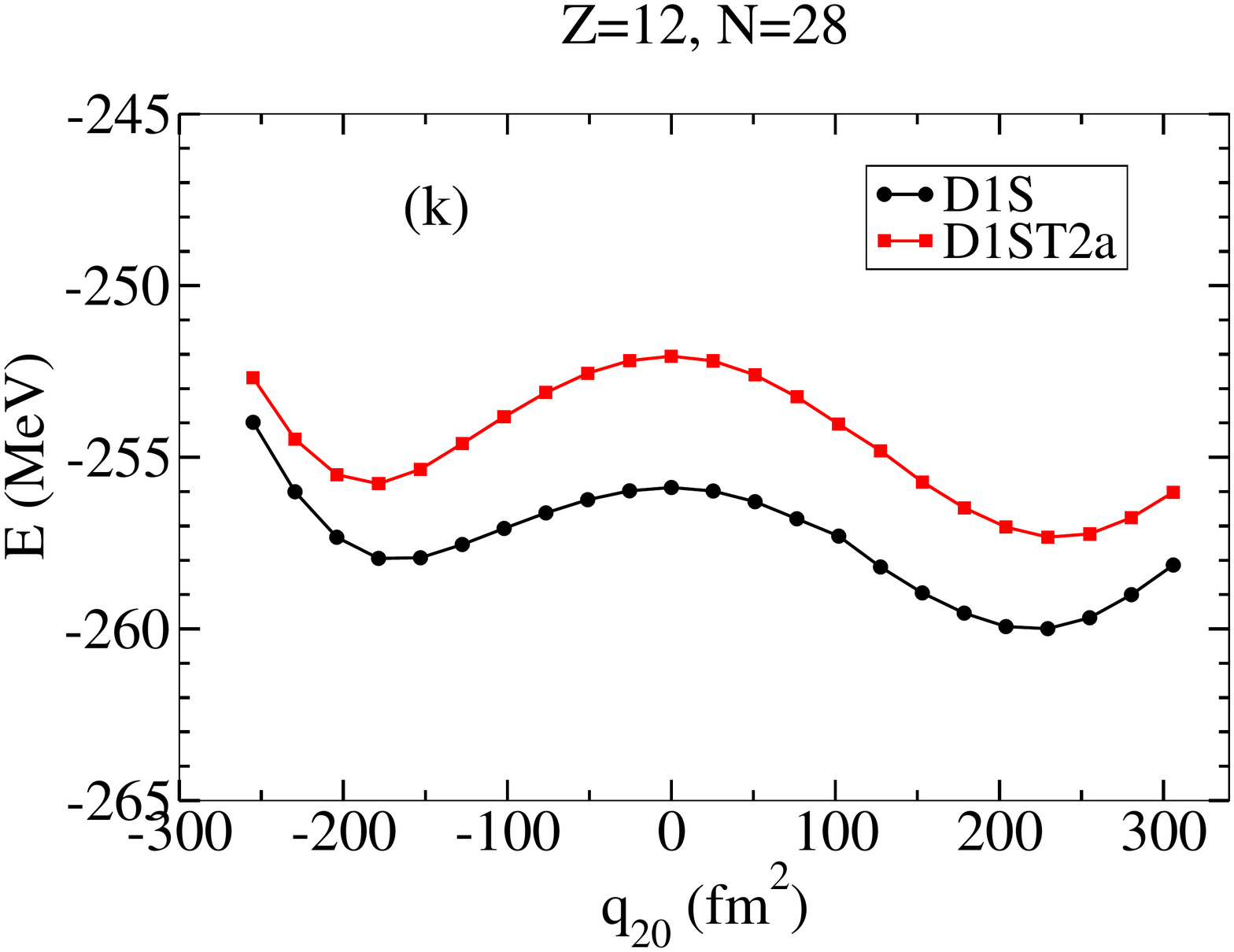} & 
\end{tabular}
\caption{Potential Energy Curves  of the $Z=12$ chain 
for the D1S interaction (circles) and the D1ST2a one (squares). } 
\label{Mgfig}
\end{figure} 
Magnesium PEC are presented in Fig.~\ref{Mgfig} from $A=20$ to $A=40$.
Some of the nuclei presented below ($^{28}$Mg, $^{30}$Mg, etc) are soft against the quadrupole moment coordinate. Therefore
beyond mean field theories would be more appropriate to describe them.
Since the tensor interaction may modify at the mean field level the PEC in an important way, we choose
to present all the even-even isotopes in the A=20 to 40 interval regardless its structure properties. For the Magnesium chain
 the contribution of the tensor term is always repulsive, except for the $N=8$ isotope for which it is negligible. 
This is due to the fact that $N=8$ is a saturated shell,
and then, tensor contributions coming from neutrons are zero. Even for the nucleus $^{32}$Mg, namely another neutron saturated shell,
we see that the contribution of tensor terms  is negligible in the minimum region. 
For these two nuclei $E_{\rm TS}(p-p)$ and $E_{\rm TS}(p-n)$ compensate each other.
Moreover another striking feature of Fig.~\ref{Mgfig} is the independence of tensor energy with respect to the quadrupole deformation. This
occurs for a majority of the Magnesium isotopes namely from $N=8$ to $N=12$ and from $N=22$ to $N=28$.
For $N=14$ the prolate ground state with D1S becomes oblate with D1ST2a. For the two next nuclei $N=16,\,18$ D1ST2a produces a two minima well whereas D1S PEC are flat.

\begin{table}[htb] \centering
 \begin{tabular}{|c|c|l|c|c|c|}
\hline
         & Exp. $Q_S(I=2)$& Method & Exp. $Q_0(K=2)$& Exp. $Q_0(K=1)$& Exp. $Q_0(K=0)$ \\
\hline
         & $-0.29(3)$       & CER    & $-1.02$ 	      & $-2.03$ 	    & $+1.02$ 	    \\
$^{24}$Mg& $-0.18(2)$       & CER, R & $-0.63$ 	      & $-1.26$ 	    & $+0.63$ 	    \\
         & $-0.07(3)$       & ES, R  & $-0.25$ 	      & $-0.49$ 	    & $+0.25$ 	    \\
\hline
         & $-0.21(2)$       & CER    &$-0.74$	      & $-1.47$ 	    & $+0.74$ 	    \\
$^{26}$Mg& $-0.14(3)$       & CER, R &$-0.49$	      & $-0.98$ 	    & $+0.49$ 	    \\
         & $-0.10(3)$       & CER    &$-0.35$	      & $-0.70$ 	    & $+0.35 $	    \\
\hline
 \end{tabular}
\caption{Experimental $Q_S(I=2)$ (eb) (from \cite{StoADN05}) for the $Z=12$ chain. Methods are specified: CER for Coulomb 
Excitation Reorientation, R for Re-evaluated data, ES for Electron Scattering. Corresponding $Q_0$ are given in eb assuming 
$K=2$, $K=1$ and $K=0$.} 
\label{TableMg} 
\end{table}

\begin{table}[htb] \centering
 \begin{tabular}{|C{2cm}|C{2.6cm}|C{2.6cm}||C{2.6cm}|C{2.6cm}|}
\hline        
    A       &    D1S gs   & D1S isomere & D1ST2a gs & D1ST2a isomere   \\
\hline 
   20       &   $ 0.00$   &             &  $ 0.00$  &                  \\
\hline 
   22       &   $+0.54$   &   $-0.22$   &  $+0.59$  &  $-0.28$         \\
\hline 
   24       &   $+0.61$   &   $-0.28$   &  $+0.61$  &  $-0.39$         \\
\hline 
   26       &   $-0.34$   &   $+0.50$   &  $+0.59$  &  $-0.39$         \\
\hline 
   28       &   $+0.46$   &   $ 0.00$   &  $+0.53$  &  $-0.31$         \\
\hline 
   30       &   $+0.15$   &   $-0.07$   &  $-0.26$  &  $+0.31$         \\
\hline 
   32       &   $ 0.00$   &   $+0.59$   &  $ 0.00$  &                  \\
\hline 
   34       &   $+0.61$   &   $-0.14$   &  $+0.51$  &  $-0.14$         \\
\hline 
   36       &   $+0.58$   &   $-0.27$   &  $+0.63$  &  $-0.27$         \\
\hline
   38       &   $+0.59$   &   $-0.34$   &  $+0.64$  &  $-0.41$         \\
\hline
   40       &   $+0.64$   &   $-0.46$   &  $+0.65$  &  $-0.48$         \\
\hline
\end{tabular}
\caption{Theoretical $Q_0$ (eb) for the $Z=12$ chain.  }  
\label{TableMg2} 
\end{table}

In Table~\ref{TableMg}  experimental spectroscopic quadrupole moments $Q_S$($I=2$) are presented for $^{24}$Mg and $^{26}$Mg. 
The methods of the  measurement are specified and the intrinsic quadrupole moment $Q_0$ is extracted using Eq.~(\ref{defQS})
for all the possible values of $K$. These latter values have to be compared to those in Table~\ref{TableMg2} in which all the 
ground states and first isomere theoretical $Q_0$ are presented from $A=20$ to $A=40$.
For both $^{24}$Mg and $^{26}$Mg the $2^+$ state from which is extracted the spectroscopic moment is the first
excited state of the yrast band. Results for $K=2$ and $K=1$ are proportional by a factor $2$ according to Eq.~(\ref{defQS}) and give an oblate deformation 
for both nuclei in each experiment. On the contrary the $K=0$ assumption changes the sign of $Q_0$ with respect to the $K=2$ expectation and gives prolate shapes. 
Following the empirical rule that we get $I=K$ for the band head we could exclude the value
$K=2$. In Table~\ref{TableMg2} we see that the D1S interaction predicts a prolate ground state for $A=24$ and an oblate one for $A=26$. On the other hand D1ST2a 
interaction  provides prolate ground states, it predicts the same shapes than the experimental results with $K=0$. 

\begin{figure}[htb] \centering
\begin{tabular}{ccc}
   \includegraphics[width=4.6cm]{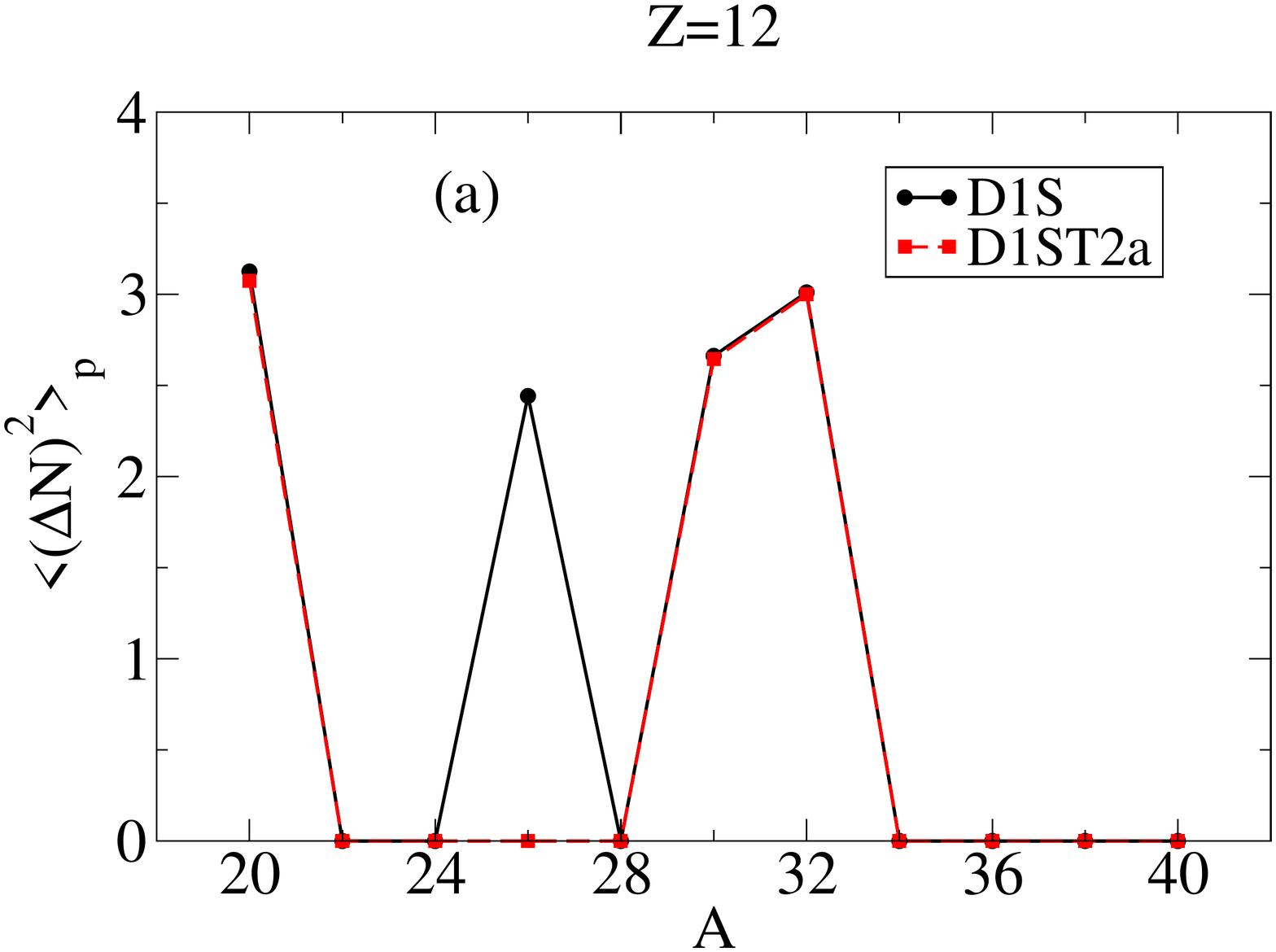}&
   \includegraphics[width=4.6cm]{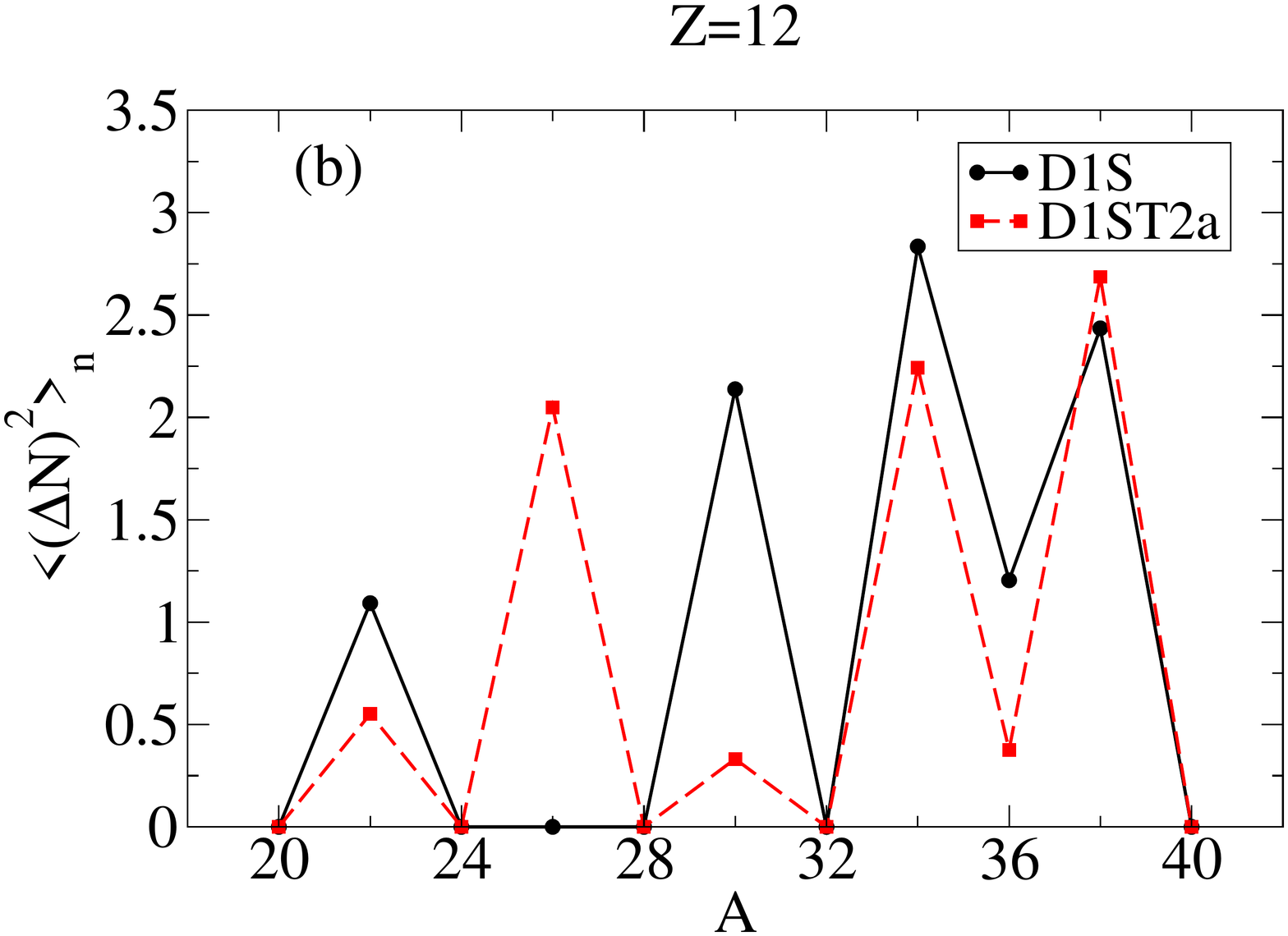}&
   \includegraphics[width=4.6cm]{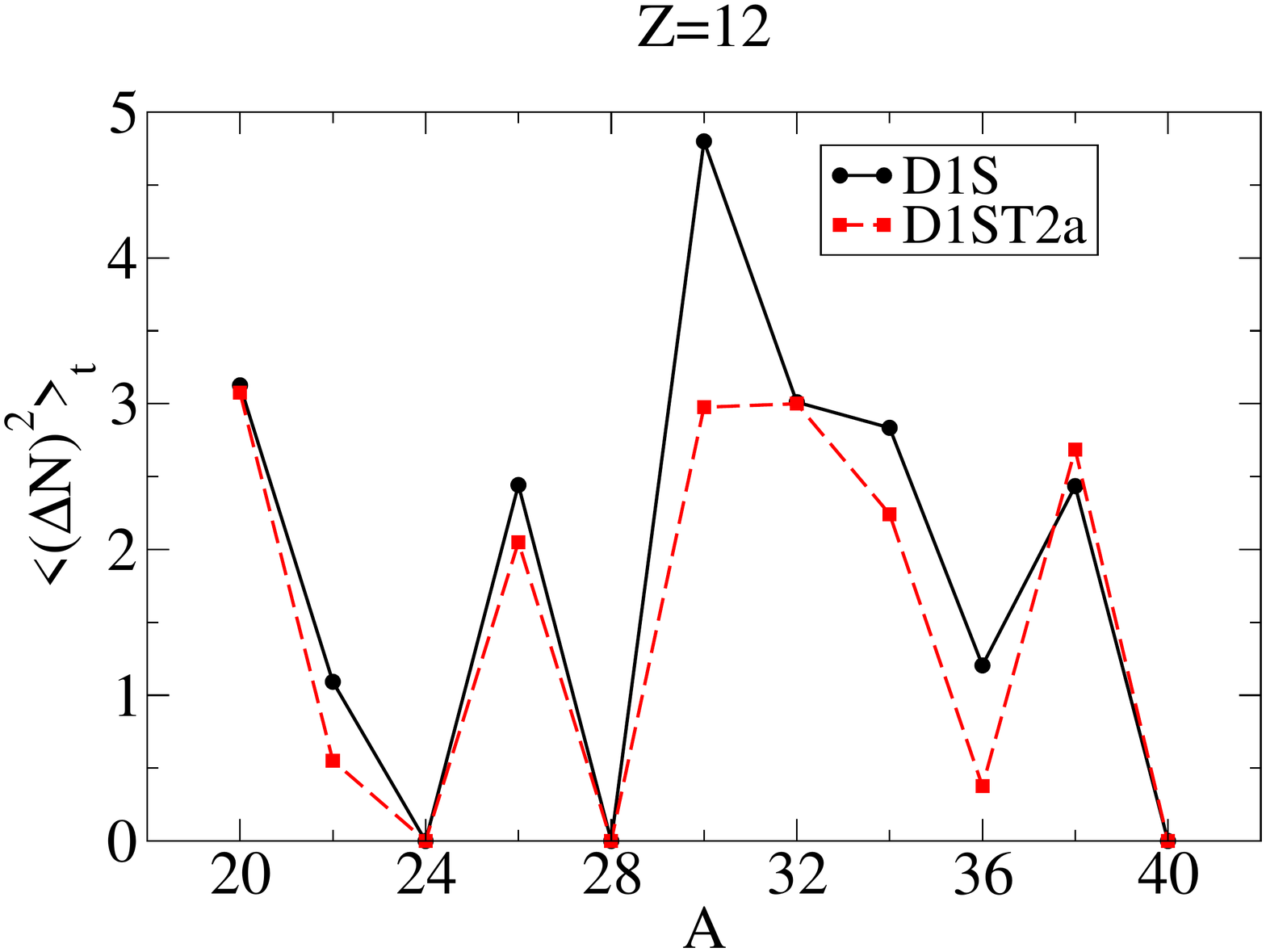} 
\end{tabular}
\caption{Proton, neutron and total particle number fluctuation for the $Z=12$ chain in the ground state. }
\label{MgDN2}
\end{figure}

In Fig.~\ref{MgDN2} we plot proton, neutron and total particle number fluctuations obtained for the ground state in the chain $Z=12$, using the D1S interaction 
(solid line) and the D1ST2a one (red dashed line).  In this case, the total fluctuation is very similar for both interactions,  with the only exception of 
$^{30}$Mg. For $^{26}$Mg the total particle number fluctuations are equal for both interactions, whereas the 
ground state locations are inverted. Here the none zero neutron contribution for the D1S interaction is compensated by the proton contribution of the 
D1ST2a one. For nuclei with non zero pairing correlations, where no inversion occurs, the D1ST2a interaction lowers the particle number fluctuation.
Taking into account that deformation does not change for the ground state, using D1S or D1ST2a interaction (except for
$^{26}$Mg and $^{30}$Mg), it seems to exist a relation between particle fluctuation and the intrinsic  
quadrupole moment (i.e. deformation) of the nuclear state: if particle fluctuations are similar with both interactions, deformation does not change.

\subsubsection{\texorpdfstring{$^{26}$}{26}Mg and \texorpdfstring{$^{30}$}{30}Mg}

\begin{figure}[htb] \centering
\begin{tabular}{ccc}
   \includegraphics[width=4.6cm]{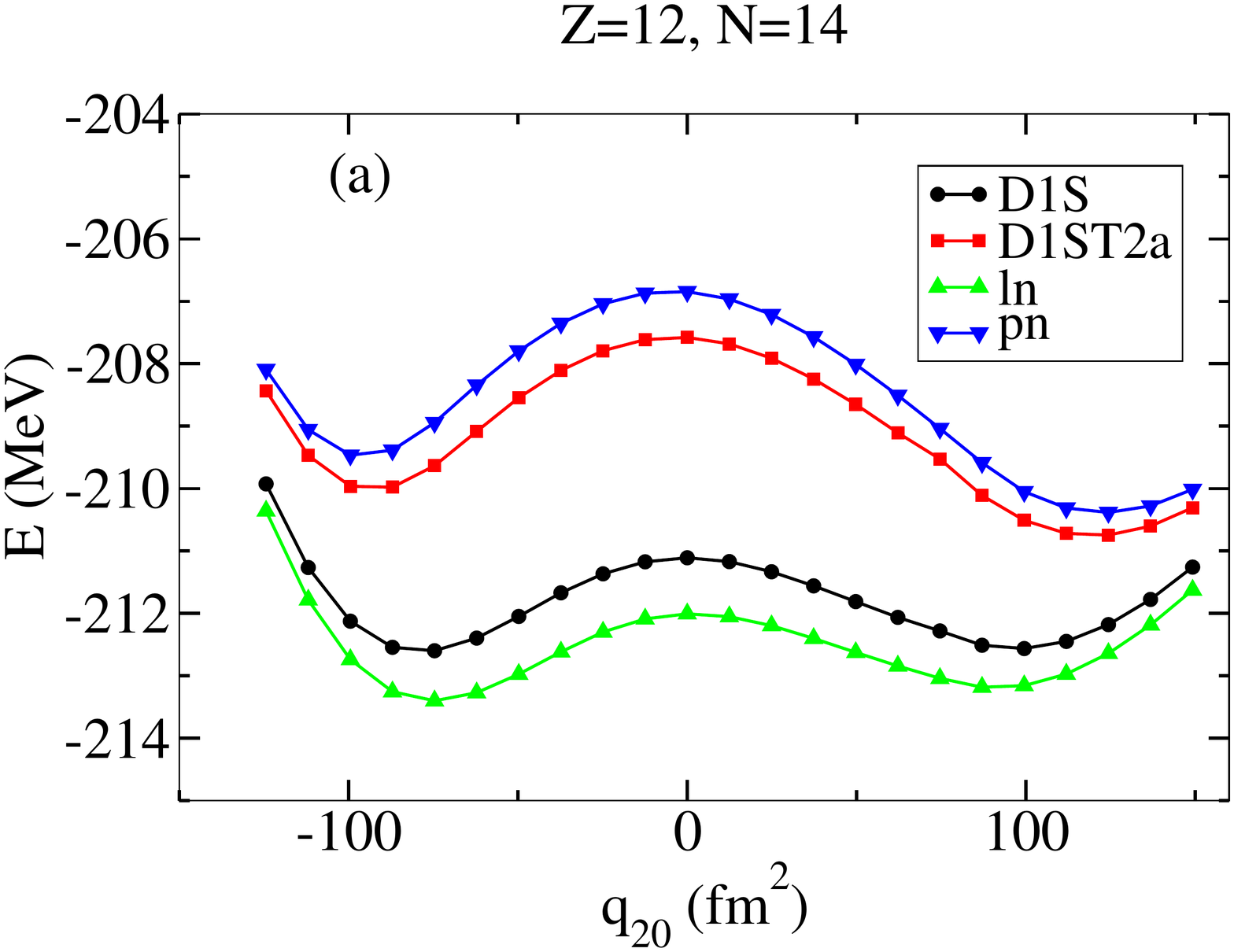} & 
   \includegraphics[width=4.6cm]{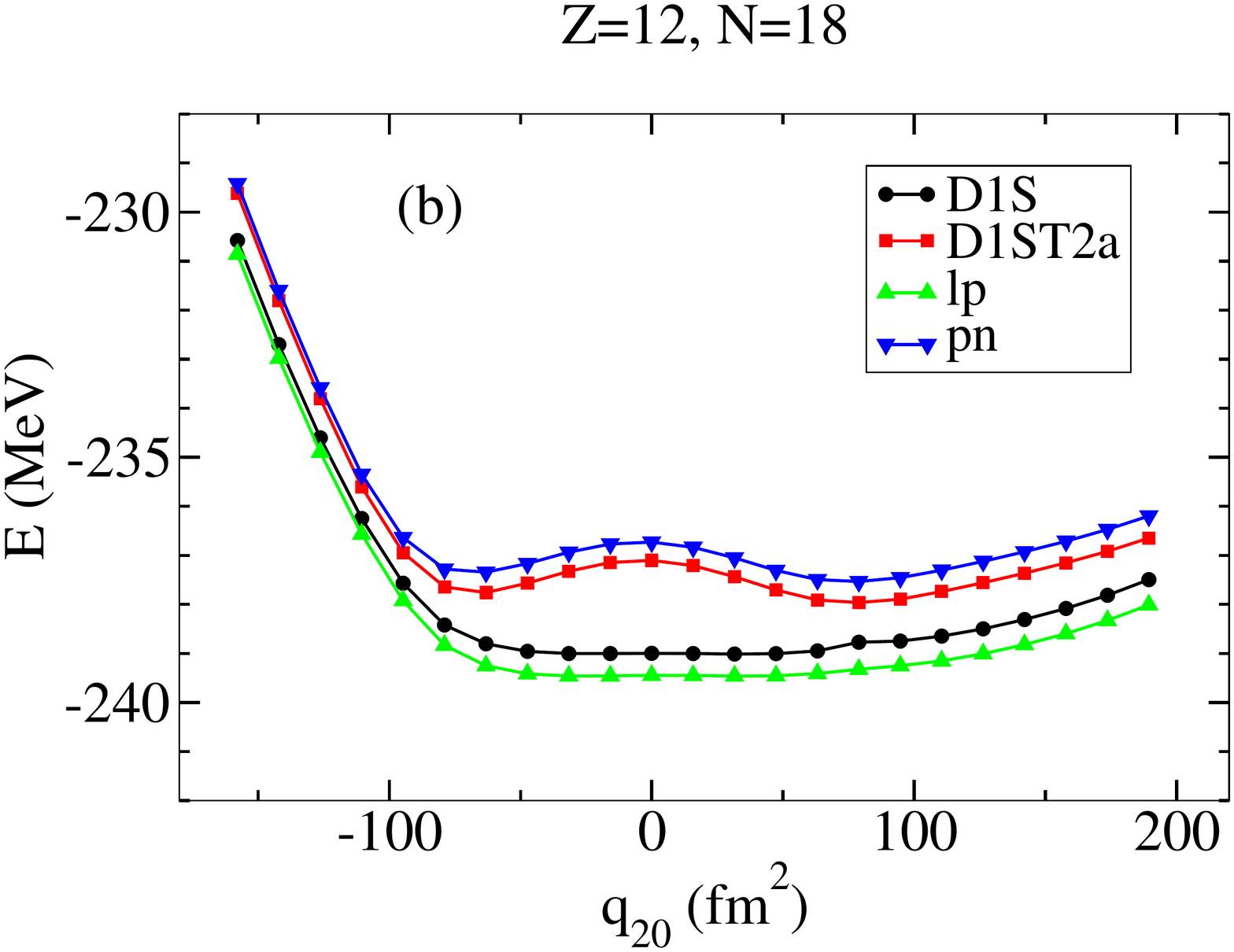} 
\end{tabular}
\caption{Like-particle and proton-neutron  contributions of the D1ST2a interaction for $^{26}$Mg and $^{30}$Mg.} 
\label{Mglnpnp}
\end{figure}

We study here in more detail two isotopes of the $Z=12$ chain. In Fig.~\ref{Mglnpnp} we show the PEC for $^{26}$Mg and $^{30}$Mg,  
using the D1S interaction (black  circles) and the D1ST2a one (red  squares).  We show too the results considering only the like-particle  part of the 
tensor interaction (green triangles up) or the proton-neutron part (blue triangles down). 
This last curve and the corresponding to the D1ST2 interaction are very close for all the nuclei, showing that
the main tensor contribution is that coming from the pn part. Then, we obtain the same conclusions than for the 
Zirconium chain: the pn part dominates the tensor term. Here both pn and lp contributions
are almost deformation independent. For $^{30}$Mg 
the D1S and lp PEC are very flat, all the states from $q_{20}=-60$ fm$^2$ to $q_{20}=+60$ fm$^2$ are almost degenerated 
and the location of the lp ground state differs ($q_{20}=+16$ fm$^2$) from the D1S one ($q_{20}=+60$  fm$^2$). Two distinct 
minima appears when adding the tensor around $q_{20}=-70$ fm$^2$ and $q_{20}=+63$ fm$^2$ for both D1ST2a and the pn part.

\begin{figure}[htb] \centering
\begin{tabular}{cc}
   \includegraphics[width=4.6cm]{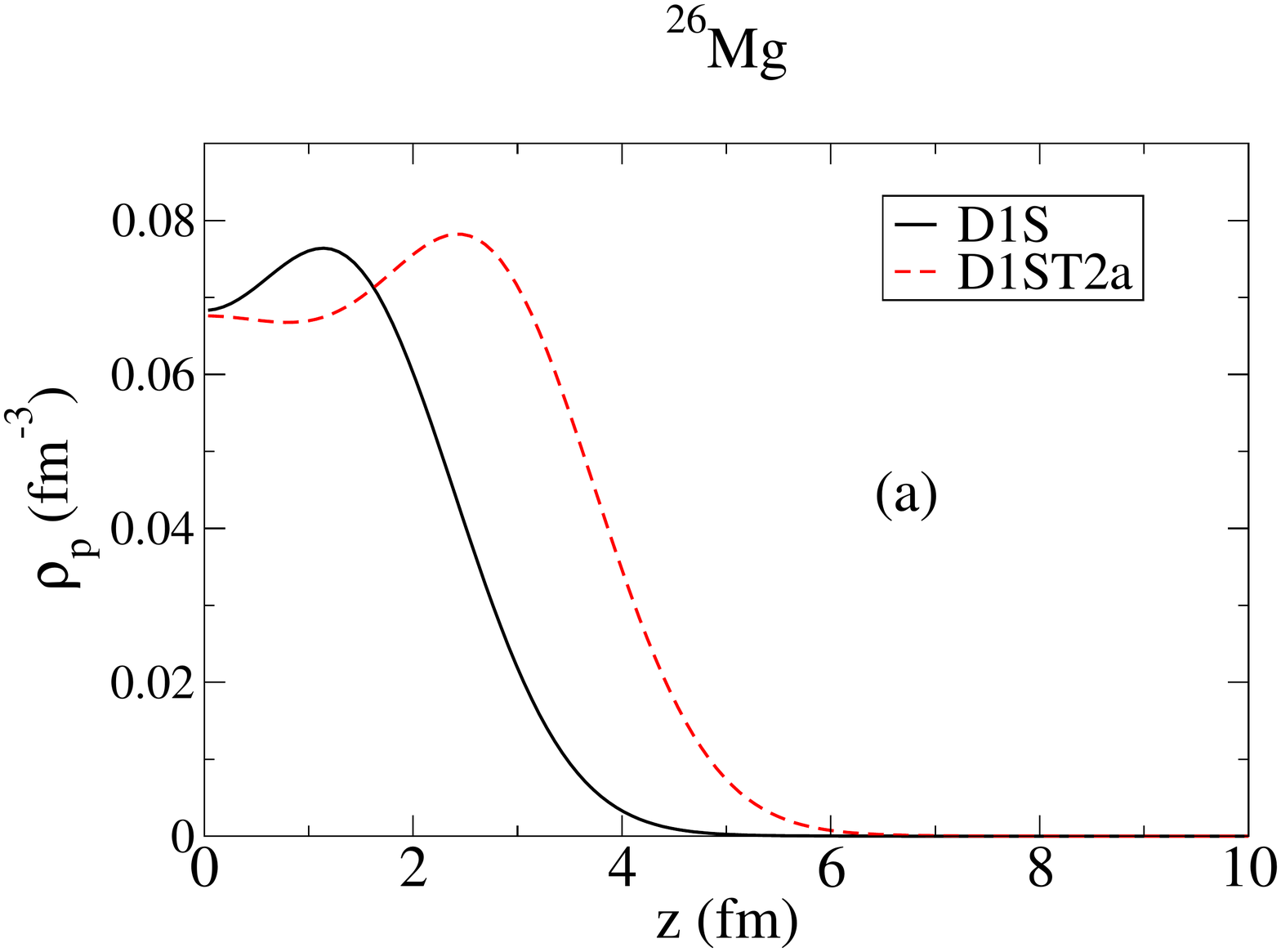} & 
   \includegraphics[width=4.6cm]{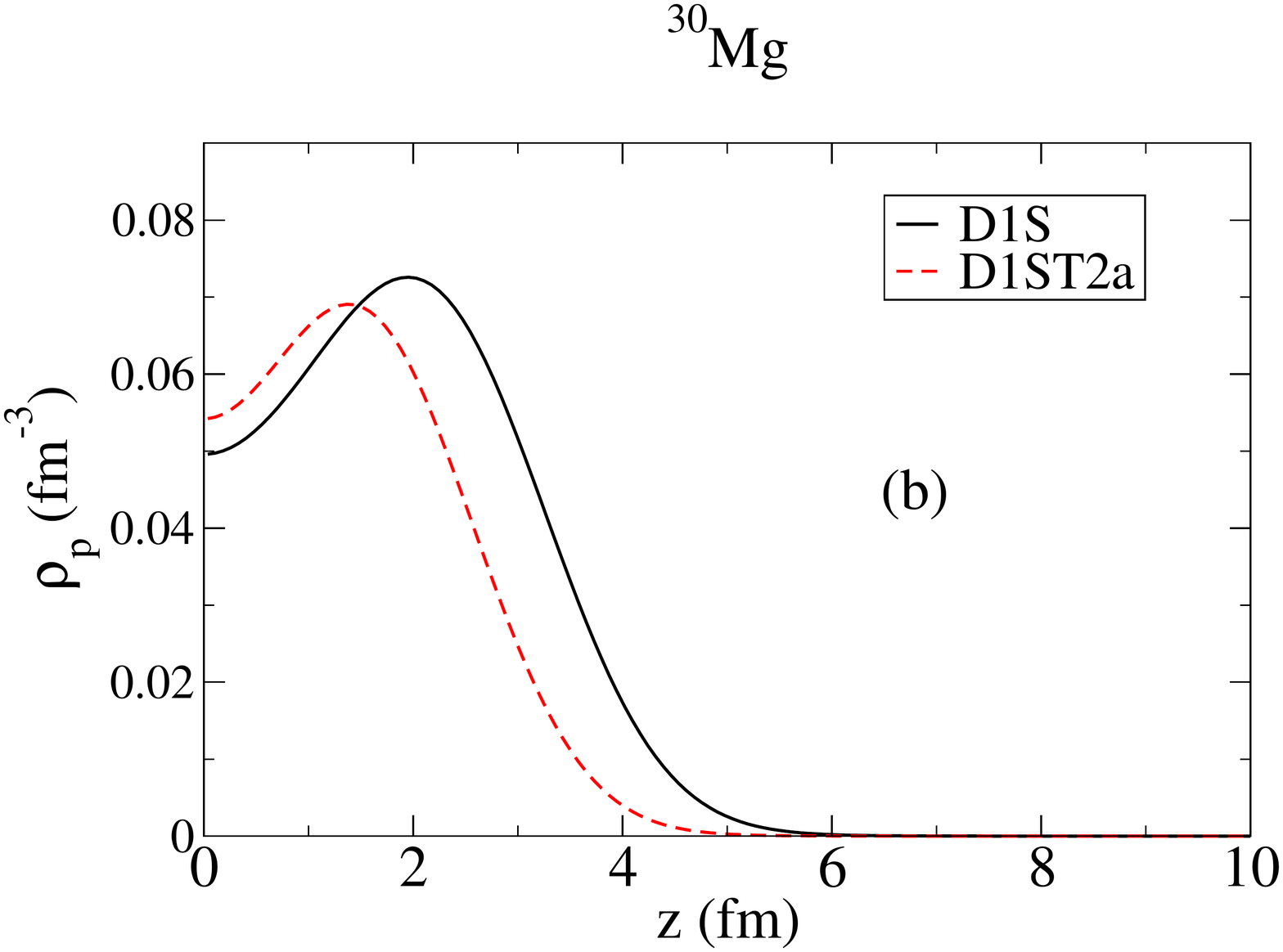} \\
   \includegraphics[width=4.6cm]{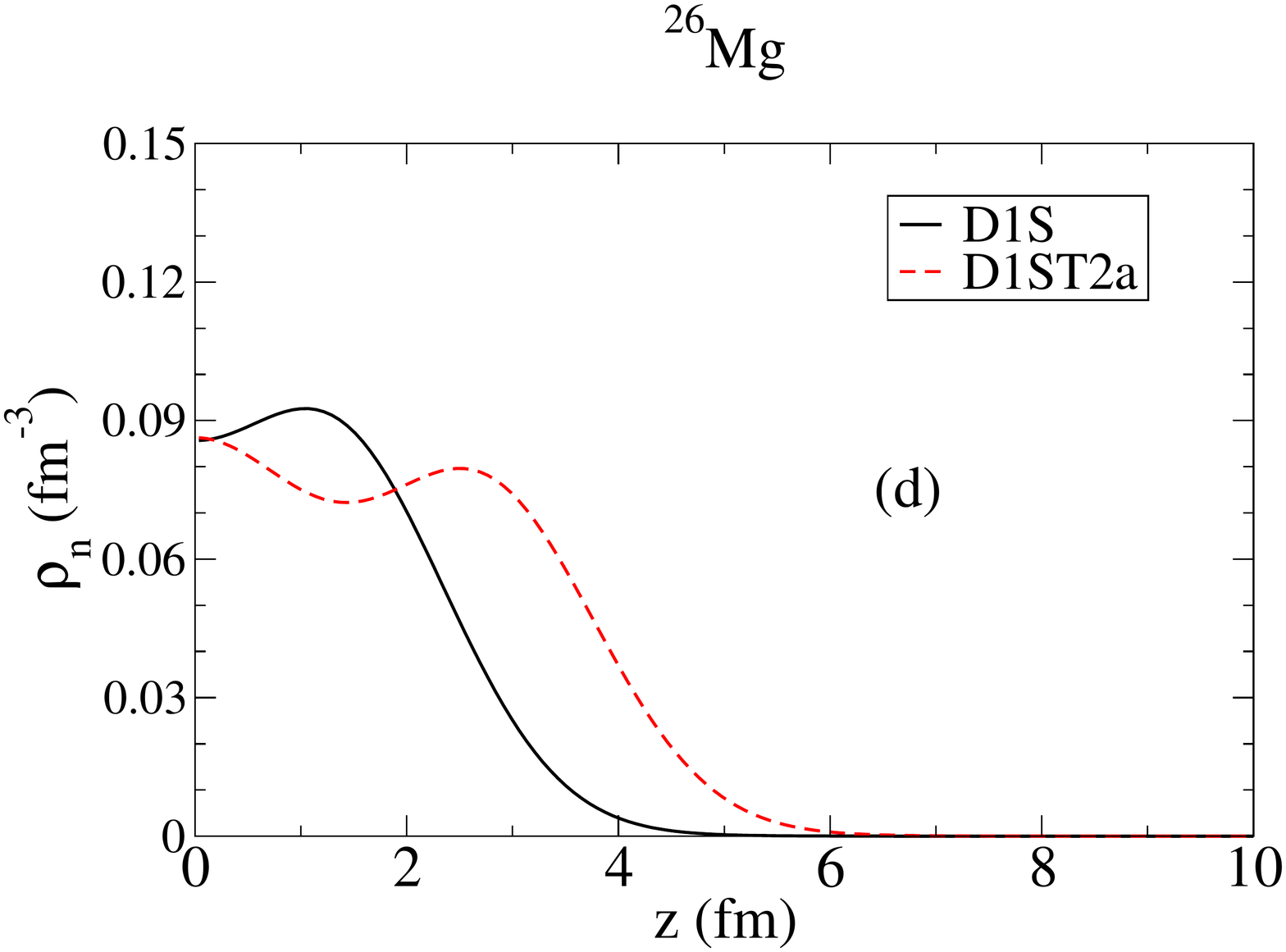} & 
   \includegraphics[width=4.6cm]{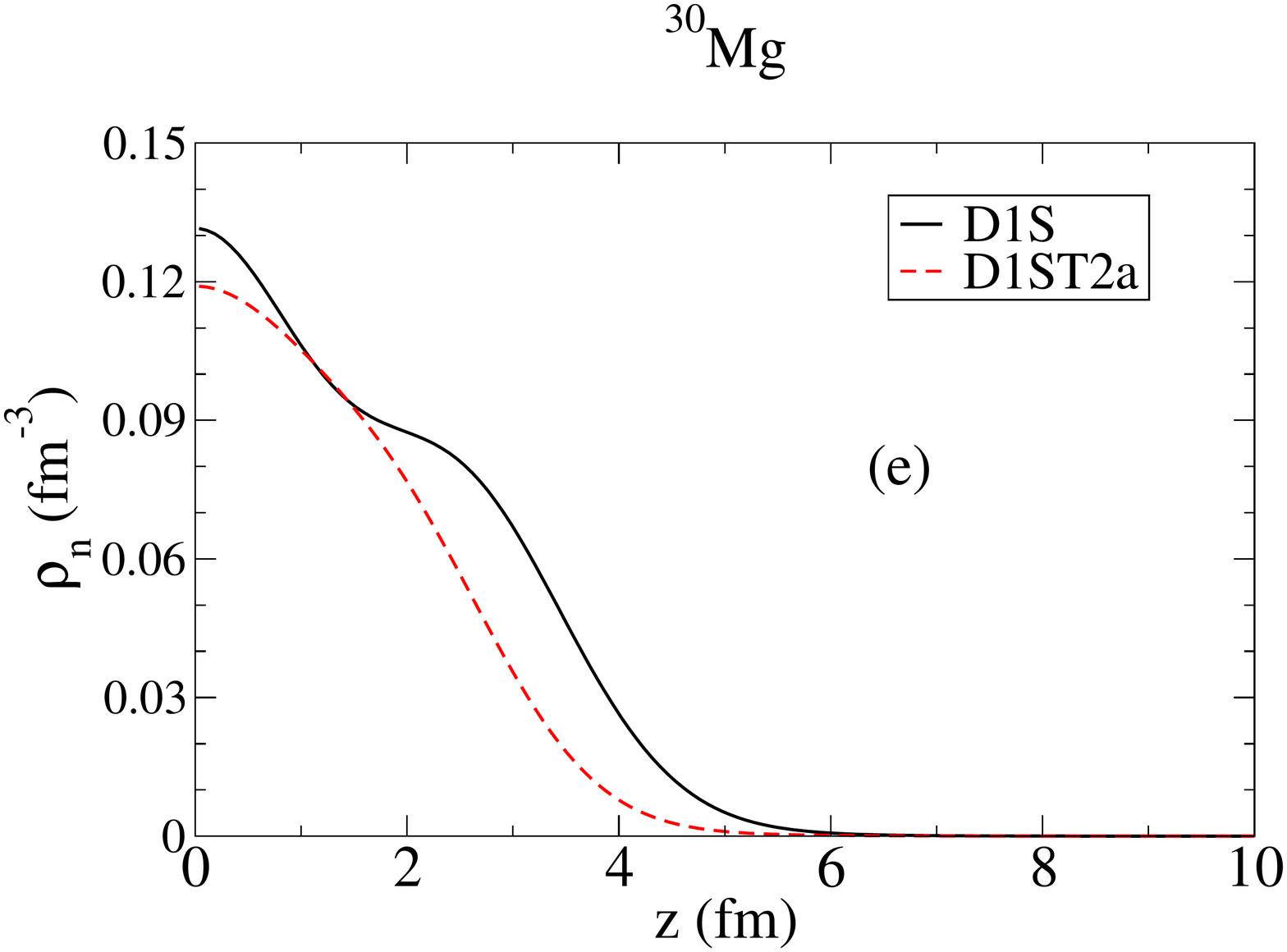} \\
\end{tabular}
\caption{Proton and neutron densities for $^{26}$Mg and $^{30}$Mg for D1S and D1ST2a interactions.} 
\label{MgDenspn}
\end{figure}

We observe in Fig.~\ref{MgDenspn} that the proton and neutron densities are modified when using  the D1ST2a interaction with respect to the result for the D1S one in 
the case of nuclei $^{26}$Mg  and  $^{30}$Mg because the ground state obtained for each interaction has different deformation characteristics. On the contrary for all 
the other Magnesium isotopes, where the D1ST2a interaction does not invert the ground state location, the densities are scarcely modified.

\clearpage

\subsection{Silicon chain}

\begin{figure}[htb] \centering
\begin{tabular}{ccc}
   \includegraphics[width=4.6cm]{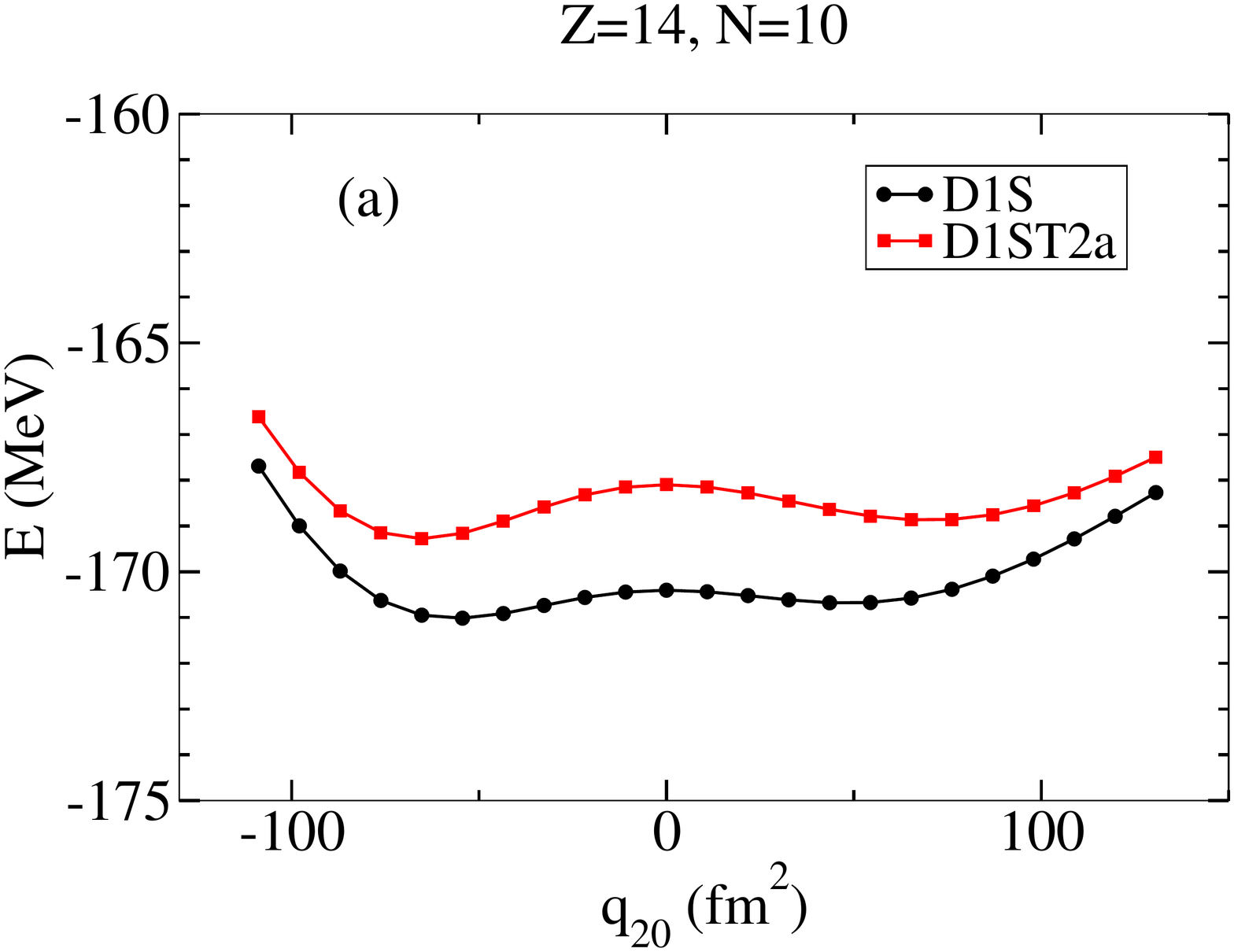} &
   \includegraphics[width=4.6cm]{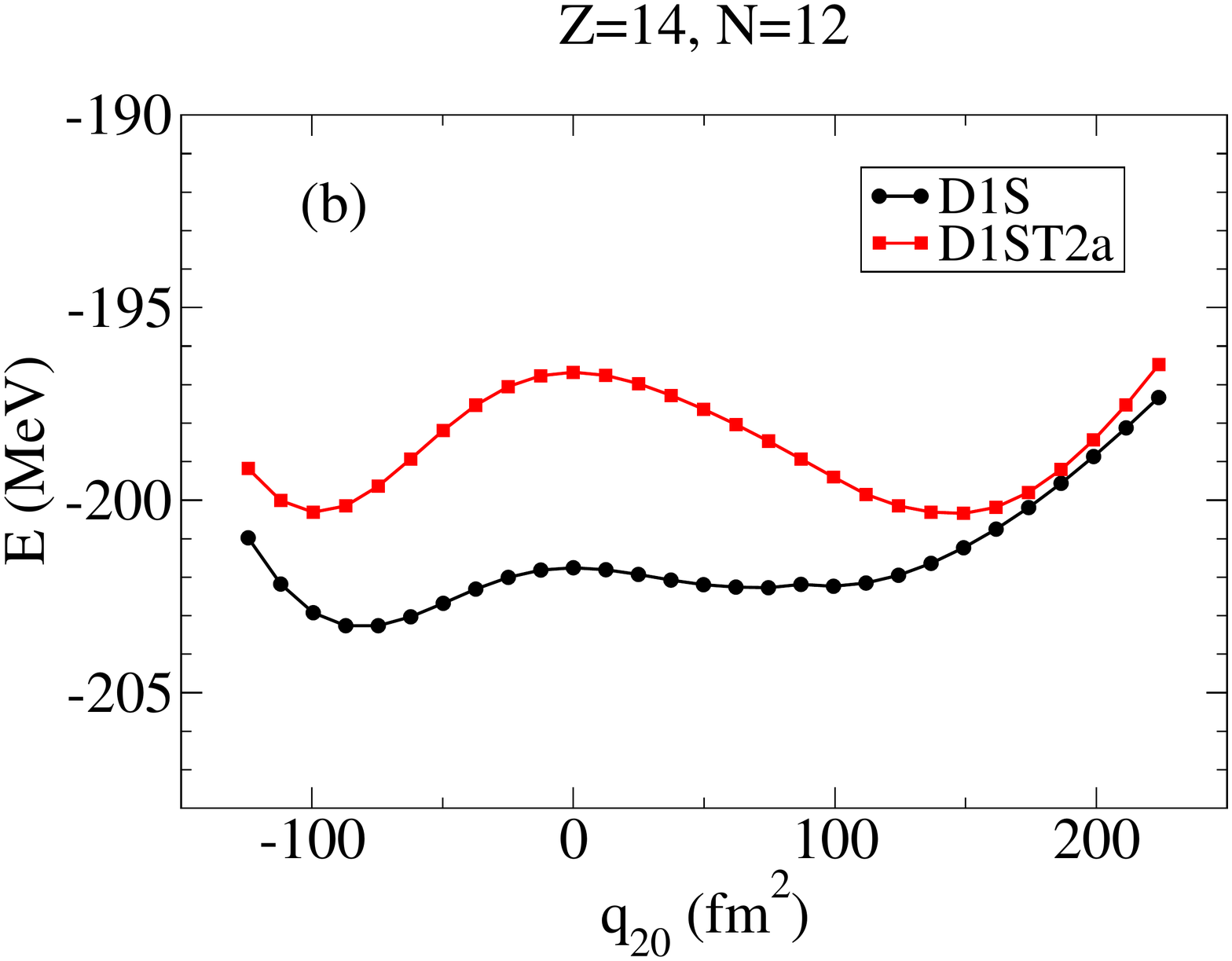} &
   \includegraphics[width=4.6cm]{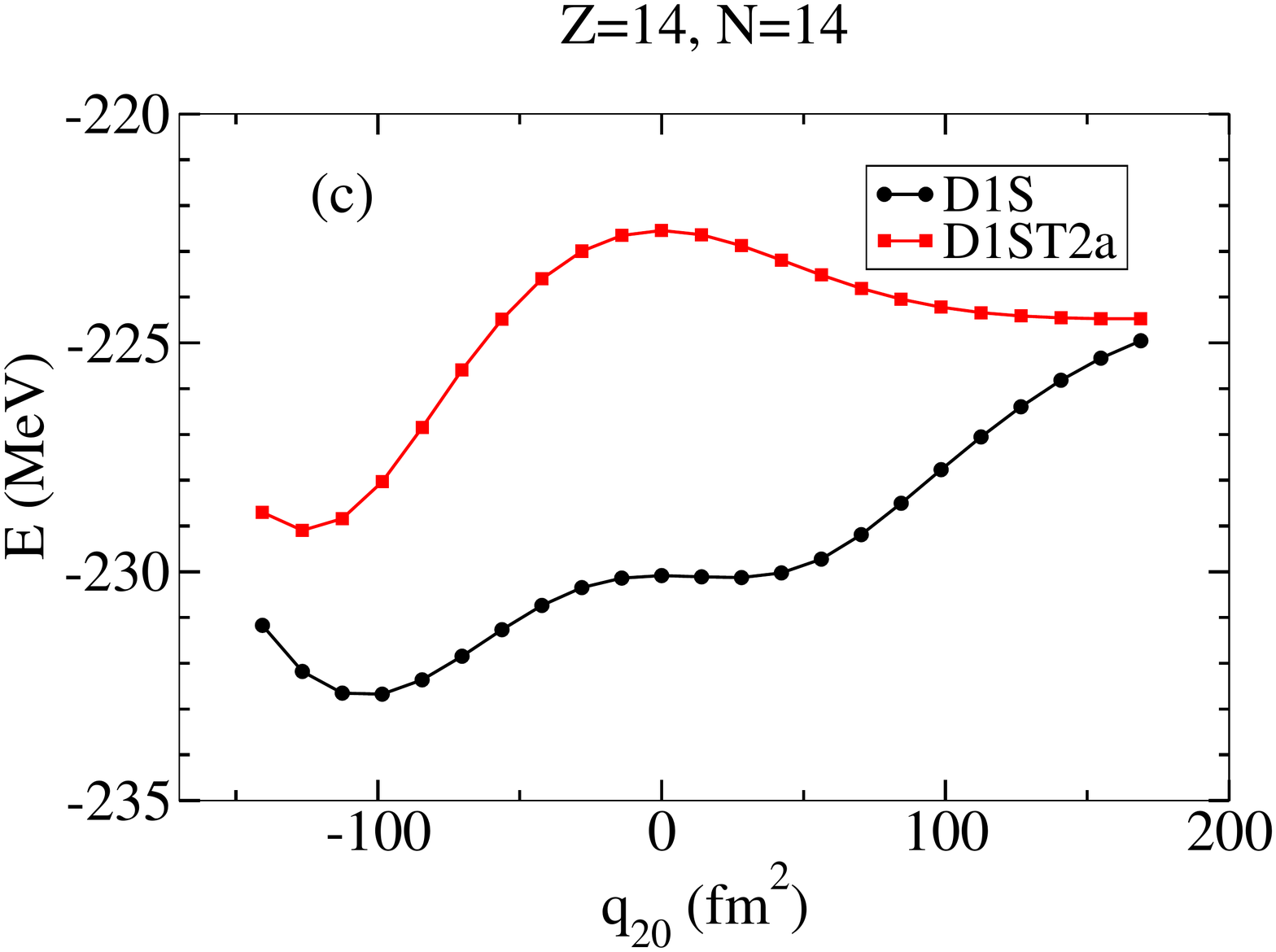} \\
   \includegraphics[width=4.6cm]{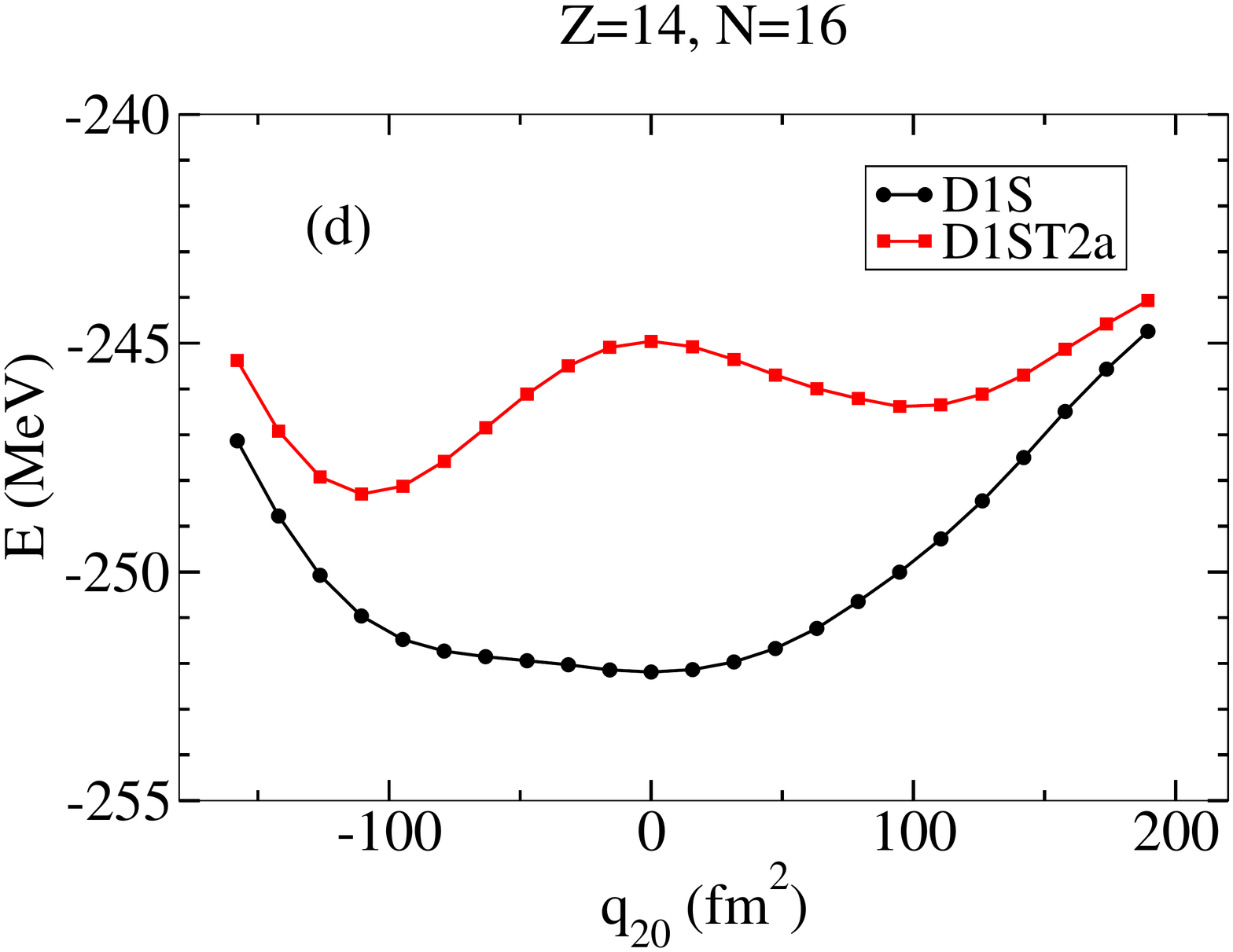} &
   \includegraphics[width=4.6cm]{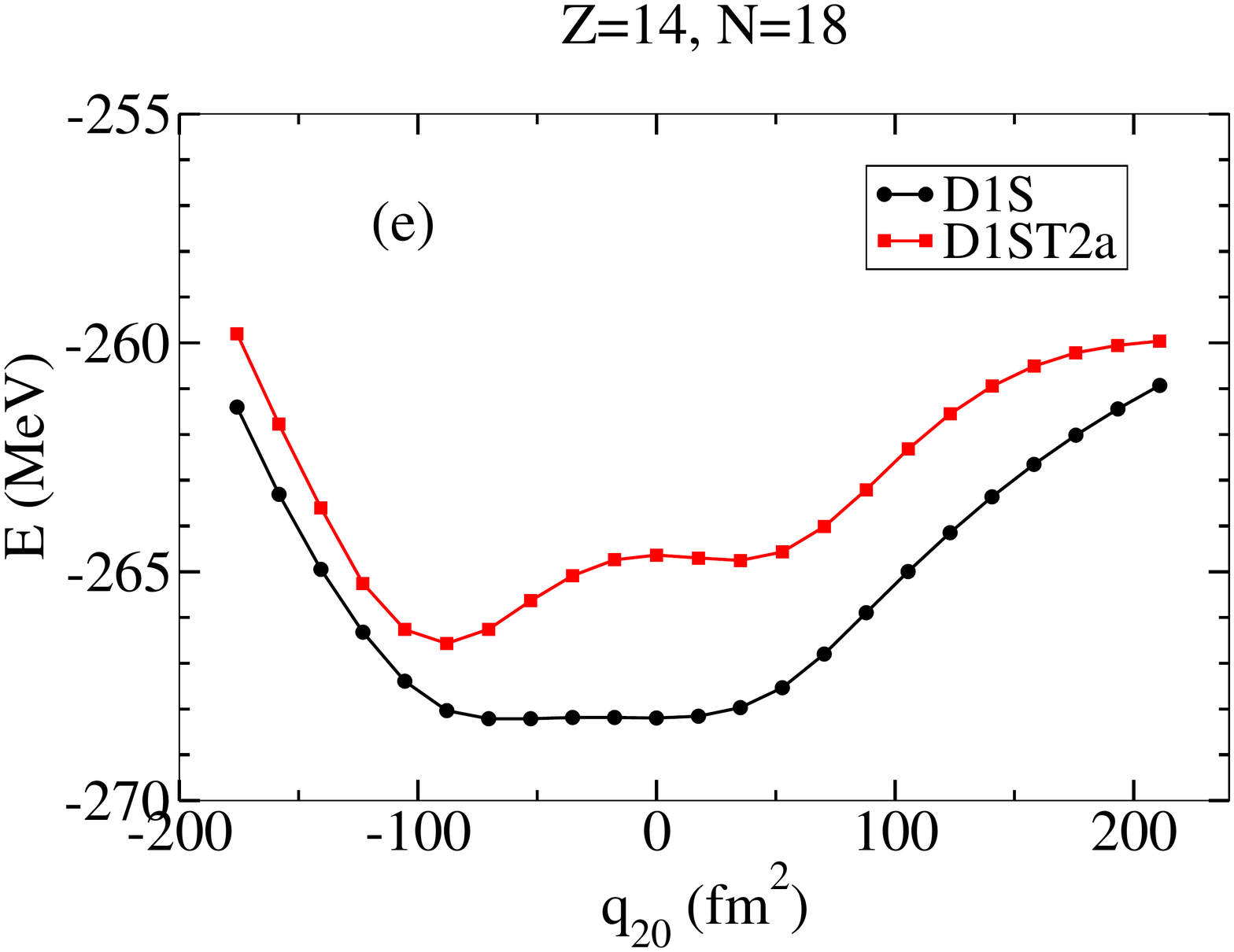} &
   \includegraphics[width=4.6cm]{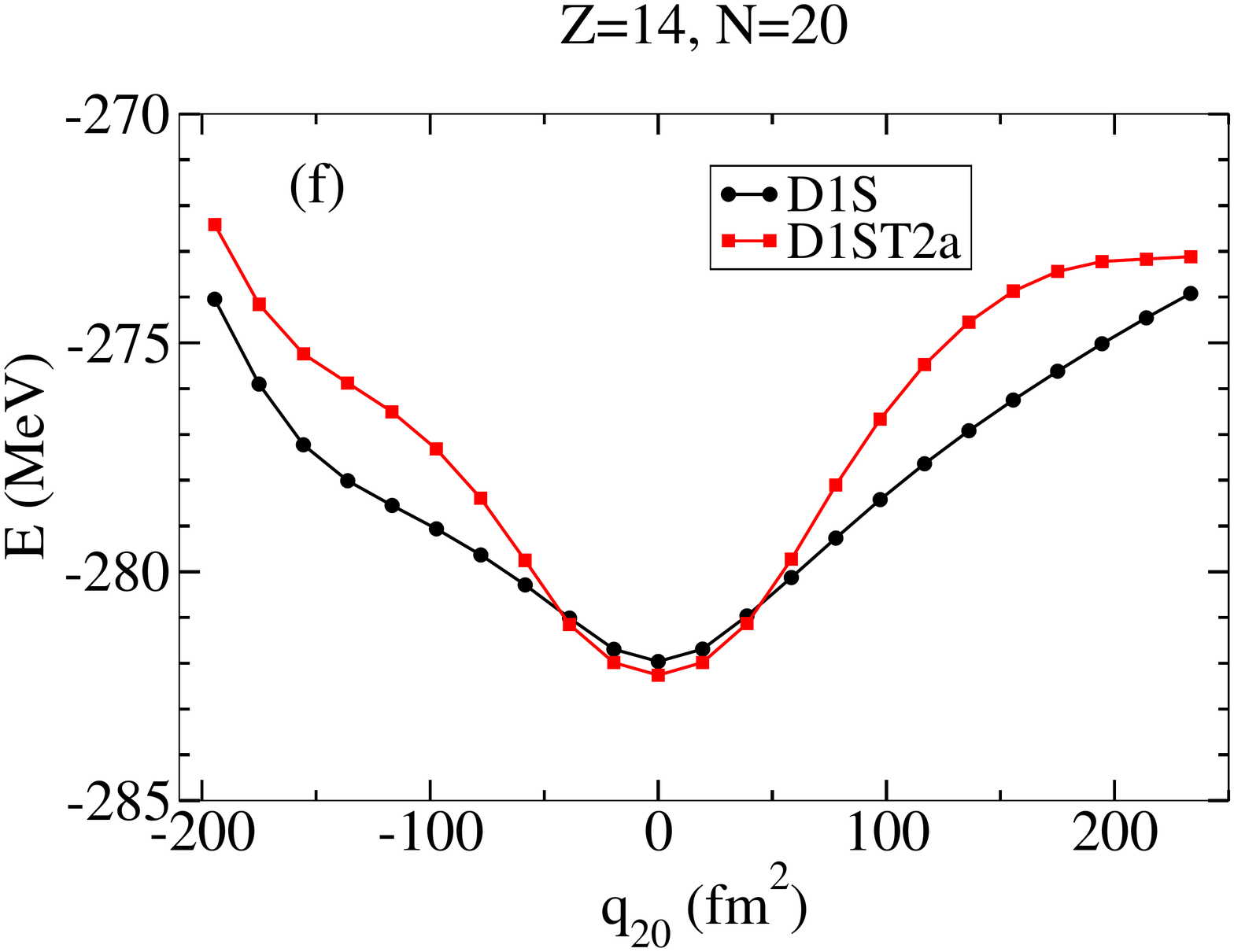} \\
   \includegraphics[width=4.6cm]{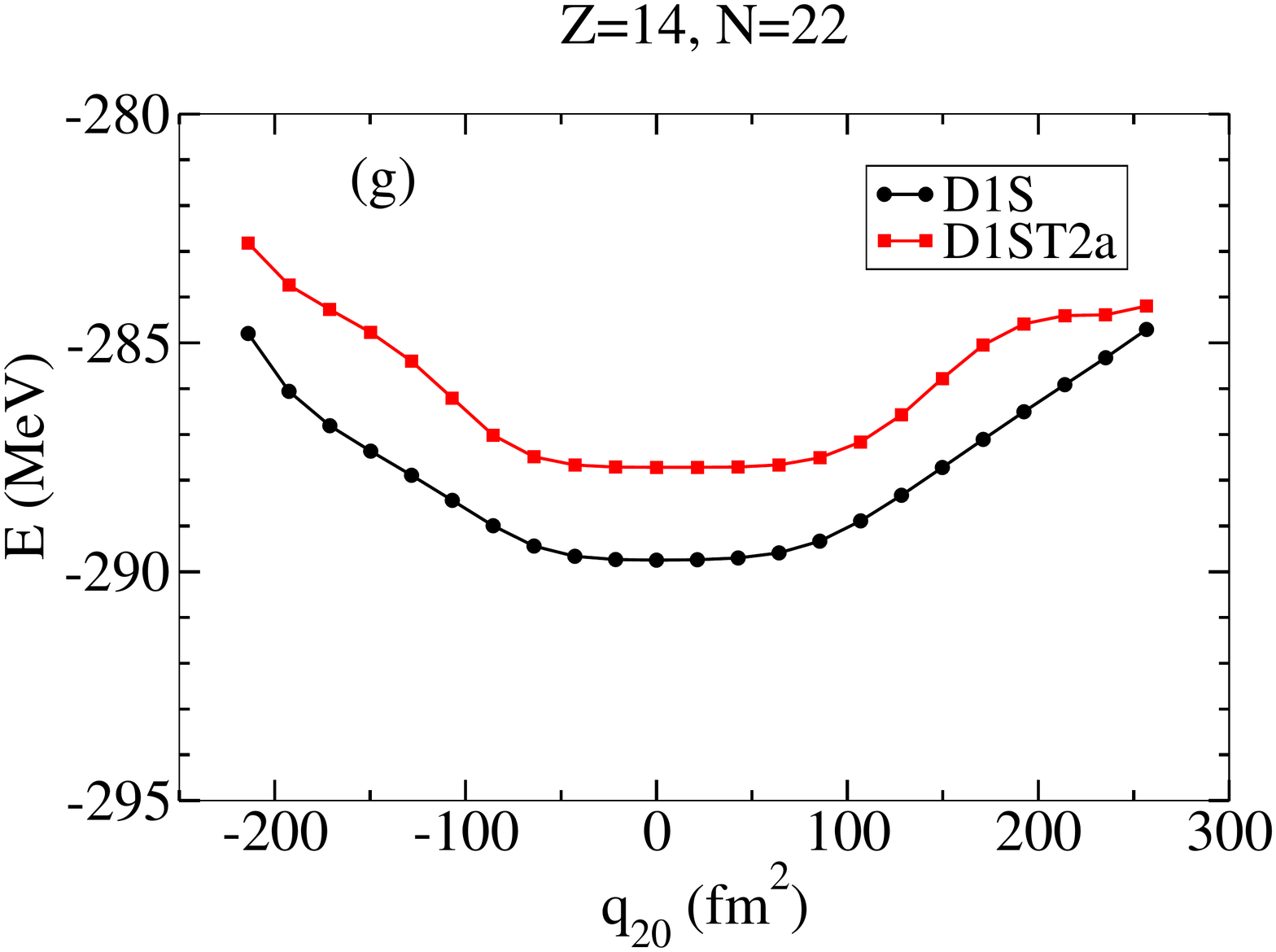} &
   \includegraphics[width=4.6cm]{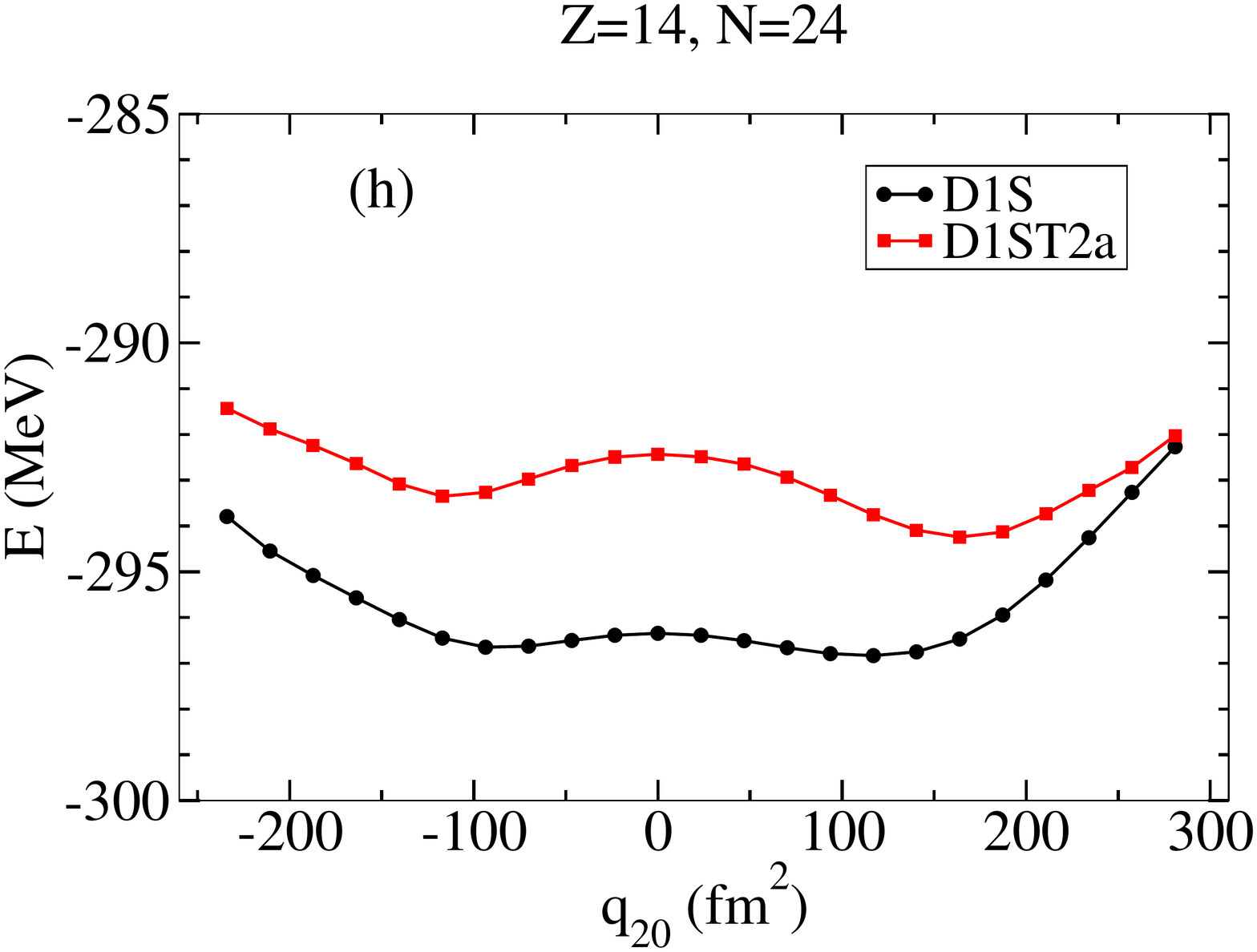} &
   \includegraphics[width=4.6cm]{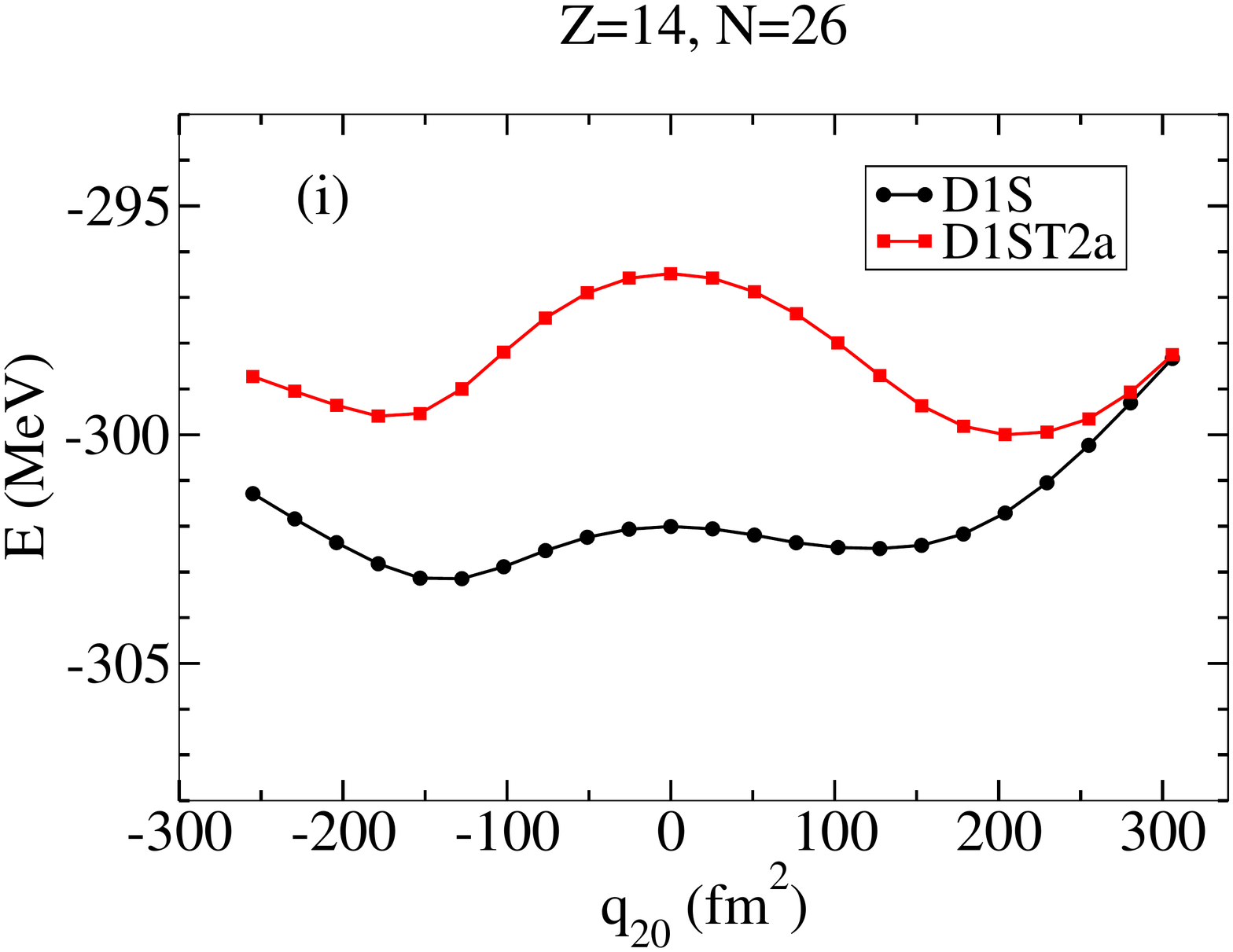} \\
   \includegraphics[width=4.6cm]{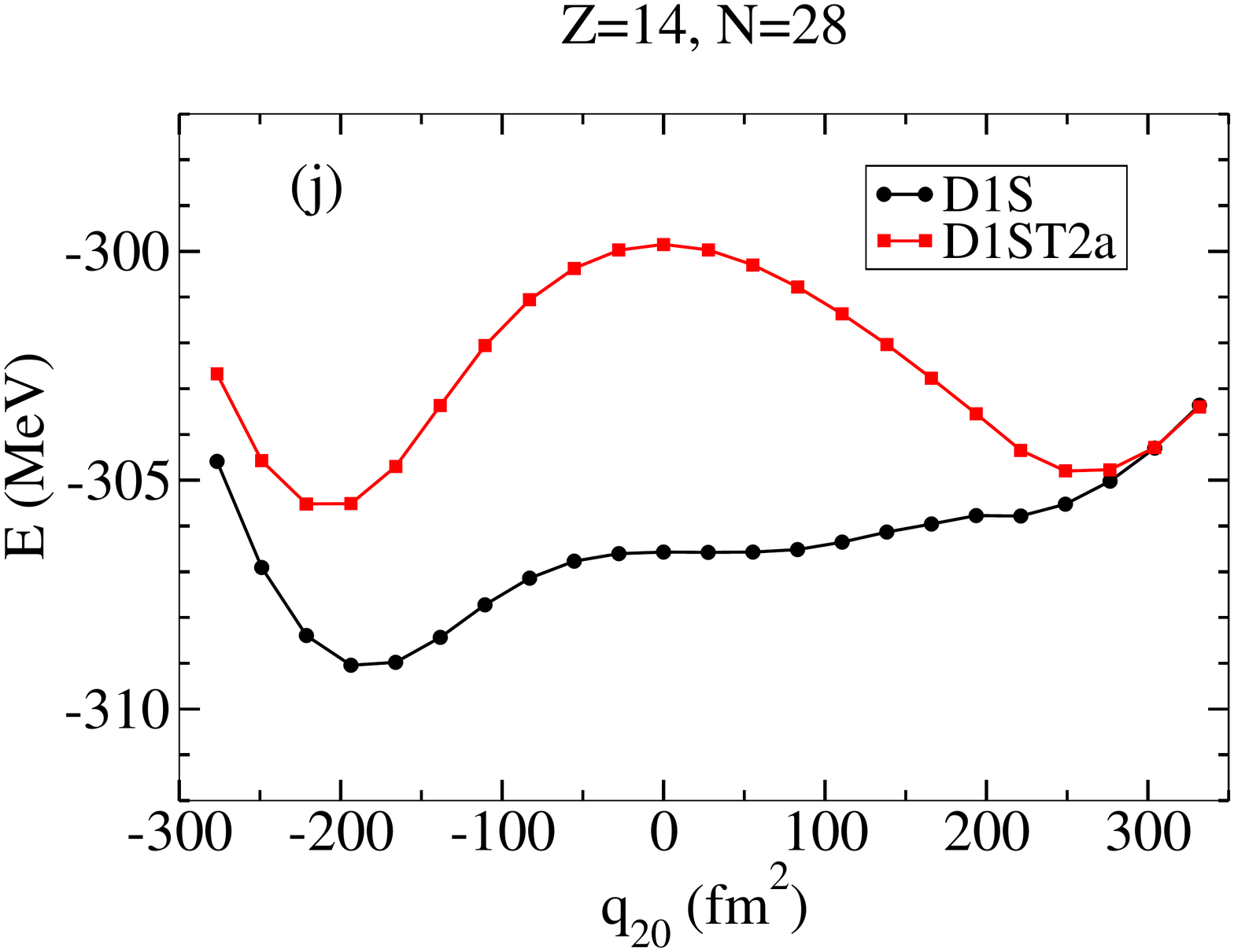} &
   \includegraphics[width=4.6cm]{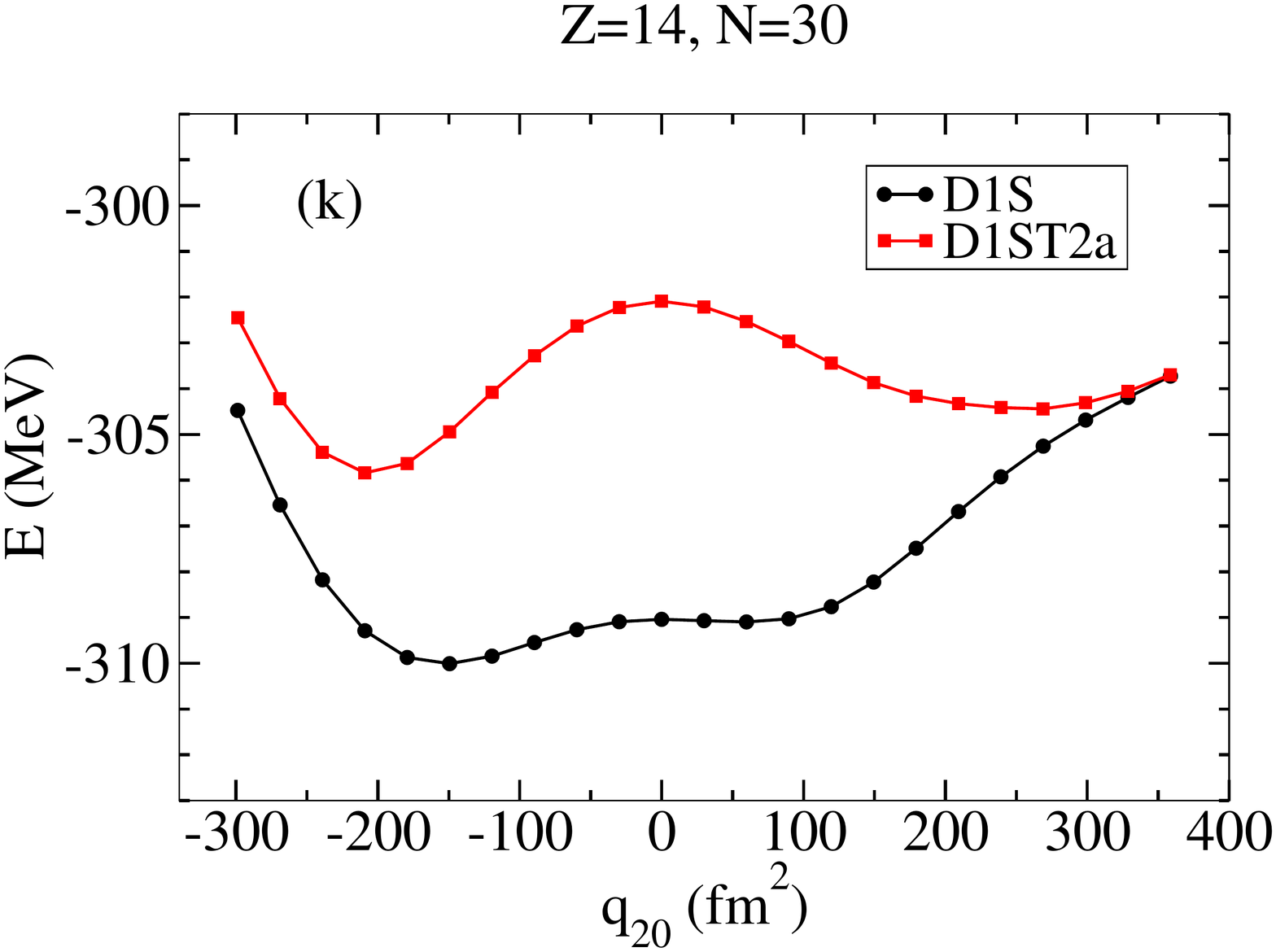} &
    \\
\end{tabular}
\caption{Potential Energy Curves  of the $Z=14$ chain 
for the D1S interaction (circles) and the D1ST2a one (squares). } 
\label{PESSi}
\end{figure}

In Fig.~\ref{PESSi} are depicted the Silicon PEC from $A=24$ to $A=44$. The intrinsic quadrupole moment
of their two first minima are presented in Table~\ref{TableSi2}.
The tensor interaction is always repulsive except for $A=34$ around $q_{20}=0$ fm$^2$ and at large
deformation for the heaviest isotopes $A\ge40$. 
For $^{34}$Si, which is neutron spin-saturated, the tensor contribution is slightly attractive and is mainly due to
its $p-p$ part ($E_{\rm TS}(p-p)=-0.792$ MeV, $E_{\rm TS}({\rm tot})=-0.675$ MeV).
In a general way the D1S and D1ST2a minima may differ along the chain. 
All the Silicon isotopes are spherical or oblate with D1S except $^{38}$Si. 
Among the eleven isotopes the D1ST2a interaction changes the nucleus shape for two of them: $^{30}$S  and $^{40}$Si. 
They are globally more deformed with the D1ST2a interaction.
For $^{30}$Si the D1S interaction gives a flat well around sphericity and the D1ST2a one predicts a well located oblate ground state ($Q_0=-0.54$ eb).
For $^{40}$Si the prolate and oblate minima are reversed comparing the two interactions and the D1ST2a interaction predicts more deformed minima.

\begin{table}[htb] \centering
 \begin{tabular}{|C{2cm}|C{2.6cm}|C{2.6cm}||C{2.6cm}|C{2.6cm}|}
\hline        
    A       &    D1S gs   & D1S isomere & D1ST2a gs & D1ST2a isomere   \\
\hline 
   24       &   $-0.35$   &   $+0.24$   &  $-0.42$  &  $+0.36$         \\
\hline 
   26       &   $-0.42$   &   $+0.37$   &  $-0.56$  &                  \\
\hline 
   28       &   $-0.50$   &   $+0.14$   &  $-0.64$  &                  \\
\hline 
   30       &   $ 0.00$   &             &  $-0.54$  &  $+0.45$         \\
\hline 
   32       &   $-0.32$   &   $ 0.00$   &  $-0.40$  &  $+0.16$         \\
\hline 
   34       &   $ 0.00$   &             &  $ 0.00$  &                  \\
\hline 
   36       &   $ 0.00$   &             &  $ 0.00$  &                  \\
\hline
   38       &   $+0.37$   &   $-0.35$   &  $+0.57$  &  $-0.44$         \\
\hline
   40       &   $-0.44$   &   $+0.39$   &  $+0.72$  &  $-0.61$         \\
\hline
   42       &   $-0.61$   &   $+0.08$   &  $-0.71$  &  $+0.83$         \\
\hline
   44       &   $-0.45$   &   $+0.16$   &  $-0.65$  &  $+0.87$         \\
\hline
\end{tabular}
\caption{Theoretical $Q_0$ (eb) for the $Z=14$ chain.  }  
\label{TableSi2} 
\end{table}

\clearpage

\begin{figure}[htb] \centering
\begin{tabular}{cc}
   \includegraphics[width=4.6cm]{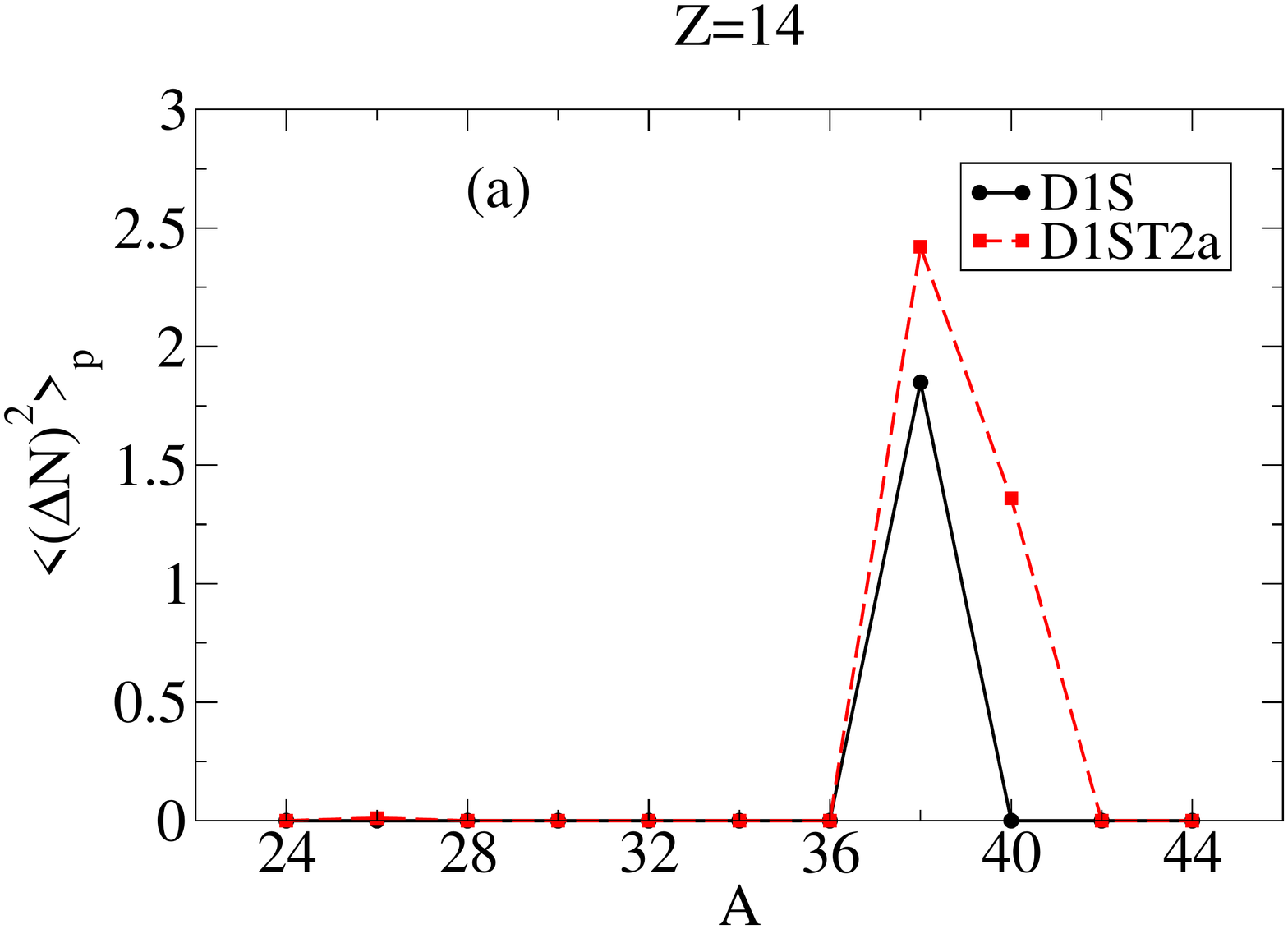} 
   \includegraphics[width=4.6cm]{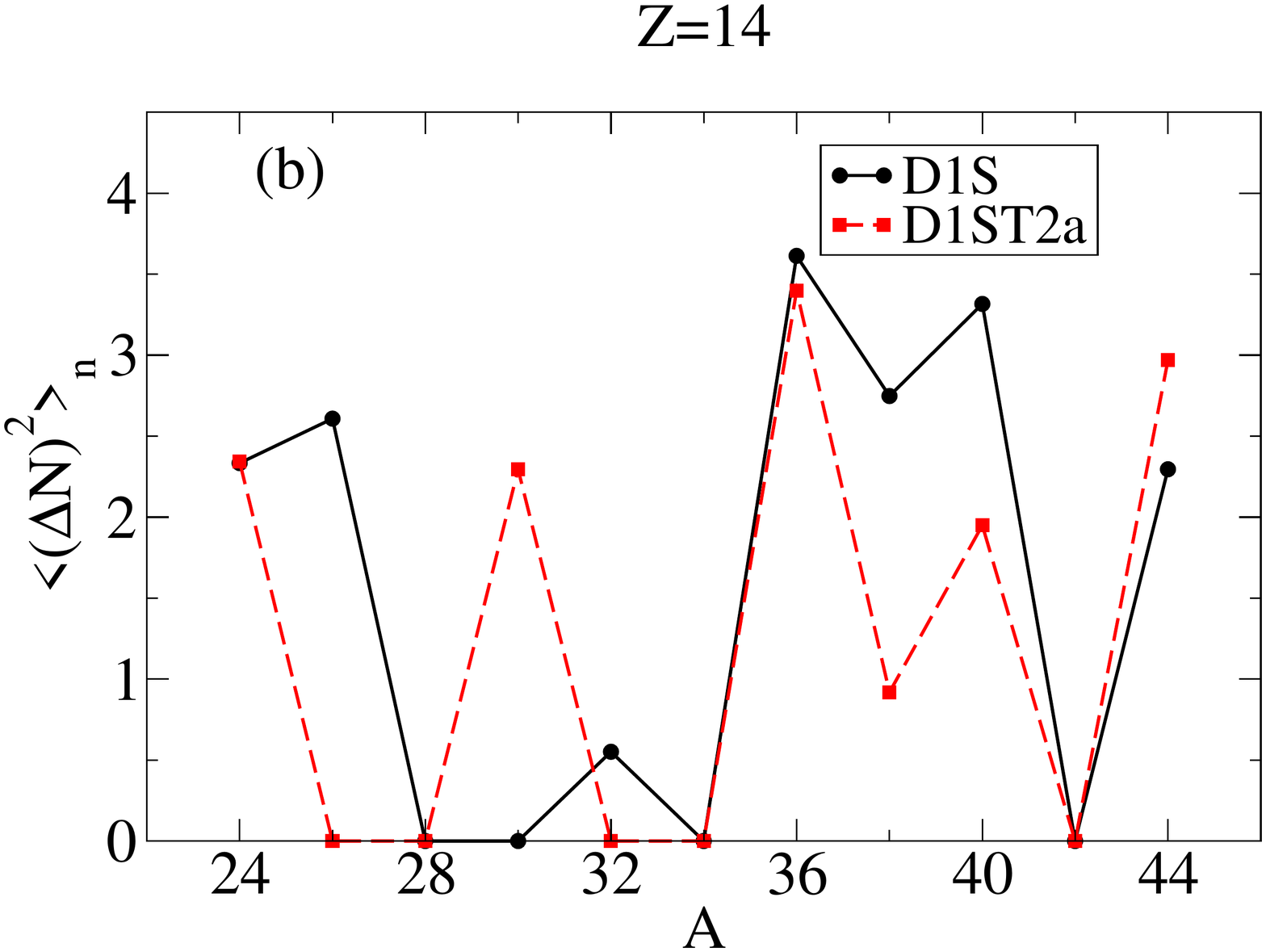}
\end{tabular}
\caption{Proton and neutron particle number fluctuation for the $Z=14$ chain.} 
\label{SiDN2}
\end{figure}

The proton pairing is zero for the spherical and oblate ground state for both interactions, as we can see in Fig.~\ref{SiDN2}.
Non zero proton number fluctuation values are associated to prolate shapes.
In the neutron side the D1ST2a curve only roughly get the same trend than the D1S one since the D1S and D1ST2a potential energy landscapes may differ a lot. 
As for the Mg chain, when both interactions give the same deformation the particle number fluctuation is most of the time smaller with the D1ST2a one.

\subsubsection{\texorpdfstring{${}^{30}$}{30}Si and \texorpdfstring{${}^{40}$}{40}Si }

\begin{figure}[htb] \centering
\begin{tabular}{cc}
   \includegraphics[width=4.6cm]{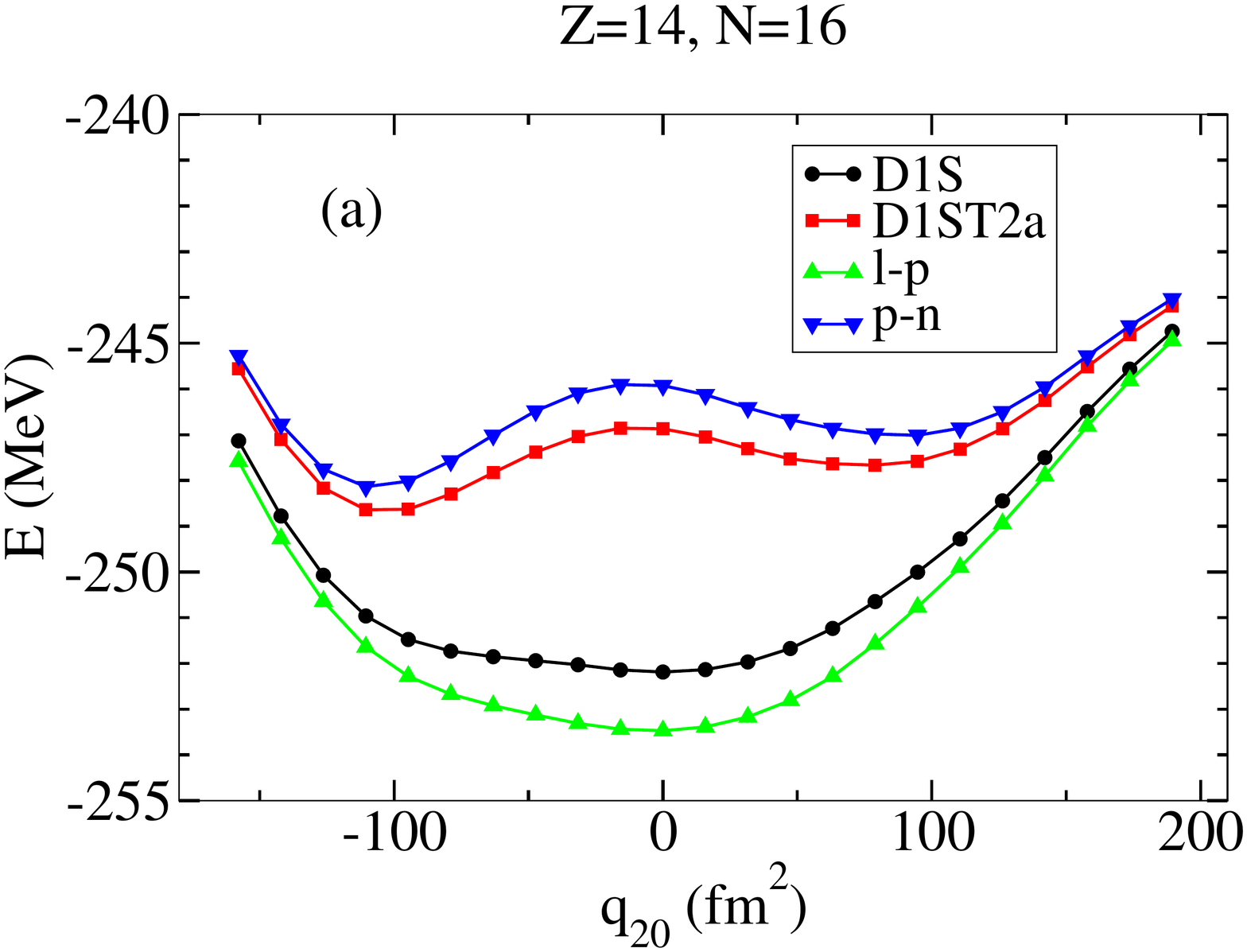} & 
   \includegraphics[width=4.6cm]{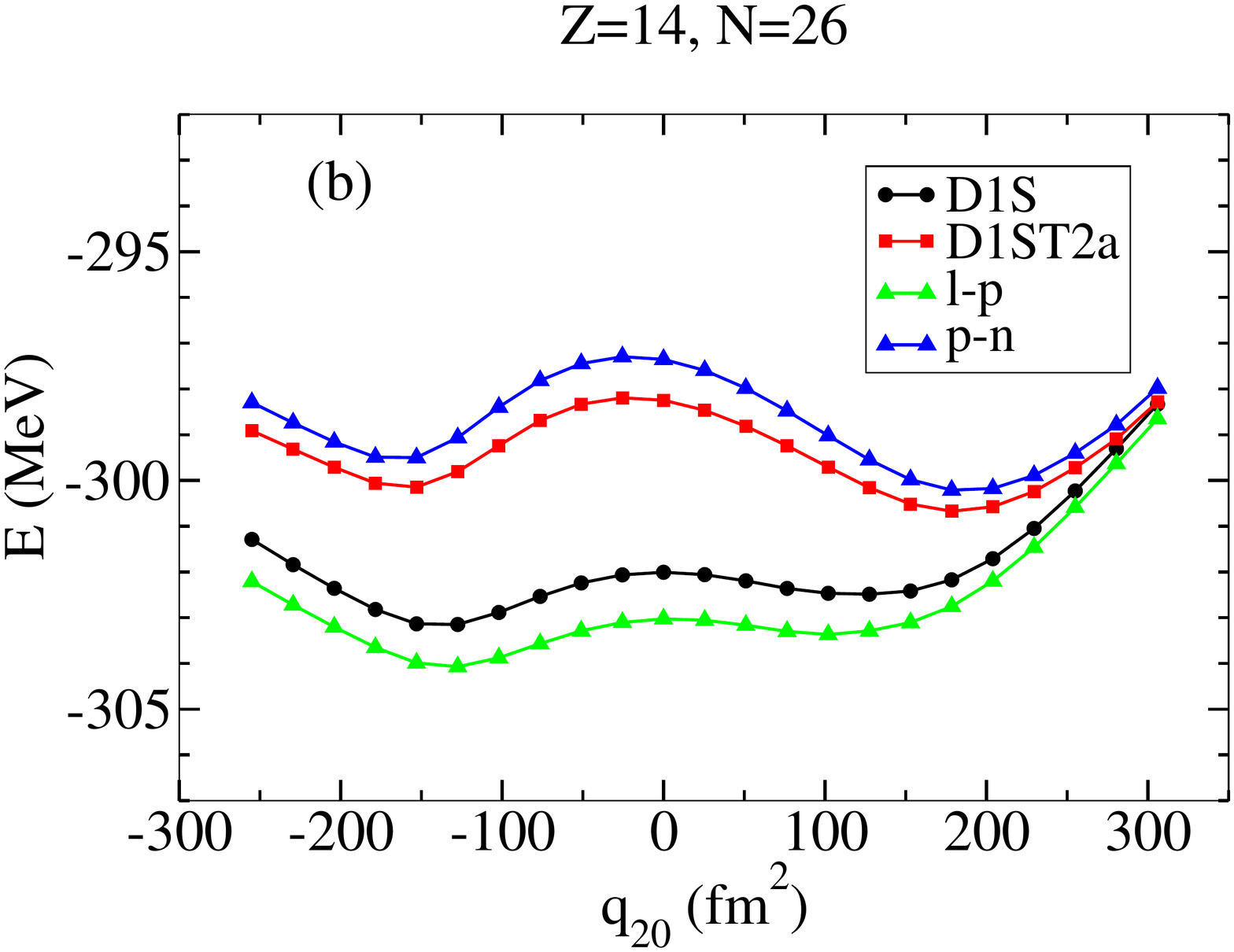}
\end{tabular}
\caption{Like-particle and proton-neutron contributions of the D1ST2a interaction for $^{30}$Si and $^{40}$Si.} 
\label{Silnpnp}
\end{figure}

In Fig.~\ref{Silnpnp} we plot the lp (green triangles up) and pn (blue triangles down) parts of the interaction D1ST2a, comparing with  the D1S (black circles) 
and the full D1ST2a (red squares) ones. As for the other chains, lp and D1S results, on one hand, and pn and D1ST2a, on the other hand, 
are very similar. This shows again that the pn part is the main contribution of the tensor part, for all the deformations. It is interesting to note that in general, 
the contribution of the tensor interaction is less in the regime of high prolate deformations.

\begin{figure}[htb] \centering
\begin{tabular}{cc}
   \includegraphics[width=4.6cm]{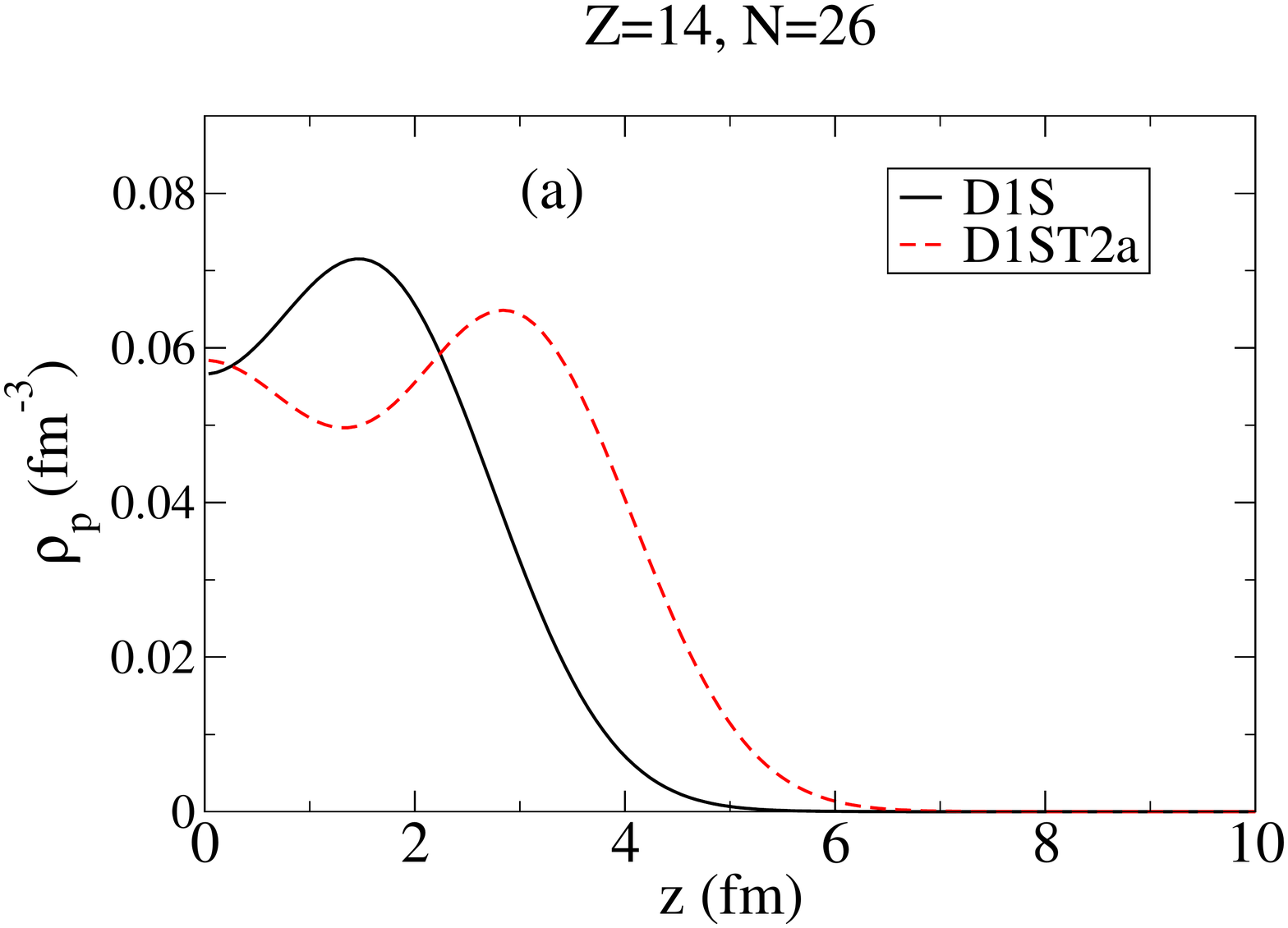} & 
   \includegraphics[width=4.6cm]{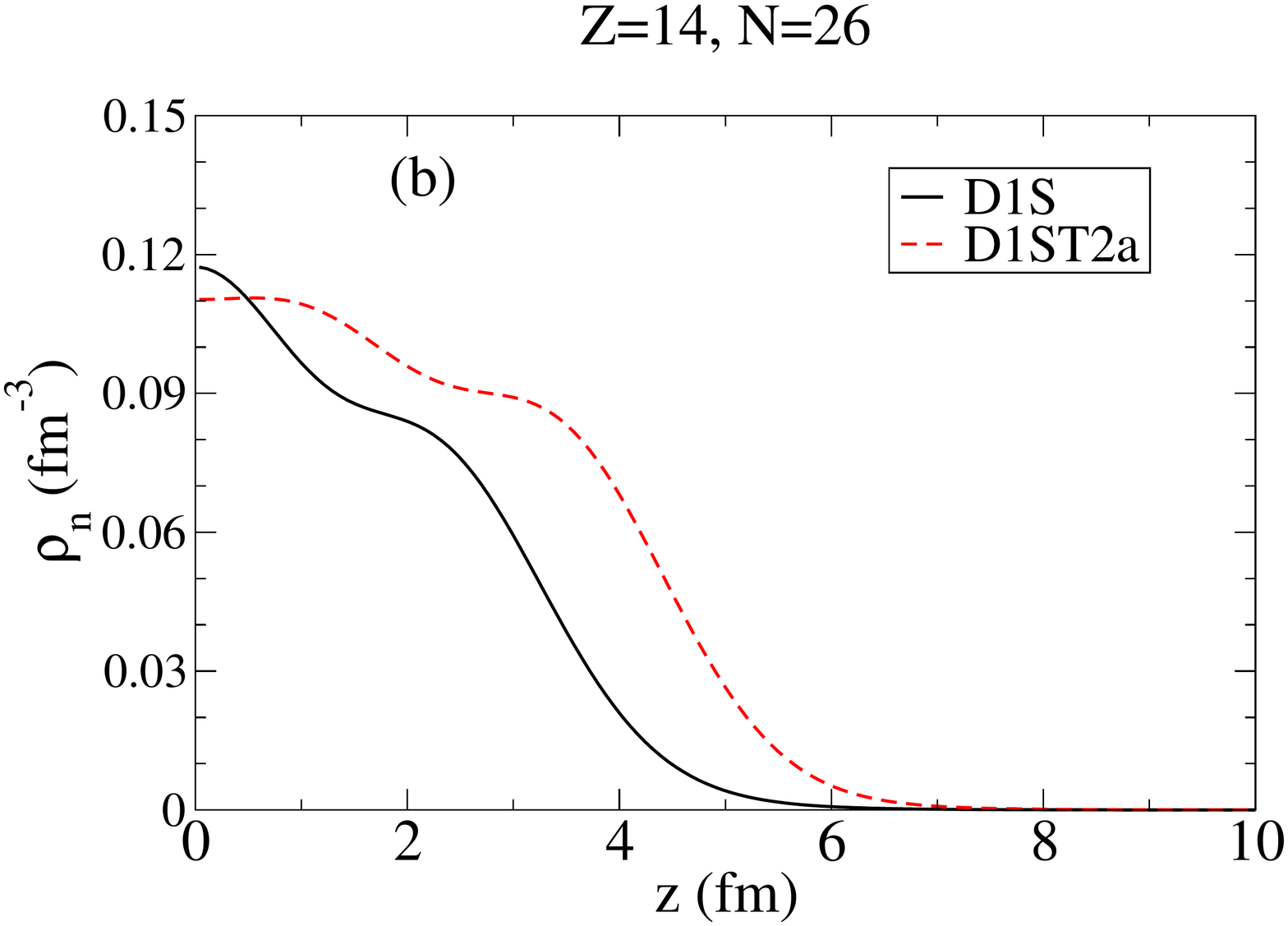}
\end{tabular}	
\caption{Proton and neutron densities for the $^{40}$Si with the D1S (black solid lines) interaction and the D1ST2a one (red dashed lines).} 
\label{SiDenspn}
\end{figure}

In Fig.~\ref{SiDenspn} we show the ground state proton and neutron densities obtained using D1S (black solid lines) and D1ST2a (red dashed lines) for 
protons (panel a) and neutrons (panel b).  The differences observed are due to the fact that the ground state is oblate for the D1S interaction and 
prolate for the D1ST2a one.

\subsubsection{Slater approximation}

\begin{figure}[htb] \centering
\begin{tabular}{ccc}
   \includegraphics[width=4.6cm]{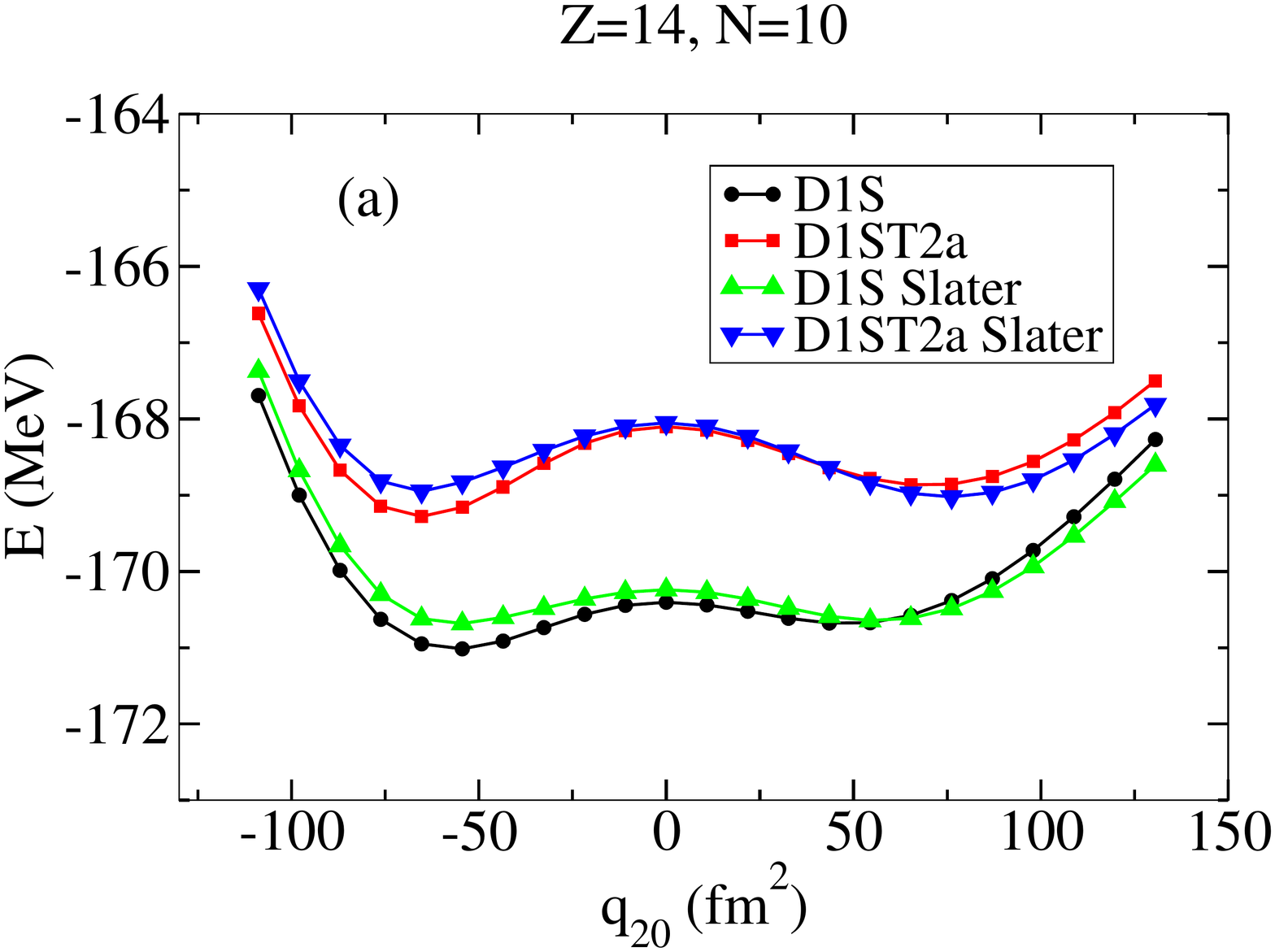} &
   \includegraphics[width=4.6cm]{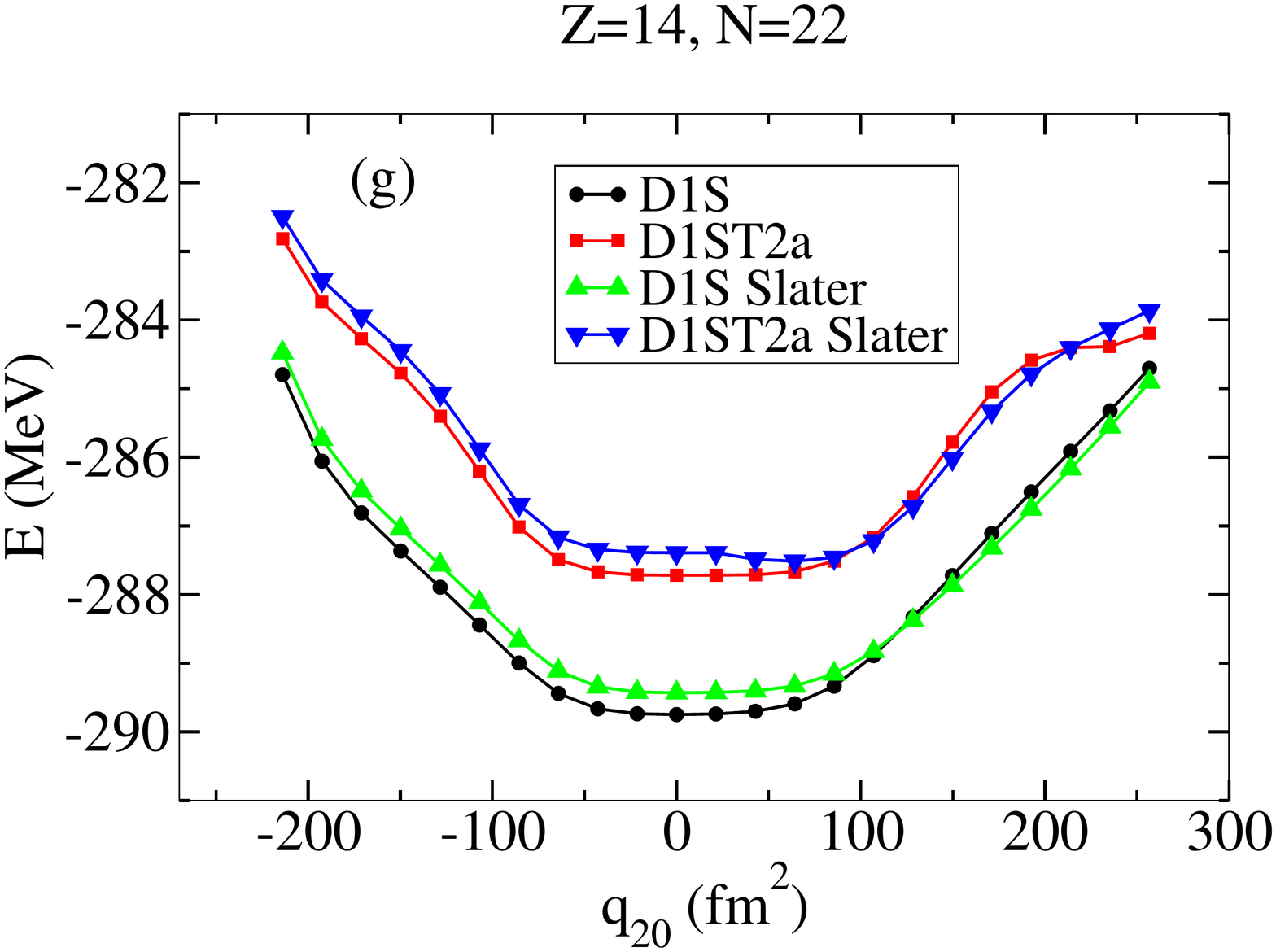} 
\end{tabular}

\caption{Potential Energy Curves for the Si isotopes for the D1S interaction (black circles),  D1ST2a (red squares),  D1S$ +$ Slater approximation 
(green triangles up) and D1ST2$+$ Slater approximation (blue triangles down).} 
\label{ZrPESslat}
\end{figure}

As it is shown before the changes of the ground state localization in PEC can be assigned to the tensor force.  
They can also result from the commonly used Slater approximation on the Coulomb term.
Alongside the tensor study we
 analyze the impact of the Slater approximation on the Coulomb term all over the Si chain. As expected it barely modifies the PEC. 
However these small differences are big enough to change the $^{24}$Si and $^{36}$Si ground states for the D1ST2a interaction. We show in Fig.~\ref{ZrPESslat} the PEC 
computed with both interactions using the 
Slater approximation for the Coulomb exchange term for these latter nuclei. It is compared with the one obtained using the 
exact form for the exchange Coulomb term (that is our standard calculation).
The Slater approximation generally reproduces the same trend than the exact calculation. 
For the first isotope the PEC depicts a two minima landscape. For the D1S interaction with exact or approximated Coulomb term  the ground state is predicted oblate.  
On the other hand considering the Coulomb term at the Slater approximation with the D1ST2a interaction the ground state moves from the oblate
minimum to the prolate one. For $^{36}$Si the Slater approximation moves the ground state location from sphericity to a prolate shape with the D1ST2a interaction.

\subsubsection{The D1ST2c parameterization}

The D1ST2a parameterization chosen for this study keeps the spin-orbit parameter of the Gogny interaction unchanged. This choice is supported by the fact
that we aim to analyze and isolate as well as possible the effect of the tensor term in the deformation.
However as already mentionned above another parameterization called D1ST2c has been published in the literature \cite{GraPRC13}. 
In this latter the adjustment of the spin-orbit and tensor parameters is done in three steps keeping all the other ones unchanged from the D1S parameterization.
First the spin-orbit parameter is set looking at the neutron splitting $1f$ in $^{40}$Ca for which the tensor force vanishes. Then the like-particle $V_{\rm T1}+V_{\rm T2}$ 
tensor parameter is determined by the fit of the neutron splitting $1f$ in $^{48}$Ca. Finally the unlike $V_{T2}$ parameter is adjusted by looking at the
same splitting in $^{56}$Ni. 
The intention of \cite{GraPRC13} is to give acceptable signs and boundaries for the tensor parameters.
It worth stressing out that these adjustments are done at the HF level.
In Table~\ref{parD1ST2ac} are reminded the parameters of interest.

\begin{table}[htb] \centering
 \begin{tabular}{|c|c|c|}
\hline        
   Parameter                &    D1ST2a   & D1ST2c         \\
\hline 
   $W_{\rm LS}$             &   $ 130.0$   & $ 103.0$      \\
\hline 
   $V_{\rm T1}+V_{\rm T2}$  &   $ -20.0$   & $ -75.0$      \\
\hline 
   $V_{\rm T2}$             &   $ 115.0$   & $  60.0$      \\
\hline
\end{tabular}

\caption{Refitted parameters (MeV) for D1ST2a and D1ST2c. }  
\label{parD1ST2ac} 
\end{table}

\begin{figure}[htb] \centering
\begin{tabular}{ccc}
   \includegraphics[width=4.6cm]{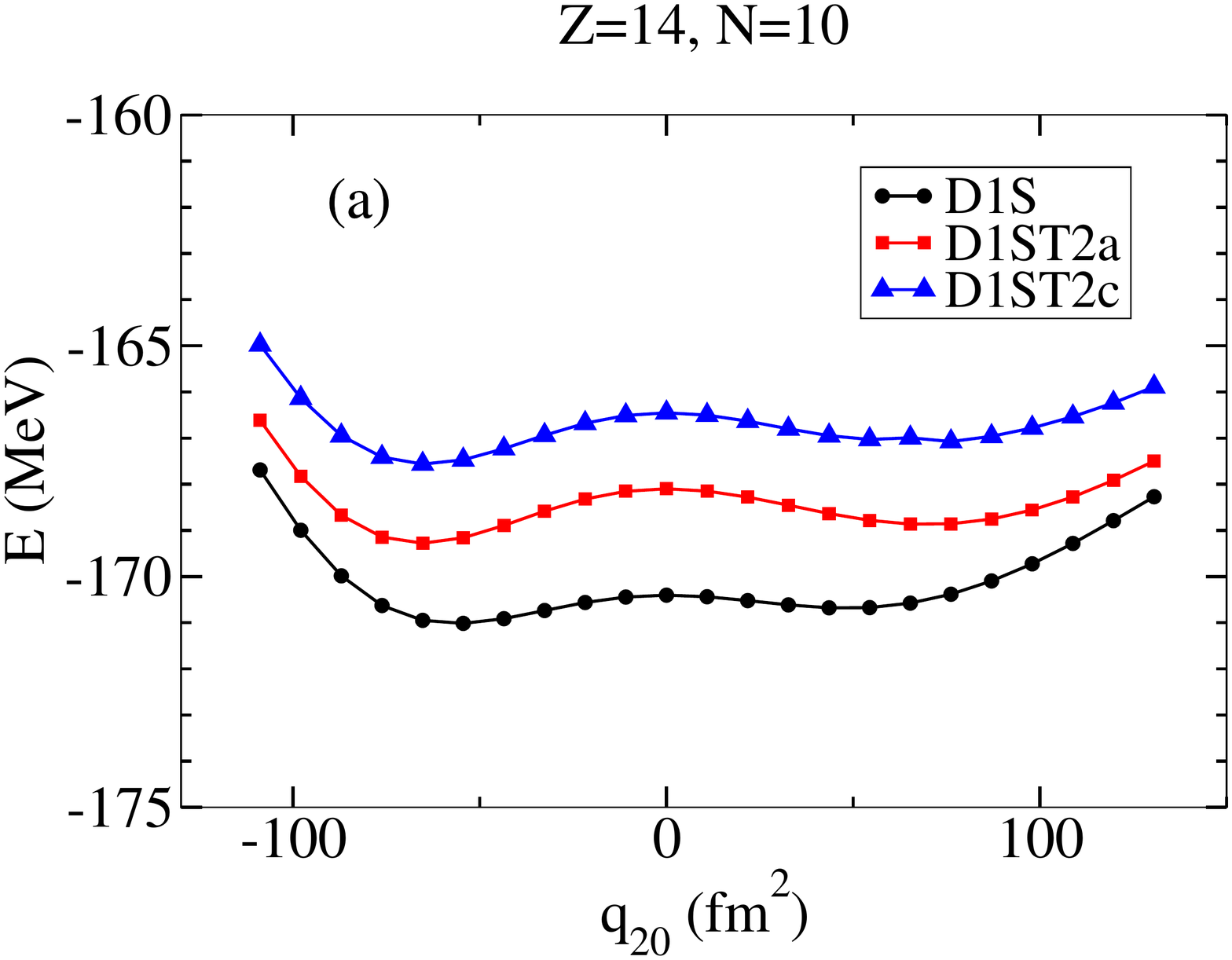} &
   \includegraphics[width=4.6cm]{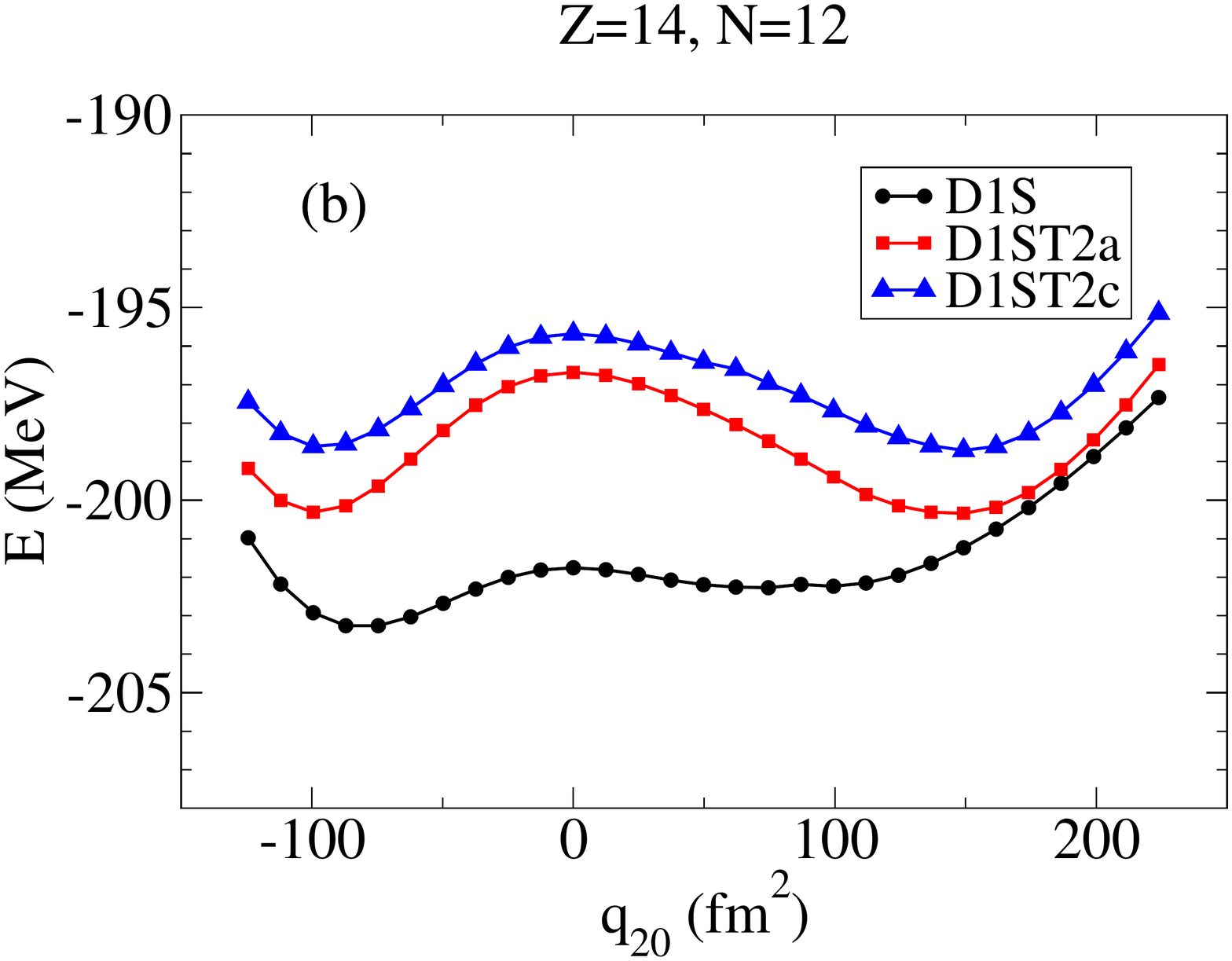} &
   \includegraphics[width=4.6cm]{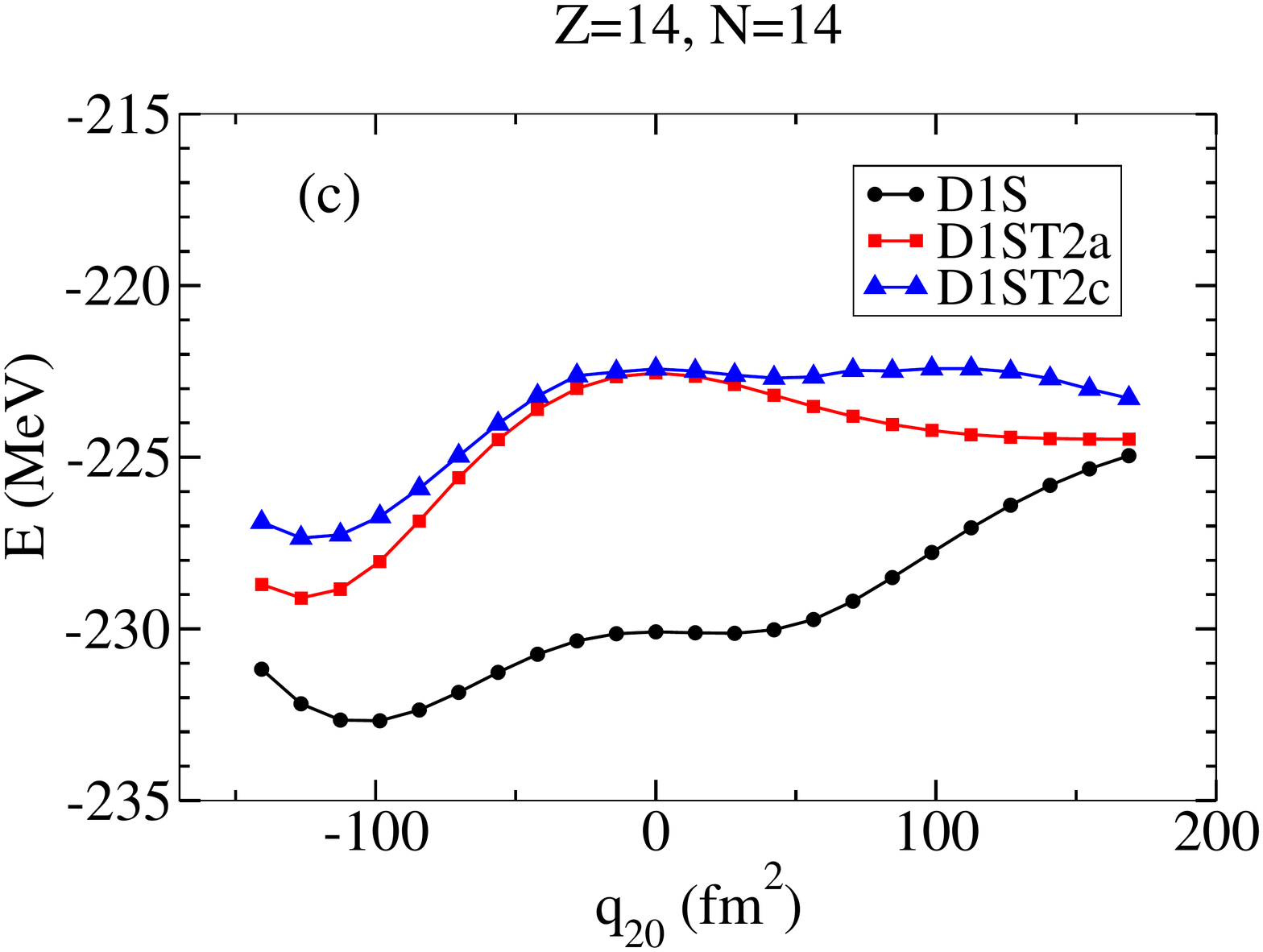} \\
   \includegraphics[width=4.6cm]{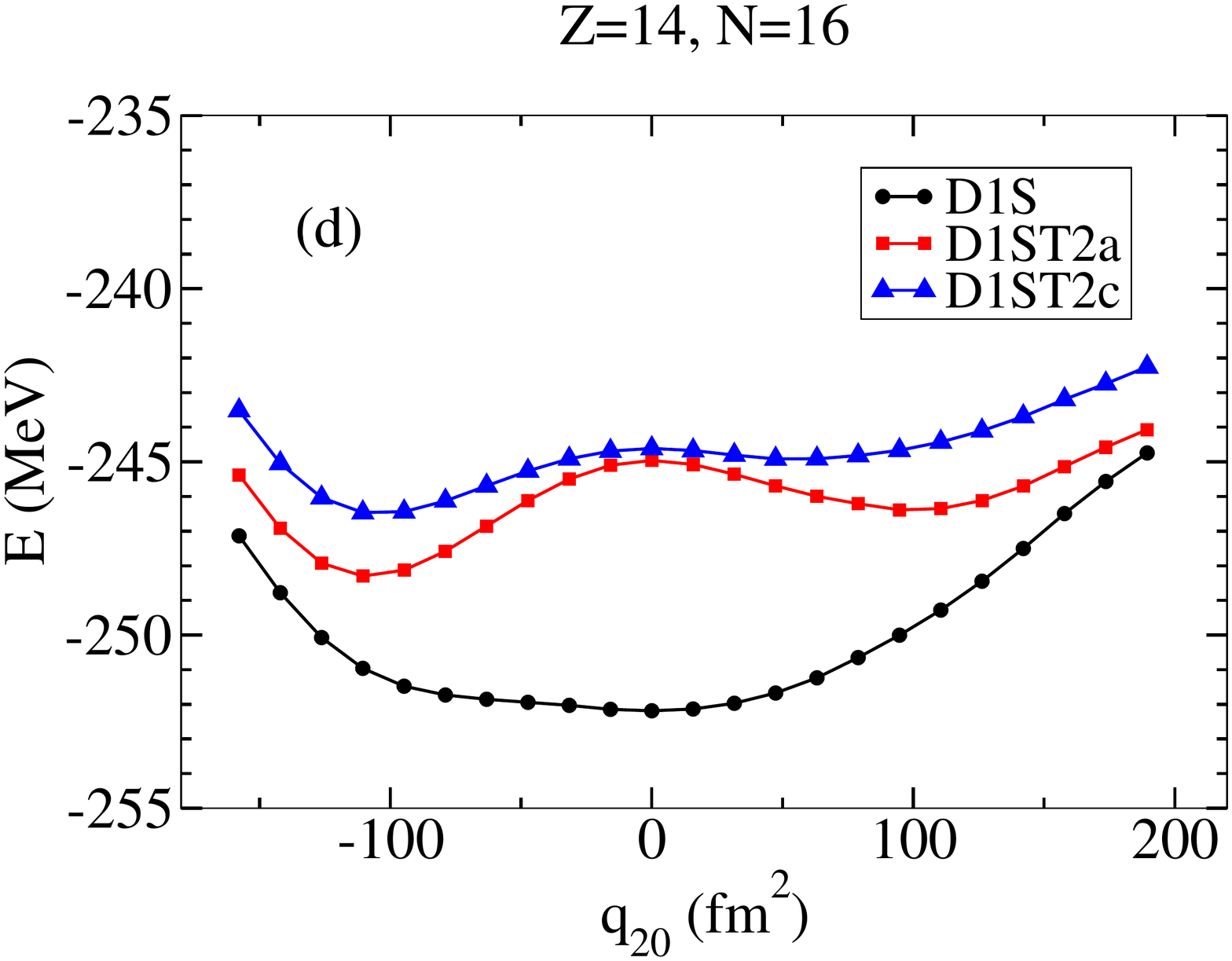} &
   \includegraphics[width=4.6cm]{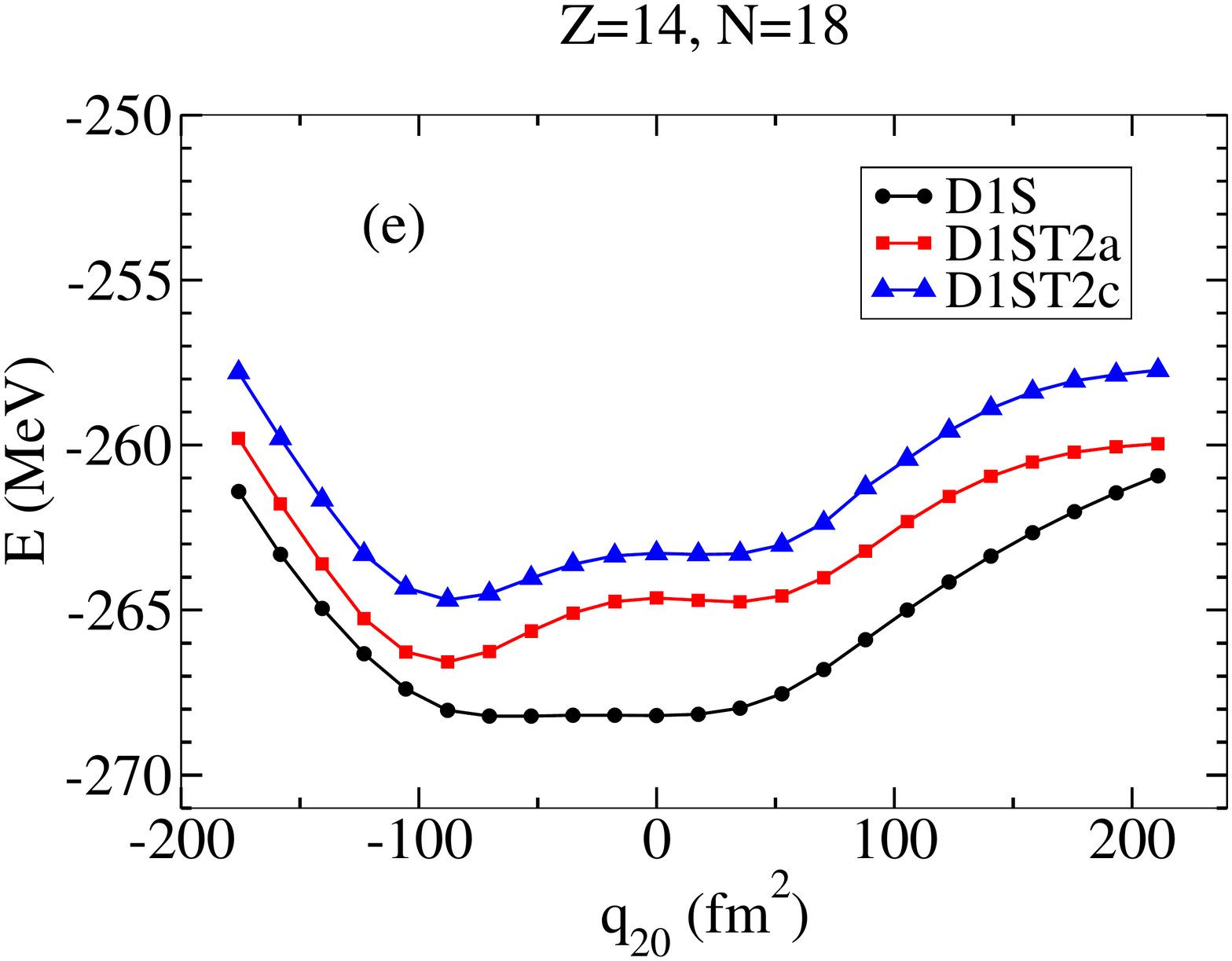} &
   \includegraphics[width=4.6cm]{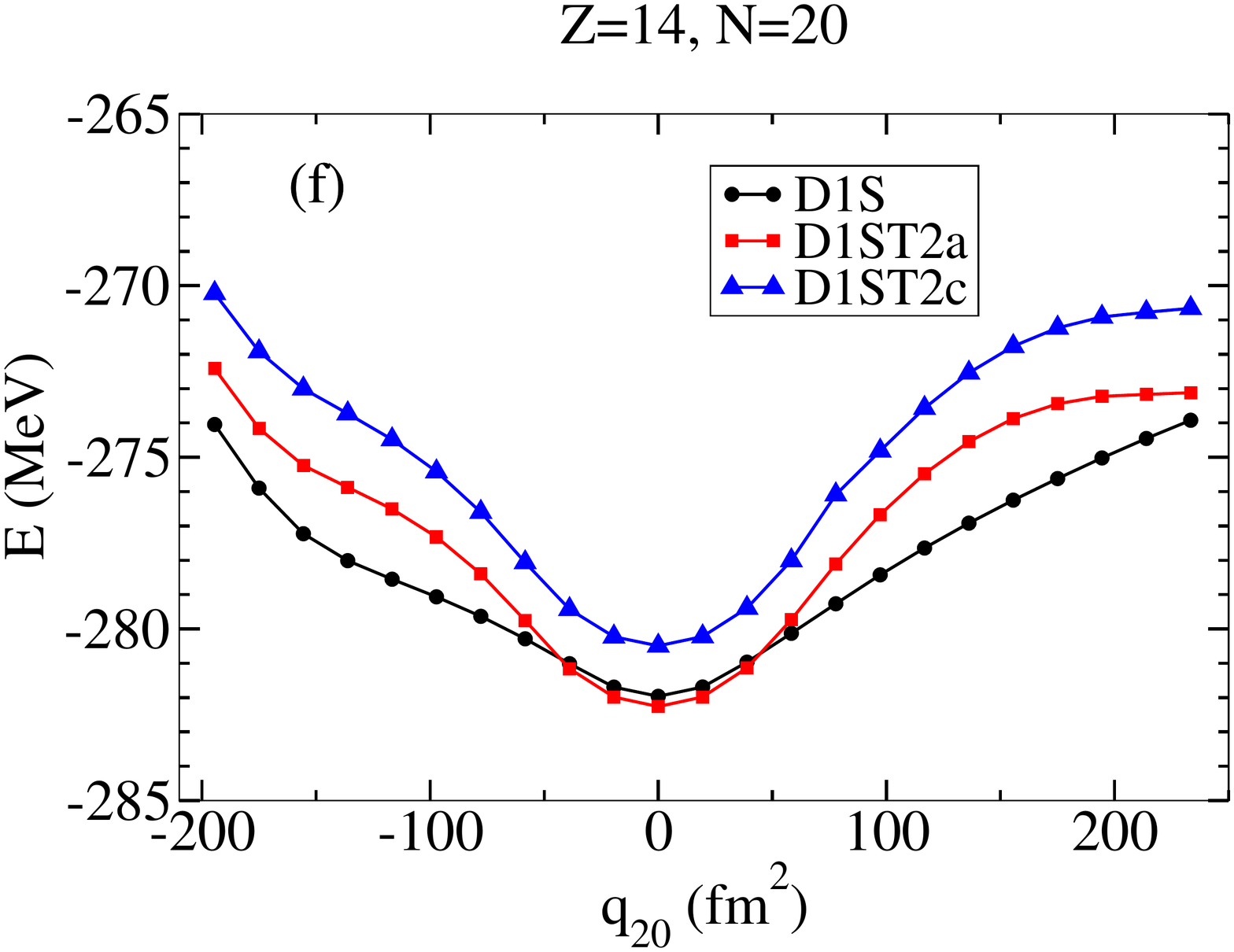} \\
   \includegraphics[width=4.6cm]{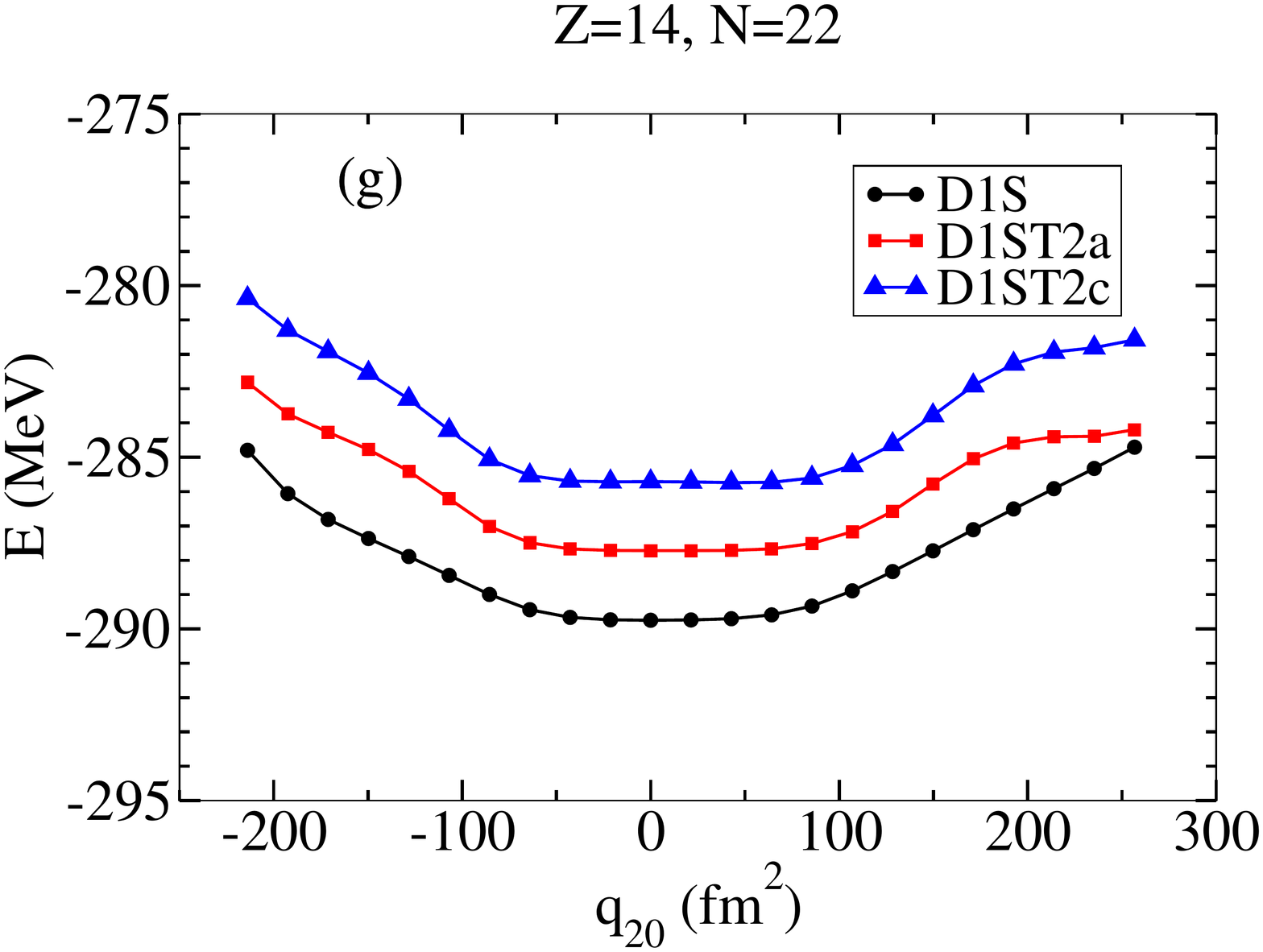} &
   \includegraphics[width=4.6cm]{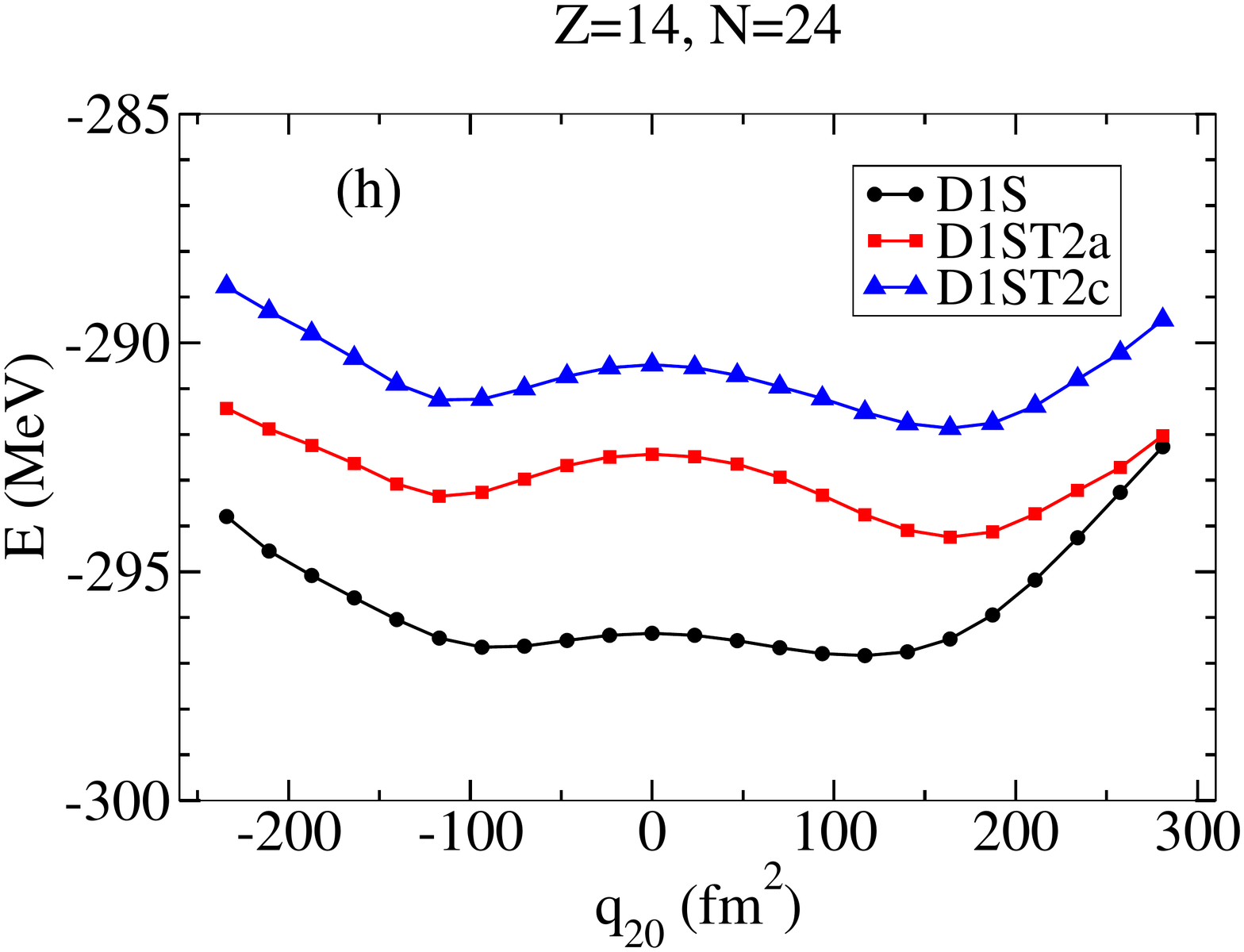} &
   \includegraphics[width=4.6cm]{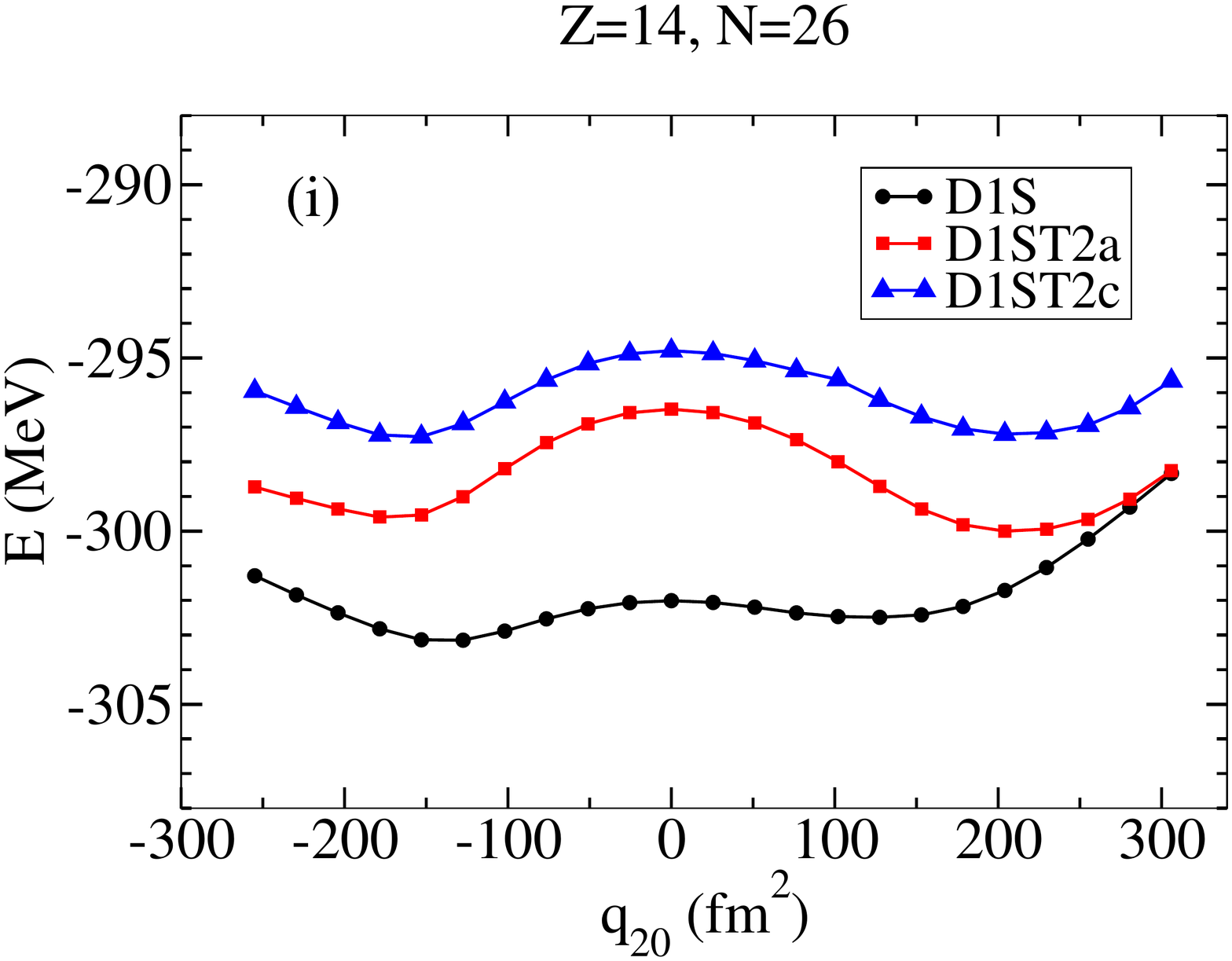} \\
   \includegraphics[width=4.6cm]{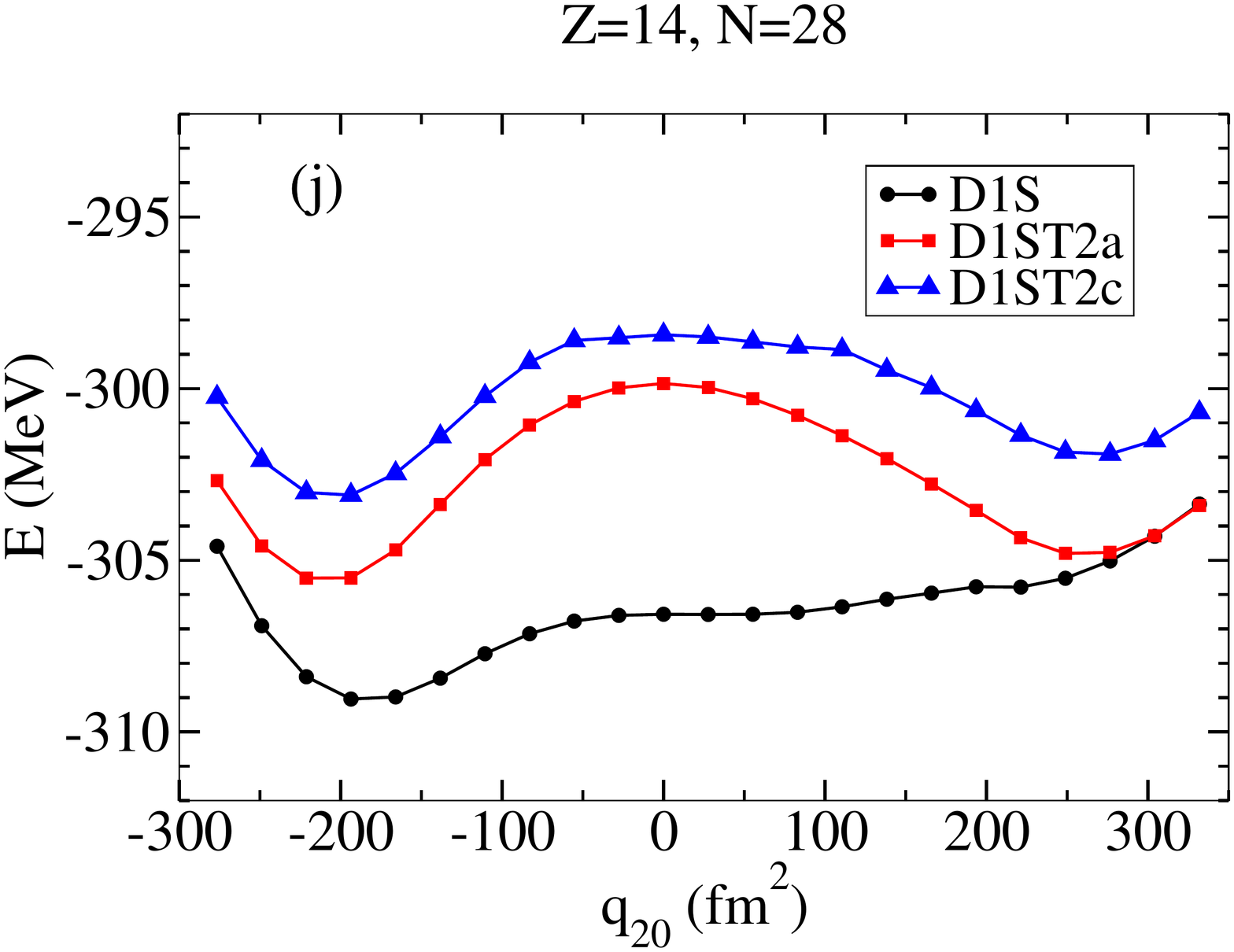} &
   \includegraphics[width=4.6cm]{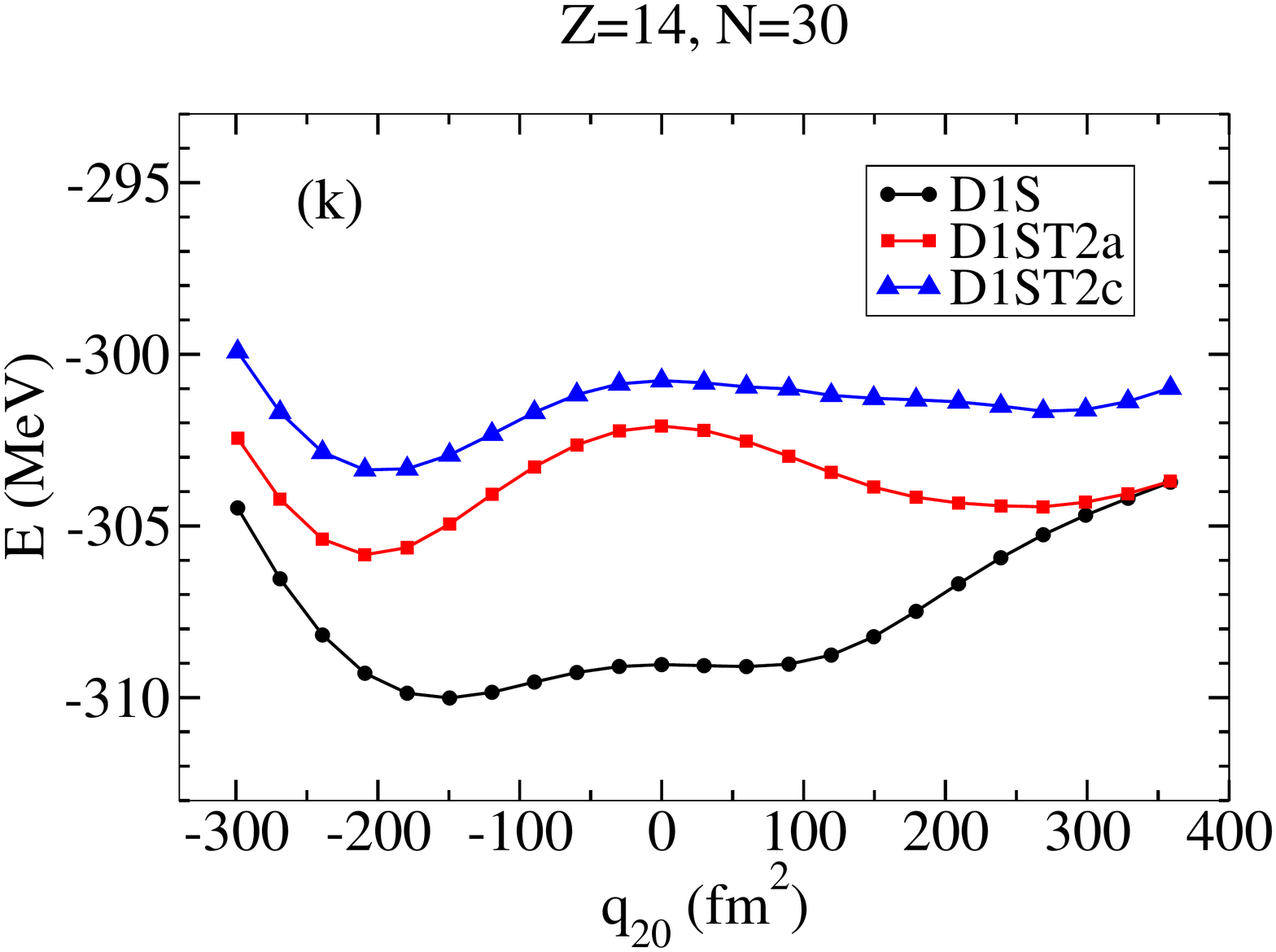} &
    \\
\end{tabular}
\caption{Potential Energy Curves  of the $Z=14$ chain 
for the D1S interaction (circles), the D1ST2a one (squares) and the D1ST2c one (triangles). } 
\label{PESSiD1ST2c}
\end{figure} 

The D1ST2a and D1ST2c PEC are compared in Fig.~\ref{PESSiD1ST2c} for the Silicon isotopes. 
The minima in the D1ST2c PEC are located at the same quadrupole deformation than the ones obtained with D1ST2a.
The nuclei are less bound with the D1ST2c parameterization, by about $2$ MeV along the Silicon chain.
The D1ST2a and D1ST2c curves get the same variations in all the PEC, except for the $^{28,30}$Si isotopes around sphericity.
For these two nuclei the valence shell, the $1d_{5/2}$ and $2s_{1/2}$ respectively, are full. The $1d_{5/2}$ subshell partner is empty and we expect the tensor 
contribution to be important according to the schematic arguments presented in Sec. \ref{TED}.
Another interesting Si isotope is the neutron SS $^{34}$Si. At sphericity the tensor energy only comes from the $p-p$ contribution and the pairing vanishes 
for both parameterizations. 
The ratio between the tensor energy from D1ST2a and from D1ST2c is about $0.239$ very close to the expected value ($0.267$) in the schematic tensor energy decomposition 
in Sec. \ref{TED}. This also applies for the spin-orbit energies: $1.281$ close to $W_{\rm LS}^{a}/W_{\rm LS}^{c}\simeq 1.262$.
The total HFB energy difference ($\simeq -1.77$ MeV) between D1ST2a and D1ST2c is not only due to the spin-orbit plus tensor energy difference ($\simeq 6.45$ MeV); 
all the contributions to the total HFB energy are rearranged by the D1ST2c fit. This is generally true all along the chain.

\clearpage

\subsection{Sulfur chain}

\begin{figure}[htb] \centering
\begin{tabular}{ccc}
   \includegraphics[width=4.6cm]{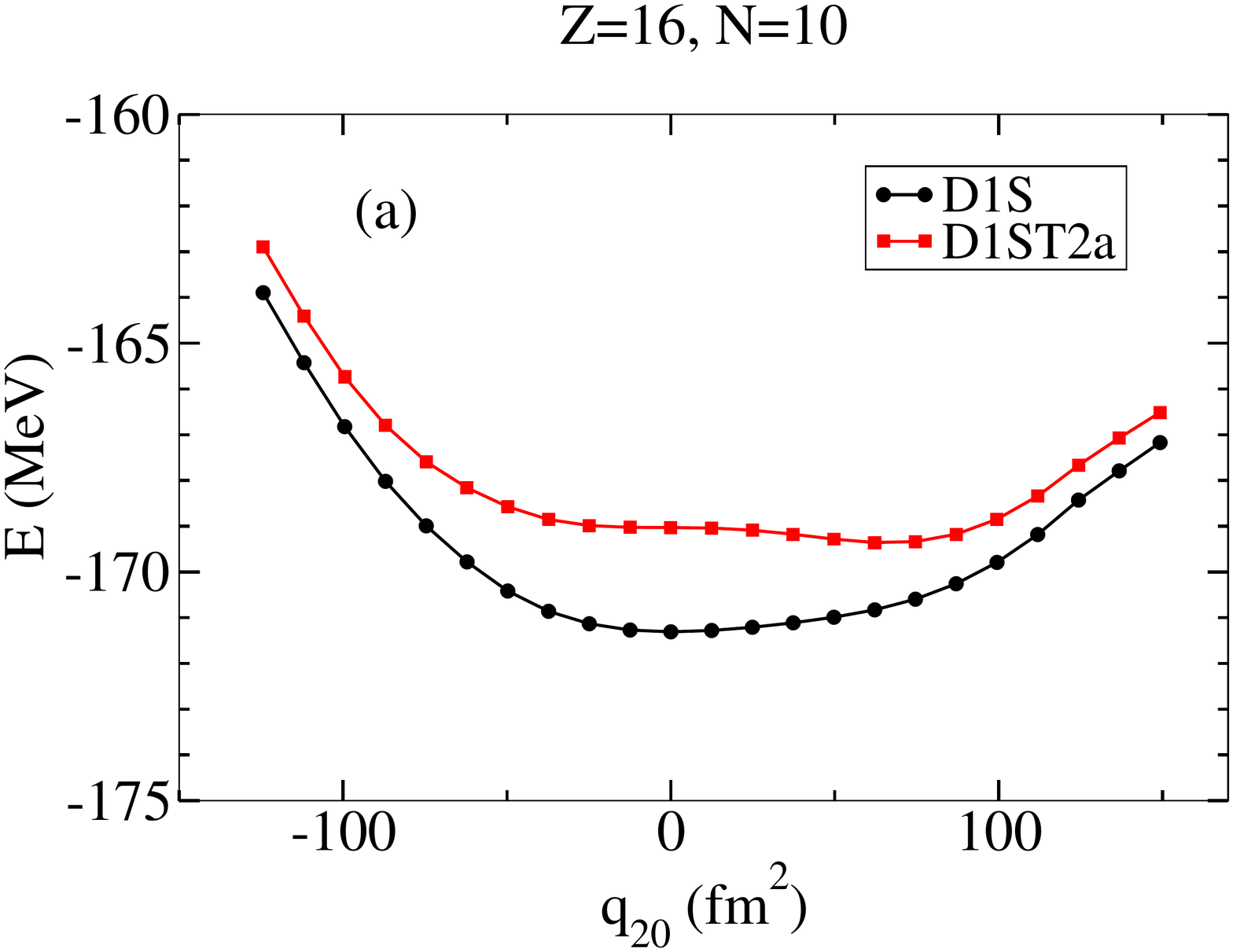} &
   \includegraphics[width=4.6cm]{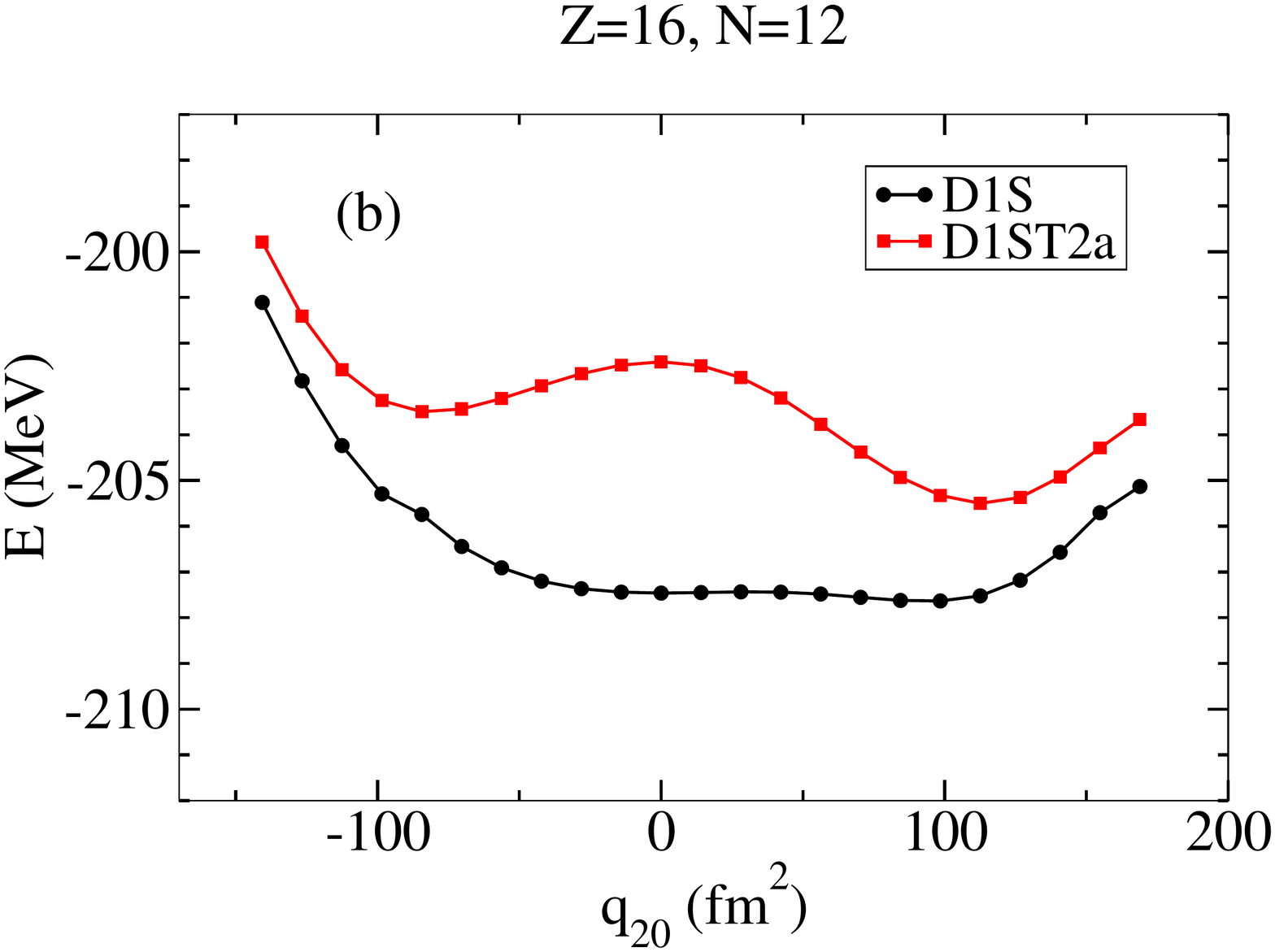} &
   \includegraphics[width=4.6cm]{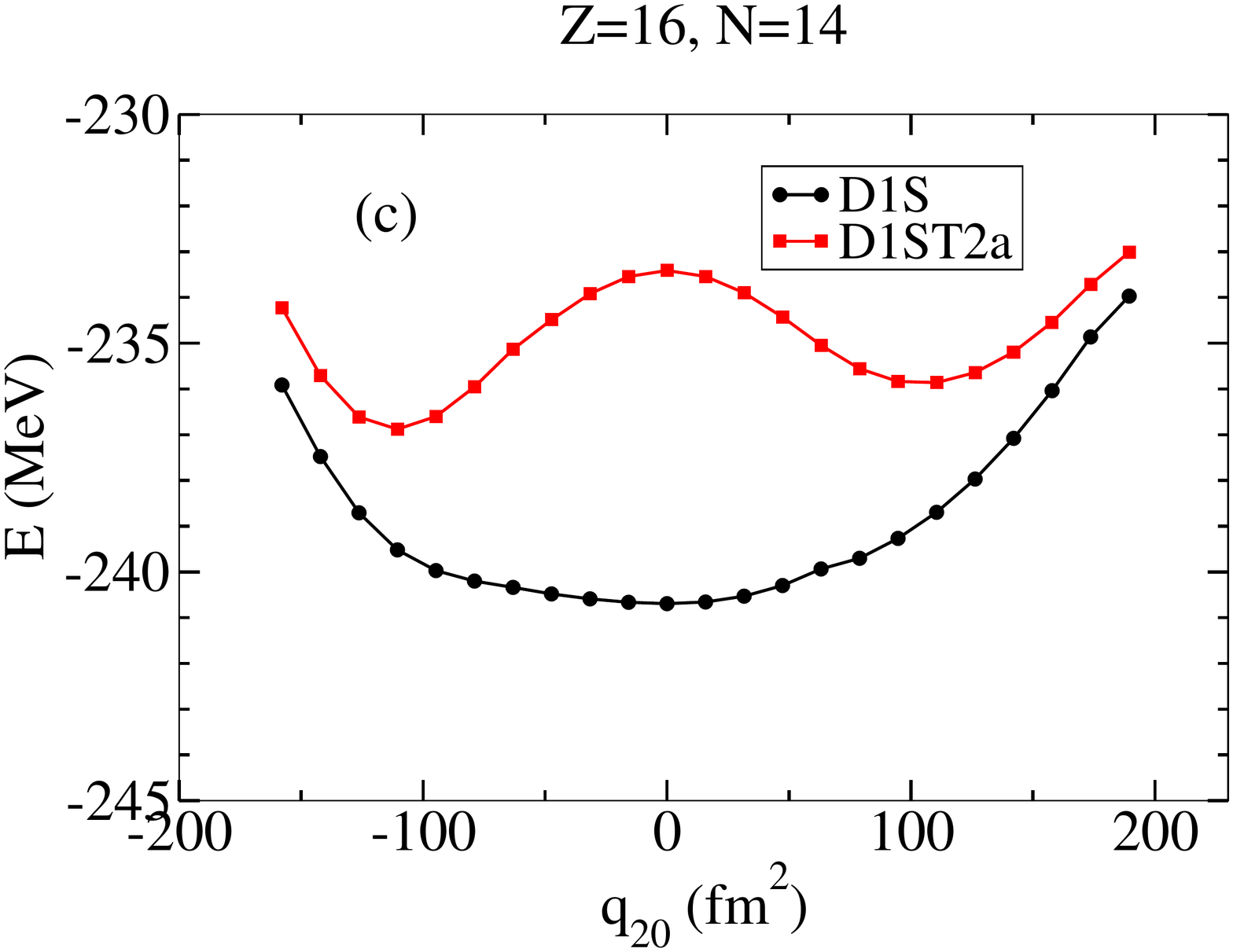} \\
   \includegraphics[width=4.6cm]{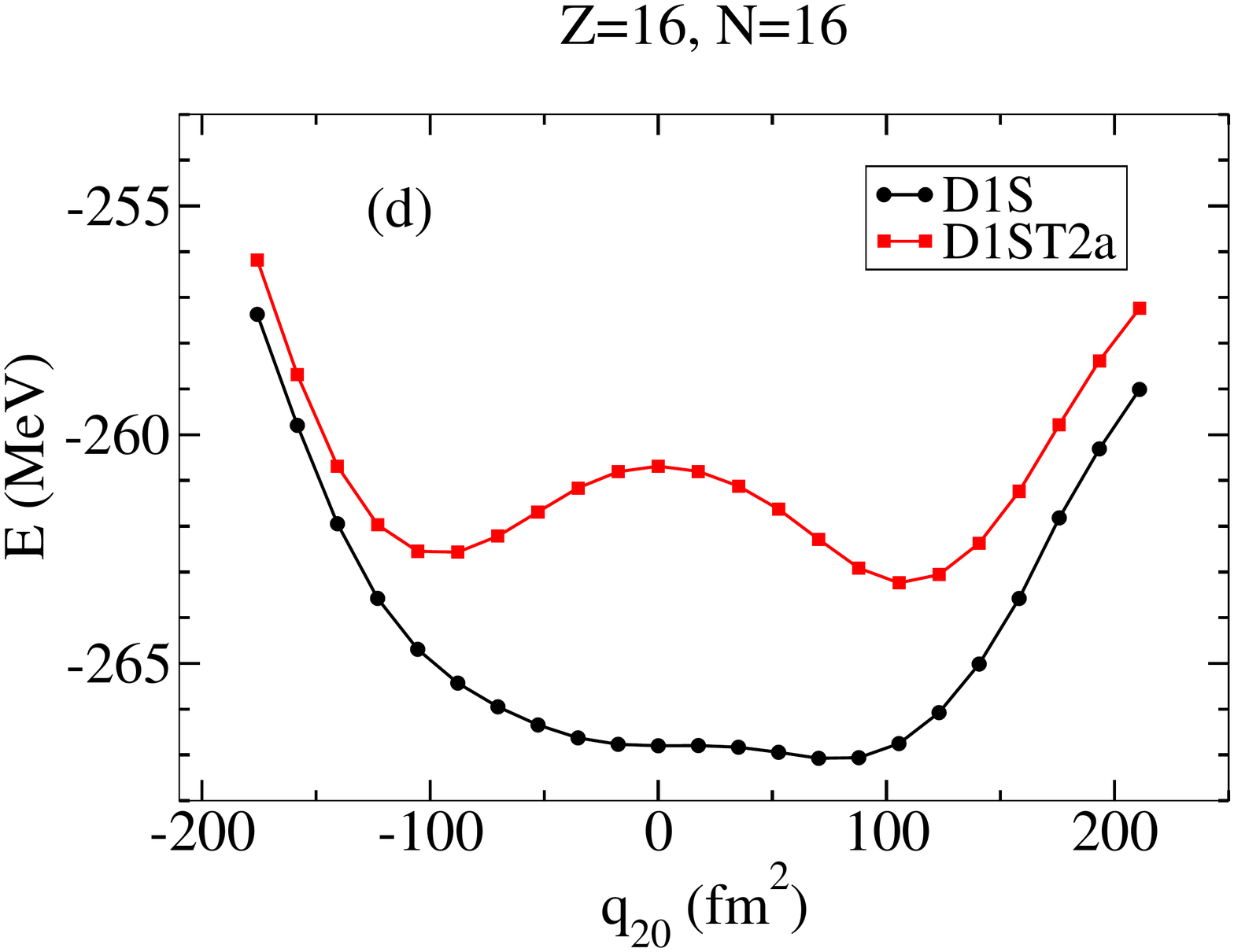} &
   \includegraphics[width=4.6cm]{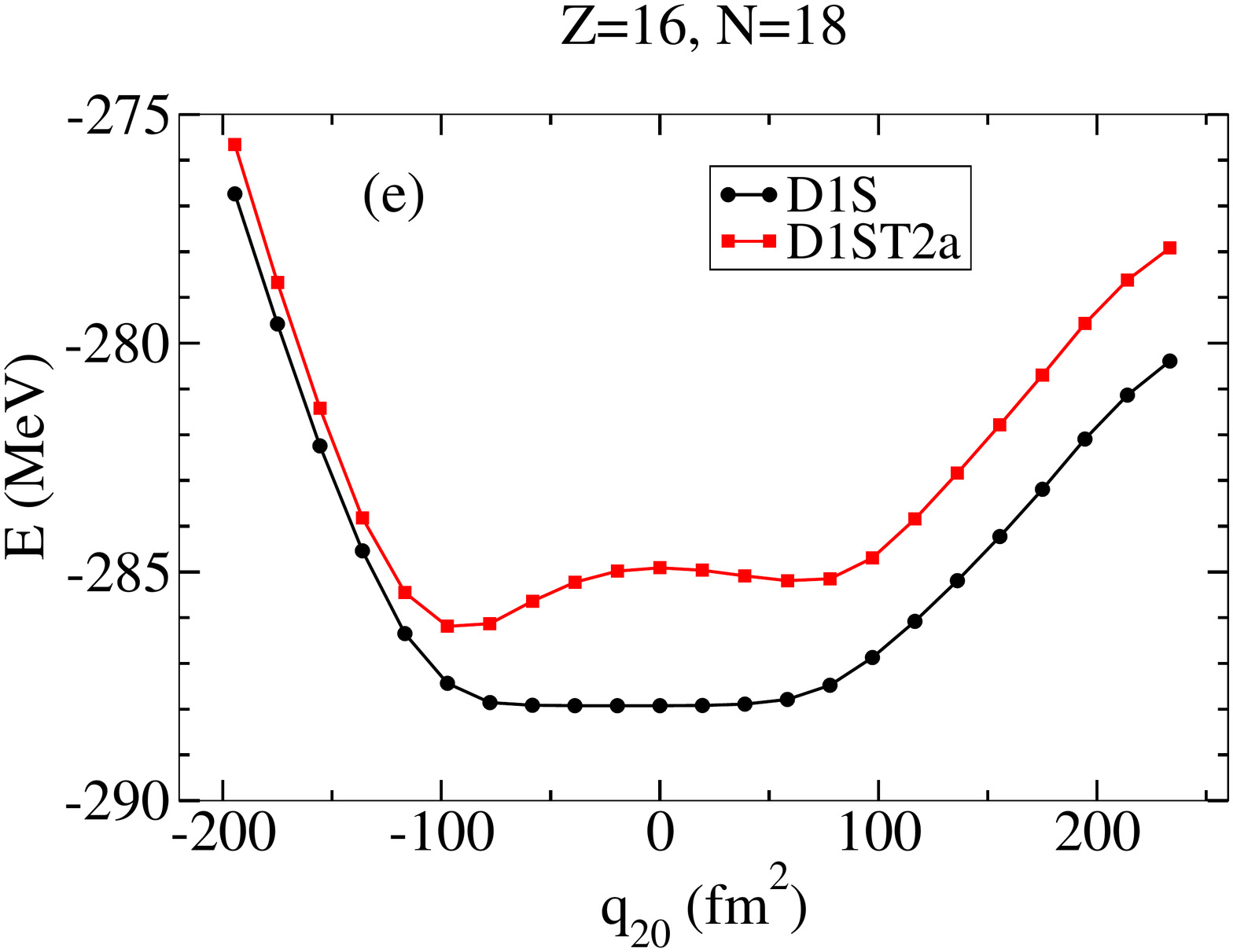} &
   \includegraphics[width=4.6cm]{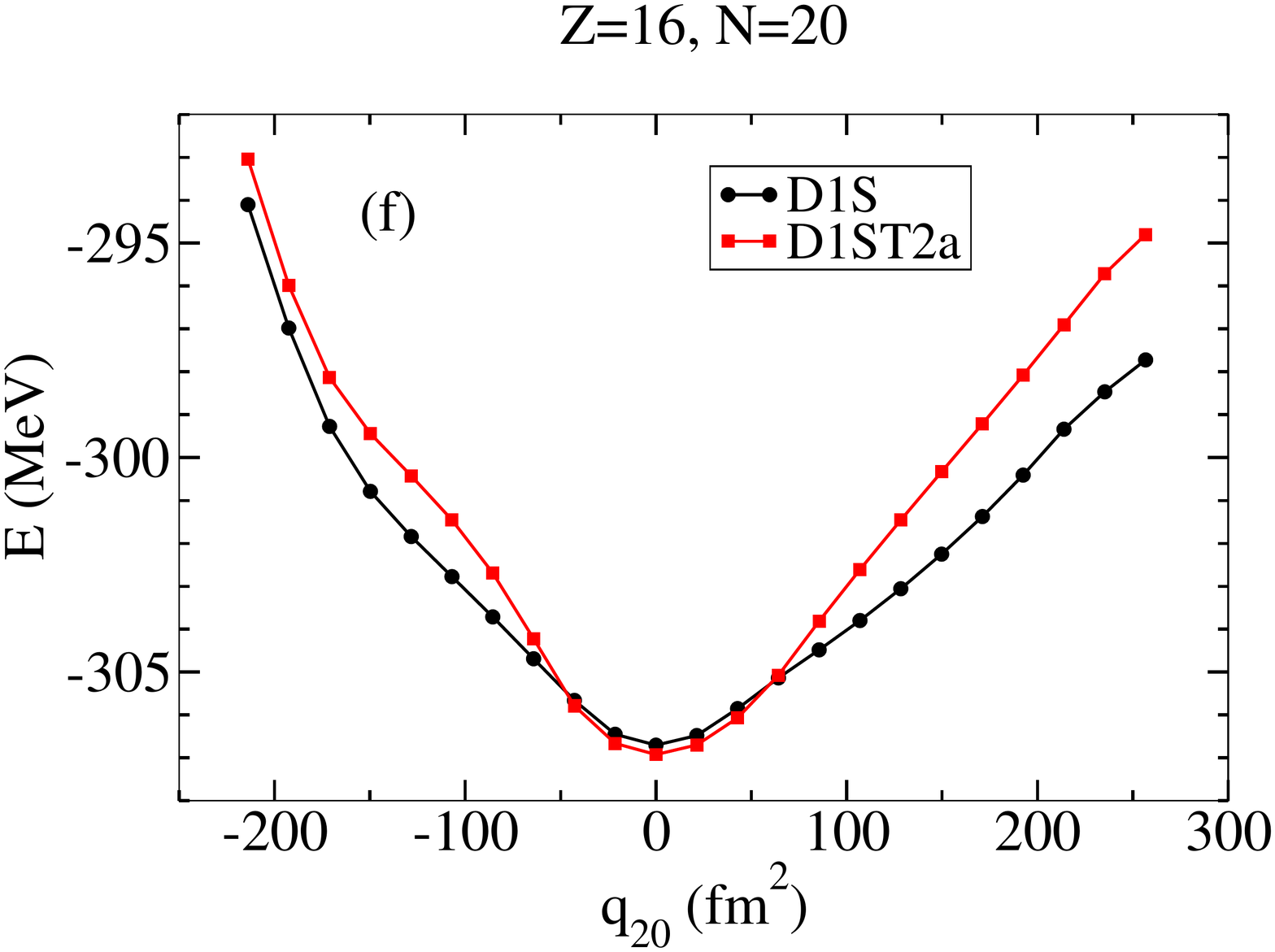} \\
   \includegraphics[width=4.6cm]{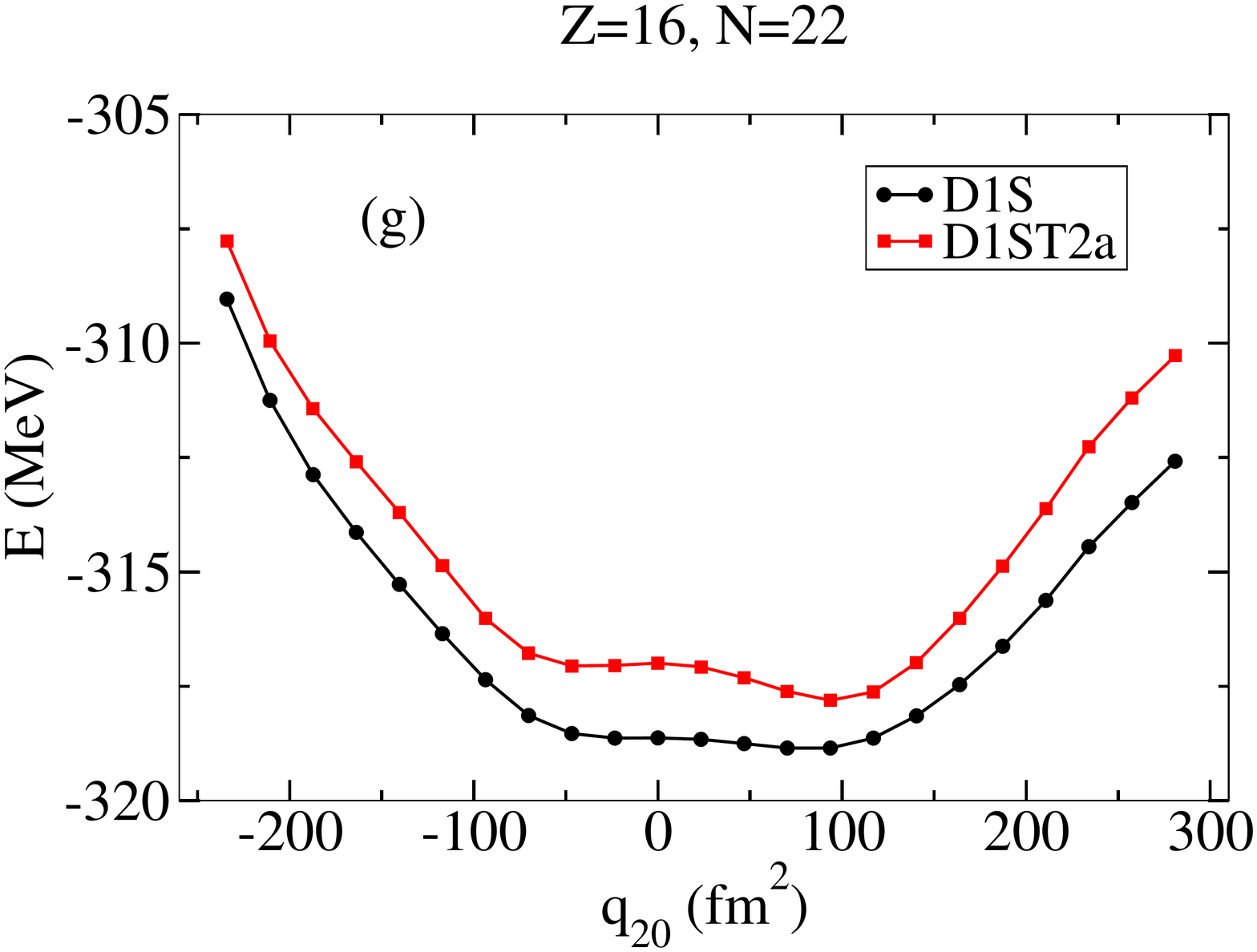} &
   \includegraphics[width=4.6cm]{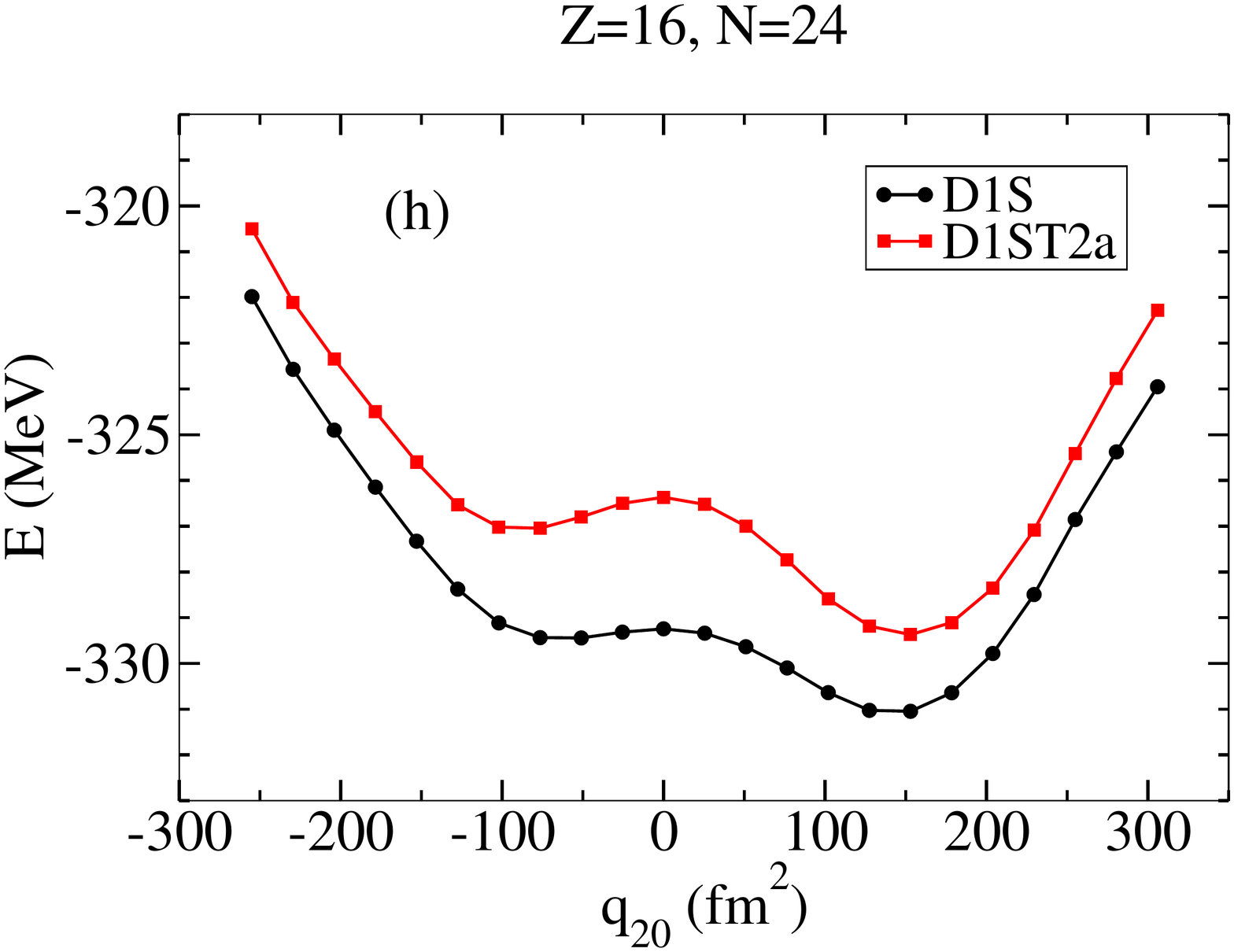} &
   \includegraphics[width=4.6cm]{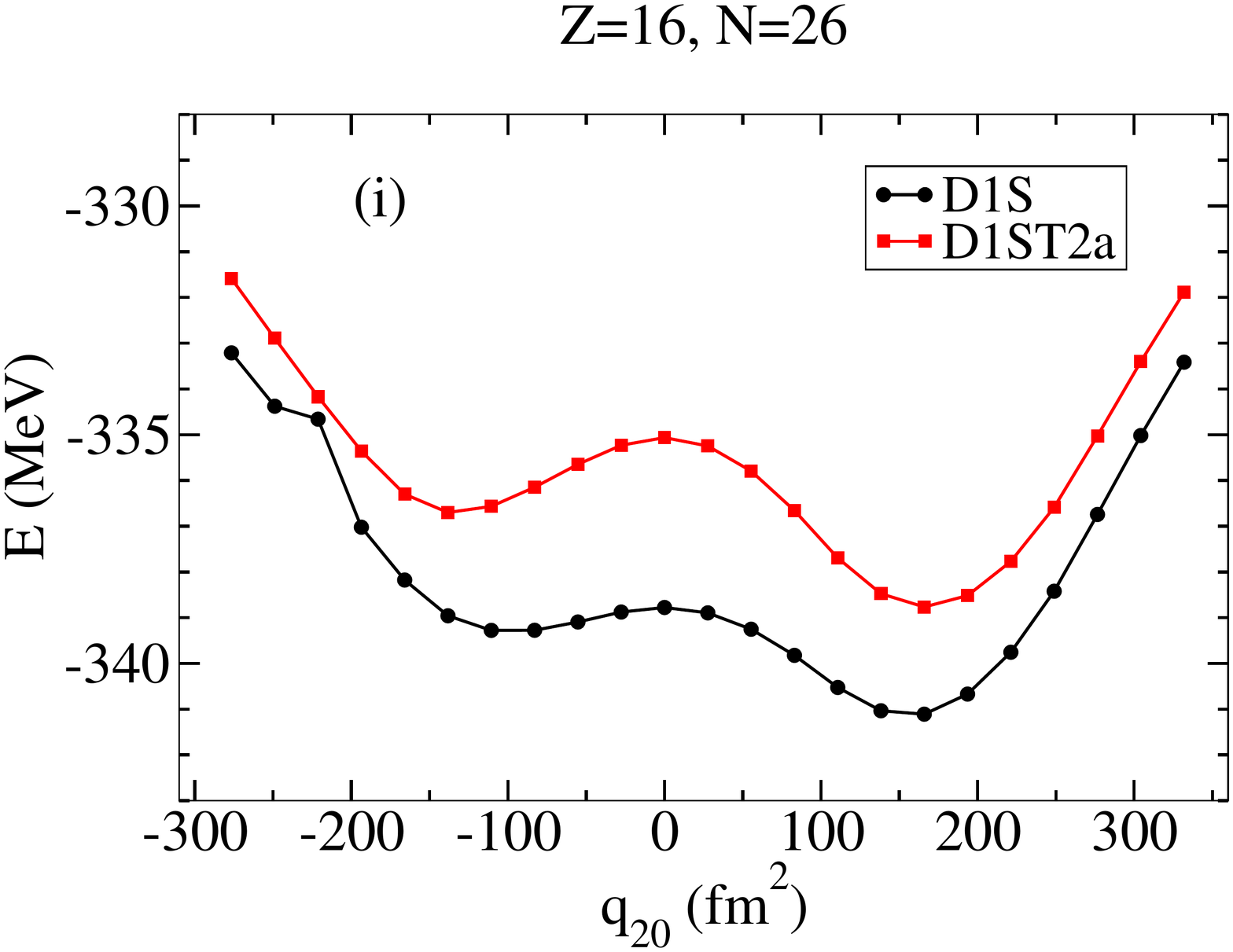} \\
   \includegraphics[width=4.6cm]{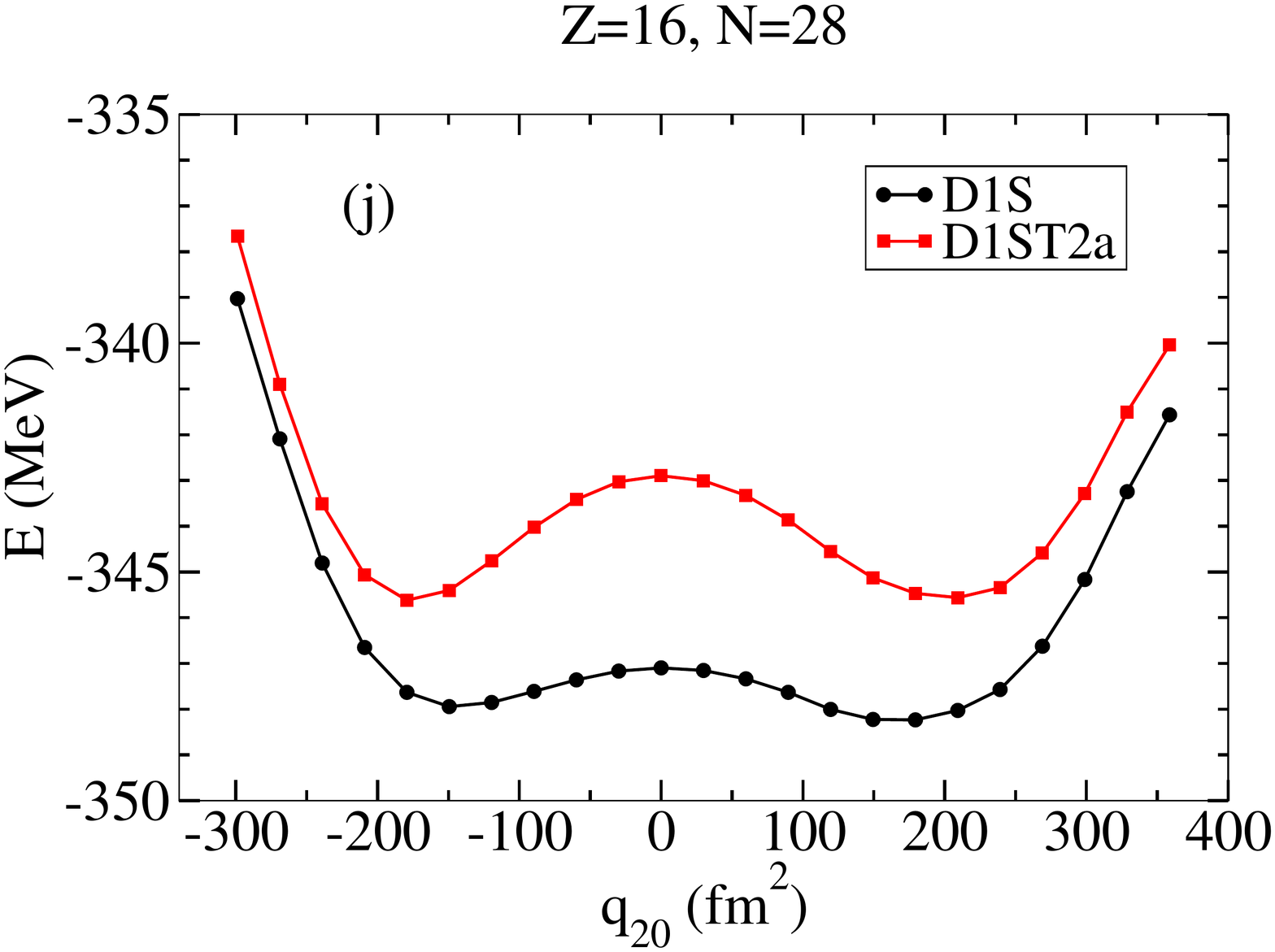} &
   \includegraphics[width=4.6cm]{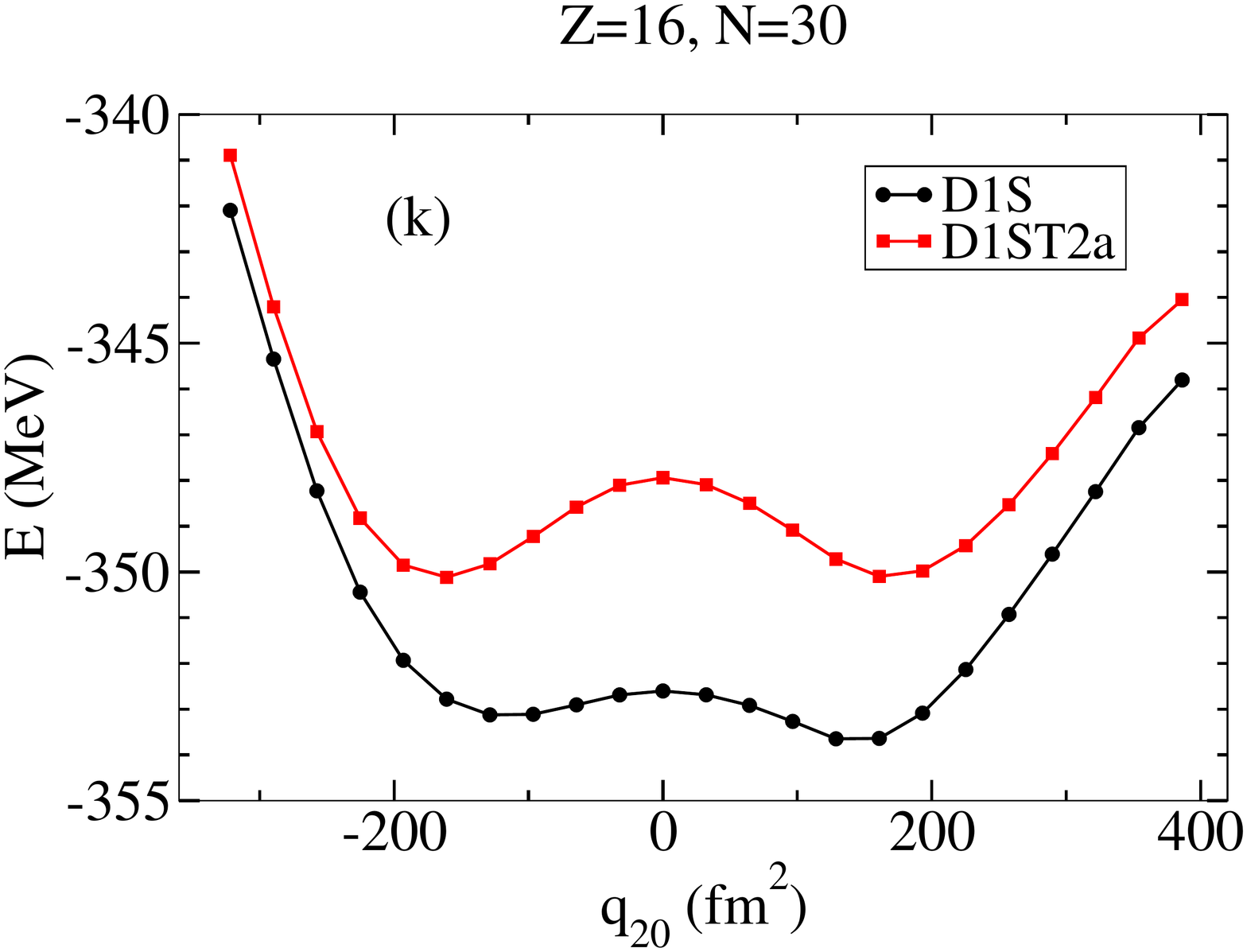} &
   \includegraphics[width=4.6cm]{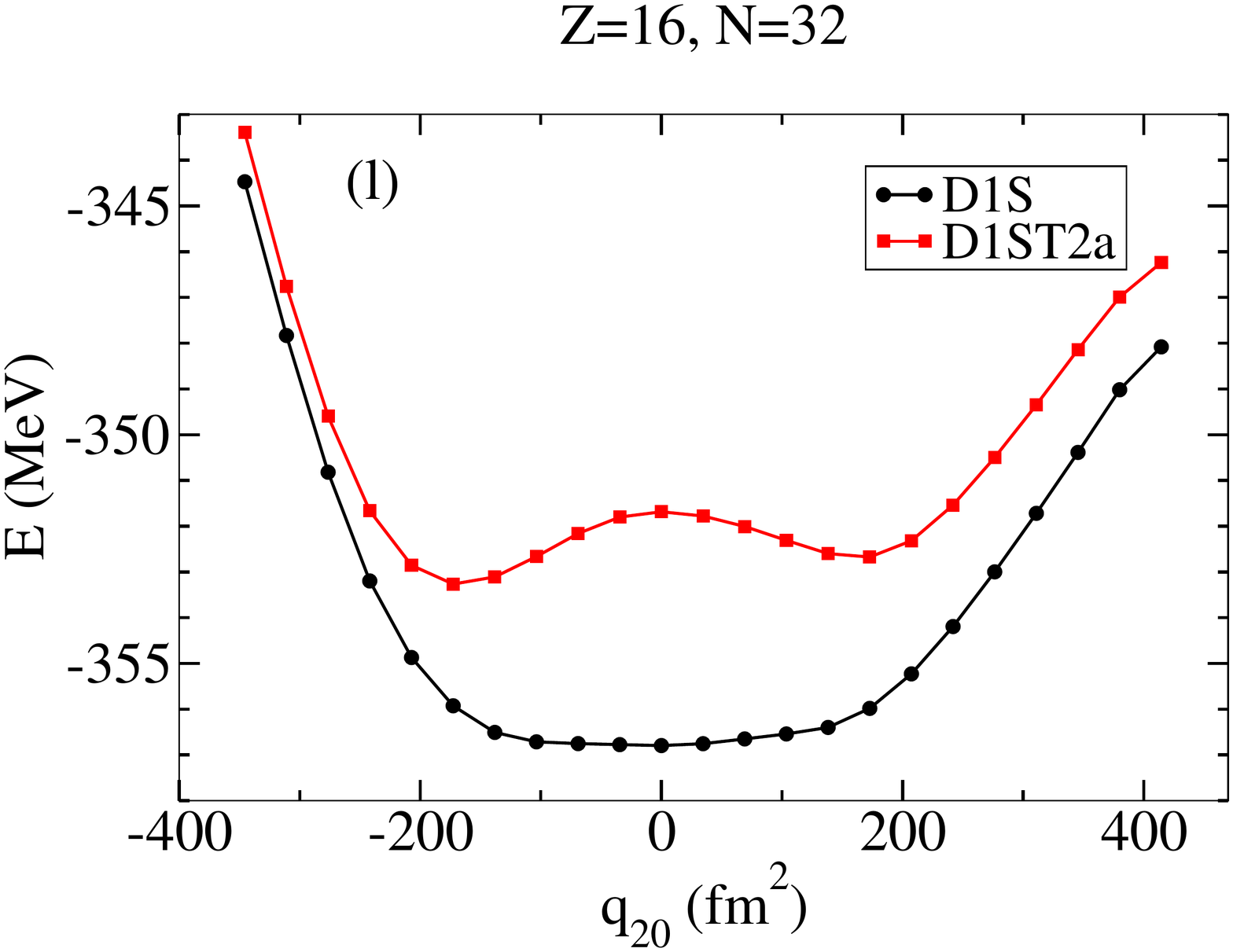} \\
   \includegraphics[width=4.6cm]{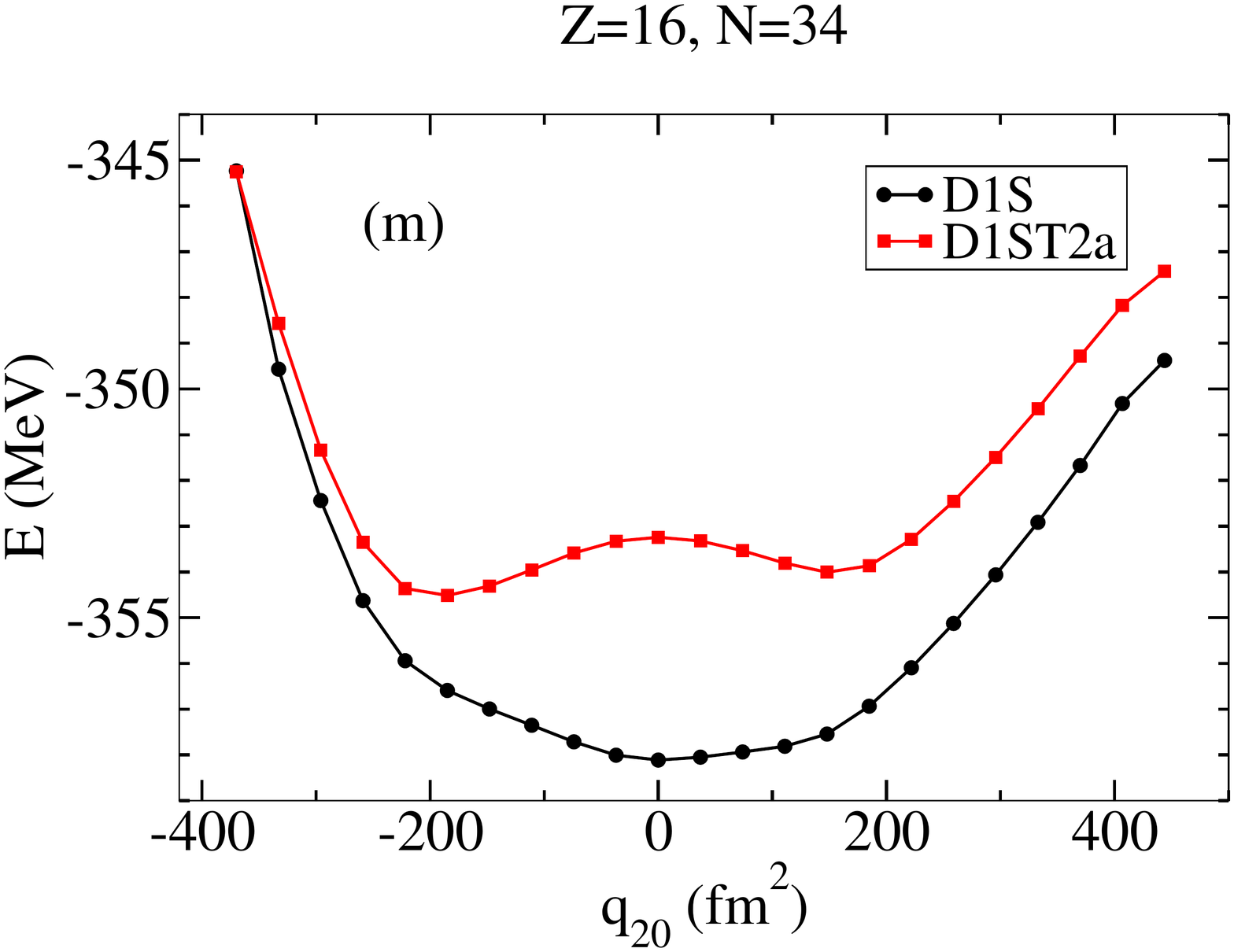} 
\end{tabular} 
\caption{Potential Energy Curves of the $Z=16$ chain 
for D1S (black circles) and D1ST2a (red squares) interactions. } 
\label{PESS}
\end{figure} 

We present the Sulfur PEC in Fig.~\ref{PESS} and the associated two first minima $Q_0$ 
obtained with D1S and D1ST2a in Table~\ref{TableS2}.
In this chain there is an important impact of the tensor term on the PEC. 
D1ST2a modifies the ground state location for several nuclei.
With the D1S interaction all the isotopes are found to be either spherical or prolate.
For the lighest isotopes (up to $A=38$) and the two last ones ($A=48,\,50$) the D1S curve describes a flat well. 
Two of them are prolate ($A=28,\,32$) even if the well is centered around their nascent second minimum at $q_{20}=0$ fm$^ 2$, 
and the rest of the them are spherical. 
For $40\le A\le 46$ the D1S PEC gives two distinct oblate and prolate minima.

\begin{table}[htb] \centering
 \begin{tabular}{|C{2cm}|C{2.6cm}|C{2.6cm}||C{2.6cm}|C{2.6cm}|}
\hline        
    A       &    D1S gs   & D1S isomere & D1ST2a gs & D1ST2a isomere   \\
\hline 
   26       &   $ 0.00$   &             &  $+0.37$  &                  \\
\hline 
   28       &   $+0.54$   &   $ 0.00$   &  $+0.61$  &  $-0.49$         \\
\hline 
   30       &   $ 0.00$   &             &  $-0.58$  &  $+0.58$         \\
\hline 
   32       &   $+0.36$   &   $ 0.00$   &  $+0.54$  &  $-0.45$         \\
\hline 
   34       &   $ 0.00$   &             &  $-0.47$  &  $+0.31$         \\
\hline 
   36       &   $ 0.00$   &             &  $ 0.00$  &                  \\
\hline
   38       &   $+0.40$   &   $-0.10$   &  $+0.40$  &  $-0.20$         \\
\hline
   40       &   $+0.57$   &   $-0.20$   &  $+0.56$  &  $-0.31$         \\
\hline
   42       &   $+0.59$   &   $-0.40$   &  $+0.59$  &  $-0.51$         \\
\hline
   44       &   $+0.60$   &   $-0.50$   &  $-0.61$  &  $+0.67$         \\
\hline
   46       &   $+0.44$   &   $-0.41$   &  $-0.54$  &  $+0.54$         \\
\hline
   48       &   $ 0.00$   &             &  $-0.55$  &  $+0.57$         \\
\hline
   50       &   $ 0.00$   &             &  $-0.56$  &  $+0.52$         \\
\hline
\end{tabular}

\caption{Theoretical $Q_0$(eb) for the $Z=16$ chain.  }  
\label{TableS2} 
\end{table}

Using the D1ST2a interaction, we observe significative changes in the PEC. It tends to eliminate the one-minimum flat well produced by D1S to a two-minima landscape. 
Two of the lightest isotopes become prolate with the D1ST2a interaction: $^ {30}$S and  $^ {34}$S. Besides, the ground state location is changed for the 
four last isotopes: all of them become oblate with the D1ST2a interaction from a prolate ($A=44,\, 46$) or spherical ($A=48,\, 50$) D1S configuration. 
Moreover, $^{36}$S is a special case because it is neutron spin-saturated. As expected here the tensor energy becomes negligible in its spherical ground state. 

\begin{figure}[htb] \centering
\begin{tabular}{ccc}
   \includegraphics[width=4.6cm]{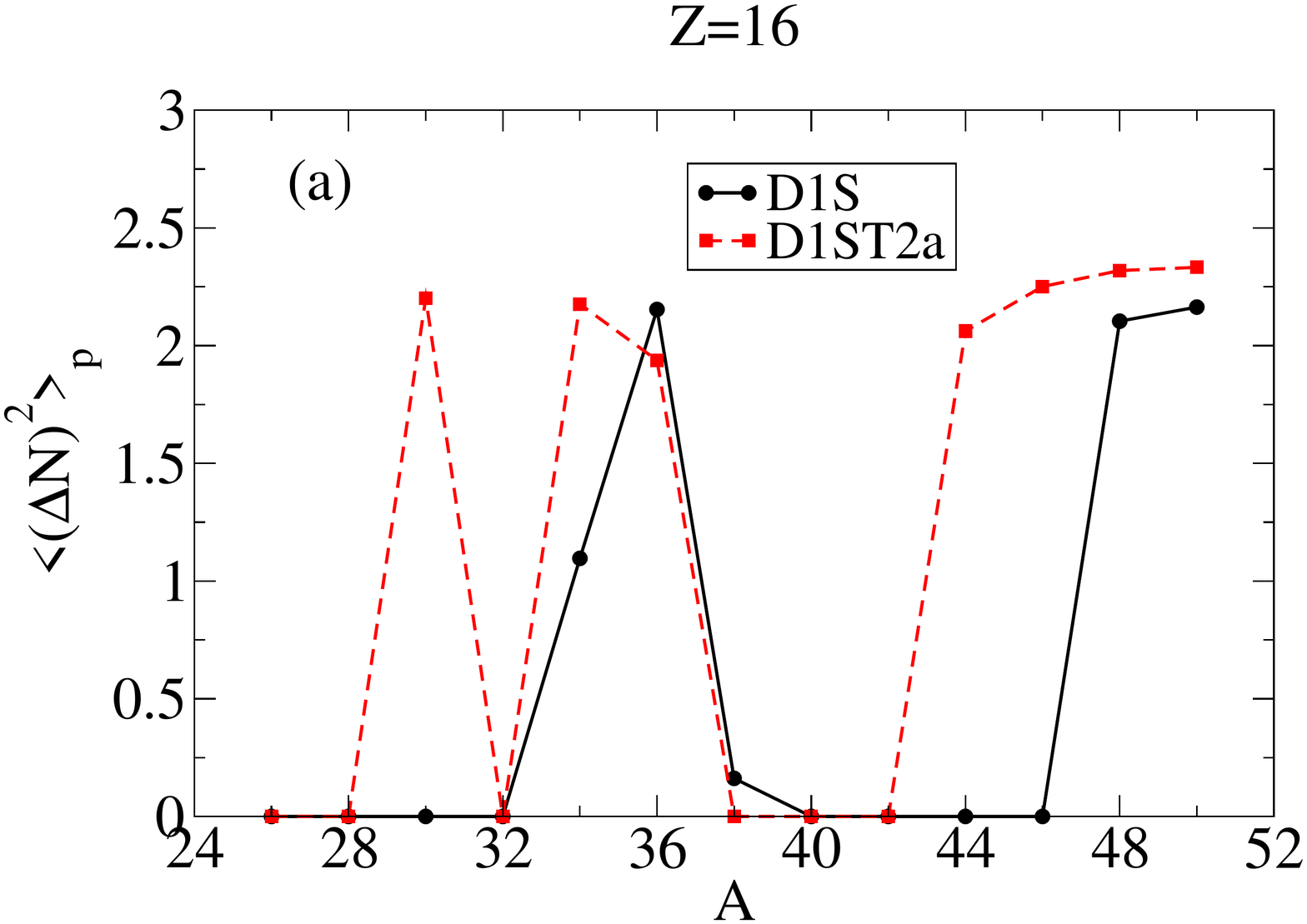} &
   \includegraphics[width=4.6cm]{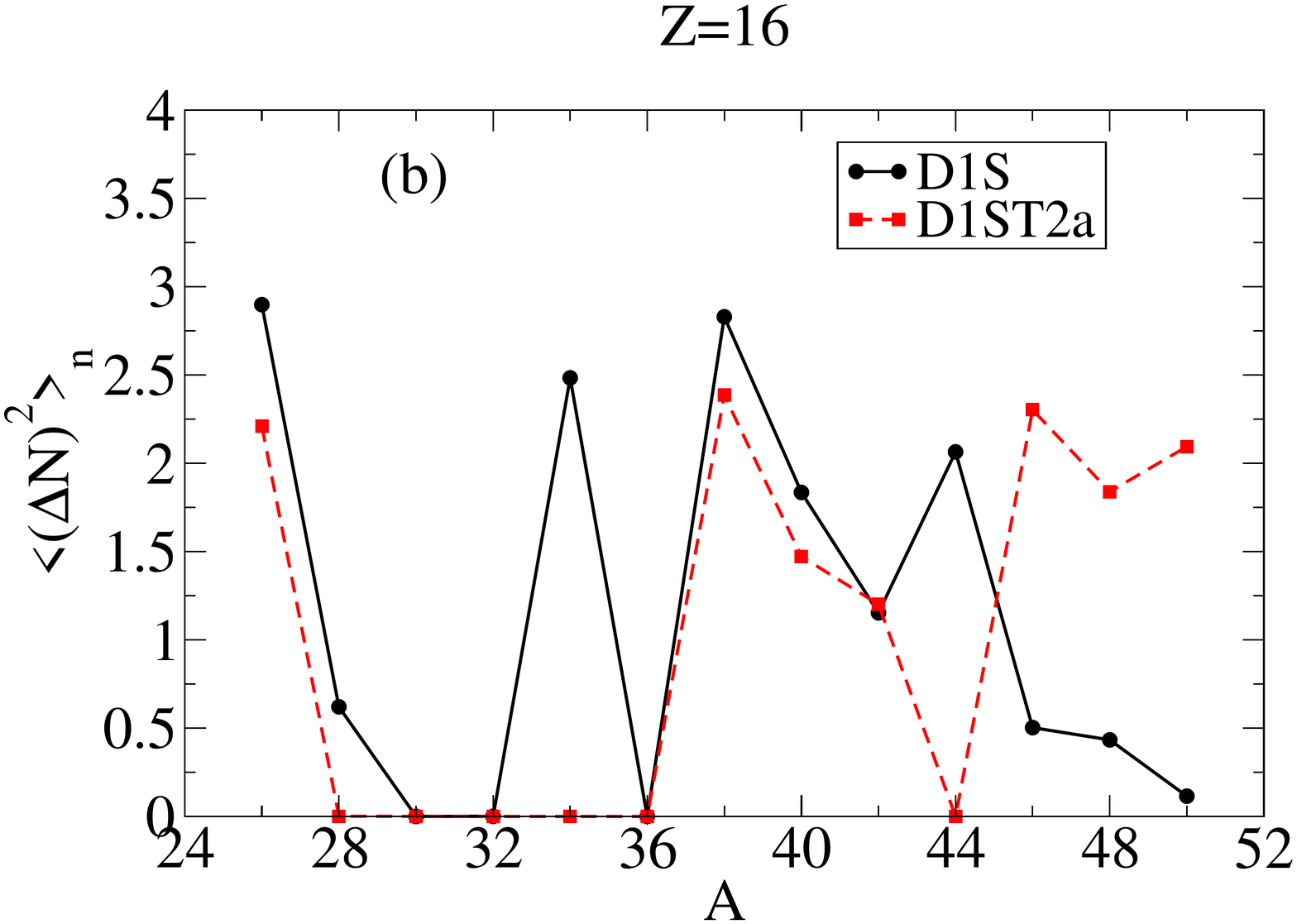} &
   \includegraphics[width=4.6cm]{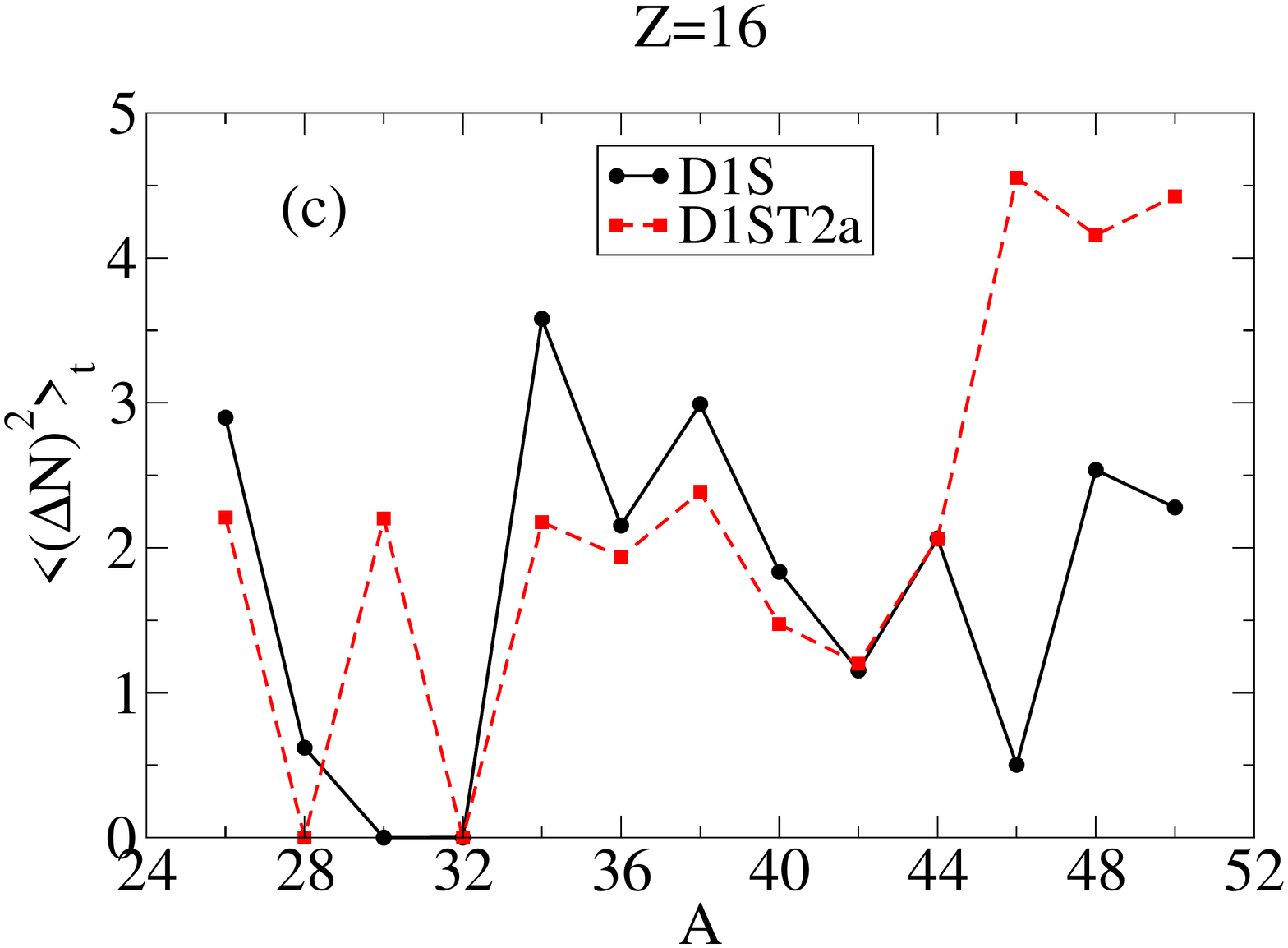} 
\end{tabular}
\caption{Proton, neutron and total particle number fluctuation for the $Z=16$ chain obtained with the D1S interaction (black circles) and with the D1ST2a one (red squares).} 
\label{SDN2}
\end{figure}

\begin{figure}[htb] \centering
\begin{tabular}{c}
   \includegraphics[width=4.8cm]{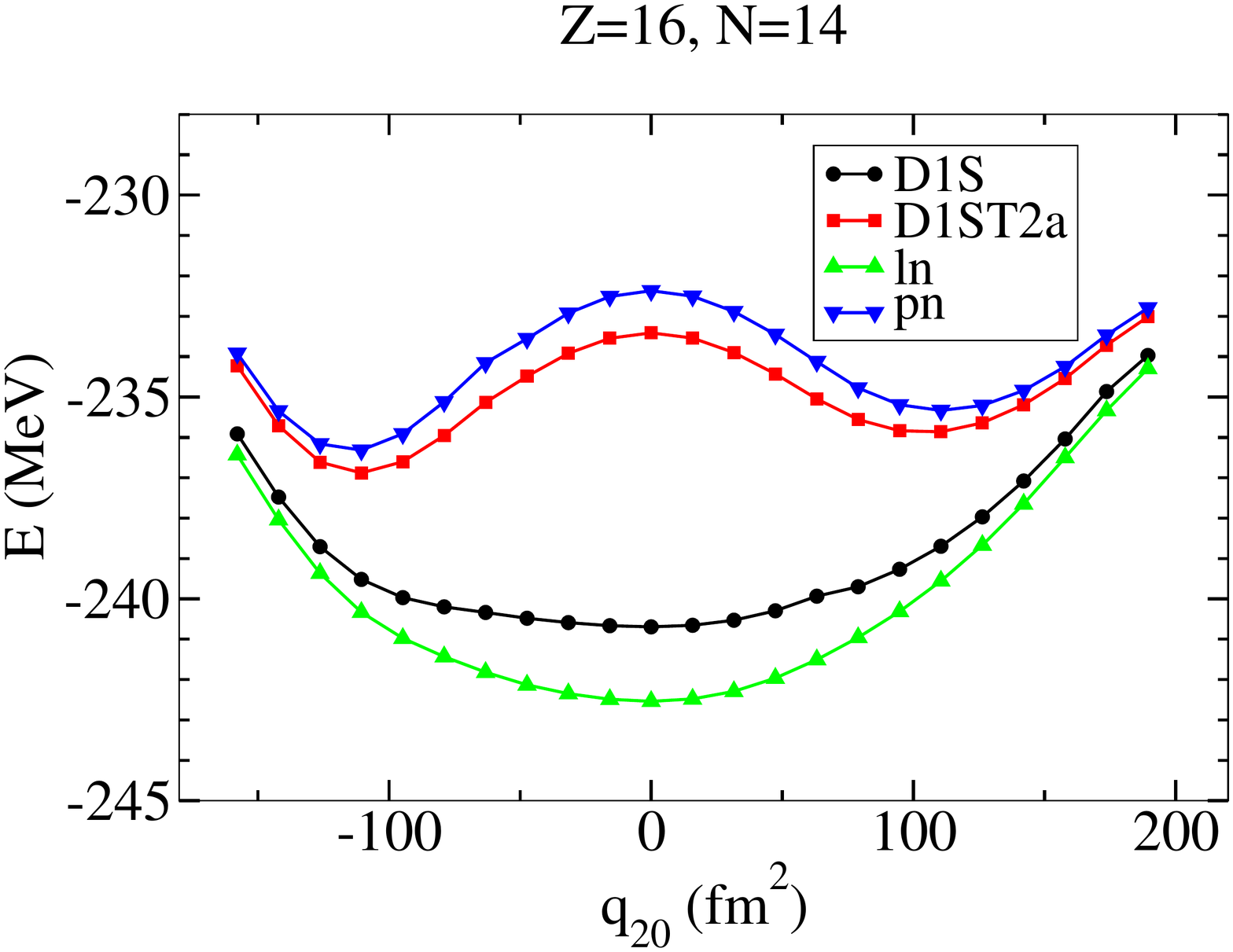} 
\end{tabular}
\caption{Like-particle and proton-neutron contributions of the D1ST2a interaction for $^{30}$S.} 
\label{Slnpnp}
\end{figure}

In Fig.~\ref{SDN2} we show the particle number fluctuations of the Sulfur isotopes. The main differences are found where the ground state location are different 
for both proton and neutron fluctuations. For the total particle number fluctuations, the behaviour is more or less the same up to $A=44$, that is, 
the D1ST2a interaction slightly lowering the fluctuations when the ground state shapes are similar, except for the nucleus $^{30}$S,  for which remains the difference 
obtained for the proton case. For this  latter nucleus, we present in Fig.~\ref{Slnpnp} the PEC, with  the lp (green triangles up)  
and pn (blue triangles down) tensor contributions separately. We see again that the main contribution is coming from the pn part, because it is very similar to the 
result using the full D1ST2a interaction (red squares).


\begin{figure}[htb] \centering
\begin{tabular}{cc}
   \includegraphics[width=4.8cm]{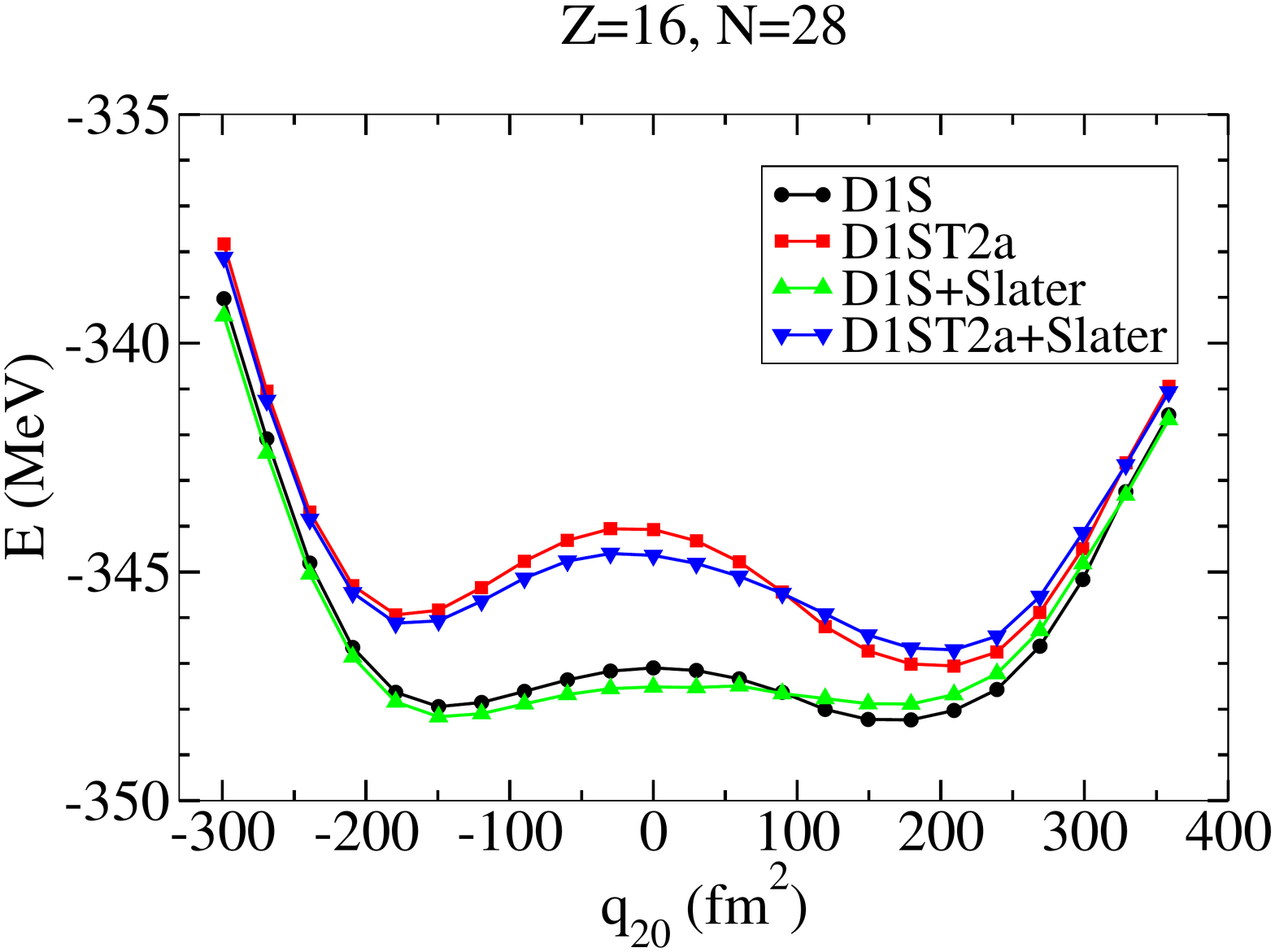} 
\end{tabular}
\caption{Potential Energy Curves for $^{44}$S  using the exact Coulomb for the D1S interaction with the exact Coulomb (black circles)  
or Slater approximation (green triangles up). The same for the D1ST2a interaction, exact Coulomb (red squares) and Slater approximation (blue triangles down).} 
\label{44SPES}
\end{figure}

\begin{figure}[htb] \centering
\begin{tabular}{cc}
   \includegraphics[width=4.6cm]{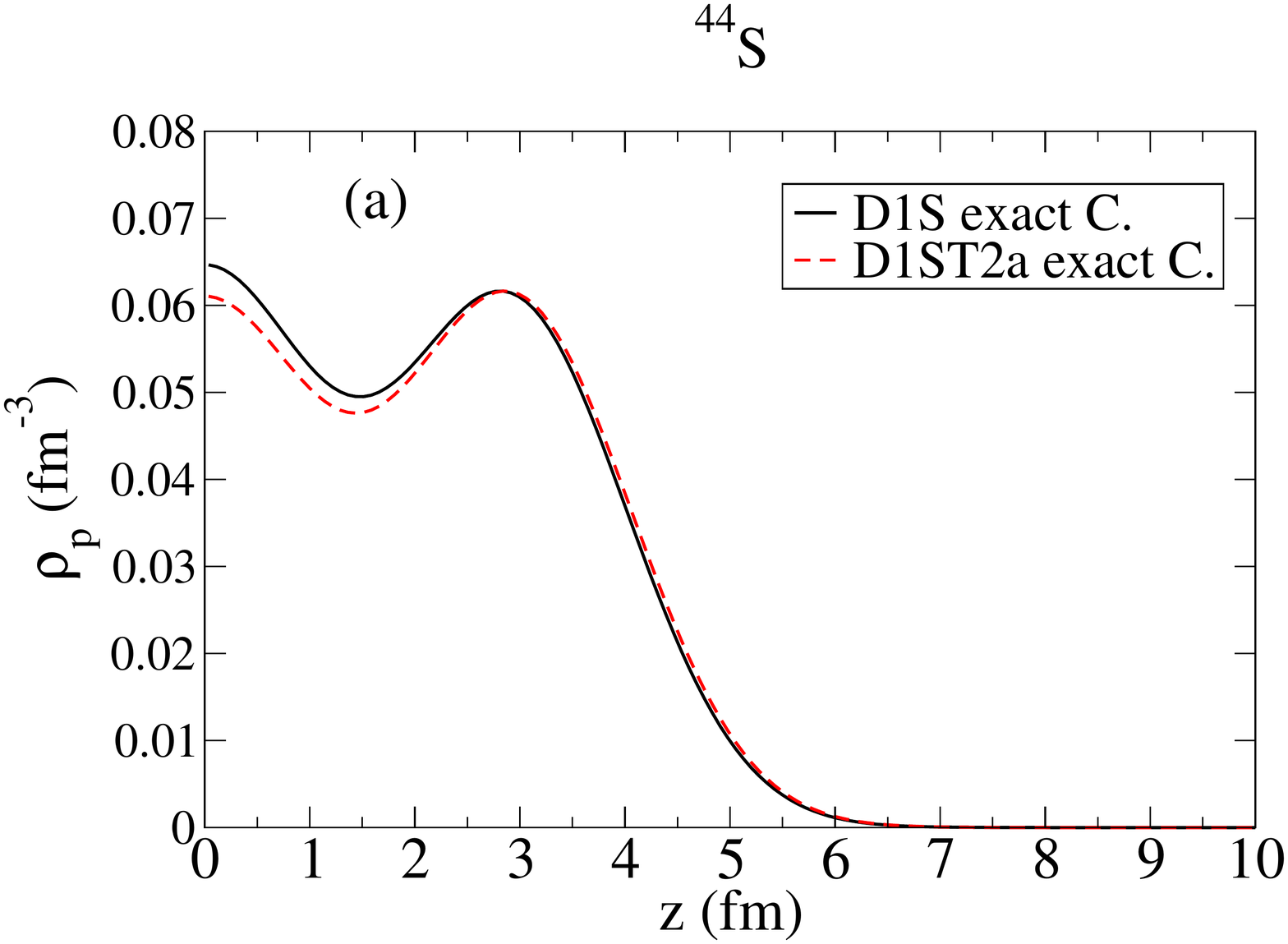} & 
   \includegraphics[width=4.6cm]{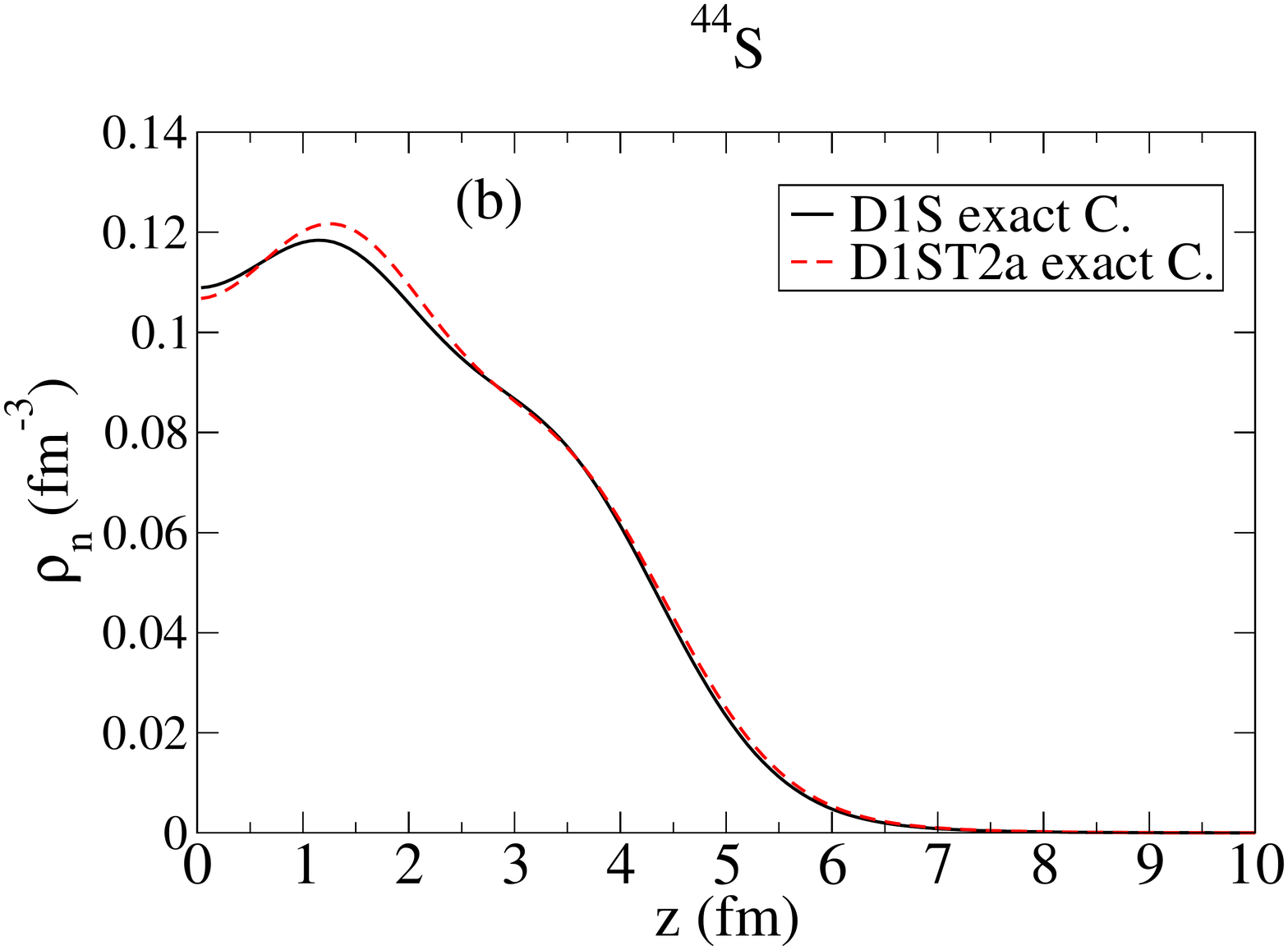} \\
   \includegraphics[width=4.6cm]{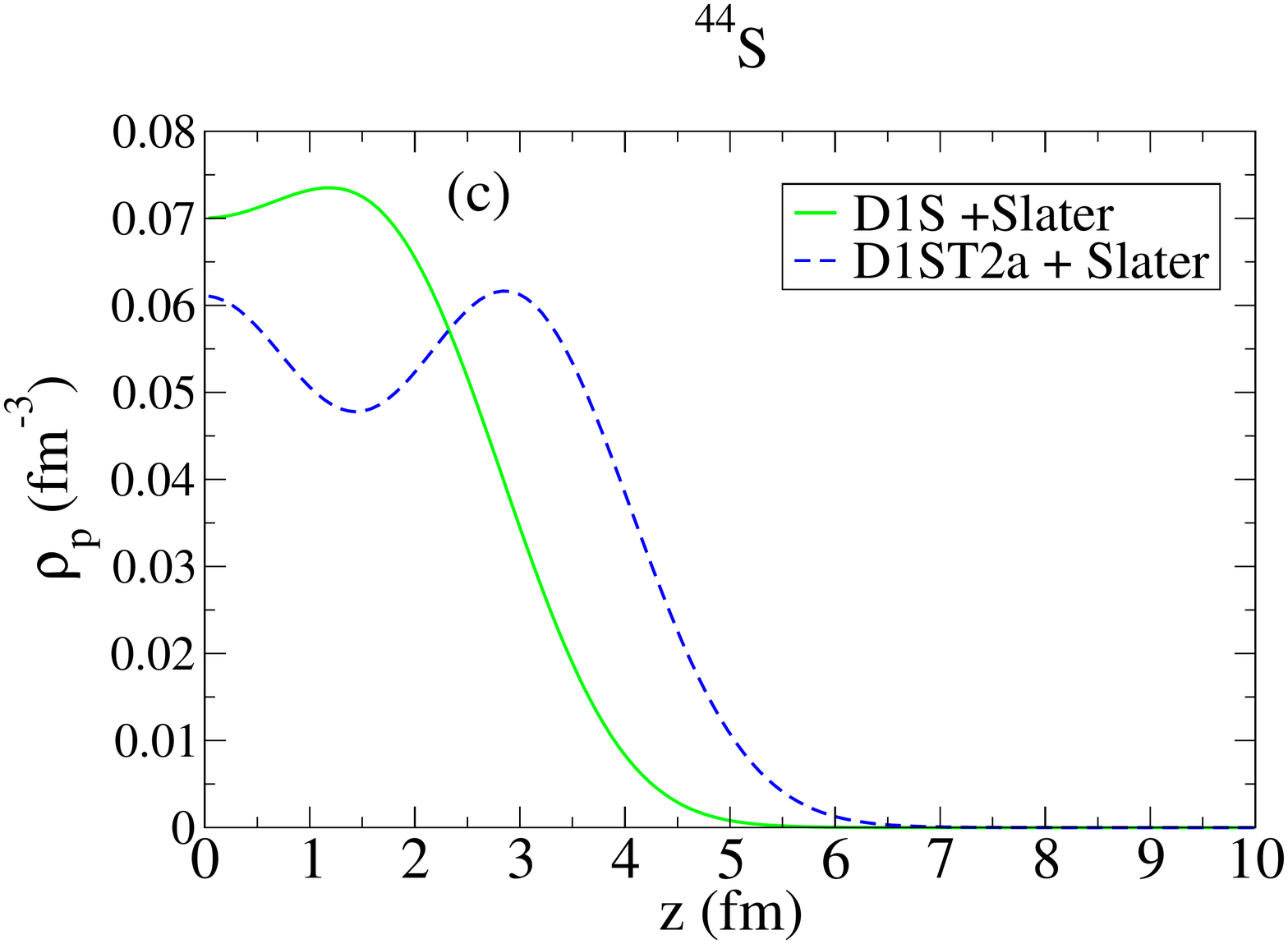} & 
   \includegraphics[width=4.6cm]{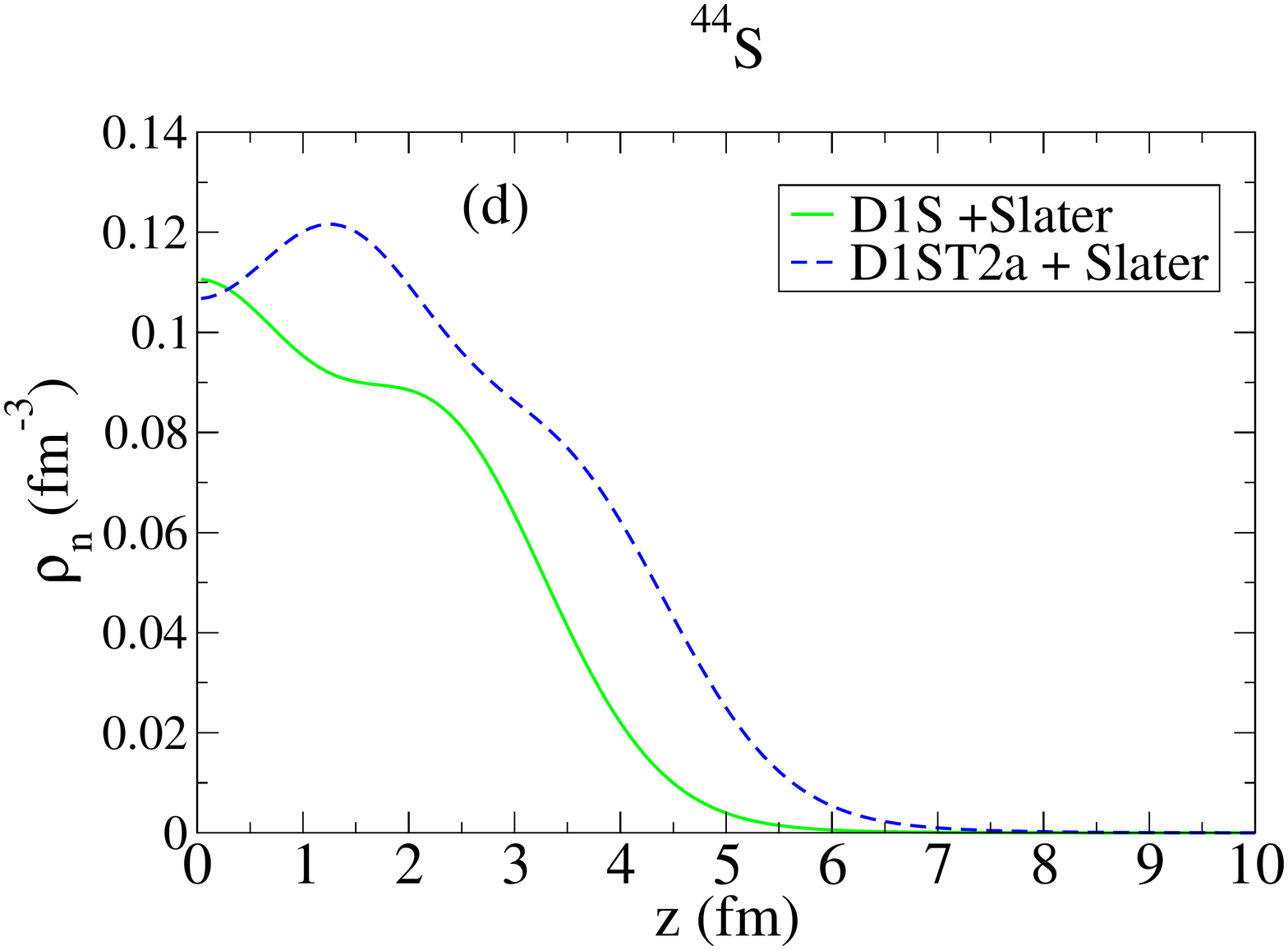} \\
\end{tabular}
\caption{Proton and neutron densities for $^{44}$S for the D1S interaction (black circles) and the D1ST2a one (red squares). 
The results for the D1S interaction using the Slater approximation for the Coulomb interaction are shown by green triangles up and those for 
the D1ST2a one by blue triangles down.} 
\label{SDenspn}
\end{figure}

In Fig.~\ref{44SPES} we show the PEC for the nucleus $^{44}$S, using both interactions and considering or not the Slater approximation for 
the exchange Coulomb term. We see that the ground state does not change in the case of the D1ST2a interaction, but considering the D1S one, the ground 
state is prolate for the exact treatment of the exchange Coulomb term and oblate when the Slater approximation is used. Then, this is the reason to 
obtain different proton and neutron distributions for both interactions when Slater approximation is considered in the calculations, as we show in 
Fig.~\ref{SDenspn}, lower panels.   


\clearpage

\subsection{Argon chain}

\begin{figure}[htb] \centering
\begin{tabular}{ccc}
   \includegraphics[width=4.6cm]{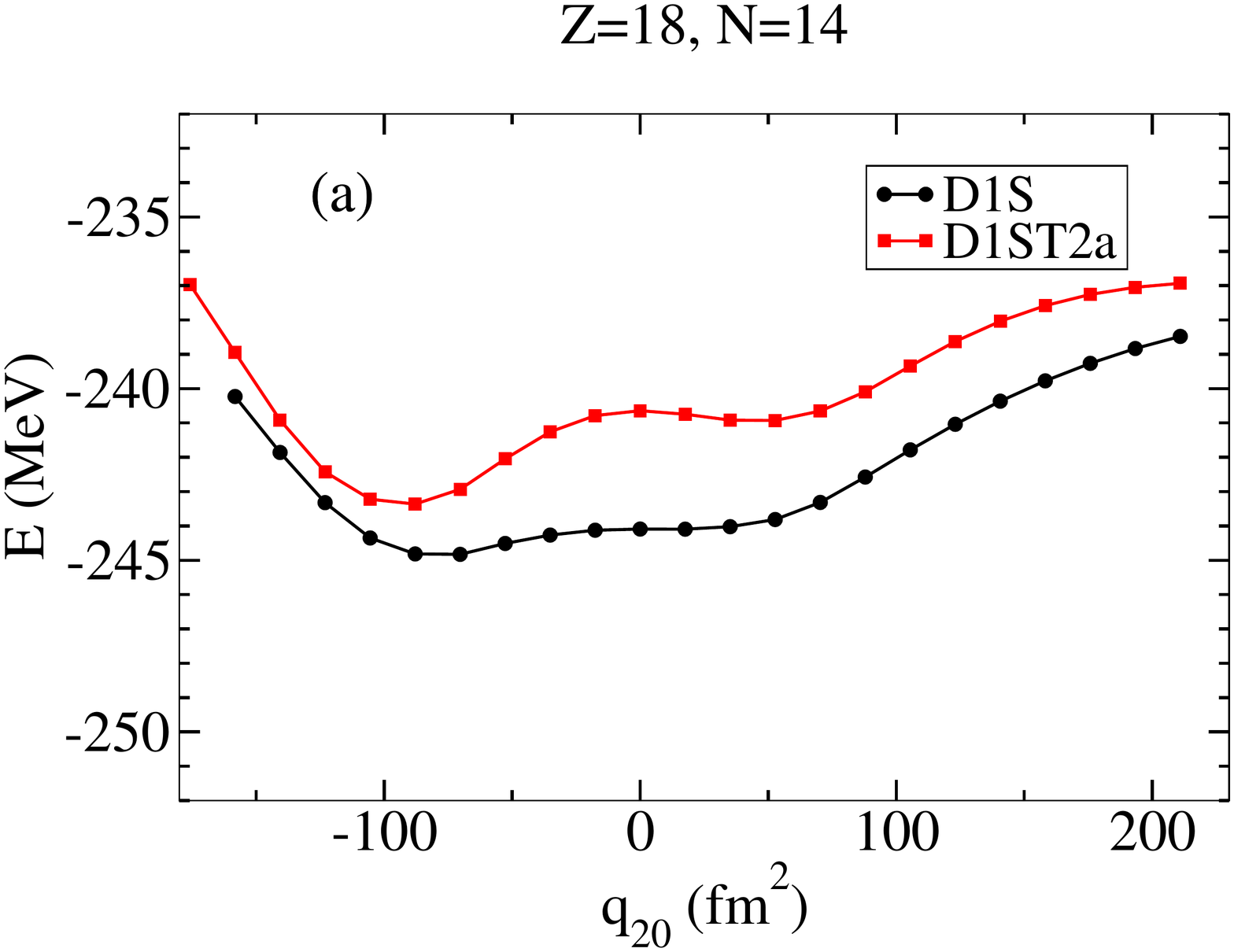} &
   \includegraphics[width=4.6cm]{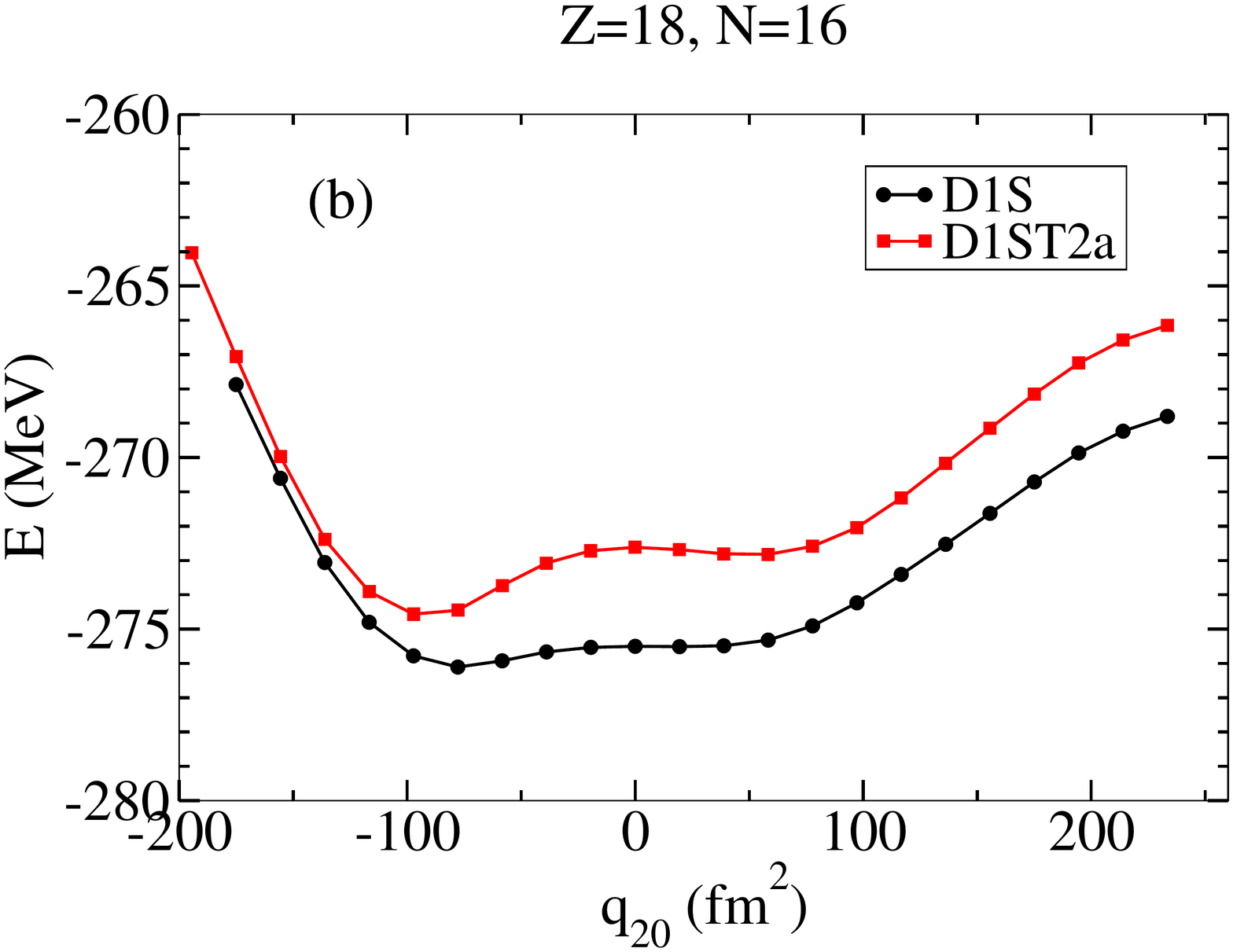} &
   \includegraphics[width=4.6cm]{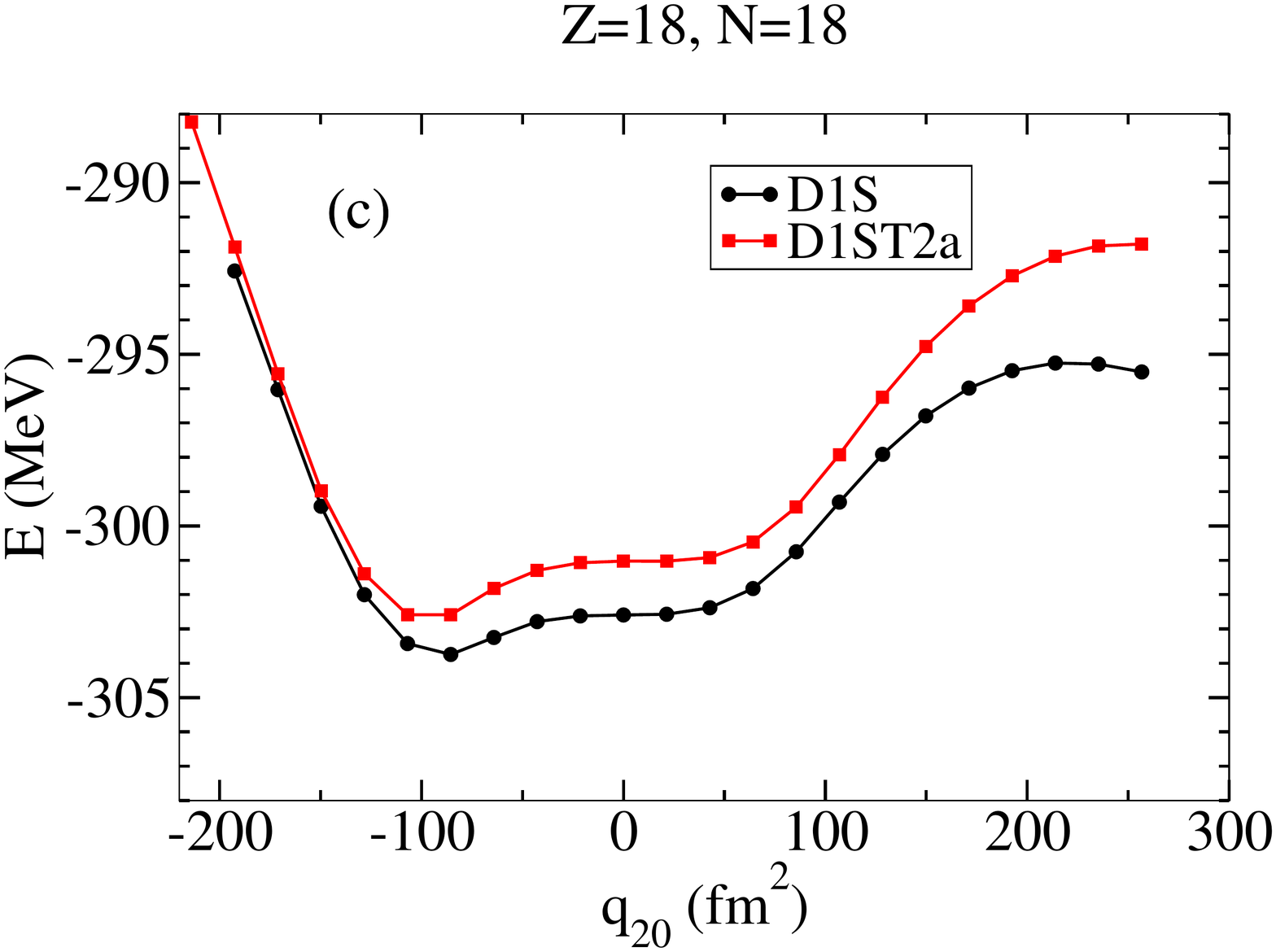} \\
   \includegraphics[width=4.6cm]{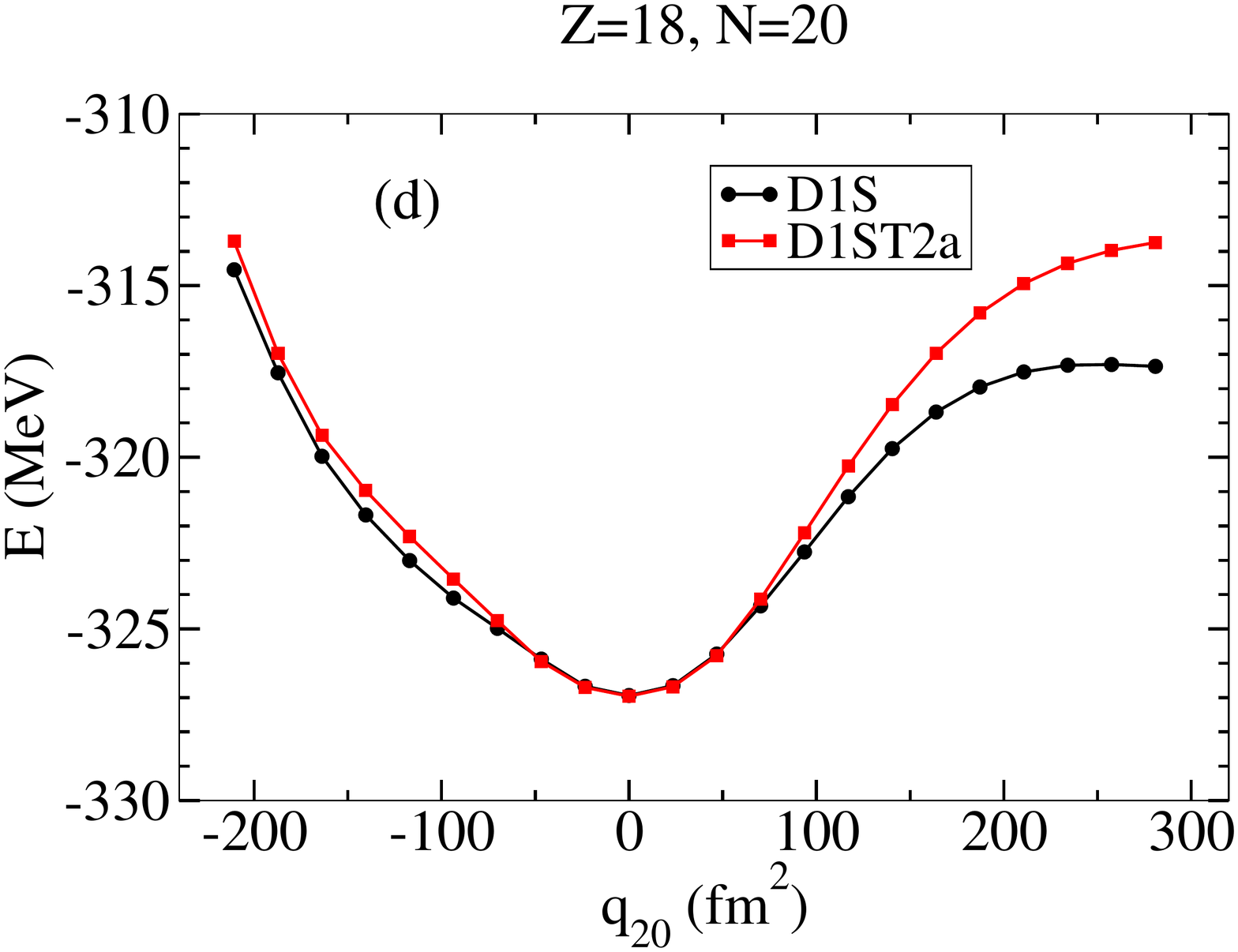} &
   \includegraphics[width=4.6cm]{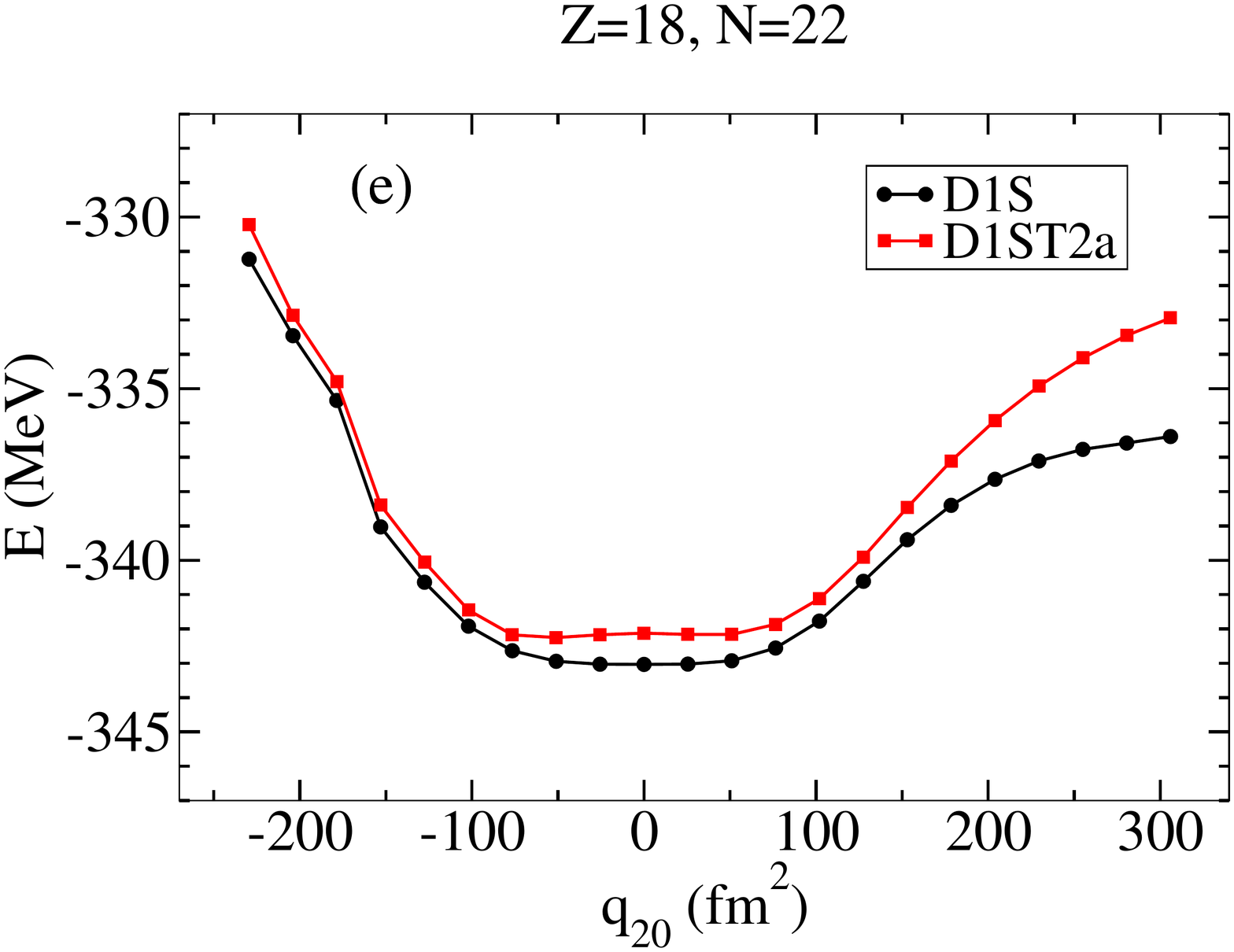} &
   \includegraphics[width=4.6cm]{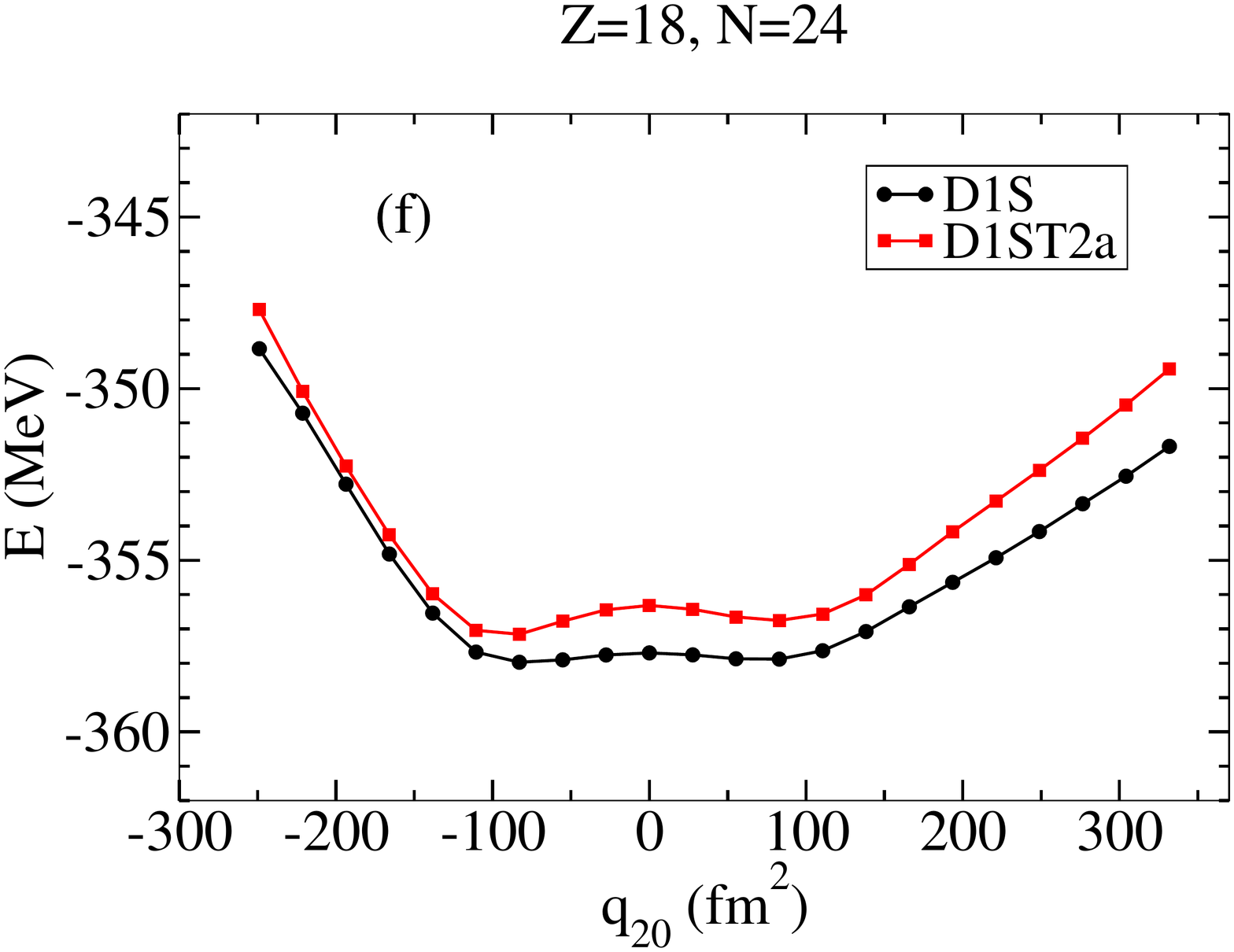} \\
   \includegraphics[width=4.6cm]{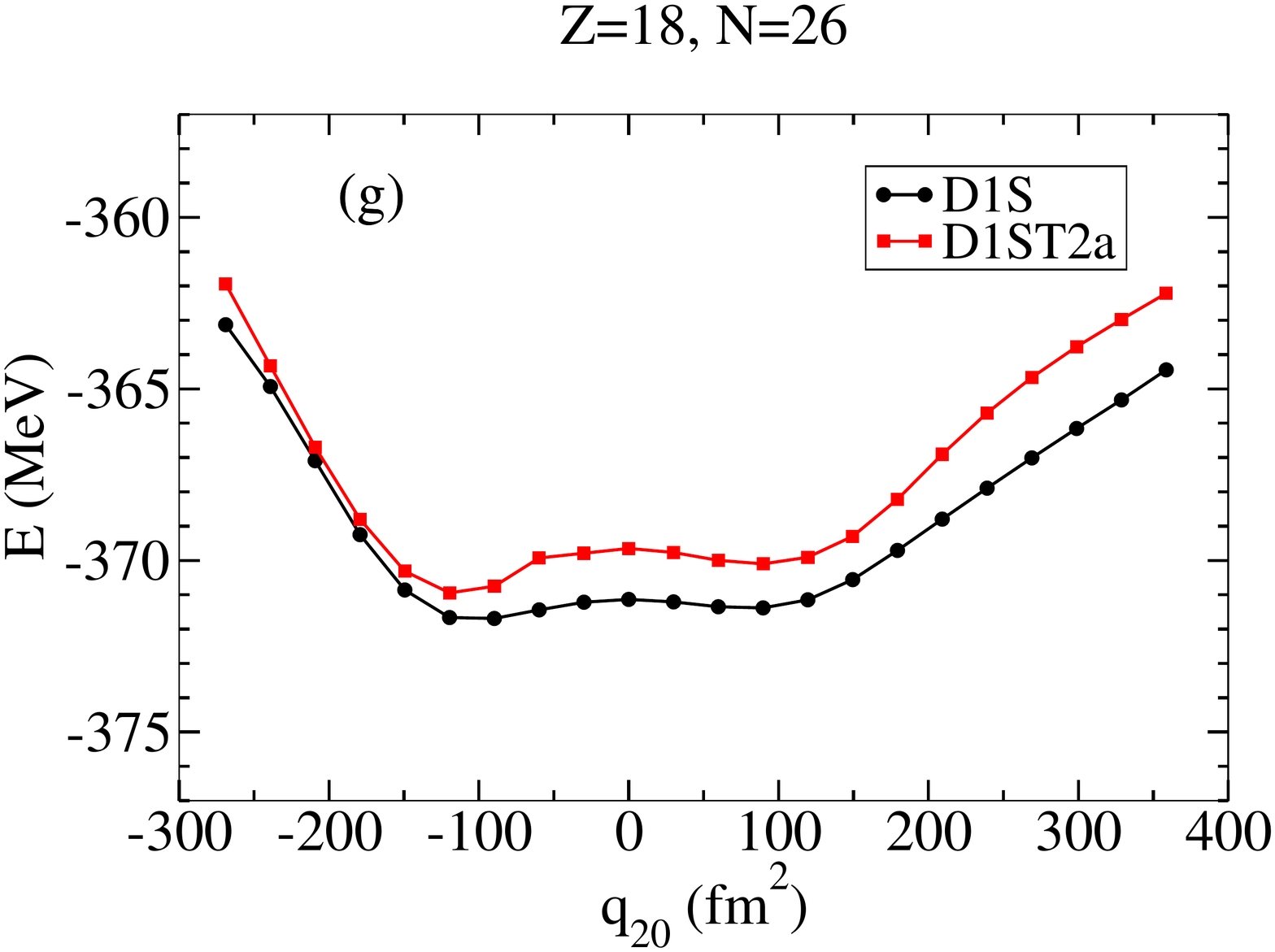} &
   \includegraphics[width=4.6cm]{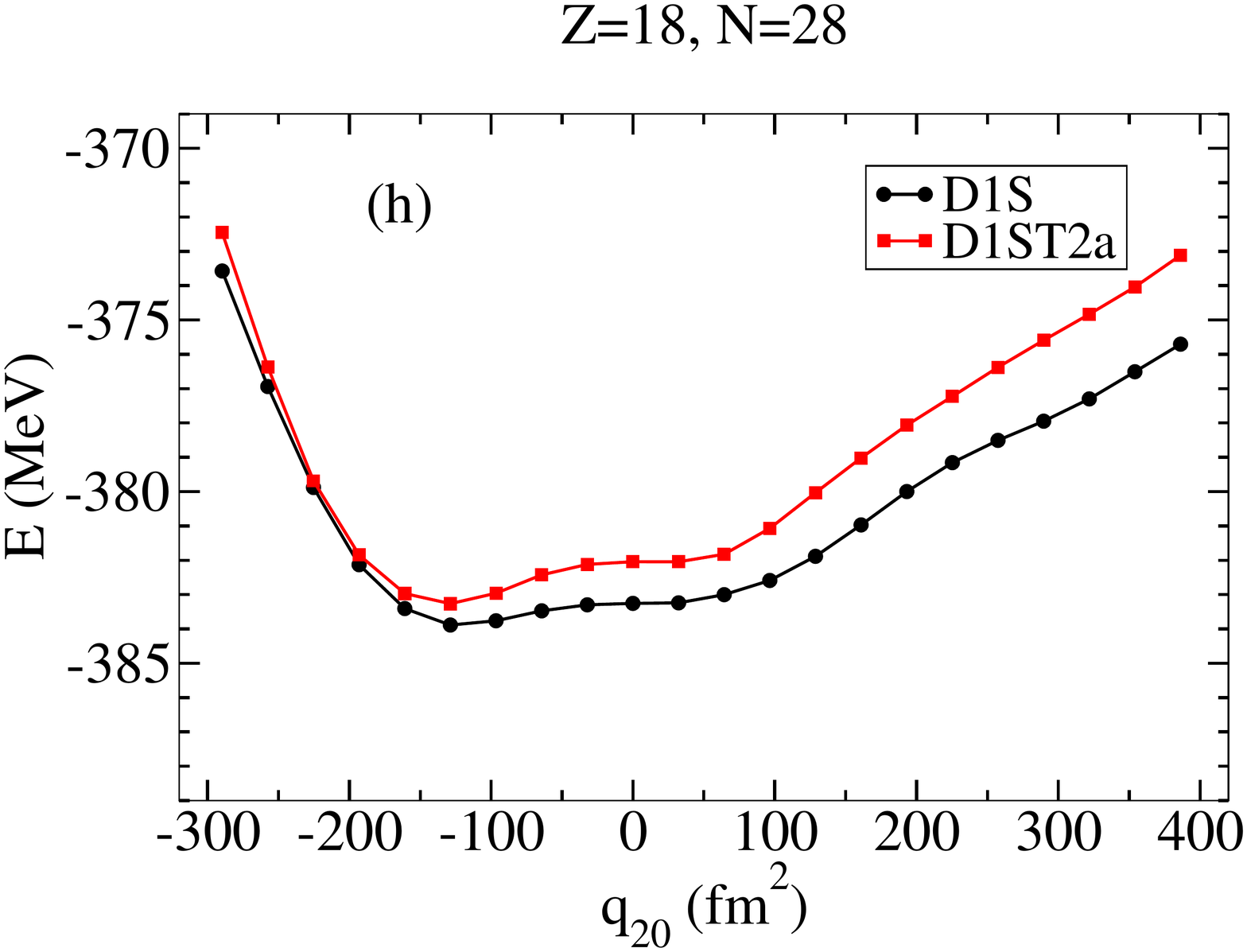} &
   \includegraphics[width=4.6cm]{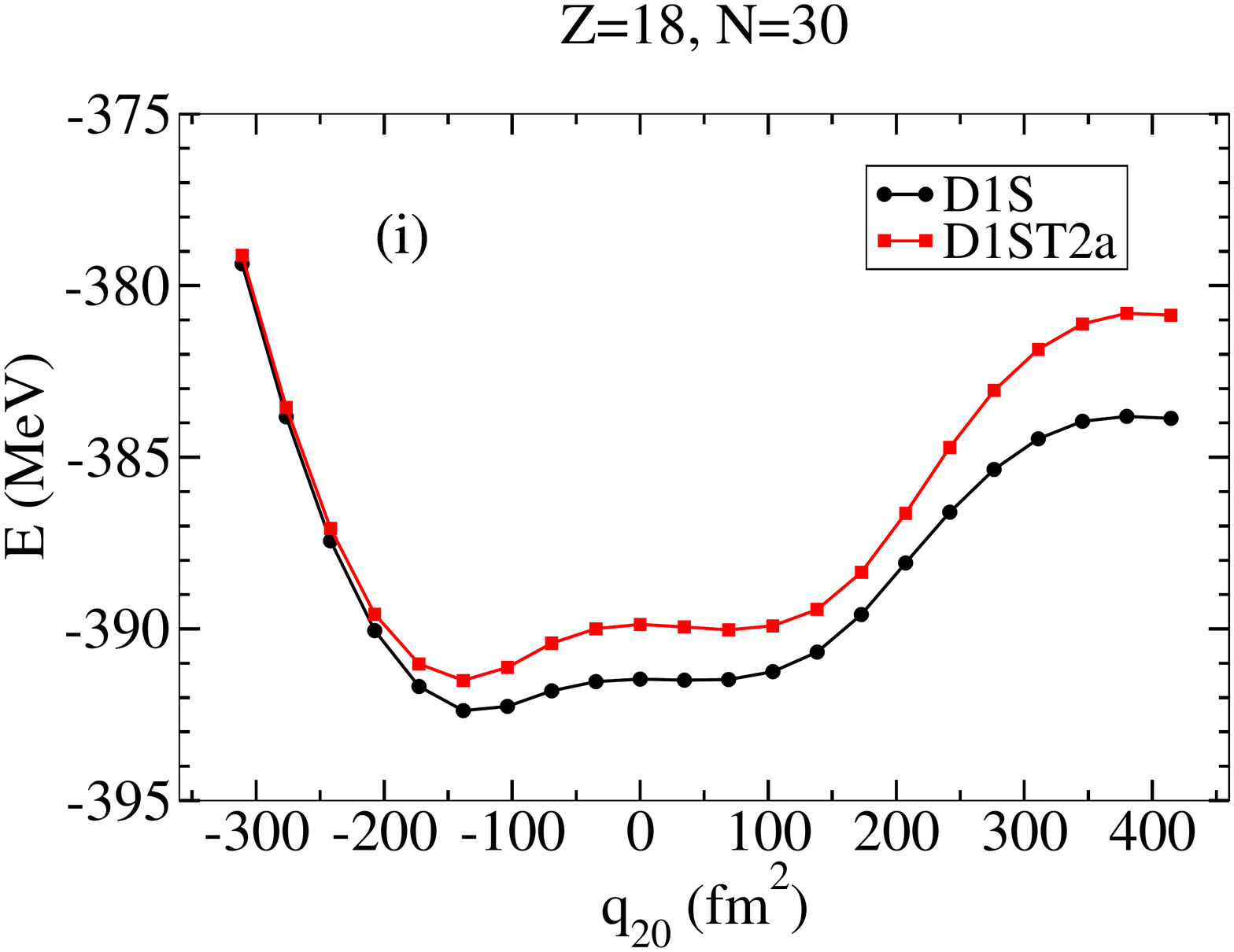} \\
   \includegraphics[width=4.6cm]{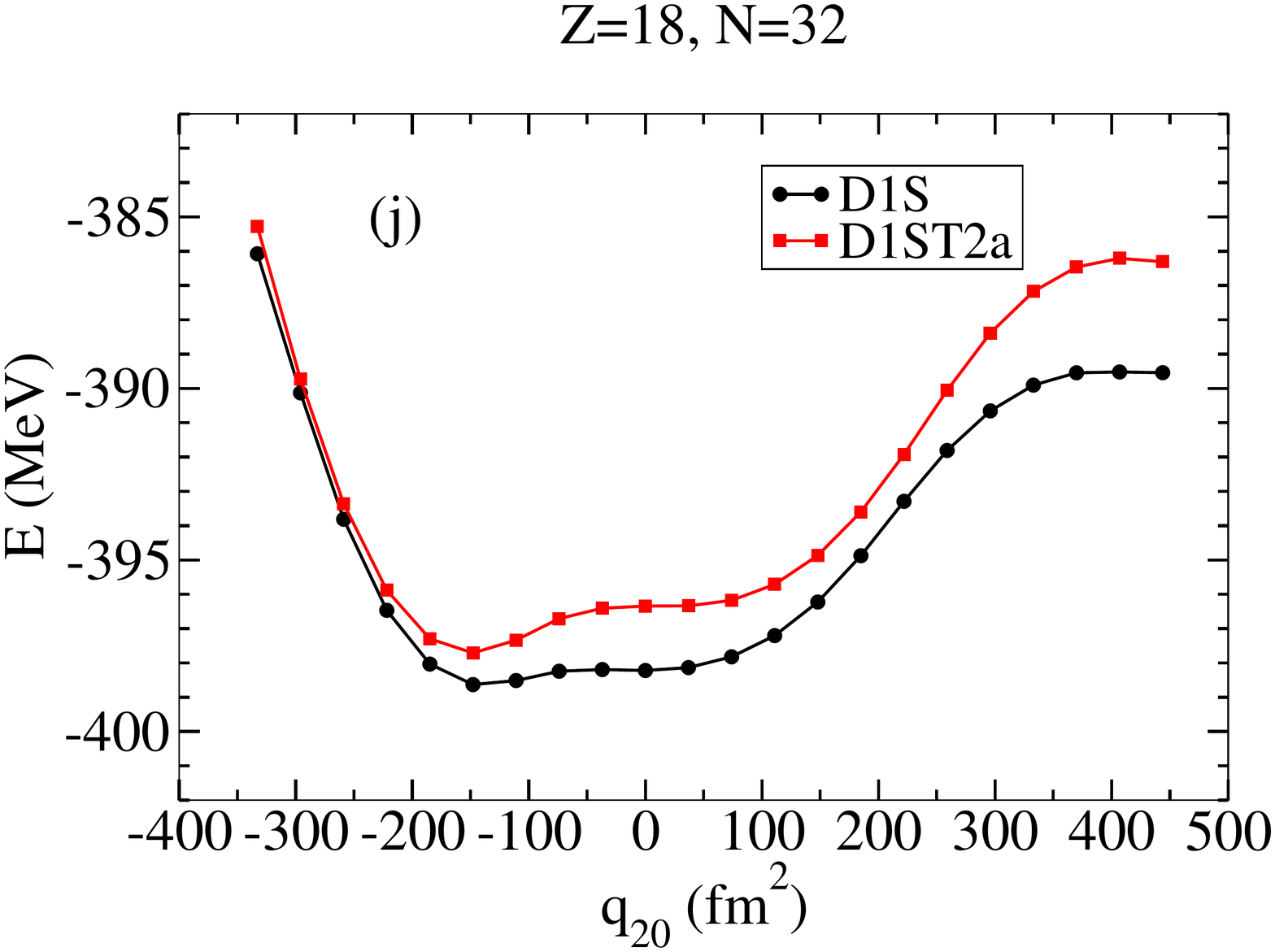} &
    & \\
\end{tabular}
\caption{Potential Energy Curves of the $Z=18$ chain 
for the D1S interaction (black circles) and the D1ST2a one (red squares). } 
\label{figAr}
\end{figure} 

In Fig.~\ref{figAr} we plot the PEC for Argon isotopes. We can see that the effect of  tensor interaction onto these curves is not very important. 
It does not change the ground state and the possible isomere.
It is worth to note the nascent two-minima well for D1ST2a in $^{40}$Ar which arises for D1S from $A\ge 42$.

\begin{table}[htb] \centering
 \begin{tabular}{|C{2cm}|C{2.6cm}|C{2.6cm}||C{2.6cm}|C{2.6cm}|}
\hline        
    A       &    D1S gs   & D1S isomere & D1ST2a gs & D1ST2a isomere   \\
\hline 
   32       &   $-0.39$   &   $+0.10$   &  $-0.49$  &  $+0.28$         \\
\hline 
   34       &   $-0.42$   &   $+0.10$   &  $-0.52$  &  $+0.29$         \\
\hline 
   36       &   $-0.43$   &             &  $-0.43$  &  $+0.11$         \\
\hline
   38       &   $ 0.00$   &             &  $ 0.00$  &                  \\
\hline
   40       &   $ 0.00$   &             &  $-0.25$  &  $+0.11$         \\
\hline
   42       &   $-0.38$   &   $+0.31$   &  $-0.37$  &  $+0.30$         \\
\hline
   44       &   $-0.38$   &   $+0.32$   &  $-0.48$  &  $+0.32$         \\
\hline
   46       &   $-0.48$   &             &  $-0.48$  &  $+0.13$         \\
\hline
   48       &   $-0.49$   &   $+0.12$   &  $-0.50$  &  $+0.23$         \\
\hline
   50       &   $-0.49$   &   $ 0.00$   &  $-0.51$  &                  \\
\hline
\end{tabular}
\caption{Theoretical $Q_0$ (eb) for the $Z=18$ chain.  }  
\label{TableAr2} 
\end{table}

For $^{36}$Ar both interactions predict an oblate ground state.
In Table~\ref{TableAr} we see that the spectroscopic quadrupole moment gives an oblate
shape for $^{36}$Ar only when $K=0$. The experimental $Q_0$ value $-0.39$ eb is then very close to the
theoretical ones (both at $-0.43$ eb). For $^{40}$Ar the experimental $Q_0$ gives an  spherical ground state, as predicted by the D1S interaction. 
With the D1ST2a one, the ground state is slightly prolate. However for this nucleus the D1ST2a well is rather flat and centered around sphericity as the D1S one.
Thus the ground state shape is sensitive to any modification of the PEC.

\begin{table}[htb] \centering
 \begin{tabular}{|c|c|c|c|c|}
\hline
          & Exp. $Q_S(I=2)$    & Exp. $Q_0(K=2)$  & Exp. $Q_0(K=1)$ & Exp. $Q_0(K=0)$ \\
\hline
$^{36}$Ar &  $+0.11(6)$        & $+0.39$          & $+0.19$ 	    & $-0.39$ 	    \\
\hline
$^{40}$Ar &  $+0.01(4)$        & $+0.04$          & $+0.07$ 	    & $-0.04$ 	    \\
\hline
 \end{tabular}
\caption{Experimental $Q_S(I=2)$ (eb) (from~\cite{StoADN05}) for the $Z=18$ chain. Corresponding $Q_0$ are given in eb assuming  $K=2$, $K=1$ and $K=0$.} 
\label{TableAr} 
\end{table}

\clearpage

\subsection{Strontium chain}
\begin{figure}[htb] \centering
\begin{tabular}{ccc}
   \includegraphics[width=4.6cm]{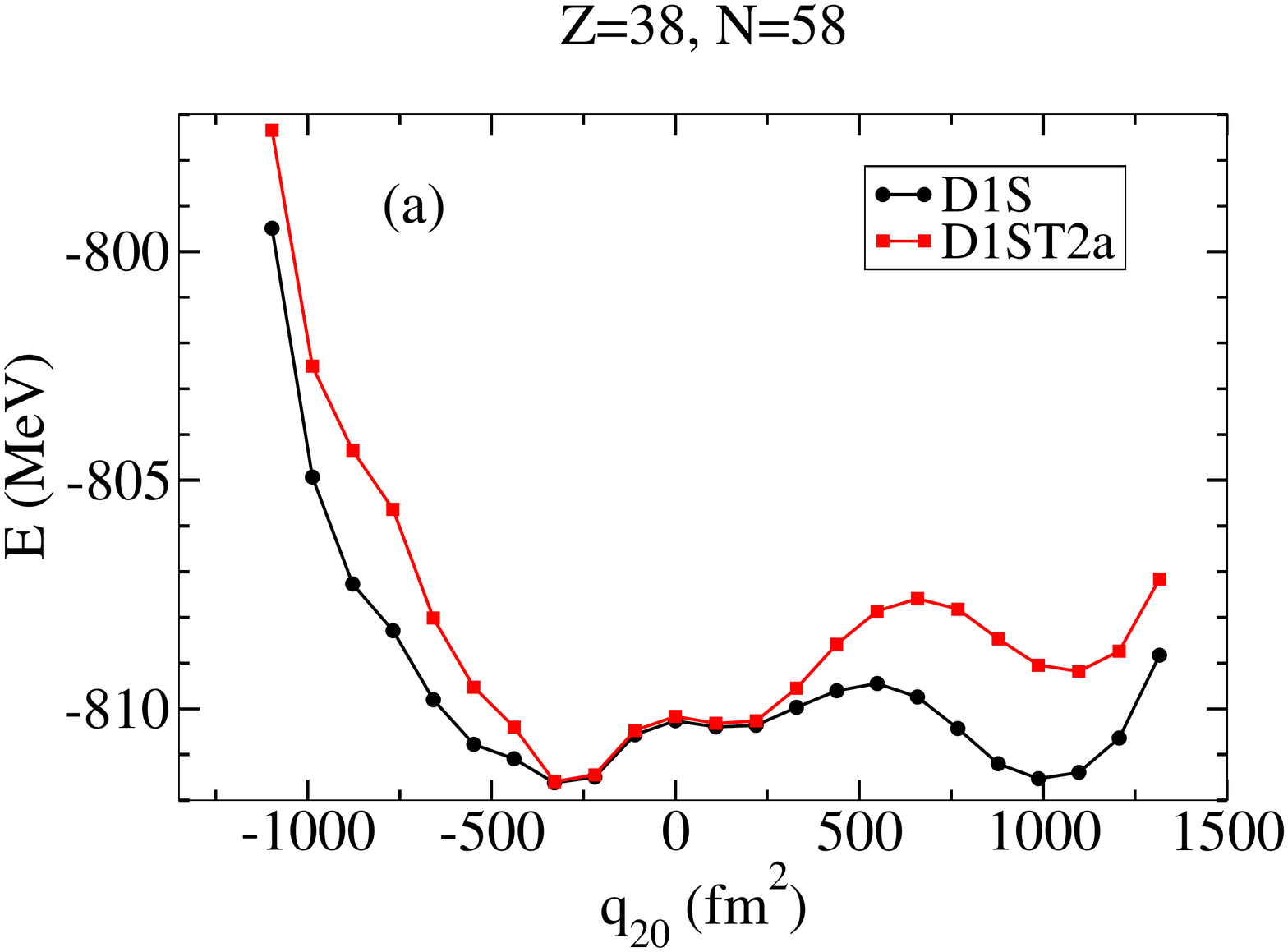} &
   \includegraphics[width=4.6cm]{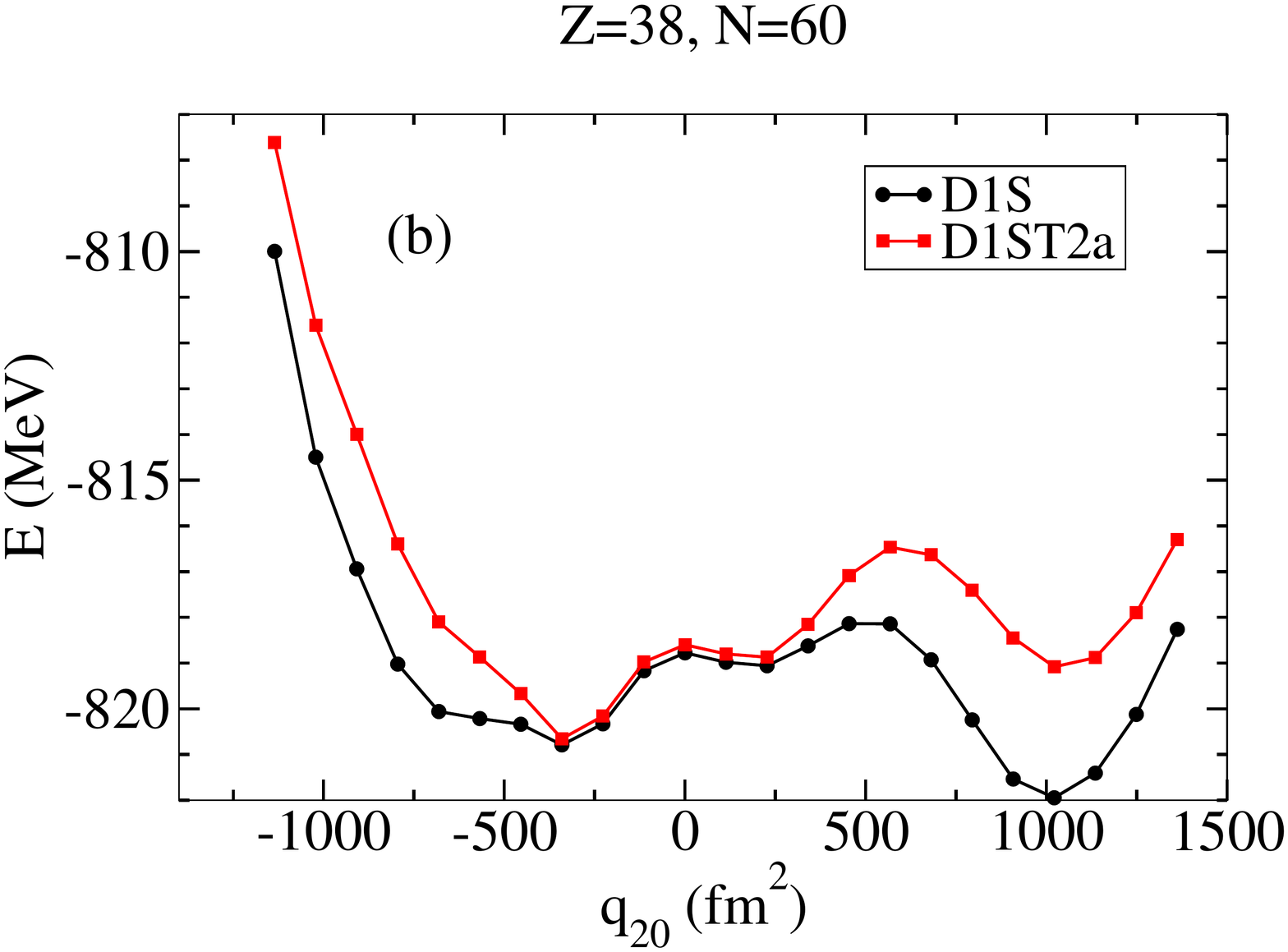} &
   \includegraphics[width=4.6cm]{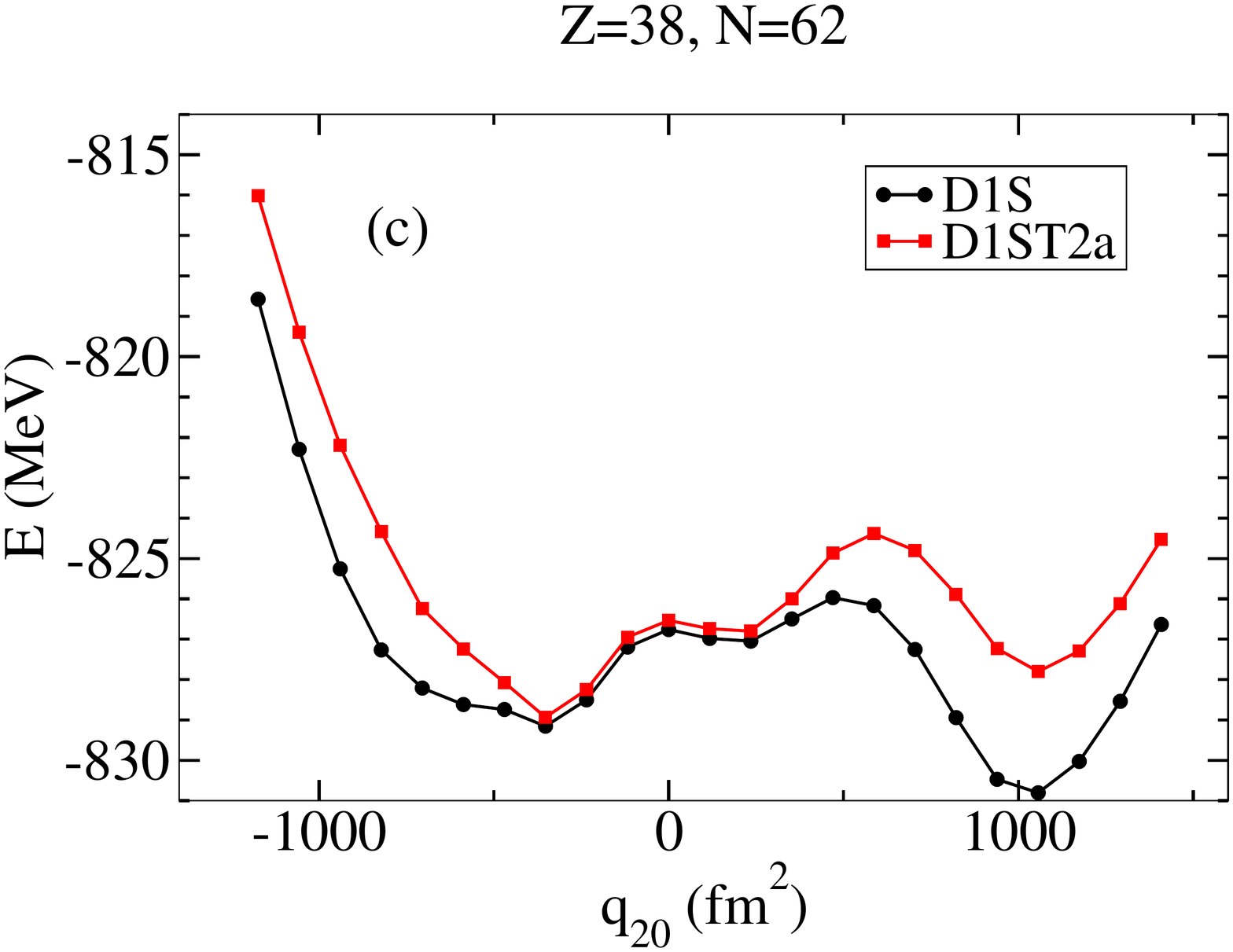} \\
   \includegraphics[width=4.6cm]{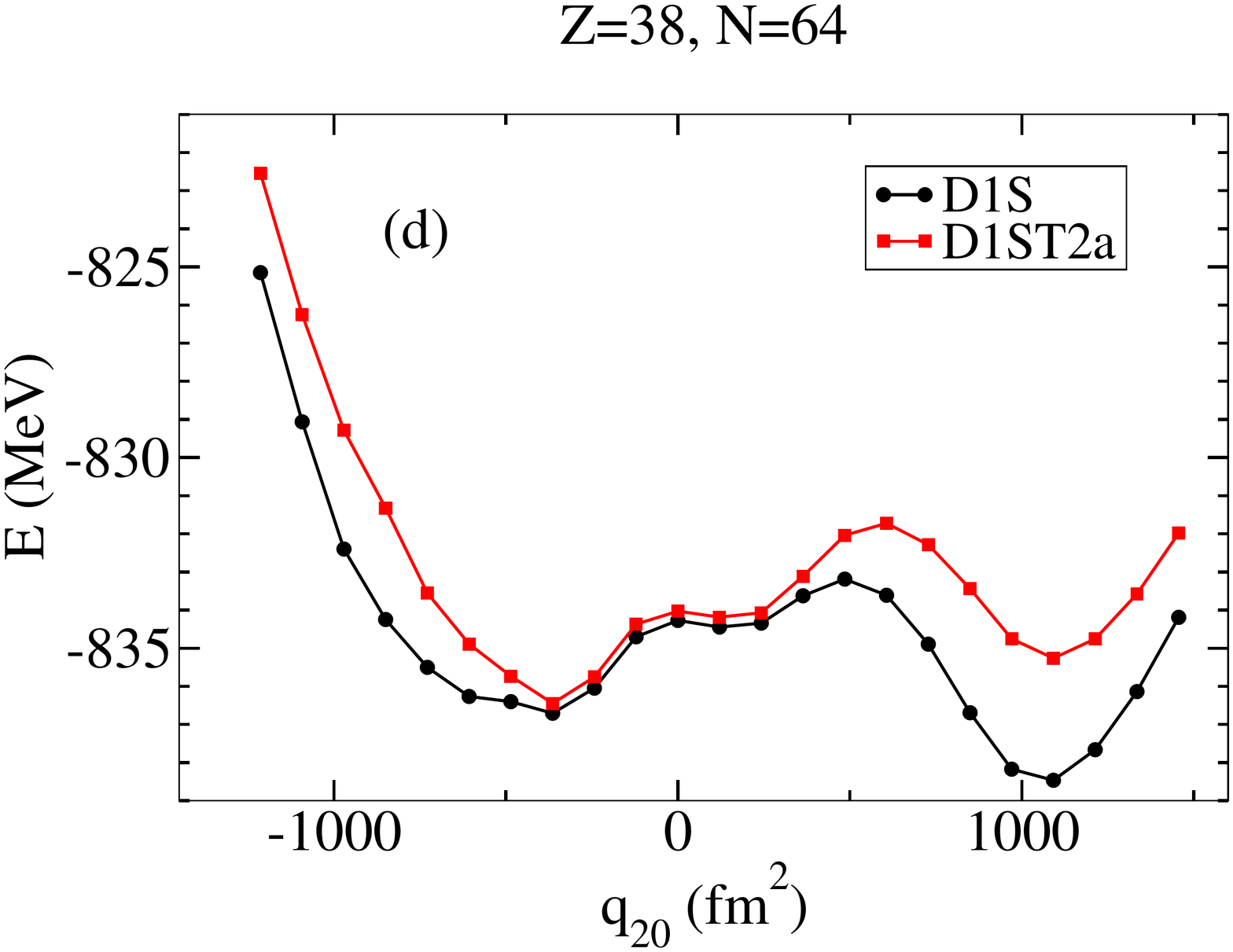} &
   \includegraphics[width=4.6cm]{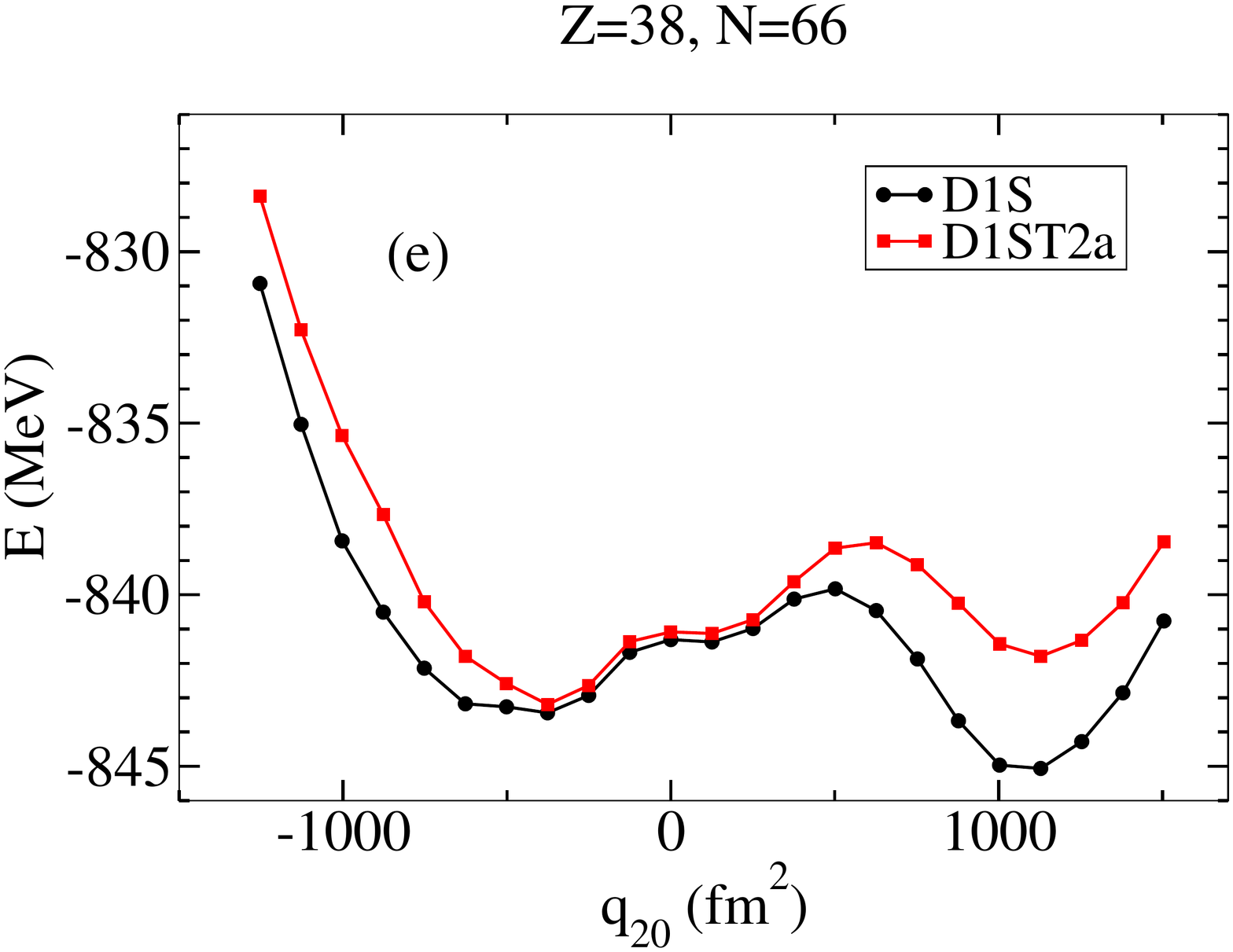} &
   \includegraphics[width=4.6cm]{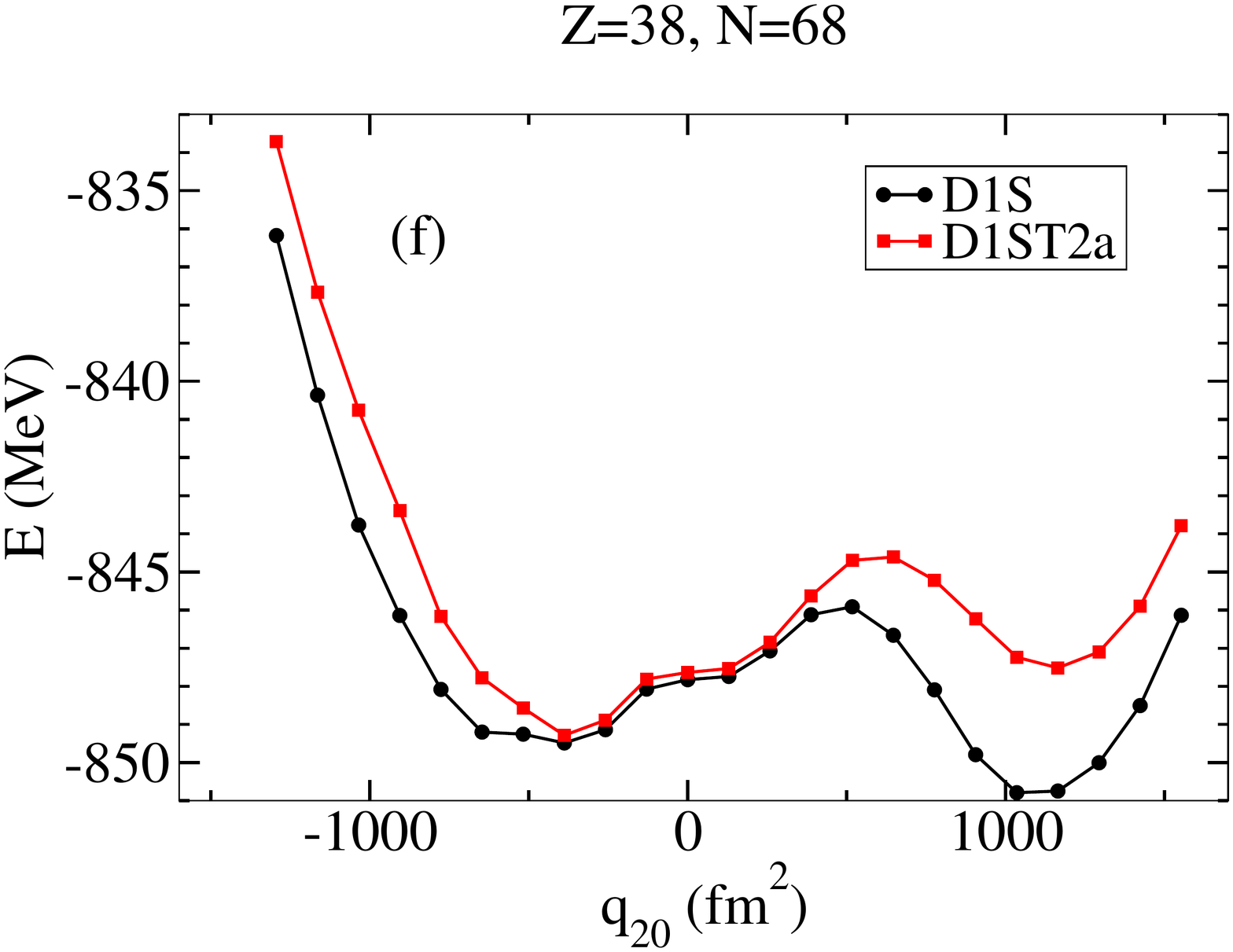} \\
   \includegraphics[width=4.6cm]{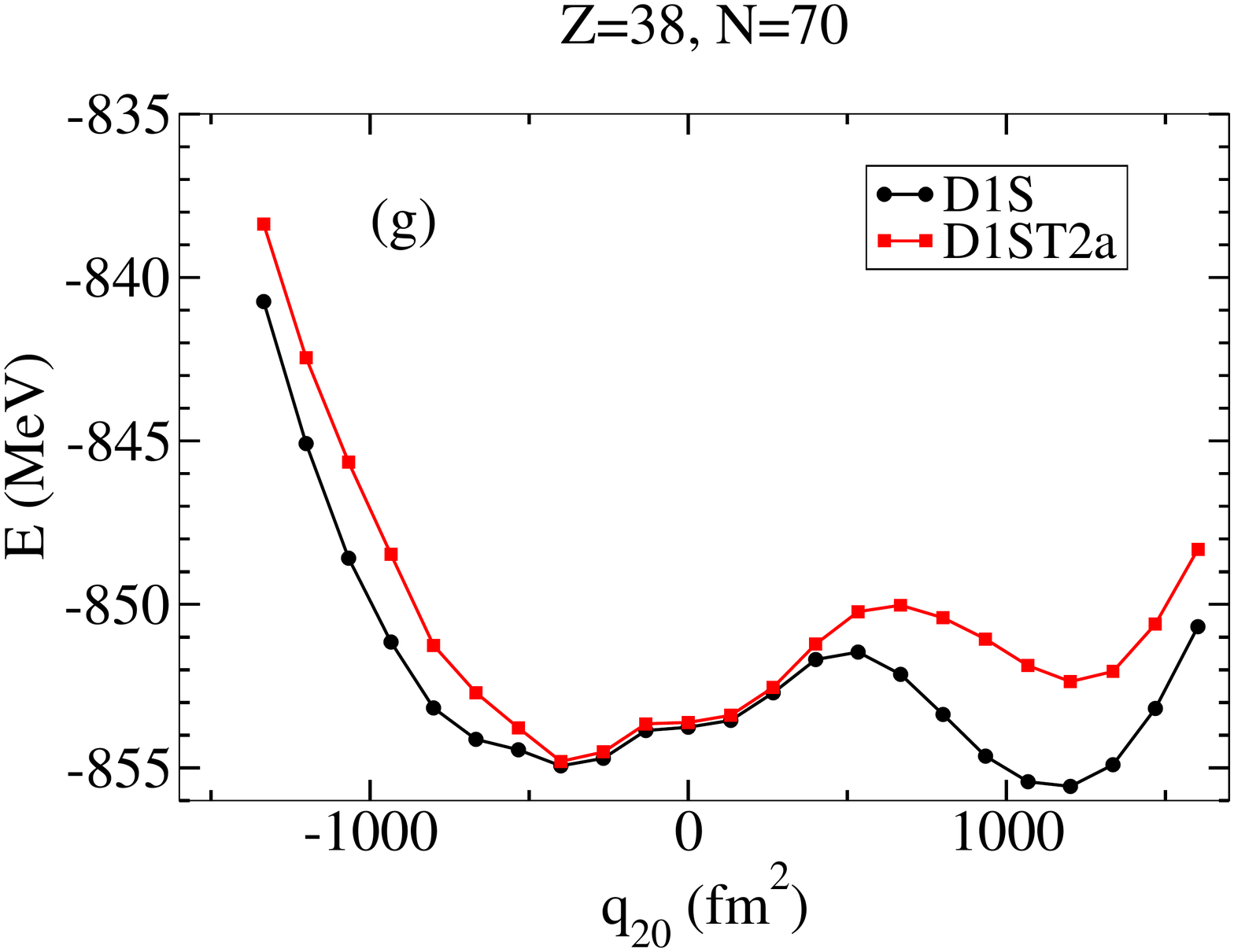} &
   \includegraphics[width=4.6cm]{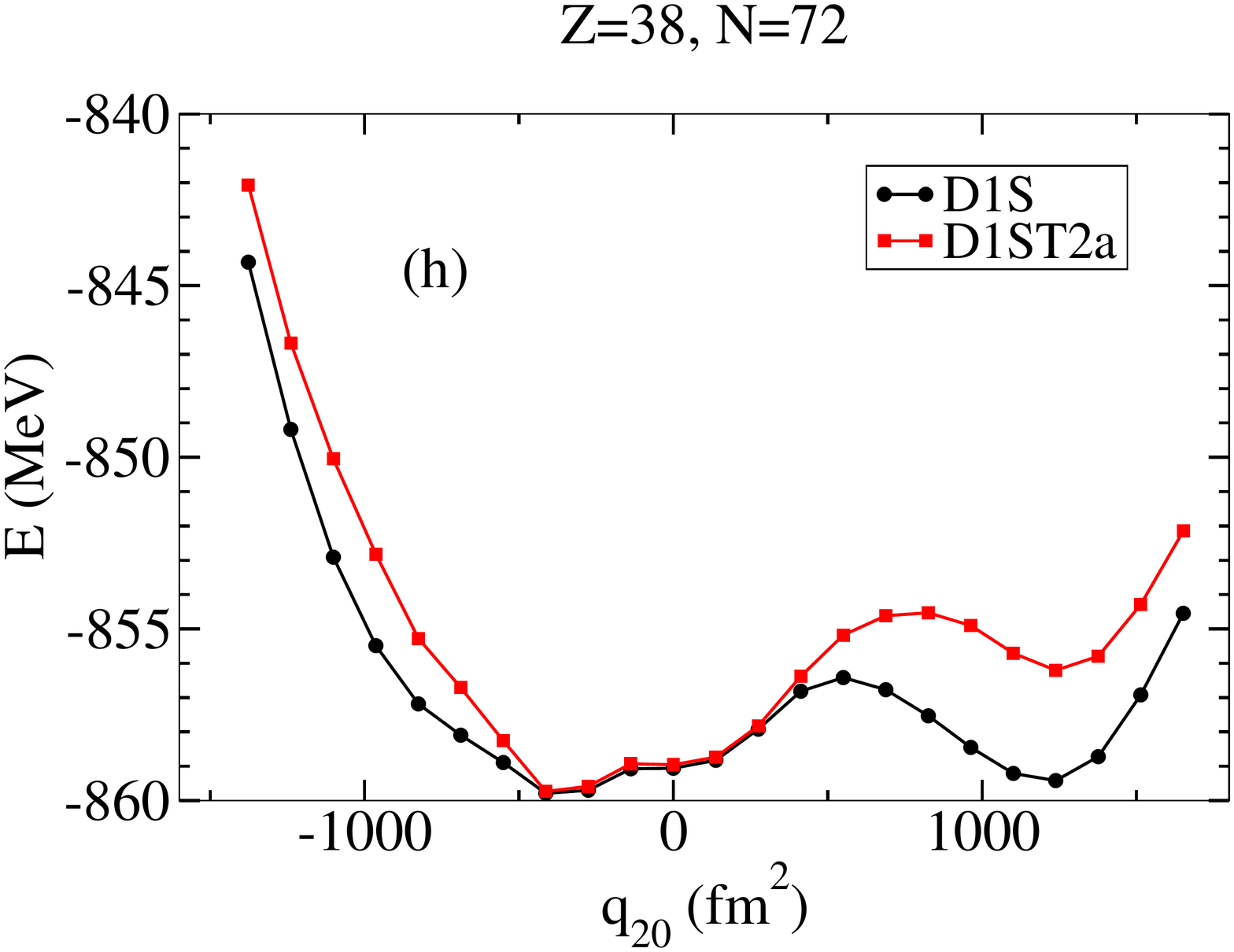} &
   \includegraphics[width=4.6cm]{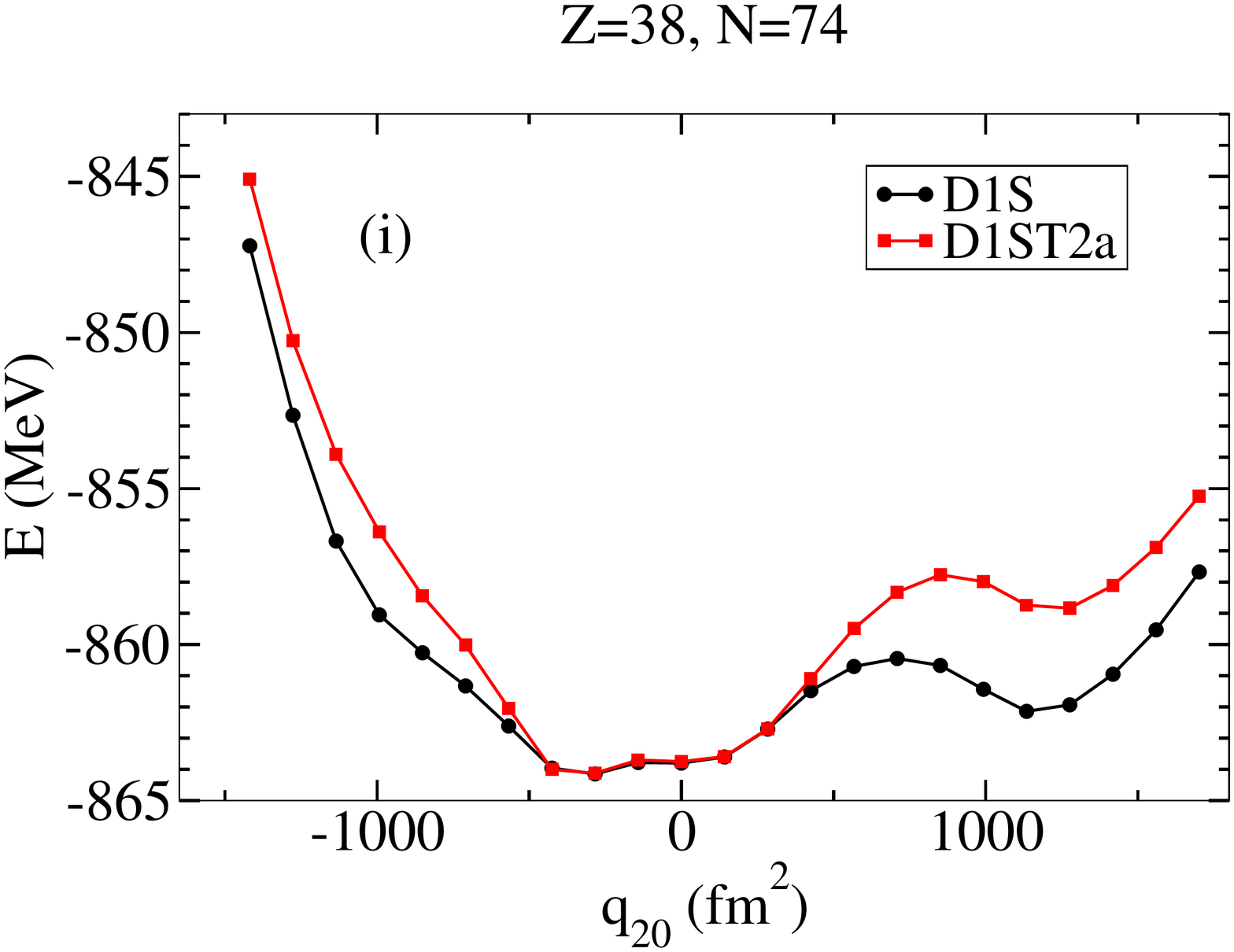} 
\end{tabular}
\caption{Potential Energy Curves of the $Z=38$ chain  for the D1S interaction (black circles) and D1ST2a one (red squares). } 
\label{PESSr}
\end{figure}

In Fig.~\ref{PESSr} are depicted the PEC for Strontium isotopes from $A=96$ to $A=112$.  
Ground states and first isomeres $Q_0$ are given in Table~\ref{TableSr2}. D1S and D1ST2a PEC get at least two minima along the chain: a prolate one and a slightly
oblate one. A nascent spherical minimum may arise for $A\le104$ or for $A\ge110$ for the D1ST2a interaction. 

The D1S curves get a pronounced prolate minimum from $A=98$ to $A=108$ which
is the ground state of these Strontium isotopes. On the contrary for the first ($A=96$) and the two last
isotopes $A=110,\,112$ the ground state is oblate. 
In a general way, for the D1ST2a interaction the prolate minimum is always high in energy, about $5$ MeV above 
the D1S minimum as for the Zr chain, whereas the energies obtained with both interactions are similar at sphericity and at the oblate minimum.
Around sphericity this chain gets a special proton SU configuration due to the mixing of the $2p$ and $1f$ subshells.
Indeed the proton subshells are filled up to the valence $1f_{5/2}$ one, including the $1f_{7/2}$ and $2p_{3/2}$ ones, and the $2p_{1/2}$ subshell is empty.
The Sr PEC can be compared to the Zr ones in Fig.~\ref{PESZr2} for which the proton $2p_{1/2}$ subshell is also full.  
Around $q_{20}=0$ fm$^{2}$ the tensor energy is small and repulsive or negligible along the Sr chain whereas it is attractive for the neutron SU Zr isotopes.
This directly highlights the impact of the $2p_{1/2}$ on the $p-n$ and $p-p$ contributions to the tensor energies.
For prolate deformations the tensor energy is strongly repulsive for both chains. In this region the tensor energy is dominated by the strongly repulsive $p-n$ contribution.
Consequently the D1ST2a interaction reverses the ground state for all the nuclei that D1S predicts to be prolate namely from $A=98$ to $A=108$.

\begin{table}[htb] \centering
 \begin{tabular}{|c|c|c|c|c|c|c|c|c|c|}
\hline
       A        &   96    &   98    &   100   &   102   &   104   &   106   &   108   &   110   &  112   \\
\hline
 D1S gs         & $-1.20$ & $+3.98$ & $+4.04$ & $+4.09$ & $+4.14$ & $+3.78$ & $+4.20$ & $-1.34$ & $-0.98$   \\
 D1S isomere    & $+3.92$ & $-1.20$ & $-1.21$ & $-1.23$ & $-1.26$ & $-1.28$ & $-1.31$ & $+4.25$ & $0.00$    \\
\hline
 D1ST2a gs      & $-1.20$ & $-1.21$ & $-1.22$ & $-1.24$ & $-1.26$ & $-1.28$ & $-1.31$ & $-1.33$ & $-0.98$   \\
 D1ST2a isomere & $+0.39$ & $+0.75$ & $+0.77$ & $+0.42$ & $+0.43$ & $+4.19$ & $+4.22$ & $0.00$  & $0.00$    \\
\hline
 \end{tabular}
\caption{Theoretical $Q_0$ (eb) for the $Z=38$ chain.  }  
\label{TableSr2} 
\end{table}

\begin{figure}[htb] \centering
\begin{tabular}{c}
   \includegraphics[width=5.0cm]{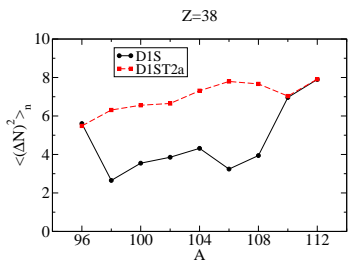} 
\end{tabular}
\caption{Proton and neutron particle number fluctuation for the $Z=38$ chain.} 
\label{SrDN2}
\end{figure}

In Fig.~\ref{SrDN2} we only have plotted the neutron particle number fluctuations for both interactions. 
There is no proton pairing for all the D1ST2a oblate ground states. The only two Strontium isotopes which get a small proton pairing at their ground states 
with the D1S interaction are A$=96$ and A$=110$, and they are the only nuclei with oblate deformation in the chain with A$=112$. 
On the other hand, for neutrons,  the fluctuation number is bigger using the D1ST2a interaction. It is interesting to note that 
this interaction produces oblate deformations for all the isotopes. This could indicate a relation between oblate deformations and pairing fluctuations.

\section{Conclusions} \label{conc}

In this paper we studied the tensor interaction in deformed nuclei within the Hartree-Fock-Bogoliubov theory.
The D1ST2a Gogny interaction used here is built from the D1S Gogny interaction to which is added a finite range
tensor term \cite{AngPRC12}. We focused our work on the Mg, Si, S, Ar, Sr and Zr chains looking at the potential energy curves with 
respect to the quadrupole deformation moment variable and ground states properties.
We found that the D1ST2a interaction is globally more repulsive than the D1S one along the PEC even if it can vanish or become attractive in some cases. 
When two PEC minima are close in energy, D1ST2a interaction can invert the minima order and thus modify the ground state shape.
Consequently the ground states densities and particle number flucuations can change drastically.
When no inversion occurs the pairing tends towards slightly weakening with the D1ST2a interaction, and the particle number fluctuation follows the same trend than 
the D1S one.

The range of the D1ST2a tensor term was chosen as the longest one of the D1S parametrizations and the two remaining parameters of the tensor term were 
adjusted on specific structure properties \cite{AngPRC12}.
Modifying these latter parameters enable us to decompose the tensor contribution to the 
HFB energy to a like-particle part and a proton-neutron part. 
We analyzed these two contributions along the PEC:  the like-particle part is attractive and 
most of the time the total tensor energy is dominated by the repulsive proton-neutron part of the tensor term.

In general the tensor contribution behaves in a different way around sphericity and well deformed shapes. 
The way the tensor term contributes to the total HFB energy is driven by the filling of the valence subshells 
and their spin-orbit partners.
Within a spin-orbit doublet there are three different contributions to the tensor energy: the one where
both particles are in the lower subshell, the one where both particles are in the higher subshell and the case
where particles are in distinct subshells. Regardless of the parameters signs the two first ones gives a positive contribution and the last one gives a negative 
contribution in such a way that when both subshells are completely filled the three contributions cancel each other.
More generally for the proton-neutron energy the sign of the contribution to the energy is given by the spins signs
assigned to the subshells within the shell model picture. For two subshells in the same position in their respective spin-orbit
doublet the tensor energy will be positive; when both spins are different, the tensor will be attractive.
As a result for SS/SS the tensor energy vanishes. 

When one of the subshells is not spin-saturated (SS/SU case) the PEC become completely different.
In this configuration the tensor energy is either attractive or negligible around sphericity.
Here the like-particle energy from the spin-saturated subshell is negligible. The proton-neutron
contribution reduces drastically from the one of its SU/SU neighbors and the SU like-particle energy
becomes the dominant attractive contribution, or at least compensate the proton-neutron one.

For SU/SU nuclei several situations may arise.
The most probable case occurs when the tensor term is the most important around spherical shape and decrease at
larger deformation (oblate and prolate sides). 
In contrast the tensor energy is almost deformation independent in some isotopes of the Mg chain. It occurs for isotopes which begin the filling  of the 
proton $p$ and $sd$ shells and such a behaviour, specific to the Mg chain, is not reproduce for the next opened or closed proton subshell (Si, S and Ar chains). 
Finally the case of the Strontium chain is different. The inversion of the $1f_{5/2}$ and $2p_{3/2}$ subshell makes the Sr looks 
more like a SS subshell than a classic SU one. 

Summarizing, we can conclude that the tensor interaction can modify in an important way the deformation properties of some nuclei. Then, it is possible to use 
observables related with deformation in order to test different interactions including tensor. One of the most important features dealing with the tensor interaction 
in effective field theory is the fact that the major effect of the bare tensor interaction is already included in the central term of the effective interaction.
In addition whereas light nuclei, such as deuteron, bring information about the bare tensor force,
the difficulty to isolate the effect of the tensor force from the other terms of the effective interaction 
arises for medium-mass or heavy nuclei.
With the present study we aim to bring additional information on the effective tensor force 
to achieve a global fit of the Gogny interaction with a complete built-in tensor force. 
We have considered the D1ST2a parameterization, for which the spin-orbit strength does not change with respect to the D1S Gogny interaction, 
because we are interested in the analyzis and the isolation of the effect of the tensor term in the deformation as well as possible.
This study is a preliminar work in order to highlight regions where tensor contributions are important.
A refit of all the parameters of the Gogny + full tensor effective interaction is required. 
An overall fit of this new effective Gogny interaction would aim to improve the deficiencies in spin-orbit splitting in exotic neutron rich nuclei.
However the tensor force effect is not the unique feature responsible for the variations in the single particle energies which makes its inclusion in the 
effective interaction delicate. 
For instance the spin-orbit coupling may be reduced by neutrons skin appearance, or
in the light nuclei the interaction between bound orbitals and continuum states can modify the energy gaps.
The D1S parameterization fails to reproduce the $N=20$ single particle energy spectra in exotic nuclei in the O, Ne, or Mg 
chains \cite{BecPRL06,UtsPRC04,TriPRL05,NeyPRL05}; one would expect some improvement from a new Gogny + tensor parameterization on that matter. 
Besides, since the tensor force may modify considerably the single particle energies, all the nuclear structure properties can be affected.
One of the main weaknesses of the D1S parameterization is its disability to correctly reproduce the binding energies all over the nuclide chart,
resulting in particular to the release of the D1M parameterization \cite{GorPRL09}. 
The present study shows that with the tensor inclusion some medium mass nuclei can become more bound for several MeV at the mean field level and 
completely change their ground state deformation.
It has also been highlighted by Pudlinger \cite{PudPRC97} that bound energies are sensitives to the tensor force in lighter nuclei.
All these features make a new global parameterization challenging.
Beforehand the importance of the tensor force on odd nuclei must be analyzed following on from these studies.
This is what we plan to do in the near future. In particular we will focus on the impact of the tensor term in the odd-even staggering of 
the binding energy.

\section{Acknowledgments}
R.N.B would like to thank L. M. Robledo for his help in dealing with the \hbox{HFBAXIAL} code and
 acknlowledges the support from CPAN (CSD2007-00042). 
Financial support from the Junta de Andaluc\'{\i}a (FQM0220),  
the Spanish Ministerio de Econom\'{\i}a y Competitividad (FPA2012-31993), 
and the European Regional Development Fund (ERDF) is gratefully acknowledged.


\section*{Appendix}
\appendix

\subsection{Decomposition of the tensor energy contribution in terms of isospin components}

The total tensor energy can be decomposed according to the isospin quantum number of the particles as:
\be
E_{\rm TS }(tot) = E_{\rm TS} (p-p) + E_{\rm TS} (n-n) + E_{\rm TS} (p-n) + E_{\rm TS} (n-p)  \, ,
\ee
where the isospin dependent energies $E_{\rm TS} (i-j)$ read:
\begin{eqnarray}
E_{\rm TS} (i-j) &=& V_{ij} \sum_{\forall a,b,c,d} \bra{ab}|V^{TS}\ket{\widetilde{cd}}  \Lambda^{(i)}_{db} \Lambda^{(j)}_{ca} \, ,
       \label{Eiso}
\end{eqnarray}
with $\bra{ab}|V^{TS}\ket{\widetilde{cd}}$ the antisymmetrized two body matrix element of the (isospin free) tensor interaction.
In the above equation $a$ and $c$ denotes quasiparticles of isospin $j$ whereas $b$ and $d$ get an isospin $i$.
The coefficient $V_{ij}$ is determined from Eq.~(\ref{parts}): $V_{ij}=V_{{\rm T}1}+V_{{\rm T}2}$ when $i=j$ and $V_{ij}=V_{{\rm T}2}$ for $i \ne j$. 
Moreover it is straightforward to show that the two unlike-particle contributions are equal: $E_{\rm TS} (p-n) = E_{\rm TS} (n-p)$.
In the following we redefine for convenience $E_{\rm TS} (p-n)$ as their sum and use the coefficients
$V_{\rm lp}\equiv V_{{\rm T}1}+V_{{\rm T}2}$ and $V_{\rm pn}\equiv 2V_{{\rm T}2}$.

In Eq.~(\ref{Eiso}) the quasiparticles indices run over all the states of the quasiparticle basis of the Fock space.
In the following we will assume that the tensor energy comes from the valence subshell and their spin-orbit partners 
in the shell model picture.
For instance in the case of $^ {40}$Ca we will consider only nucleons from the $1d_{3/2}$ and $1d_{1/2}$ subshells. Furthermore
for this nucleus there is no pairing and the density matrices are identity in the $\{1d_{3/2}, 1d_{5/2}\}$ subspace: 
$\Lambda_{ca}^{(j)}=\delta_{ca}$ for $a,c \in \{1d_{3/2},1d_{5/2}\}$. Thus the energy becomes:
\begin{eqnarray}
E_{\rm TS} &=& V_{ij} \sum_{a,b \in 1d_{3/2},1d_{5/2}} \bra{ab}|V_{\rm TS}\ket{\widetilde{ab}}  \, .
\end{eqnarray}
Defining $X_{\alpha,\beta}$ as the summation of the matrix elements over the subshells $\alpha$ and $\beta$:
\begin{eqnarray}
X_{\alpha,\beta} = \sum_{a \in \alpha} \sum_{b \in \alpha} \bra{ab}| V_{\rm TS}\ket{\widetilde{ab}} \, ,
\end{eqnarray}
the different components of $^{40}$Ca are:
\begin{eqnarray}
\begin{cases}
E_{\rm TS}(p-p) = V_{\rm lp} \left (  X_{1d_{3/2},1d_{3/2}} + X_{1d_{5/2},1d_{5/2}} + 2\, X_{1d_{5/2},1d_{3/2}}\right )  \, , \\
E_{\rm TS}(n-n)= V_{\rm lp} \left (  X_{1d_{3/2},1d_{3/2}} + X_{1d_{5/2},1d_{5/2}} + 2\, X_{1d_{5/2},1d_{3/2}} \right )  \, , \\
E_{\rm TS}(p-n)= V_{\rm pn} \left (  X_{1d_{3/2},1d_{3/2}} + X_{1d_{5/2},1d_{5/2}} + 2\, X_{1d_{5/2},1d_{3/2}} \right )  \, .
\end{cases}
\label{decCa}
\end{eqnarray}
Thus we get the relations
\begin{eqnarray}
 E_{\rm TS}(p-p) & = & E_{\rm TS} (n-n)  \, , \\
 E_{\rm TS} (p-n) & =  & {\displaystyle \frac{V_{\rm lp}}{V_{\rm pn}}} E_{\rm TS} (p-p) .
 \end{eqnarray}
These latter properties are in agreement with the numerical numbers  of   $^{40}$Ca in Table~\ref{TSenergy}. This analysis can be 
done for all types of nuclei and we will follow in this appendix to work at the Hartree-Fock level for all nuclei. 

In the case of SS/SU nuclei with the same spin-orbit partners for neutron and proton (such as $^{34}$Si),
we get for a neutron spin-saturated subshells ($\alpha$ and $\beta$ full) and a proton spin-unsaturated subshell ($\alpha$ full and $\beta$ empty)  configuration:
\begin{eqnarray}
\begin{cases}
 E_{\rm TS} (p-p) = V_{\rm lp} \, X_{\alpha,\alpha} \, , \\
 E_{\rm TS} (n-n) = V_{\rm lp} \left ( X_{\alpha,\alpha} + X_{\beta,\beta} +2 X_{\alpha,\beta} \right ) \, ,  \\
 E_{\rm TS} (n-p) = V_{\rm pn} \left ( X_{\alpha,\alpha} + X_{\alpha,\beta}  \right ) \, ,
\end{cases}
\end{eqnarray}
which leads to
\begin{eqnarray}
\begin{cases}
 X_{\alpha,\alpha} = E_{\rm TS} (p-p) / V_{\rm lp}  \, , \\
 X_{\alpha,\beta } = E_{\rm TS} (p-n) / V_{\rm pn} - E_{\rm TS} (p-p) / V_{\rm lp}  \, , \\
 X_{\beta ,\beta } = E_{\rm TS} (n-n) / V_{\rm lp} + E_{\rm TS} (p-p) / V_{\rm lp} - E_{\rm TS}(p-n)/V_{\rm pn} \, .
\end{cases}
\end{eqnarray}
In the pf shell the $1f_{7/2}/1f_{5/2}$ and $2p_{3/2}/2p_{1/2}$ partners are mixed, 
that is to say the $2p_{3/2}$ subshell is located between the $1f_{7/2}$ and $1f_{5/2}$ ones. 
As a result we encounter in that region configurations where more than two subshells have to
be taken into account to understand the tensor energy contributions. For instance for $^{76}$Sr
both proton and neutron subshells are full up to the $1f_{5/2}$ and the tensor energies are:
\begin{eqnarray}
  E_{\rm TS}(p-p)  &  =&   V_{\rm lp}  \left [ X_{1f_{7/2},1f_{7/2}} + X_{1f_{5/2},1f_{5/2}} + X_ {2 p_{3/2},2p_{3/2}}  
   \right . \nonumber \\
 & &  \left .  +  2 \, \big(X_{1f_{7/2},1f_{5/2}}  + X_{1f_{7/2},2p_{3/2}}  + X_{1f_{5/2},2p_{3/2}}\big)  \right ] \, , \nonumber \\
  E_{\rm TS} (n-n)  & = &  E_{\rm TS} (p-p)  \, , \\
  E_{\rm TS} (p-n)  & =  & {\displaystyle \frac{V_{\rm pn}}{V_{\rm lp}} }  E_{\rm TS} (p-p) \nonumber \, .
\end{eqnarray}
There are several ways to determine most of the $X_{\alpha,\beta}$. For example the $X_{1f_{7/2},1 f_{7/2}}$
is directly given by $E_{\rm TS}(p-p)$  or $E_{\rm TS}(n-n)$ for \  $^{56}$Ni or $E_{\rm TS}(p-p)$ for  \ $^{66}$Ni or 
$E_{\rm TS} (n-n)$ for  \ $^{48}$Ca and so on.
Most of the energies mixes several $X_{\alpha,\beta}$ which means that several energy contributions are necessary to 
determine some specific $X_{\alpha,\beta}$ in most cases.
The different ways to calculate the $X_{\alpha,\beta}$ from different nuclei gives slightly different numerical values. The average
value is then chosen as the reference value. This latter is presented in Table~\ref{TableX} for all the 
subshells from the $1d_{5/2}$ subshell to the $2p_{1/2}$ one except the $2s$ subshell. In effect the study of nuclei 
$^{30}$Si, $^{32}$S and $^{36}$S shows that its contributions to the tensor energies are always small 
and are consequently neglected.
By way of example, some $X_{\alpha,\beta}$ values are presented at the end of this Appendix.

\begin{table}[htb] \centering
 \begin{tabular}{|c|c|c|c|c|c|c|}
\hline
 $X$           & $1d_{5/2}$  & $1d_{3/2}$ & $1f_{7/2}$ & $1f_{5/2}$& $2p_{3/2}$  & $2p_{1/2}$   \\
\hline
 $1d_{5/2}$    &   $+0.0410$ &  $-0.0426$ &  $+0.0385$ & $-0.0270$ &  $+0.0037$  &              \\
\hline
 $1d_{3/2}$    &             &  $+0.0443$ &  $-0.0378$ & $+0.0275$ &  $-0.0048$  &              \\
\hline
 $1f_{7/2}$    &             &            &  $+0.0574$ & $-0.0547$ &  $-0.0008$  &  $+0.0014$     \\
\hline
 $1f_{5/2}$    &             &            &            & $+0.0539$ &  $-0.0023$  & $+0.0016$     \\
\hline
 $2p_{3/2}$    &             &            &            &           &  $+0.0112$  &  $-0.0123$      \\
\hline
 $2p_{1/2}$    &             &            &            &           &             &  $+0.0139$      \\
\hline
 \end{tabular}
\caption{ Average $X_{\alpha,\beta}$ for selected subshells from Hartree-Fock calculations.   }
\label{TableX} 
\end{table}
Several conclusions can be drawn from Table~\ref{TableX}.
The first one deals with the sign of the $X_{\alpha,\beta}$. The sign of the $X_{\alpha,\alpha}$ type
is always positive.
Within spin-orbit partners subshells the $X_{\alpha,\beta}$ crossed term ($\alpha \ne \beta$) is always negative.
In a more general way, the sign of the $X_{\alpha,\beta}$ is determined by the location of the subshells
$\alpha$ and $\beta$ in the level scheme of the shell model picture: 
when both spin quantum numbers $s(\alpha)$ and $s(\beta)$ are equal, $X_{\alpha,\beta}$ is positive
and when $s(\alpha) \neq s(\beta)$, $X_{\alpha,\beta}$   is negative.
The only exception in Table~\ref{TableX} concerns the smallest $X$ value: $X_{1f_{7/2}, 2p_{3/2}}=-0.0008$.

The second noticeable feature in Table~\ref{TableX} deals with the orders of magnitude.
For a given subshell $\alpha$, the $X_{\alpha,\alpha}$ quantity is about $\sim 0.01$ and rises with the numbers
of nucleons in the subshell from $+0.0112$ for 2p$_{3/2}$ up to $+0.0574$ for $1f_{7/2}$.
Moreover for two spin-orbit partners $(\alpha, \, \beta)$ we get $X_{\alpha,\alpha} \simeq X_{\beta,\beta}$
and $X_{\alpha,\alpha}\simeq - X_{\alpha,\beta}$. 
As a result the sum of the tensor energy contributions from spin-orbit partner subshells are negligible:
\begin{eqnarray}
  X_{\alpha,\alpha}+X_{\beta,\beta} + 2 \, X_{\alpha,\beta} = \begin{cases}
                                                      + 0.0001\, \mbox{ for } \alpha, \beta = 1d_{5/2}, 1d_{3/2} \, , \\
                                                      + 0.0019\, \mbox{ for } \alpha, \beta = 1f_{7/2}, 1f_{5/2} \, , \\
                                                      +0.0005\, \mbox{ for } \alpha, \beta = 2p_{3/2}, 2p_{1/2} \, .\\ 
                                                     \end{cases}
\end{eqnarray}
This can be correlated to the fact that when both subshells are full and are non valence subshells their contributions
to the tensor energy is neglected in our assumption.

Finally as the number of nucleons is small for the $2p_{3/2}$ and $2p_{1/2}$ subshells we get small crossed 
$X_{\alpha, \beta}$ terms involving one of these subshells and one of the biggest ones.

\subsection{Explicit calculation of the \texorpdfstring{$X_{\alpha,\beta}$}{Xab} }
\label{appX}

In this section we give a few examples of the way the different Hartree-Fock averaged $X$ are calculated. 
They are determined as the average of the $X$ extracted from the tensor energies of several nuclei.
\begin{eqnarray}
\begin{cases}
  E_{\rm TS}(p-p)/V_{\rm lp} = +0.0421 & \hspace{1 cm} \mbox{($^{28}$Si)} \, ,  \\
  E_{\rm TS}(n-n)/V_{\rm lp} = +0.0441 & \hspace{1cm} \mbox{($^{28}$Si)}  \, , \\
  E_{\rm TS}(p-n)/V_{\rm pn} = +0.0431 & \hspace{1cm} \mbox{($^{28}$Si)}  \, ,  \\
  E_{\rm TS}(p-p)/V_{\rm lp} = +0.0396 & \hspace{1cm} \mbox{($^{34}$Si)}  \, , \\
  E_{\rm TS}(p-p)/V_{\rm lp} = +0.0361 & \hspace{1cm} \mbox{($^{42}$Si)}  \, . \\
 \end{cases}
\end{eqnarray}
Thus we get $X_{1d_{5/2},1d_{5/2}} =+0.0410$.
\begin{eqnarray}
\begin{cases}
  E_{\rm TS}(p-p)/V_{lp} = +0.0548  & \hspace{1cm} \mbox{($^{56}$Ni)} \, ,   \\  
  E_{\rm TS}(n-n)/V_{lp} = +0.0580  & \hspace{1cm} \mbox{($^{56}$Ni)}  \, , \\
  E_{\rm TS}(p-n)/V_{pn} = +0.0563  & \hspace{1cm} \mbox{($^{56}$Ni)}  \, , \\
  E_{\rm TS}(n-n)/V_{lp} = +0.0624  & \hspace{1cm} \mbox{($^{48}$Ca)} \, ,  \\
  E_{\rm TS}(p-p)/V_{lp} = +0.0538  & \hspace{1cm} \mbox{($^{60}$Ni)} \, ,  \\
  E_{\rm TS}(p-p)/V_{lp} = +0.0517  & \hspace{1cm} \mbox{($^{66}$Ni)} \, ,  \\
  E_{\rm TS}(n-n)/V_{lp} = +0.0598  & \hspace{1cm} \mbox{($^{42}$Si)}  \, , \\
  E_{\rm TS}(n-n)/V_{lp} = +0.0621  & \hspace{1cm} \mbox{($^{44}$S )}  \, . \\
 \end{cases}
\end{eqnarray}
We get $X_{1_{f7/2},1f_{7/2}} =+0.0574$. Moreover $X_{1{\rm d}_{5/2},1{\rm d}_{3/2}}$ is directly given by $^{34}$Si:
\begin{eqnarray}
 X_{1{\rm d}_{5/2},1{\rm d}_{3/2}} = E_{\rm TS}(p-n)/V_{pn} - X_{1{\rm d}_{5/2},1{\rm d}_{5/2}} = -0.0426   & \hspace{1cm} \mbox{($^{34}$Si)} \, .
\end{eqnarray}

Besides we calculate the contributions of the $2s$ subshell to the tensor energy in the $sd$ shell.
To do so the $2s$ subshell is explicitly considered in $^{30}$Si and $^{36}$S. 
As it shown below the $X$ involving the $2s$ subshell are negligible.

\begin{eqnarray}
 X_{2{ s},1{ d}_{5/2}} &=& E_{\rm TS}(p-n)/V_{pn} - X_{1{ d}_{5/2},1{ d}_{5/2}} \hspace{2.5cm}= -0.0001   \hspace{1cm} \mbox{($^{30}$Si)}  \, , 
\\
 X_{2{ s},1{ d}_{3/2}} &=& \Big( E_{\rm TS}(n-n)/V_{lp} - 2X_{2{ s},1{ d}_{5/2}}  - 2X_{1{ d}_{3/2},1{ d}_{5/2}}\nonumber\\ 
 &&\hspace{1cm}- X_{1{ d}_{5/2},1{ d}_{5/2}}  
 - X_{1{ d}_{3/2},1{ d}_{3/2}} -X_{2{ s},2{ s}} \Big)/2 = +0.0006   \hspace{1cm} \mbox{($^{36}$S)}  \, . 
\end{eqnarray}

\section*{References}


\end{document}